\documentstyle[12pt]{article}
\input epsf.sty
\def\ZZZ{{\hbox{ Z\kern-1.6mm Z}}}

\newcommand{\eps}{\epsilon}
\newcommand{\ra}{\rangle}
\newcommand{\la}{\langle}

\newcommand{\vp}{\varphi}

\newcommand{\tl}{\wt\lambda}
\newcommand{\dt}{(\vec \nabla T)^2}
\newcommand{\hp}{{\wh\Phi}}
\newcommand{\hq}{{\wh Q_B}}
\newcommand{\he}{{\wh\eta_0}}
\newcommand{\ha}{{\wh{A}}}
\newcommand{\lllb}{\Bigl\langle\Bigl\langle}
\newcommand{\rrrb}{\Bigr\rangle\Bigr\rangle}
\newcommand{\tf}{\wt f}

\newcommand{\VV}{{\cal V}}
\newcommand{\BB}{{\cal B}}

\newcommand{\GG}{{\cal G}}
\newcommand{\AAA}{{\cal A}}
\newcommand{\FF}{{\cal F}}
\newcommand{\HH}{{\cal H}}

\newcommand{\CC}{{\cal C}}
\newcommand{\OO}{{\cal O}}
\newcommand{\QQ}{{\cal Q}}
\newcommand{\PP}{{\cal P}}
\newcommand{\EE}{{\cal E}}
\newcommand{\LL}{{\cal L}}
\newcommand{\lll}{\langle\langle}
\newcommand{\rrr}{\rangle\rangle}
\newcommand{\square}{\Box}
\newcommand{\half}{{1\over 2}}
\newcommand{\wt}{\widetilde}
\newcommand{\wh}{\widehat}
\newcommand{\wc}{\check}

\newcommand{\bd}{\bar{\rm D}}
\newcommand{\RR}{{\cal R}}
\newcommand{\NN}{{\cal N}}
\newcommand{\TT}{{\cal T}}

\newcommand{\al}{\alpha}

\newcommand{\onk}{\omega^{(N)}_{\vec k_\perp}}

\newcommand{\be}{\begin{equation}}
\newcommand{\ee}{\end{equation}}
\newcommand{\ben}{\begin{eqnarray}\displaystyle}
\newcommand{\een}{\end{eqnarray}}
\newcommand{\refb}[1]{(\ref{#1})}
\newcommand{\p}{\partial}
\newcommand{\sectiono}[1]{\section{#1}\setcounter{equation}{0}}

\def\one{{\hbox{ 1\kern-.8mm l}}}
\def\zero{{\hbox{ 0\kern-1.5mm 0}}}

\begin{document}
{}~
{}~
\hfill\vbox{\hbox{hep-th/0410103}
}\break

\vskip .6cm

\centerline{{\Large \bf Tachyon Dynamics in Open String
Theory}\footnote{Based on lectures given at the 2003 and 2004 ICTP Spring
School, TASI 2003, 2003 Summer School on Strings, Gravity and 
Cosmology at Vancouver, 2003 IPM String
School at Anzali, Iran, 2003 ICTP Latin American School at Sao
Paolo, 2004 Nordic meeting at Groningen and 2004 Onassis Foundation 
lecture at Crete.}}

\medskip

\vspace*{4.0ex}

\centerline{\large \rm
Ashoke Sen}

\vspace*{4.0ex}

\centerline{\large \it Harish-Chandra Research
Institute}

\centerline{\large \it  Chhatnag Road, Jhusi,
Allahabad 211019, INDIA}

\centerline{E-mail: ashoke.sen@cern.ch,
sen@mri.ernet.in}

\vspace*{5.0ex}

\centerline{\bf Abstract} \bigskip

In this review we describe our current understanding of the properties of
open string tachyons on an unstable D-brane or brane-antibrane system in
string theory. The various string theoretic methods used for this study
include techniques of two dimensional conformal field theory, open string
field theory, boundary string field theory, non-commutative solitons etc.
We also describe various attempts to understand these results using field
theoretic methods. These field theory models include toy models like
singular potential models and $p$-adic string theory, as well as more
realistic version of the tachyon effective action based on
Dirac-Born-Infeld type action. Finally we study closed string background
produced by the `decaying' unstable D-branes, both in the critical string
theory and in the two dimensional string theory, and describe the open
string completeness conjecture that emerges out of this study. According
to this conjecture the quantum dynamics of an unstable D-brane system is
described by an internally consistent quantum open string field theory
without any need to couple the system to closed strings. Each such system
can be regarded as a part of the `hologram' describing the full string
theory.

\vfill \eject

\baselineskip=16pt

\tableofcontents

\sectiono{Introduction} \label{s1}

This introductory section is divided into two parts. In section
\ref{smot} we give a brief motivation for studying the tachyon dynamics in
string theory. Section \ref{sorg} summarizes the organisation of the
paper.

\subsection{Motivation} \label{smot}

Historically, a tachyon was defined as a particle that travels faster than
light. Using the relativistic relation $v=p/\sqrt{p^2 + m^2}$ between
the velocity $v$, the spatial
momentum
$p$ and mass $m$ of a particle we see that
for real $p$ a
tachyon must have negative mass$^2$. Clearly neither of these
descriptions
makes a convincing case for the tachyon.

Quantum field theories offer a much better insight into the role of
tachyons. For this consider a scalar field $\phi$ with conventional
kinetic term, and a potential
$V(\phi)$ which has an extremum at the origin. If we carry out
perturbative quantization of the scalar field by expanding the potential
around $\phi=0$, and ignore the cubic and higher order terms in the
action, we find a particle like state with mass$^2=V''(0)$. For $V''(0)$
positive this describes a particle with positive mass$^2$. But for
$V''(0)<0$ we have a particle with negative mass$^2$, {\it i.e.} a
tachyon!

In this case however the existence of the tachyon has a clear
physical interpretation. For $V''(0)<0$, the potential $V(\phi)$
has a maximum at the origin, and hence a small displacement of
$\phi$ away from the origin will make it grow exponentially in
time. Thus perturbation theory, in which we treat the cubic and
higher order terms in the potential to be small, breaks down. From
this point of view we see that the existence of a tachyon in a
quantum field theory is associated with an instability of the
system which causes a breakdown of the perturbation theory. This
interpretaion also suggests a natural remedy of the problem. We
simply need to expand the potential around a new point in the
field space where it has a minimum, and carry out perturbative
quantization of the theory around this point. This in turn will
give a particle with positive mass$^2$ in the spectrum.

Unlike quantum field theories which provide a second
quantized description of a particle, conventional formulation of string
theory uses a first quantized formalism. In this formulation the spectrum
of single `particle' states in the theory are obtained by quantizing the
vibrational modes of a single string. Each such state is characterized by
its
energy $E$ and momentum $p$ besides  other quantum numbers, and
occasionally one finds states for which $E^2-p^2<0$. Since $E^2-p^2$ is
identified as the mass$^2$ of a particle, these states correspond to
particles of negative mass$^2$, {\it i.e.} tachyons.

The simplest
example
of such a tachyon appears in the $(25+1)$ dimensional bosonic string
theory. This theory has closed strings as its fundamental excitations, and
the lowest mass$^2$ state of this theory turns out to be tachyonic.
One might suspect that this tachyon
may have the same origin as in a quantum field theory, {\it i.e.} we may
be carrying out perturbation expansion around an unstable point, and that
the tachyon may be removed once we expand the theory about a stable
minimum of the potential.
Unfortunately, the first quantized description of string theory
does not allow us
to test this hypothesis. In particular, whether the closed string tachyon
potential  in the
bosonic string theory has a stable minimum still remains an unsolved
problem, and many people believe that this theory is inconsistent due to
the presence of the tachyon in its spectrum.
Fortunately various versions of superstring theories, defined in (9+1)
dimensions, have tachyon free closed string
spectrum. These theories are the starting points of most
attempts at constructing a unified theory of nature.

Besides closed strings, some string theories also contain open
string excitations with appropriate boundary conditions at the two
ends of the string. According to our current understanding, open
string excitations exist only when we consider a theory in the
presence of soliton like configurations known as
D-branes\cite{9510017,9611050,0007170}. Conversely, inclusion of open 
string
states in the spectrum implies that we are quantizing the theory
in the presence of a D-brane. To be more specific, a D-$p$-brane is a
$p$-dimensional extended object, and in the presence of such a
brane lying along a $p$-dimensional hypersurface $S$, the theory
contains open string excitations whose ends are forced to move
along the surface $S$. In the presence of $N$ D-branes (not
necessarily of the same kind) the spectrum contains $N^2$
different types of open string, with each end lying on one of the
$N$ D-branes. The physical interpretation of these open string
states is that they represent quantum excitations of the system of
D-branes.

It turns out that in some cases the spectrum of open string states
on a system of D-$p$-branes also contains tachyon. This happens
for example on D-$p$-branes in bosonic string theory for any $p$,
and D-$p$-branes in type IIA / IIB superstring theories for odd /
even values of $p$. Again, from our experience in quantum field
theory one would guess that the existence of the open string
tachyons represents an instability of the D-brane system whose
quantum excitations they describe. The natural question that
arises then is: is there a stable minimum of the tachyon potential
around which we can quantize the theory and get sensible results?

Although our understanding of this subject is still not complete,
last several years have seen much progress in answering this question.
These notes are designed to primarily review the main developments in this
subject.

\subsection{Organisation of the review} \label{sorg}

This review is organized as follows. In section \ref{s2} we give a
summary of the main results reviewed in this article.
In sections \ref{s3} - \ref{snc} we analyze
time independent classical solutions involving the open string tachyon 
using various techniques. Section \ref{s3} uses
the correspondence between two dimensional conformal field
theories and classical solutions of the equations of motion in
open string field theory. Section \ref{s4} is based on direct analysis of 
the equations
of motion of open string field theory. In sections \ref{s7} and
\ref{snc} we discuss application of the methods of boundary string
field theory and non-commutative field theory respectively. In section 
\ref{s10} we construct and analyze the
properties of time dependent solutions involving the tachyon. In
section \ref{s5.3} we describe an effective field theory which
reproduces qualitatively some of the results on time independent
and time dependent classical solutions involving the tachyon.
Section \ref{s5} is devoted to the discussion of other toy models,
{\it e.g.} field theories with singular potential and $p$-adic
string theory, which exhibit some of the features of the static
solutions involving the open string tachyon. In section \ref{s9}
we study the effect of closed string emission from the time
dependent rolling tachyon background on an unstable D-brane. In
section \ref{s11} we apply the methods discussed in this review to
study the dynamics and decay of an unstable D0-brane in two
dimensional string theory, and compare these results with exact
description of the system using large $N$ matrix models. Finally
in section \ref{sopenclosed} we propose an open string
completeness conjecture and generalized holographic principle which
explain some of the results of sections \ref{s9} and \ref{s11}.

Throughout this paper we work in the units:
 \be \label{econv1}
\hbar = c =\alpha' = 1\, .
 \ee
Thus in this unit the fundamental string tension is $(2\pi)^{-1}$.
Also our convention for the space-time metric will be
$\eta_{\mu\nu}={\rm diag}(-1,1,\ldots 1)$.

Before concluding this section we would like to caution the reader that 
this review 
does not cover all aspects of tachyon condensation. For example we do not 
address open string tachyon condensation on D$p$-D$p'$ brane system or 
branes at angles\cite{9704006,9703217}. We also do not review various 
attempts to 
find possible cosmological applications of the open string 
tachyon\cite{9812483,0105204,0105203,0107058,0204008}; nor do we 
address issues 
involving 
closed 
string tachyon condensation\cite{0108075}. We refer the reader to the 
original papers and their citations in spires database for learning these 
subjects.

Finally we would like to draw the readers' attention to many other
reviews where different aspects of tachyon condensation have been
discussed. A partial list includes 
refs.\cite{9908126,9908144,0005029,0102076,0102085,0106195,
0301094,0311017,
0405064}. For some early studies in open string tachyon dynamics, 
see \cite{bar1,bar2,bar3}.

\sectiono{Review of Main Results}
\label{s2}

In this section we summarize the main results reviewed in this article. 
The derivation of these results will be discussed in the rest of this 
article.

\subsection{Static solutions in superstring theory} \label{s2.1}

We begin our discussion by reviewing the properties of D-branes in
type IIA and IIB superstring theories. D$p$-branes are by definition
$p$-dimensional extended objects on which fundamental open strings can
end. It is well known\cite{dai,leigh,9407031} that type IIA/IIB
string theory contains BPS D$p$-branes for even / odd
$p$,
and that these D-branes carry Ramond-Ramond (RR) charges\cite{9510017}.
These
D-branes are oriented, and have definite mass per unit $p$-volume known as
tension. The tension of a BPS D$p$-brane in type IIA/IIB string theory is
given by:
 \be \label{e2.1}
\TT_p = (2\pi)^{-p} \, g_s^{-1}\, ,
 \ee
where $g_s$ is the closed string coupling constant. The BPS D-branes are
stable, and all the open string modes living on such a brane have
mass$^2\ge 0$. Since these branes are
oriented, given a specific BPS D$p$-brane, we shall call a D$p$-brane
with opposite orientation an
anti-D$p$-brane, or a $\bd p$-brane. The D0-brane in type IIA string
theory also has an anti-particle known as $\bd$0-brane, but we cannot
describe it as a D0-brane with reversed orientation.

Although a BPS D$p$-brane does not have a negative mass$^2$ (tachyonic) 
mode, if we consider a
coincident BPS D$p$-brane - $\bd p$-brane pair, then the open string
stretched
from the brane to the anti-brane (or vice-versa) has a tachyonic
mode\cite{9403040,9511194,9604091,9604156,9612215}. This is due to the 
fact 
that 
the GSO projection rule for these open 
strings is opposite of that for open strings whose both ends lie on the 
brane (or the anti-brane). As a result the ground state in the 
Neveu-Schwarz (NS) sector, which is normally removed from the spectrum by 
GSO projection, now becomes part of the spectrum, giving rise to a 
tachyonic mode. Altogether there are 
two tachyonic modes in the spectrum, -- one from the open string 
stretched from the brane to the anti-brane and the other from the open 
string stretched from the anti-brane to the brane.
The mass$^2$ of each of these tachyonic modes is given by 
 \be \label{e2.3}
m^2 = -{1\over 2}\, .
 \ee

Besides the stable BPS D$p$-branes, type II string theories also
contain in their spectrum unstable, non-BPS
D-branes\cite{9805019,9806155,9808141,9809111,9901014}.
The simplest way to define these D-branes in IIA/IIB string theory is to
begin with a coincident BPS D$p$ -- $\bd p$-brane pair in type IIB/IIA
string theory, and then take an orbifold of the theory by $(-1)^{F_L}$,
where ${F_L}$ denotes the contribution to the space-time fermion
number from the left-moving sector of the world-sheet. 
Since the RR fields are odd
under $(-1)^{F_L}$, all the RR fields of type IIB/IIA theory
are
projected out by the $(-1)^{F_L}$ projection. The twisted sector states
then give us back the  RR fields of type IIA/IIB theory. Since
$(-1)^{F_L}$ reverses the sign of the RR charge, it
takes a BPS D$p$-brane to a $\bd p$-brane and vice versa. As a result its 
action on the
open string states on a D$p$-$\bd p$-brane system is to
conjugate the Chan-Paton
factor by the exchange operator
$\sigma_1$. Thus modding out the D$p$ - $\bd p$-brane by $(-1)^{F_L}$
removes
all open string states with Chan-Paton factor $\sigma_2$ and $\sigma_3$
since these
anti-commute with $\sigma_1$, but keeps the open string states with 
Chan-Paton factors $I$ and $\sigma_1$. This gives us a non-BPS 
D$p$-brane\cite{9812031}.

The non-BPS D-branes
have precisely those dimensions which BPS D-branes do not have. Thus
type IIA string theory has non-BPS D$p$-branes for odd $p$ and type IIB
string theory has non-BPS D$p$-branes for even $p$. These branes are
unoriented and carry a fixed mass per unit $p$-volume, given by
 \be \label{e2.2}
\wt\TT_p = \sqrt 2\, (2\pi)^{-p} \, g_s^{-1}\, .
 \ee
The most important feature that distinguishes the non-BPS D-branes from
BPS D-branes is that the spectrum of open strings on a non-BPS D-brane
contains a single mode of negative mass$^2$ besides infinite number of
other modes of
mass$^2\ge 0$. This tachyonic mode can be identified as a particular 
linear combination of the two tachyons living on the original 
brane-antibrane pair that survives the $(-1)^{F_L}$ projection, and has 
the same mass$^2$ as given in \refb{e2.3}.
Another important feature that distinguishes a BPS D$p$-brane from a
non-BPS D$p$-brane is that unlike a BPS D$p$-brane which is charged
under the RR $(p+1)$-form gauge field of string theory, a
non-BPS D-brane is neutral under these gauge fields.
Various other properties of
non-BPS D-branes have been reviewed in \cite{9904207,9905006,9908126}.

Our main goal will be to understand the dynamics of these tachyonic modes. 
This however is not a simple task.
The dynamics of open strings living on a D$p$-brane is described by a
$(p+1)$ dimensional (string) field theory, defined such that the free
field quantization of the field theory reproduces the spectrum of open
strings on the D$p$-brane, and the S-matrix elements computed from this
field theory reproduce the S-matrix elements of open string theory on the
D-brane. On a non-BPS D-brane the existence of a single scalar tachyonic
mode shows that the corresponding open string field theory must contain a
real scalar field $T$ with mass$^2=-1/2$, whereas the same reasoning shows
that open string field theory associated with a coincident
brane-anti-brane system must contain two real scalar fields, or
equivalently one complex scalar field $T$ of mass$^2=-1/2$. However these
fields have non-trivial coupling to all the infinite number of other
fields in open string field theory, and hence one cannot study the
dynamics of these tachyonic modes in isolation. Furthermore since the
$|mass^2|$ of the tachyonic modes is of the same order of magnitude as
that of the other heavy modes of the string, one cannot work with a simple
{\it low energy} effective action obtained by integrating out the other
heavy
modes of the string. This is what makes the analysis of the tachyon
dynamics non-trivial. Nevertheless, it is convenient to state the results
of the analysis in terms of an effective action
$S_{eff}(T, \ldots)$ obtained by
formally integrating out all the positive mass$^2$ fields.
This is what we shall
do.\footnote{At this stage we would like to remind the reader that
our analysis will
be only at the level of classical open string field theory,
and hence integrating
out the heavy fields simply amounts to eliminating them by their equations
of motion.
} Here $\ldots$ stands for all the massless bosonic fields, which in
the case of
non-BPS D$p$-branes include one gauge field and $(9-p)$
scalar fields associated
with the transverse coordinates.
For D$p$-$\bd p$ brane pair the massless fields
consist of two $U(1)$ gauge fields and $2(9-p)$ transverse scalar fields.

First we shall state two properties of $S_{eff}(T,\ldots)$ which are trivially
derived from the analysis of the tree level S-matrix:
 \begin{enumerate}
\item For a non-BPS D-brane the tachyon effective action has a $Z_2$
symmetry under $T\to -T$, wheras for a brane-anti-brane system the tachyon
effective action has a phase symmetry under $T\to e^{i\alpha}T$.

\item Let $V(T)$ denote the tachyon effective potential, defined such that for
space-time independent field configuration,
and with all the massless fields set to
zero, the tachyon effective action $S_{eff}$
has
the form:
 \be \label{e2.4}
-\int d^{p+1} x \, V(T)\, .
 \ee
In that case $V(T)$ has a maximum at $T=0$. This is a straightforward
consequence of the fact that the mass$^2$ of the field $T$ is
given by $V''(T=0)$, and this is known to be negative. We shall
choose the additive constant in $V(T)$ such that $V(0)=0$.
\end{enumerate}
The question that we shall be most interested in is whether $V(T)$ has a
(local) minimum, and if it does, then how does the theory behave around this
minimum? The answer to this question is summarized in the following three
`conjectures'
\cite{9805019,9805170,9807138,
9808141,9911116,9812031}:\footnote{Although
initially these properties were
conjectured, by now
there is sufficient evidence for these conjectures
so that one can refer to them as
results rather than conjectures.}
 \begin{enumerate}
\item $V(T)$ does have a pair of
global minima at $T=\pm T_0$ for the non-BPS D-brane, and a one parameter
($\alpha$) family of
global minima at $T=T_0 e^{i\alpha}$ for the brane-antibrane
system. At this minimum
the
tension of the original D-brane configuration
is exactly canceled by the negative
contribution of the potential $V(T)$. Thus
 \be \label{e2.5}
V(T_0) + \EE_p = 0\, ,
 \ee
where
 \be \label{e2.6}
\EE_p = \cases{\wt \TT_p \quad \hbox{for non-BPS D$p$-brane} \cr
2\TT_p \quad \hbox{for D$p$-$\bd p$ brane pair} }\, .
 \ee
Thus the total energy density vanishes at the minimum of the
tachyon potential. This
has been illustrated in Fig.\ref{f1}.

 \begin{figure}[!ht]
 \begin{center}
\leavevmode
\epsfysize=5cm
\epsfbox{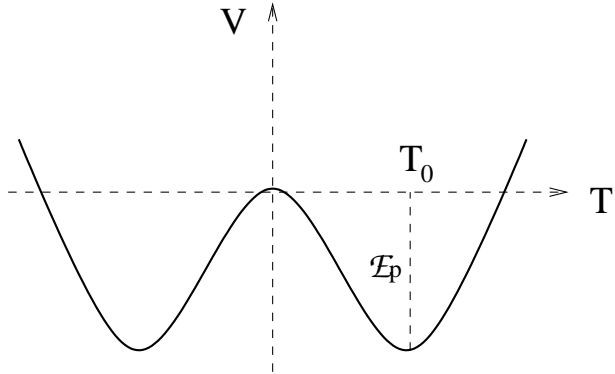}
\end{center}
\caption{The tachyon potential on an unstable D-brane in
superstring theories. The tachyon potential on a
brane-antibrane system is obtained by revolving this diagram about the 
vertical axis.
} \label{f1}
\end{figure}

\item Since the total energy density vanishes at $T=T_0$, and
furthermore, neither
the non-BPS D-brane nor the brane-antibrane system carries any RR charge,
it is
natural to conjecture that the configuration $T=T_0$ describes the vacuum
without any
D-brane.
This in turn implies that in perturbation theory we should not get any
physical
open string states by quantizing the theory
around the minimum of the potential, since open string states live only on
D-branes. This is counterintuitive, since in conventional
field theories the
number
of perturbative physical states do not change as we go from
one extremum of the
potential to another extremum.

 \begin{figure}[!ht]
 \begin{center}
\leavevmode
\epsfysize=5cm
\epsfbox{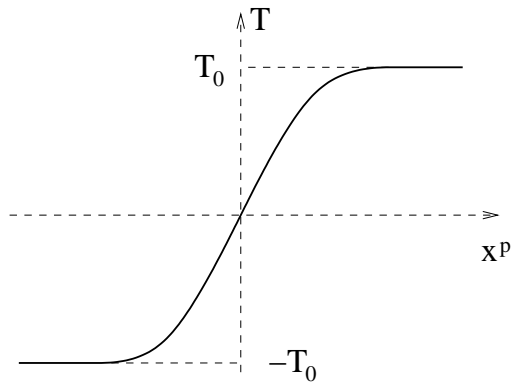}
\end{center}
\caption{The kink solution on a non-BPS D-brane.} \label{f2}
\end{figure}

\item Although there are no perturbative physical states
around the minimum of the
potential, the equations of motion derived from the tachyon effective action
$S_{eff}(T,\ldots)$ does have non-trivial time independent classical
solutions. It is conjectured that these
solutions represent lower dimensional D-branes. Some examples are
given
below:
 \begin{enumerate}
\item The tachyon effective
action on a non-BPS D$p$-brane admits a classical kink
solution as shown in Fig.\ref{f2}. This solution depends on
only one of the spatial
coordinates, labeled by $x^p$ in the figure, such that $T$ approaches $T_0$ as
$x^p\to\infty$ and
$-T_0$ as $x^p\to -\infty$, and interpolates between these two
values around $x^p=0$.
Since the total energy density vanishes for $T=\pm T_0$, we
see that for the above
configuration the energy density is concentrated around a
$(p-1)$ dimensional subspace $x^p=0$. This kink
solution describes a BPS
D-$(p-1)$-brane in the same theory\cite{9812031,9812135}.

\item There is a similar
solution on a brane-antibrane system, where the imaginary
part of the tachyon field
is set to zero, and the real part takes the form given in
Fig.\ref{f2}. This is not a
stable solution, but describes a non-BPS D-$(p-1)$-brane
in the same theory\cite{9805019,9808141}.
\item Since the tachyon field $T$
on a D$p$-$\bd p$-brane system is a complex field,
one can also construct a vortex solution where $T$ is a
function of two of the
spatial coordinates (say $x^{p-1}$ and $x^p$) and takes the form:
 \be \label{e2.7}
T = T_0 \, f(\rho) e^{i\theta}\, ,
 \ee
where
 \be \label{e2.8}
\rho = \sqrt{(x^{p-1})^2 + (x^p)^2}, \qquad \theta = \tan^{-1}(x^p/x^{p-1})\,,
 \ee
are the polar coordinates on the $x^{p-1}$-$x^p$ plane and the function 
$f(\rho)$ has the property:
 \be \label{e2.9}
f(\infty)=1, \qquad f(0)=0\, .
 \ee
Thus the potential energy associated with the solution vanishes as
$\rho\to\infty$. Besides the tachyon the solution also
contains an accompanying
background gauge field which makes
the covariant derivative of the tachyon fall off
sufficiently fast for large $\rho$ so
that the net energy density is concentrated
around the $\rho=0$ region. This gives a
codimension two soliton solution. This solution describes a BPS
D-$(p-2)$-brane in the same
theory\cite{9808141,0003124}.

\item If we take a coincident pair of non-BPS D-branes,
then the D-brane effective field theory
around $T=0$ contains a U(2) gauge field,
and there are four tachyon states represented by a $2\times 2$ hermitian
matrix
valued scalar field transforming in the
adjoint representation of this gauge group.
The $(ij)$ component of the matrix represents the tachyon in the open
string sector beginning on the $i$-th D-brane and ending on the $j$-th
D-brane. A family of minima of the tachyon potential can be found by
beginning with the configuration $T=T_0\pmatrix{1 & 0 \cr 0 & -1}$ which
represents the tachyon on the first D-brane at its minimum $T_0$ and the
tachyon on the second D-brane at its minimum $-T_0$, and then making an
SU(2) rotation. This gives a family of  minima of the form $T=T_0\, \hat
n.\vec \sigma$, where $\hat n$ is a unit vector and $\sigma_i$ are the
Pauli matrices.   At any of these minima of the tachyon potential
the SU(2) part of the gauge group is broken to U(1) by the vacuum
expectation value of
the tachyon. 

This theory contains a 't Hooft - Polyakov monopole
solution\cite{thooft,polyakov} which depends on three of the spatial 
coordinates $\vec x$, and
for which
the asymptotic form of the tachyon and the SU(2) gauge field strengths
$F^a_{\mu\nu}$
are given by:
 \be \label{ethooft}
T(\vec x) \simeq T_0 \, {\vec\sigma.\vec x \over |\vec x|}\, , \qquad
F^a_{ij}(\vec
x) \simeq
\epsilon^{aij} {x^a\over |\vec x|^3}\, .
 \ee
The energy density of this solution is concentrated around $\vec x=0$ and
hence
this gives
a codimension 3 brane. This solution describes
a BPS D-$(p-3)$-brane in the same
theory\cite{9812135,0003124}.

\item If we consider a system of two D$p$-branes and two $\bd p$-branes, all
along the same plane, then the
D-brane world-volume theory has an $U(2)\times U(2)$
gauge field, and a $2\times 2$
matrix valued complex tachyon field $T$, transforming in
the (2,2) representation of the gauge group. The $(ij)$ component of the
matrix represents the tachyon field coming from the open string with ends
on
the $i$-th D-brane and the $j$-th $\bd$-brane. In this case the minimum
of the tachyon potential where the 11 component of the tachyon
takes value $T_0 e^{i\alpha}$ and the 22 component of the tachyon takes
value $T_0 e^{i\beta}$ corresponds to $T=T_0\pmatrix{e^{i\alpha} & 0\cr 0
& e^{i\beta}}$. A family of minima may now be found by making arbitrary
U(2) rotations from the left and the right. This gives $T=T_0\, U$ with
$U$ being an arbitrary $U(2)$ matrix.

Let $A_\mu^{(1)}$ and $A_\mu^{(2)}$
denote the gauge fields in the two SU(2) gauge groups. Then we can construct a
codimension 4 brane solution where the fields depend on four of the spatial
coordinates, and have the asymptotic behaviour:
 \be \label{e2.10}
T \simeq T_0 \, U(x^{p-3}, x^{p-2}, x^{p-1}, x^p), \qquad A_\mu^{(1)}
\simeq {i} \p_\mu U U^{-1}, \qquad  A_\mu^{(2)}\simeq 0 \, ,
 \ee
where $U$ is an SU(2)
matrix valued function of four spatial coordinates, corresponding to the
identity map
(winding number one map) from
the surface $S^3$ at spatial infinity to the SU(2) group
manifold. This describes a
BPS D-$(p-4)$-brane in the same theory \cite{9808141,0003124}.

\end{enumerate}

Quite generally if we begin
with sufficient number of non-BPS D9-branes in type IIA
string theory, or D9-$\bd 9$-branes in type IIB string theory, we can
describe any lower dimensional D-brane as classical solution in this
open string field theory \cite{9810188,9812135,0003124}. This has led to a
classification of
D-branes using
a branch of mathematics  known as 
K-theory\cite{9810188,9812135,9812226,9901042,
9902160,9904153,
9905034,
9907140,
9908091,9908121,
9910109,9912279,0001143,
0002023,0006223,
0007175,0010007,
0012164,0108100,
0111151,
0111169,0212059,
0304018}.

\end{enumerate}

\subsection{Time dependent solutions in superstring theory} \label{s2.2}

So far we  have only discussed time independent solutions of the tachyon
equations of
motion. One could also ask questions about time dependent solutions. In
particular,  given that the tachyon potential on a non-BPS D$p$-brane or a
D$p$-$\bd p$ pair has the form given in
Fig.\ref{f1}, one
could ask:  what happens if we displace the tachyon from the maximum of
the potential
and let it  roll down towards its minimum?\footnote{For simplicity in this
section we shall only describe spatially homogeneous time dependent
solutions, but more general solutions which depend on both space and time
coordinates can also be studied\cite{0207105,0212248}.} If $T$ had been an
ordinary scalar field
then the answer  is simple: the tachyon field $T$ will simply oscillate
about the
minimum $T$ of the  potential, and in the absence of any dissipative force
(as is the
case at the classical level) the oscillation will continue for ever.
The energy density $T_{00}$ will remain constant during this oscillation,
but other
components of the energy-momentum  tensor, {\it e.g.} the pressure
$p(x^0)$, defined through $T_{ij}=p(x^0)\, \delta_{ij}$ for $1\le i,j \le
p$,
will oscillate
about their average value.
However for the
case of the string theory tachyon  the answer is different and somewhat
surprising\cite{0203211,0203265}.
It turns out that for the rolling tachyon solution on an unstable
D-brane the  energy density on the brane remains constant as in
the case of
a usual scalar field, but the pressure, instead of oscillating about an average
value, goes to zero asymptotically. More precisely, the non-zero
components of $T_{\mu\nu}$ take the form\footnote{The energy momentum
tensor
$T_{\mu\nu}$ is confined to the plane of the original D-brane, and
hence all expressions for
$T_{\mu\nu}$ are accompanied by a $\delta$-function in the transverse
coordinates which we shall denote by $\delta(x_\perp)$.
This factor may occasionally be
omitted for brevity. Also, only the
components of the stress tensor along the world-volume of the brane are
non-zero, {\it i.e.} $T_{\mu\nu}\ne 0$ only for $0\le \mu, \nu \le p$.}
 \ben \label{e2.11}
&& T_{00}=\EE \, \delta(x_\perp), \qquad T_{ij} = p(x^0) \, \delta_{ij},
\qquad 1\le i,j\le p\nonumber \\
&& p(x^0) = -\EE_p \, \tf(x^0)\, \delta(x_\perp)\, ,
 \een
where $\EE$ is a constant labelling the energy density on the
brane, $\EE_p$ is given by \refb{e2.6}, $\delta(x_\perp)$ denotes a
delta-function in the coordinates transverse to the brane and the function
$\tf(x^0)$ vanishes as $x^0\to \infty$. In order to give the 
precise form of $\wt f(x^0)$ we need to consider two
different
cases:
 \begin{enumerate}
\item $\EE\le \EE_p$: In this case we  can label the solution by
a parameter
$\tl$ defined through the relation:
 \be \label{e2.12}
T_{00} = {\EE_p} \, \cos^2(\pi \tl) \, \delta(x_\perp).
 \ee
$T_{00}$ includes the contribution from the tension of the D-brane(s) as
well as the tachyon kinetic and potential energy.
Since the total energy density available to the system is less than $\EE_p$,
-- the energy density at the maximum of the tachyon potential describing the
original brane configuration, -- at some instant of time during its
motion the tachyon is expected to come to rest at some point away from
the maximum of the
potential. We can  choose this instant of time as
$x^0=0$. The function $\tf(x^0)$ in this case takes the form:
 \be \label{e2.13}
\tf(x^0) = {1\over 1 + e^{\sqrt 2 x^0} \sin^2(\tl\pi)} + {1 \over
1 + e^{-\sqrt 2 x^0} \sin^2(\tl\pi)} - 1\, .
 \ee
{}From this we see that as $x^0\to \infty$, $\tf(x^0)\to 0$. Thus the
pressure
vanishes asymptotically.

Note that for $\tl={1\over 2}$ both $T_{00}$ and $p(x^0)$ vanish
identically. Thus this solution has the natural interpretation as the
tachyon being placed at the minimum of its potential. The solution for
$\tl={1\over 2}+\epsilon$ is identical to the one at $\tl={1\over
2}-\epsilon$; thus the inequivalent set of solutions are obtained by
restricting $\tl$ to the range $[-{1\over 2}, {1\over 2}]$.

\item $\EE\ge \EE_p$: In this case we can label the solutions by
a parameter
$\tl$ defined through the relation:
 \be \label{e2.14}
T_{00} = {\EE_p} \cosh^2(\pi\tl) \, \delta(x_\perp)\, .
 \ee
Since the total energy density available to the system is
larger than $\EE_p$,
at some instant of time during its
motion the tachyon is expected to pass the point $T=0$ where the potential
has a maximum.
We can choose our initial condition such that at
$x^0=0$ the tachyon is at the maximum of the potential and has a non-zero
velocity. The function $\tf(x^0)$ in this case takes the
form:\footnote{This result can be trusted only for $|\tl|\le
\sinh^{-1}1$.}
 \be \label{e2.15}
\tf(x^0) = {1\over 1 + e^{\sqrt 2 x^0} \sinh^2(\tl\pi)} + {1 \over
1 + e^{-\sqrt 2 x^0} \sinh^2(\tl\pi)} - 1\, .
 \ee
Since as $x^0\to \infty$, $\tf(x^0)\to 0$, the pressure
vanishes asymptotically.

\end{enumerate}

The energy momentum tensor $T_{\mu\nu}$ given above is computed by 
studying the coupling of the D-brane to the graviton coming from the 
closed string sector of the theory. Besides the graviton, there are other 
massless 
states in superstring theory, and a D-brane typically couples to these
massless fields as well. We can in particular consider the sources $Q$ and 
$J^{(p)}_{\mu_1\ldots
\mu_p}$  produced by the D-brane for 
the 
dilaton
$\Phi_D$ and RR $p$-form gauge fields $C^{(p)}_{\mu_1\ldots
\mu_p}$ respectively. It turns out that as the
tachyon rolls down on a non-BPS D-$p$ brane or a D$p$-$\bd
p$-brane pair stretched along the $(x^1,x^2,\ldots x^p)$
hyperplane, it produces a source for the dilaton field of the
form:
 \be \label{edilcharge}
Q(x^0) = \EE_p \wt f(x^0)\, \delta(x_\perp)\, ,
 \ee
where $\wt f(x^0)$ is the same function as defined in \refb{e2.13} and
\refb{e2.15}.
Furthermore a rolling tachyon on a non-BPS D-$p$-brane produces
an RR $p$-form source of the form\cite{0204143}:
 \be \label{errcharge}
J^{(p)}_{1\ldots p} \propto \sin(\tl\pi) \left[ {e^{x^0/\sqrt 2} \over 1 +
\sin^2(\tl\pi) e^{\sqrt 2 x^0} } - {e^{-x^0/\sqrt 2} \over 1 +
\sin^2(\tl\pi) e^{-\sqrt 2 x^0} } \right] \, \delta(x_\perp)\, ,
 \ee
for the case $\EE\leq \EE_p$, and
 \be \label{errchargep}
J^{(p)}_{1\ldots p} \propto \sinh(\tl\pi) \left[ {e^{x^0/\sqrt 2} \over 1
+
\sinh^2(\tl\pi) e^{\sqrt 2 x^0} } + {e^{-x^0/\sqrt 2} \over 1 +
\sinh^2(\tl\pi) e^{-\sqrt 2 x^0} }\right] \, \delta(x_\perp)\, ,
 \ee
for the case $\EE\geq\EE_p$. The sources for other massless fields
vanish
for this solution.

The assertion that around
the tachyon vacuum there are no physical open string states implies
that there is no small oscillation of finite frequency around the minimum
of the
tachyon potential. The lack of oscillation in the pressure is consistent
with this
result.
However the existence of classical solutions with arbitrarily
small energy density (which can be achieved by taking $\tl$ close to 1/2 in
\refb{e2.12}) indicates that quantization of open string field
theory around
the tachyon vacuum does give rise to non-trivial quantum states which in the
semi-classical limit are described by the solutions that we have found.

\subsection{Static and time dependent solutions in bosonic string theory}
\label{s2.3}

Bosonic string theory in (25+1) dimensions has D$p$-branes for all
integers $p\le
25$ with tension\cite{polbook}
 \be \label{e2.16}
\TT_p = g_s^{-1} \, (2\pi)^{-p}\, ,
 \ee
where $g_s$ as usual denotes the closed string coupling constant and we
are using
$\alpha'=1$ unit.
The spectrum of open strings on each of these D-branes
contains a
single
tachyonic state with mass$^2=-1$, besides infinite number of other states of
mass$^2\ge 0$. Thus among the infinite number of fields appearing in the
string
field theory on a D$p$-brane, there is a scalar field $T$ with negative
mass$^2$.
If as in the case of superstring theory we denote by $S_{eff}(T,\ldots)$ the
effective action obtained by integrating out the fields with positive
mass$^2$, and by
$V(T)$ the effective potential for the tachyon obtained by restricting to
space-time
independent field configurations and setting the massless fields to zero,
then
$V(T)$ will have a maximum at $T=0$. Thus we can again ask: does the
potential
$V(T)$ have a (local) minimum, and if it does, how does the open string
field theory
behave around this minimum?

Before we go on to answer these questions, let us recall that
bosonic string theory also has a tachyon in the closed string
sector, and hence the theory as it stands is inconsistent. Thus
one might wonder why we should be interested in studying
D$p$-branes in bosonic string theory in the first place. The
reason for this is simply that 1) although closed string tachyons
make the quantum open string field theory inconsistent due to
appearance of closed strings in open string loop diagrams,
classical open string field theory is not directly affected by the
closed string tachyon, and 2) the classical tachyon dynamics on a
bosonic D$p$-brane has many features in common with that on a
non-BPS D-brane or a brane-antibrane pair in superstring theory,
and yet it is simpler to study than the corresponding problem in
superstring theory. Thus studying tachyon dynamics on a bosonic
D-brane gives us valuable insight into the more relevant problem
in superstring theory.

We now summarise the three conjectures describing the static properties of the
tachyon effective action on a bosonic D$p$-brane\cite{9902105,9911116}:

 \begin{figure}[!ht]
 \begin{center}
\leavevmode
\epsfysize=5cm
\epsfbox{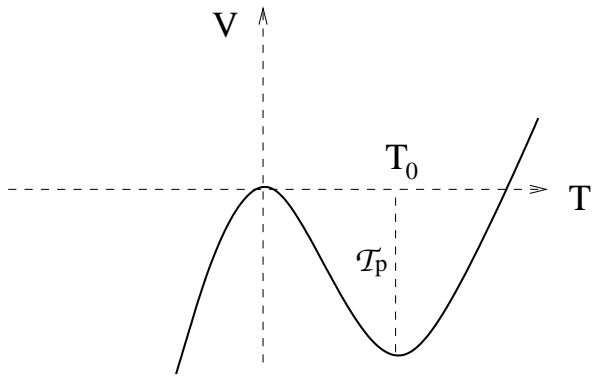}
\end{center}
\caption{The tachyon effective potential on a D$p$-brane in bosonic
string theory.
} \label{f3}
\end{figure}

 \begin{enumerate}

\item The tachyon effective potential $V(T)$ has a {\it local} minimum at
some value $T=T_0$,
and at this minimum the tension $\TT_p$ of the original D-brane is
exactly canceled
by the negative value $V(T_0)$ of the potential. Thus
 \be \label{e2.17}
V(T_0) + \TT_p = 0\, .
 \ee
The form of the potential has been shown in Fig.\ref{f3}. Note that unlike
in the case of superstring theory, in this case the tachyon potential does
not have a global minimum.

\item Since the total energy density vanishes at $T=T_0$, it is
natural to identify the configuration $T=T_0$ as the vacuum without any
D-brane.
This in turn implies that there are no physical perturbative open strings
states
around the minimum of the potential, since open string states live only on
D-branes.

 \begin{figure}[!ht]
 \begin{center}
\leavevmode
\epsfysize=5cm
\epsfbox{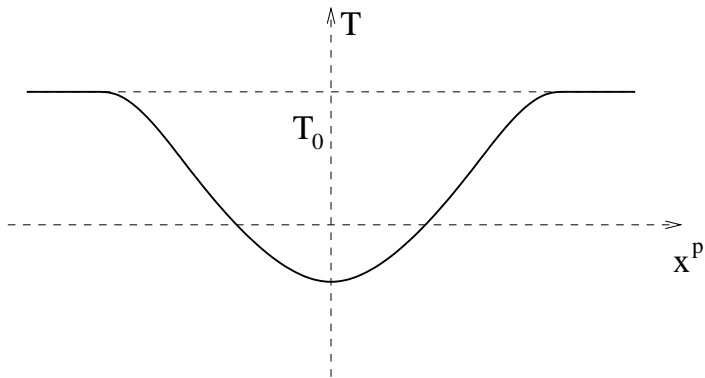}
\end{center}
\caption{The lump solution on a D$p$-brane in bosonic string theory.}
\label{f4}
\end{figure}

\item Although there are no perturbative physical states around the
minimum of the
potential, the equations of motion derived from the tachyon effective action
$S_{eff}(T,\ldots)$ does have non-trivial time independent classical
{\it lump} solutions of various codimensions. A codimension $q$ lump
solution on a
D$p$-brane, for which $T$ depends on $q$ of the spatial coordinates and
approaches
$T_0$ as any one of these  $q$ coordinates goes to infinity,
represents a
D-$(p-q)$-brane of bosonic string theory. An example of a
codimension 1
lump solution has been shown in Fig.\ref{f4}.

\end{enumerate}

This summarises the properties of time independent solutions,
but one can also ask
about time dependent solutions. In particular we can ask: what happens if we
displace the tachyon from the maximum of its potential and let it roll?
Unlike in
the case of superstrings, in this case the potential (shown in
Fig.\ref{f3}) is not
symmetric around $T=0$,
and hence we expect different behaviour depending on whether we have
displaced the
tachyon to the left (away from the local minimum) or right (towards the
local
minimum). As in the case of superstring theory, the energy density
on the brane remains
constant
during the motion, but the pressure along the brane evolves in time:
 \be \label{e2.18}
p(x^0) = - \TT_p \, \tf(x^0)\, \delta(x_\perp) \, .
 \ee
In order to specify the form of $\tf(x^0)$ we consider two cases
separately.

 \begin{enumerate}

\item $T_{00} = \EE\delta(x_\perp)$, $\EE\leq \TT_p$: In this case we can
parametrize $T_{00}$ as:
 \be \label{e2.19}
T_{00} = {\TT_p} \, \cos^2(\pi\tl) \, \delta(x_\perp)\, ,
 \ee
and choose the origin of the time coordinate $x^0$ such that at $x^0=0$ 
the tachyon has zero
velocity and is displaced from $T=0$ by a certain amount
determined by the
parameter $\tl$.
Then
the function $\tf(x^0)$ appearing in \refb{e2.18} is given
by\cite{0203211,0203265}:
 \be \label{e2.20}
\tf(x^0)={1\over 1 + e^{x^0} \sin(\tl\pi)} + {1 \over
1 + e^{-x^0} \sin(\tl\pi)} - 1\, .
 \ee
Note that $\pm \tl$ gives the same $T_{00}$ but different $\tf(x^0)$. This
is due to
the
fact that positive sign of $\tl$ corresponds to displacing the tachyon
towards the
local minimum of the potential, whereas negative value of $\tl$
corresponds to
displacing $T$ towards the direction in which the potential is unbounded
from
below. As we can see from \refb{e2.20}, for positive $\tl$ the function
$\tf(x^0)$
approaches zero as $x^0\to\infty$, showing that the system evolves to a
pressureless
gas.
In particular, for $\tl={1\over 2}$,
 \be \label{elhalf}
\wt f(x^0) = 0\, .
 \ee
Thus $T_{\mu\nu}$ vanishes identically, and we can identify this solution
to be the one where the tachyon is placed at the local minimum of the
potential. On the other hand, for negative $\tl$, $\tf(x^0)$ blows up at
 \be \label{e2.21}
x^0 = \ln{1\over |\sin(\tl\pi)|} \equiv t_c\, .
 \ee
This shows that if we displace the tachyon towards the direction in which
the
potential is unbounded from below, the system hits a singularity at a
{\it finite
time.}

\item $T_{00}  = \EE\delta(x_\perp)$, $\EE\geq \TT_p$: In this case we
can parametrize $T_{00}$ as:
 \be \label{e2.22}
T_{00} = {\TT_p} \, \cosh^2(\pi\tl) \, \delta(x_\perp)\, .
 \ee
Then for an appropriate choice of the origin of the time coordinate $x^0$
the function $\tf(x^0)$ appearing in \refb{e2.18} is given
by\cite{0203211,0203265}:
 \be \label{e2.23}
\tf(x^0) = {1\over 1 + e^{x^0} \sinh(\tl\pi)} + {1 \over
1 - e^{-x^0} \sinh(\tl\pi)} - 1\, .
 \ee
This equation is expected to be
valid only for $|\tl|\le \sinh^{-1} 1$.
Again we see that $\pm \tl$ gives the same $T_{00}$ but different
$\tf(x^0)$. Positive
sign of
$\tl$
corresponds to pushing the tachyon towards the
local minimum of the potential, whereas negative value of $\tl$
corresponds to
pushing $T$ towards the direction in which the potential is unbounded from
below. For positive $\tl$ the function $\tf(x^0)$
approaches zero as $x^0\to\infty$, showing that the system evolves to a
pressureless
gas. On the other hand, for negative $\tl$, $\tf(x^0)$ blows up at
 \be \label{e2.24}
x^0 = \ln{1\over |\sinh(\tl\pi)|}\, .
 \ee
This again shows that if we displace the tachyon towards the direction in
which the
potential is unbounded from below, the system hits a singularity at a finite
time.

\end{enumerate}

Bosonic string theory also has a massless dilaton field and we can
define the dilaton charge density as the source that couples to this
field. As in the case of superstring theory, a rolling tachyon on
a D-$p$-brane of bosonic string theory produces a source for the
dilaton field
 \be \label{ebosdil}
Q = \TT_p \, \wt f(x^0)\, \delta(x_\perp)\, ,
 \ee
with $\wt f(x^0)$ given by eq.\refb{e2.20} or \refb{e2.23}.

\subsection{Coupling to closed strings and the open string completeness 
conjecture} \label{ssummary3}

So far we have discussed the dynamics of the open string tachyon at the
purely classical level, and have ignored the coupling of the
D-brane to closed strings. Since D-branes act as sources for various 
closed
string fields, a time dependent open string field
configuration such as the rolling tachyon solution acts as a time
dependent source for closed string fields, and produces closed string
radiation. This can be computed using the standard techniques. For
unstable Dp-branes with all $p$ directions wrapped on circles, one 
finds that the total energy carried by 
the closed string radiation is infinite\cite{0303139,0304192}. However 
since the initial D$p$-brane has 
finite energy it is appropriate to regulate this divergence by putting an 
upper cut-off on the energy of the emitted closed string. A natural choice 
of this cut-off is the initial
energy
of the D-brane. In that case one finds that

\begin{enumerate}
\item All the energy of the D-brane is radiated away into closed strings
even though any single closed string mode carries a small ($\sim g_s$)
fraction of the D-brane energy.

\item Most of the energy is carried by closed strings of mass $\sim
1/g_s$.

\item The typical momentum
carried by these closed strings along directions transverse to the
D-brane is of
order $\sqrt{1/g_s}$, and the typical winding charge carried by these
strings
along directions tangential to the D-brane is also of order
$\sqrt{1/g_s}$.

\end{enumerate}

{}From the first result one would tend to conclude that the effect of 
closed string
emission should invalidate the classical open string results on the
rolling tachyon system discussed earlier. There are however some
surprising coincidences:
\begin{enumerate}
\item The tree level open string analysis tell us that the final system
associated with the rolling tachyon configuration has
zero pressure. On the other hand closed string emission results tell us
that the final closed strings have momentum/mass and winding/mass ratio of
order $\sqrt{g_s}$ and hence pressure/ energy density ratio of order
$g_s$. In the $g_s\to 0$
limit this vanishes. Thus it appears that the classical open string
analysis correctly predicts the equation of state of the final system of
closed strings into which the system decays.

\item The tree level open string analysis tells us that the final system 
has zero
dilaton charge. By analysing the properties of the closed string radiation
produced by the decaying D-brane one finds that these closed strings also
carry zero dilaton charge. Thus the classical open string analysis
correctly captures the properties of the final state closed strings
produced during the D-brane decay.

\end{enumerate}

These results (together with some generalizations which will be 
discussed briefly in section \ref{sph}) suggest that the classical open 
string theory already
knows about the properties of the final state closed strings produced by
the decay of the D-brane\cite{0305011,0306137}. This can be formally
stated as an open string completeness
conjecture according to which the complete dynamics
of a
D-brane is captured by the quantum open string theory without any need to
explicitly consider the coupling of the system to closed
strings.\footnote{Previously this was called the open-closed string
duality conjecture\cite{0306137}. However since there are many different
kinds of open-closed string duality conjecture, we find the name open
string completeness conjecture more appropriate. In fact the proposed
conjecture is not a statement of equivalence between the open and closed
string description since the closed string theory could have many more
states which are not accessible to the open string theory.} Closed strings
provide
a dual description of the system. This does not imply that any arbitrary
state in string theory can be described in terms of open string theory on
an unstable D-brane, but does imply that all the quantum states required 
to
describe the dynamics of a given D-brane
are contained in the open string theory associated with that D-brane.

At the level of critical string theory one cannot prove this conjecture.
However it turns out that this conjecture has a simple realization in a
non-critical two dimensional string theory. This theory has two equivalent
descriptions: 1) as a regular string theory in a somewhat complicated
background\cite{DAVID,DISKAW} in which the world-sheet dynamics of the
fundamental string is described by the direct sum of a free scalar field
theory and
the Liouville theory with central charge 25, and 2) as a theory of free
non-relativistic
fermions moving under a shifted inverted harmonic oscillator
potential $-{1\over 2} \, q^2 + {1\over g_s}$\cite{GROMIL,BKZ,GINZIN}.
Although in the free fermion description the potential is unbounded from
below, the ground state of the system has all the negative energy states
filled, and hence the second quantized theory is well defined. The map
between these two theories is also known. In particular the closed string
states
in the first
desciption are related to the quanta of the scalar field obtained by
bosonizing the
second quantized fermion field in the second
description\cite{DASJEV,SENWAD,GROSSKLEB}.

In the regular string theory description the theory also has an unstable
D0-brane with a tachyonic mode\cite{0101152}. The classical properties of
this tachyon
are identical to those discussed in section \ref{s2.3} in the context of
critical bosonic string theory. In particular one
can construct time dependent solution describing the rolling of the
tachyon away from the maximum of the potential. Upon taking into account
possible closed string emission effects one finds that as in the case of
critical string theory, the D0-brane decays completely into closed
strings\cite{0305159}.

By examining the coherent closed string field configuration produced in
the D0-brane decay, and translating this into the fermionic description
using the known relation between the closed string fields and the
bosonized fermion, one discovers that the radiation produced by `D0--brane
decay' precisely corresponds to a single fermion excitation in the theory.
This suggests that the D0-brane in the first description should be 
identified as the single fermion
excitation in the second
description of the theory\cite{0304224,0305159,0305194}. Thus its
dynamics is described by that of
a single particle moving under the inverted harmonic oscillator potential 
with a lower-cutoff on the energy at the fermi level due to Pauli
exclusion principle.

Given that the dynamics of a D0-brane in the first description is 
described by an open string
theory, and that in the second description a D0-brane is identified with 
single 
fermion excitation, we can conclude that the open 
string theory for the D0-brane must be equivalent to
the single particle
mechanics with potential $-{1\over 2}q^2 + {1\over 
g_s}$, 
with an additional 
constraint $E\ge 0$.
A consistency check of this proposal is that the
second derivative of the inverted harmonic oscillator potential at the
maximum precisely matches the negative mass$^2$ of the open string tachyon
living on the D0-brane\cite{0305159}.
This `open string theory' clearly
has the ability to describe the complete dynamics of the D0-brane 
\i.e.\ the single fermion 
excitations. It is possible but not necessary to describe the system in
terms of the closed string field, \i.e.\ the scalar field obtained by 
bosonizing the second quantized
fermion field. This is in complete accordance 
with the open 
string
completeness
conjecture proposed earlier in the context of critical string theory.

\sectiono{Conformal Field Theory Methods} \label{s3}

In this section we shall analyze time independent solutions involving the 
open string tachyon using the
well known correspondence between classical solutions of
equations of
motion of string theory, and two dimensional (super-)conformal field
theories
(CFT).
A D-brane configuration in a space-time background is
associated
with a two dimensional conformal field theory on an infinite
strip (which can be conformally mapped to a disk or
the upper
half plane) describing propagation of open string excitations on the 
D-brane. Such conformal field theories are known as boundary conformal
field
theories (BCFT) since they are defined on surfaces with
boundaries. The
space-time background in which the D-brane lives determines the bulk
component of
the CFT, and associated with a particular D-brane configuration we have
specific
conformally invariant boundary conditions / interactions involving 
various fields of this
CFT.  Thus for example for a D$p$-brane in flat space-time we have
Neumann boundary condition on the $(p+1)$ coordinate fields tangential to 
the
D-brane world-volume and Dirichlet boundary condition on the coordinate
fields transverse to the D-brane. Different classical solutions
in the open string field theory describing the dynamics of a D-brane are
associated
with different conformally invariant boundary interactions in this
BCFT.
More specifically, if we add to the original world-sheet action a boundary
term
 \be \label{ex21}
\int dt \, V(t) \, ,
 \ee
where $t$ is a parameter labelling the boundary of the world-sheet and $V$
is a boundary vertex operator in the world-sheet theory, then for a
generic $V$ the conformal invariance of the theory is broken. But
for
every $V$ for which we have a (super-)conformal field theory, there is an
associated solution of the classical open string field equations.
Thus we can
construct solutions of equations of motion of open string field theory by
constructing appropriate conformally invariant boundary
interactions in
the BCFT
describing the original D-brane configuration. This is the approach we
shall take in
this section.

In the rest of the section we shall outline the logical steps
based on this approach which lead to the results on time
independent solutions described in section \ref{s2}.

\subsection{Bosonic string theory} \label{s3.1}

We begin with a space-filling D25-brane of bosonic string theory. We 
shall show
that as stated in the third conjecture in section \ref{s2.3}, we can 
regard the D24-brane as a codimension 1 lump solution of the
tachyon
effective action
on the D25-brane.
This is done
in two steps:
 \begin{enumerate}

\item First we find the conformally invariant BCFT associated with the
tachyon
lump
solution on a D25-brane. This is done by
finding a series of marginal deformations that connects the $T=0$
configuration
on the D25-brane to the tachyon lump solution.
\item Next we show that this BCFT is identical to that describing a
D-24-brane.
This is done by following what happens to the original BCFT describing
the D25-brane
under this series of marginal
deformations.
\end{enumerate}
Thus we first need to find a series of marginal deformations connecting the
$T=0$ configuration to the tachyon lump solution on the D-25-brane. This
is done as
follows:
 \begin{enumerate}
\item Let us choose a specific direction $x^{25}$ on which the lump
solution will eventually depend. For
simplicity
of notation we shall define $x\equiv x^{25}$. We first compactify $x$ on a
circle of radius $R$. The resulting world-sheet theory is conformally
invariant for every $R$.
This configuration has energy per unit 24-volume given by:
 \be \label{e3.1}
2\pi R \TT_{25} = R \TT_{24}\, .
 \ee
In deriving \refb{e3.1} we have used \refb{e2.16}.

\item At $R=1$ the {\it boundary operator} $\cos X$ becomes exactly marginal
\cite{9402113,9404008,9811237,9902105}.
A simple way to see this is as follows. For $R=1$ the bulk CFT has an
enhanced
$SU(2)_L\times SU(2)_R$ symmetry. The $SU(2)_{L,R}$ currents
$J^a_{L,R}$ ($1\le a\le
3$) are:
 \be \label{eb3}
J^3_L = i \bar\p X_L, \quad J^3_R = i \p X_R, \quad J^1_{L,R} = \cos (2
X_{L,R}) \quad
J^2_{L,R}
= \sin(2 X_{L,R})\, ,
 \ee
where $X_L$ and $X_R$ denote the left and right moving components
of $X$ respectively:\footnote{In our convention left and right
refers to the anti-holomorphic and holomorphic components of the
field respectively.}
 \be \label{eb4}
X = X_L + X_R\, .
 \ee
For $\alpha'=1$ the fields $X_L$ and $X_R$ are normalized so that
 \be \label{exnorm}
\p X_R(z) \p X_R(w) \simeq -{1\over 2(z-w)^2}\, , \qquad \bar\p
X_L(\bar z) \bar\p X_L(\bar w) \simeq -{1\over 2(\bar z-\bar
w)^2}\, .
 \ee
For definiteness let us take the open string world-sheet to be the
upper half plane with the real axis as its boundary. The Neumann
boundary condition on $X$ then corresponds to:
 \be \label{eb1}
X_L=X_R \quad \to \quad J^a_L=J^a_R\, \quad \hbox{for $1\le a\le 3$}
 \ee
on the real axis.
Using eqs.\refb{eb3}-\refb{eb1}
the boundary operator $\cos X$ can be regarded as the
restriction of $J^1_{L}$ (or
$J^1_R$) to the real axis.
Due to SU(2) invariance of the CFT we can now describe
the theory in terms of a new free scalar field $\phi$, related to $X$ by
an
SU(2) rotation, so that
 \be \label{eb2}
J^1_L = i \bar\p \phi_L, \qquad J^1_R = i \p\phi_R\, .
 \ee
Thus in terms of $\phi$ the boundary operator $\cos
X=\cos(2X_L)=\cos(2X_R)$ is proportional to the restriction of
$i\bar\partial\phi_L$ (or $i\p\phi_R$) at the boundary. This is
manifestly an exactly marginal operator, as it corresponds to
switching on a Wilson line along $\phi$.

 \begin{figure}[!ht]
 \begin{center}
\leavevmode
\epsfysize=5cm
\epsfbox{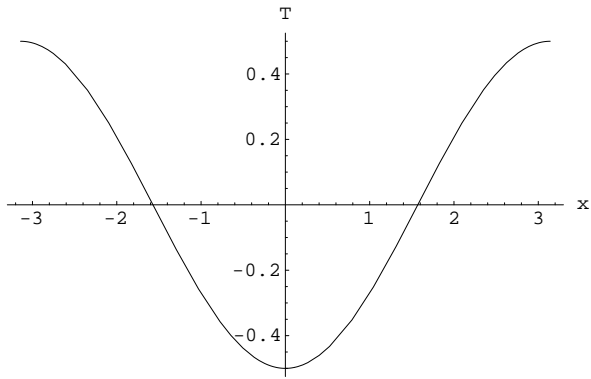}
\end{center}
\caption{The plot of $T(x) = -\al \cos x$ for $\al=.5$. This looks like a
lump
centered around $x=0$, but the height, being equal to $\al$, is
arbitrary.}
\label{f5}
\end{figure}

Due to exact marginality of the operator $\cos X$, we can
switch on a conformally invariant
perturbation of the form:
 \be \label{e3.2}
- \al \int dt \cos (X(t)) = -i\alpha \int dt \, \bar\p
\phi_L\, ,
 \ee
where $\al$ is an arbitrary constant and $t$ denotes a parameter labelling 
the boundary of the world-sheet.
From the
target space view-point switching on a perturbation proportional to
$-\cos X$
amounts to giving the tachyon field a vev proportional to $-\cos x$.
This in turn
can be interpreted as the creation of a lump centered at $x=0$ (see
Fig.\ref{f5}). At this
stage however the amplitude $\al$ is arbitrary, and hence the lump has
arbitrary
height. Since the boundary perturbation \refb{e3.2} is marginal, the
energy of
the configuration stays constant during this deformation at its initial
value at
$R=1$, $\al=0$. Using \refb{e3.1} we get the energy per unit 24-volume to
be
$\TT_{24}$, {\it i.e.} the energy density of a D-24-brane!

 \begin{figure}[!ht]
 \begin{center}
\leavevmode
\epsfysize=5cm
\epsfbox{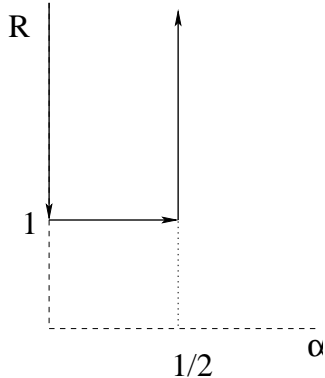}
\end{center}
\caption{Marginal flow in the $R-\al$ plane.
} \label{f6}
\end{figure}

\item Since we are interested in constructing a lump solution at 
$R=\infty$ we
need to
now take the radius back to
infinity. However for a generic $\al$, as soon as we switch on a radius
deformation,
the boundary operator $\cos {X\over R}$ develops a one point function
\cite{9902105}:
 \be \label{e3.3}
\left\la \cos {X(0)\over R} \right\ra_{R;\al} \propto (R-1) \sin(2\pi \al) 
\, ,
 \ee
for $R\simeq 1$.
This
indicates that
the configuration fails to satisfy the open string field equations and 
hence no longer describes a BCFT.\footnote{This is related to the fact 
that near $\alpha=0$ the operator $\cos(X/R)$ has dimension $1/R^2$ and 
hence deformation by $\alpha\int dt\, \cos(X(t)/R)$ does not give a BCFT 
for generic $\alpha$ and $R$.}  However if $\al=0$ or $1/2$, then the
one point function vanishes, not only for $R\simeq 1$ but for all values
of $R$
\cite{9902105}. Thus for these values of $\al$ we can get a BCFT for
arbitrary
$R$. For
$\al=0$ the resulting configuration is a D-25-brane, whereas for
$\al=1/2$ we can
identify the configuration as a tachyon lump solution on a D-25-brane for
any value of $R\ge 1$.
\end{enumerate}
The motion in the $R-\al$ plane as we follow the three step
process
has been
shown in Fig.\ref{f6}.

It now remains to show that the BCFT constructed this way with
$\al=1/2$, $R=\infty$ describes a D-24-brane. For this we need to
follow the fate of the BCFT under the three step
deformation that takes us from $(\alpha=0, R=\infty)$ to $(\alpha= 
1/2, R=\infty)$. In the first step, involving reduction of $R$ from
$\infty$ to 1, the D25-brane remains a D25-brane. The second
step,-- switching on the perturbation \refb{e3.2}, -- does
introduce non-trivial boundary interaction. It follows from the
result of \cite{9402113,9404008,9811237,9902105} that at $R=1$ the
BCFT at $\al=1/2$ corresponds to putting a Dirichlet boundary
condition on the coordinate field $X$. A simple way to see this is
as follows. Since the perturbation $\int dt \cos(X(t))$ is
proportional to $\int dt J^1_L(t)$, the effect of this
perturbation on any closed string vertex operator in the interior
of the world-sheet will be felt as a rotation by $2\pi\al$ in the 
$SU(2)_L$
group about the 1-axis.\footnote{This explains the periodicity of 
\refb{e3.3}
under $\al\to \al +1$.} For $\al=1/2$ the angle of rotation is
precisely $\pi$ and hence
it changes $X_L$ to $-X_L$ in any closed string vertex operator
inserted in the bulk. By redefining $-X_L$ as $X_L$ we can ensure
that the closed string vertex operators remain unchanged, but as a
result of this redefinition the boundary condition on $X$ changes
from Neumann to Dirichlet:
 \be \label{efx1}
X_L = - X_R\, .
 \ee
Thus
we can conclude that when probed by closed
strings, the
perturbed BCFT at $\al=1/2$ behaves as if we have Dirichlet boundary
condition on
$X$.

Since all other fields $X^\mu$ for $0\le\mu\le 24$ remain unaffected by
this
deformation, we see that the
BCFT at $R=1$, $\al=1/2$ indeed describes a D-24-brane with its transverse
direction compactified on a circle of radius 1. The subsequent fate of
the
BCFT under the radius deformation that takes us from $(\alpha=1/2, R=1)$ 
to $(\alpha=1/2, R=\infty)$ then follows the fate of a D24-brane
under such a
deformation, {\it i.e.} the D24-brane remains a D24 brane as the radius
changes. Thus
the final BCFT at $R=\infty$ describes a D24-brane in non-compact
space-time.

This establishes that a lump solution on a D-25-brane describes a
D-24-brane. Note
that the argument goes through irrespective of the boundary condition on
the
coordinates $X^1,\ldots X^{24}$; thus the same analysis shows that a
codimension 1
lump on a D$p$-brane describes a D-$(p-1)$ brane for any value of $p\ge
1$. Repeating this
procedure $q$-times
we can also establish that a codimension $q$ lump on a D$p$-brane
describes a
D-$(p-q)$-brane.

This establishes the third conjecture of section \ref{s2.3}. This in turn
indirectly
proves conjectures 1 and 2 as well. To see how conjecture 1 follows
from
conjecture 3, we note that D24-brane, and hence the lump solution, has a
finite
energy per unit 24-volume. This
means that the lump solution must have vanishing energy density as
$x^{25}\to\pm
\infty$, since otherwise we would get infinite energy per unit 24-volume by
integrating the energy density in the $x^{25}$ direction.
Thus if $T_0$ denotes the value to which $T$ approaches as
$x^{25}\to\pm\infty$,
then the total energy density must vanish at $T=T_0$. Furthermore $T_0$
must be a
local extremum of the potential in order for the tachyon equation of
motion to be
satisfied as $x^{25}\to\pm\infty$. This shows
the existence of a local extremum of the potential where the total energy
density
vanishes, as stated in conjecture 1.

To see how the second conjecture arises, note that D24-brane and hence the 
lump solution supports
open strings
with ends moving on the $x^{25}=0$ plane. This means that if we go far 
away from
the lump
solution in the $x^{25}$ direction, then there are no physical open string
excitations in this region. Since the tachyon field configuration in this
region is
by definition the $T=T_0$ configuration, we arrive at the second
conjecture that
around $T=T_0$ there are no physical open string excitations.

Finally we note that our analysis leading to the BCFT associated with the
tachyon
lump solution is somewhat indirect. One could ask if in the diagram shown
in
Fig.\ref{f6} it is possible to go from $\al=0$ to $\al=1/2$ at any value
of $R>1$
directly, without following the circuitous route of first going down to
$R=1$ and
coming back to the desired value of $R$ {\it after switching on the 
$\alpha$-deformation}. It turn out that it is possible to do this, but 
not via marginal
deformation. For a generic value of $R$, we need to perturb the BCFT
describing
D25-brane by an operator
 \be \label{e3.4}
-\al \int dt \, \cos (X(t) / R) \, ,
 \ee
which has dimension $R^{-2}$ and hence is a relevant operator for $R>1$.
It is known
that under this relevant perturbation the original BCFT describing the
D25-brane
flows into another BCFT corresponding to putting Dirichlet boundary
condition on
$X$ coordinate \cite{9406125,0003101}. In other words, although for a
generic $\al$ the
perturbation
\refb{e3.4} breaks conformal invarinance, for a specific value of $\al$
corresponding to the infrared fixed point, \refb{e3.4} describes a new BCFT
describing the D-24-brane.\footnote{Of course this will, as usual, also
induce flow in various other coupling constants labelling the boundary
interactions. The precise description of these flows will depend on the
renormalization scheme, and we could choose a suitable renormalization
scheme where the other coupling constants do not flow. In space-time
language, this amounts to {\it integrating out} the other fields.} By the
usual correspondence between equations of motion
in string theory and two dimensional BCFT, we would then conclude that
open string
field theory on a D-25-brane compactified on a circle of radius $R>1$ has
a
classical solution describing a D24-brane. This classical solution can be
identified
as the lump.

The operator $-\cos(X/R)$ looks ill defined in the $R\to\infty$ limit, but
the
correct procedure is to expand this in powers of $X/R$ and keep the leading
term. This amounts to perturbing the D25-brane BCFT by an operator
proportional to
 \be \label{e3.5}
\int dt X(t)^2\, .
 \ee
In the infrared this perturbation takes us to the BCFT of a D24-brane,
localized at
$x=0$. This result is useful in the study of tachyon
condensation in
boundary string field
theory\cite{0009103,0009148,0010108}, and will be made use of in section
\ref{s7}.

\subsection{Superstring theory} \label{s3.3}

In this section we shall generalize the analysis of section \ref{s3.1} to
superstring theory showing that various classical solutions involving the
tachyon on D9-$\bar{\rm D}9$-brane pair of IIB or non-BPS D9-brane of IIA
represent lower dimensional D$p$-branes. For definiteness we shall
illustrate in detail the representation of a non-BPS D8-brane of IIB as a
kink solution on a D9-$\bar{\rm D}9$-brane pair\cite{9808141} (case 3(b)
in section \ref{s2.1}), and then
briefly
comment on the other cases.

The tachyon state on a D9-$\bar{\rm D}9$-brane system comes from open
strings with one leg on the D9-brane and the other leg on the $\bar{\rm
D}9$-brane. Thus the corresponding vertex operator will carry an
off-diagonal Chan-Paton factor which we can take to be the Pauli matrix
$\sigma_1$ or $\sigma_2$. This gives rise to two real tachyon fields $T_1$
and $T_2$ on the world-volume of this D-brane system, which can be
combined into a complex tachyon $T=T_1+i T_2$. Thus in this convention the
coefficients of $\sigma_1$ and $\sigma_2$ represent the real and
imaginary parts respectively of the complex
tachyon field.
We shall show that a tachyonic kink involving the real part $T_1$ of $T$,
with the imaginary part set to zero, represents a non-BPS D8-brane.

The vertex operator of the tachyon $T_1$ carrying
momentum $k$ in the $-1$ picture\cite{FMS1,FMS} is given by:\footnote{A
brief review of bosonization of superconformal ghosts, picture changing 
and  physical open string vertex operators in superstring
theory is given at the beginning of section \ref{s4.4}.}
 \be
\label{efh2} V_{-1}(k) = c \, e^{-\phi_g} \, e^{ik\cdot X} \otimes
\sigma_1\, ,
  \ee
where $\phi_g$ is the bosonic field arising out of bosonization of
the $ \beta$-$\gamma$ ghost system\cite{FMS}. As usual this vertex
operator is inserted at the boundary of the world sheet. The
on-shell condition is
 \be \label{efh3} k^2 =
{1\over 2}\, ,
  \ee
showing that the corresponding state has mass$^2=-{1\over 2}$. The
same vertex operator in the $0$ picture is
 \be \label{efh4} V_0 = -
\sqrt 2\, c \, k\cdot\psi \, e^{ik\cdot X} \otimes \sigma_1\, ,
  \ee
where $\psi^\mu$ is boundary value of the world-sheet
superpartners $\psi^\mu_L$ or $\psi^\mu_R$ of $X^\mu$. On the
boundary $\psi^\mu_L$ and $\psi^\mu_R$ are equal. The fields
$\phi_g$ and $\psi^\mu$ are normalized so that their left and
right-moving components satisfy the operator product expansion:
 \ben \label{eopephipsi}
&& \p \phi_{gR}(z) \p \phi_{gR}(w) \simeq -{1\over (z-w)^2}\, ,
\qquad \bar\p \phi_{gL}(\bar z) \bar\p \phi_{gL}(\bar w) \simeq
-{1\over (\bar
z-\bar w)^2}\, , \nonumber \\
&& \psi^\mu_R(z) \psi^\nu_R(w) \simeq {\eta^{\mu\nu}\over z-w}, \qquad 
\psi^\mu_L(\bar z)
\psi^\nu_L(\bar w) \simeq {\eta^{\mu\nu}\over \bar z-\bar w}\, .
 \een

In order to show that the kink solution involving this tachyon represents
a
non-BPS D8-brane,
we proceed exactly as in the case of bosonic string theory, \i.e. we
first find the BCFT associated with the tachyonic
kink solution on the D9-$\bar{\rm D}9$-brane pair, and then show that this
BCFT is identical to that of a non-BPS D8-brane. In order to find the
BCFT associated with the tachyonic
kink, we need to identify a series of steps which take us
from the $T=0$ configuration to the tachyonic kink configuration. This is
done as follows\cite{9808141}:
 \begin{enumerate}
\item
We first compactify the $x\equiv x^9$ direction into a circle of radius
$R$. We would like to take the radius to an appropriate critical value
(analog of $R=1$ in the bosonic case) where we can create a kink solution
via a marginal boundary deformation. However,
in order to create a single kink on a circle, we need to have a
configuration where the tachyon is anti-periodic along the circle.
This can be achieved by switching on
half a unit of Wilson line along the circle associated with one of the
branes, since the tachyon field is charged under this gauge field. This is
a boundary marginal deformation.

 \begin{figure}[!ht]
 \begin{center}
\leavevmode
\epsfysize=5cm
\epsfbox{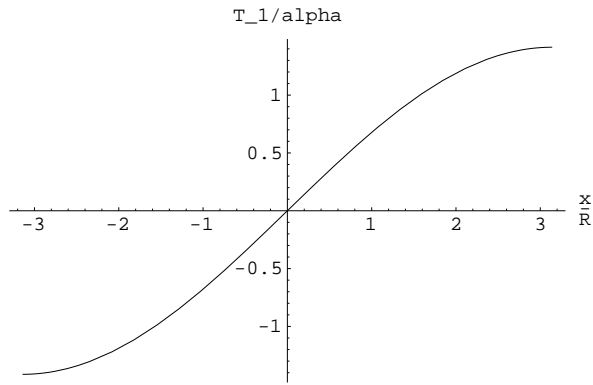}
\end{center}
\caption{The plot of $T_1(x)/\alpha$ vs. $x/R$ in the range $(-\pi,
\pi)$.
} \label{fh1}
\end{figure}

\item
In the presence of this Wilson line,
a configuration of the form:
 \be \label{efh5}
T_1(x) = \sqrt 2\,  \al\, \sin{x\over 2R}\, ,
 \ee
is an allowed configuration and
has the shape of a kink solution (see Fig.\ref{fh1}).
Here the normalization factor of $\sqrt 2$ has been chosen for
convenience.
The corresponding vertex operators in the $-1$ and the 0 pictures are
given by, respectively,
 \be \label{efh6}
V_{-1} = c e^{-\phi_g} \sin{X\over 2R} \otimes \sigma_1\, ,
 \ee
and
 \be \label{efh7}
V_{0} = {i\over \sqrt 2 R} c \, \psi \, \cos{X\over 2R} \otimes
\sigma_1\, ,
 \ee
where $X\equiv X^9$, $\psi\equiv \psi^9$.
Thus switching on a background tachyon of the form \refb{efh5}
corresponds to adding a world sheet perturbation:
 \be \label{efh8}
i\, {\al\over R} \, \int dt \, \psi \, \cos{X\over 2R} \otimes
\sigma_1\, .
 \ee
We want to find a critical value of $R$ for which \refb{efh8} represents
an exactly marginal deformation so that \refb{efh5} describes a solution 
to 
the classical open string field equations for any $\alpha$.
The perturbing operator in \refb{efh8} has dimension ${1\over 4
R^2}+{1\over 2}$.
At $R=1/\sqrt 2$ this operator has dimension 1. Furthermore, one can show
that at this radius it becomes
an exactly marginal operator\cite{9808141,0108102,0108238}. To see this,
note that at this radius
we can fermionize the space-time boson $X$ and then rebosonise the
fermions as follows:
 \be \label{efh9}
X=X_L + X_R\, ,
 \ee
 \be \label{efh10}
e^{\pm i\sqrt 2 X_L} \sim {1\over \sqrt 2} \, (\xi_L \pm i \eta_L)\, ,
\qquad
e^{\pm i\sqrt 2 X_R} \sim {1\over \sqrt 2} \, (\xi_R \pm i \eta_R)\, ,
 \ee
 \be \label{efh11}
e^{\pm i\sqrt 2 \phi_{L\atop R}}\sim {1\over \sqrt 2} \, (\xi_{L\atop R} 
\pm 
i
\psi_{L\atop R}),
\qquad
e^{\pm i\sqrt 2 \phi'_{L\atop R}}\sim {1\over \sqrt 2} \, (\eta_{L\atop R} 
\pm i
\psi_{L\atop R})\,
,
 \ee
 \ben \label{efh11.5}
&& \psi_R \xi_R = \sqrt 2 \p\phi_R, \quad \eta_R \xi_R = \sqrt 2
\p X_R,
\quad \psi_R \eta_R = \sqrt 2 \p \phi'_R, \nonumber \\
&& \psi_L \xi_L = \sqrt 2 \bar\p\phi_L, \quad \eta_L \xi_L = \sqrt
2 \bar\p X_L, \quad \psi_L \eta_L = \sqrt 2 \bar \p \phi'_L,
 \een
where $\xi_L$, $\eta_L$ are left-chiral Majorana fermions,
$\xi_R$, $\eta_R$ are right-chiral Majorana fermions,
$\phi=\phi_L+\phi_R$, $\phi'=\phi'_L+\phi'_R$ are free bosons and
$\sim$ in \refb{efh10}, \refb{efh11} denotes equality up to
cocycle factors.\footnote{For a discussion of the cocycle
factors in this case, see ref.\cite{0003124}.} $\xi$, $\eta$, $\phi$, 
$\phi'$ are
normalized so that
 \ben \label{exinorm}
&& \xi_R(z) \xi_R(w) \simeq {1\over z-w} \simeq \eta_R(z)\eta_R(w),
\qquad \xi_L(\bar z) \xi_L(\bar w) \simeq {1\over \bar z-\bar w}
\simeq \eta_L(\bar z)\eta_L(\bar w)\, , \nonumber \\
&& \p\phi_R(z) \p\phi_R(w) \simeq - {1\over 2(z-w)^2}, \qquad
\bar\p\phi_L(\bar z) \bar\p\phi_L(\bar w) \simeq - {1\over 2(\bar 
z-\bar w)^2}, \nonumber \\
&& \p\phi'_R(z) \p\phi'_R(w) \simeq - {1\over 2(z-w)^2}, \qquad
\bar\p\phi'_L(\bar z) \bar\p\phi'_L(\bar w) \simeq - {1\over 2(\bar
z-\bar w)^2} \, .
 \een
Eq.\refb{efh10} defines $\xi_{L,R}$ and $\eta_{L,R}$ in terms of
$X_{L,R}$, and eq.\refb{efh11} defines $\phi_{L,R}$ and
$\phi'_{L,R}$ in terms of $\xi_{L,R}$, $\eta_{L,R}$ and
$\psi_{L,R}$. \refb{efh11.5} follows from \refb{efh10},
\refb{efh11} and the usual rules for bosonization. The boundary
conditions on various fields, following from Neumann boundary
condition on $X$ and $\psi$, and eqs.\refb{efh10} -
\refb{efh11.5}, are:
 \be \label{efh12}
X_L=X_R, \quad \psi_L=\psi_R, \quad \xi_L=\xi_R, \quad \eta_L=\eta_R,
\quad \phi_L=\phi_R, \quad \phi'_L=\phi'_R\, .
 \ee
Using eqs.\refb{efh9}-\refb{efh12} the boundary perturbation
\refb{efh8} takes the form:
 \be \label{efh13}
i\, \sqrt 2 \, {\al } \,  \int \, dt \, \bar\p\phi_L(t) \otimes
\sigma_1\, .
 \ee
This corresponds to switching on a Wilson line along the $\phi$ direction 
and is clearly an exactly marginal deformation. Thus for any
value of the coefficient $\al$ in \refb{efh13} (and hence in \refb{efh8})
we get a boundary
CFT.

\item In order to create a kink solution in the non-compact
theory. we need to take the radius $R$ back to infinity after
switching on the deformation \refb{efh8}. However here we
encounter an obstruction; for generic $\alpha$ the boundary vertex 
operator \refb{efh8}
develops a one point function on the upper half
plane\cite{9808141} for $R\ne 1/\sqrt 2$:
 \be \label{efh14}
\la\psi\cos{X\over 2R}(0) \otimes \sigma_1\ra_{UHP} \propto (R-1)
\sin(2\pi\al)\, .
 \ee
This is a reflection of the fact that for $R\ne 1/\sqrt 2$ the operator
$\psi\cos(X/2R)$ is no longer marginal. This shows that in order
to take the radius back to $\infty$, $\al$ must be fixed at 0 or
$1/2$. $\alpha=0$ gives us back the original D-brane system,
whereas $\alpha=1/2$ gives the kink solution.
\end{enumerate}

To summarize, in order to construct the tachyon kink solution on a 
D9-$\bd9$ brane pair we first compactify the $x$ direction on a circle of 
radius $1/\sqrt 2$ and swich on half unit of Wilson line along $x$ on one 
of the branes, then switch on the deformation \refb{efh8} and take 
$\alpha$ to 1/2, and finally take $R$ back to infinity. This gives the 
construction of the tachyonic kink configuration 
on a D9-$\bar{\rm D}9$-brane pair as a BCFT. It now remains to show
that this BCFT actually describes a non-BPS D8-brane, {\it i.e.}
the effect of this deformation is to change the Neumann boundary
condition on $X$ and $\psi_L$, $\psi_R$ to Dirichlet boundary
condition. This is done by noting that \refb{efh13} is the contour
integral of an anti-holomorphic current, and hence the effect of this on
a correlation function can be studied by deforming the contour and
picking up residues from various operators. Using eqs.\refb{efh9} - 
\refb{efh11} it is easy to see that for
$\alpha=1/2$, the exponential of \refb{efh13} transforms the left-moving 
fermions $\xi_L$,
$\eta_L$ and $\psi_L$ as:
 \be \label{efh16}
\xi_L\to-\xi_L, \qquad \eta_L\to \eta_L, \qquad \psi_L\to -\psi_L\, .
 \ee
In terms of the original fields, this induces a transformation
$\bar \p X_L\to   - \bar\p X_L$, $\psi_L\to -\psi_L$ on every vertex 
operator inserted
in the interior of the world-sheet. The right-moving world-sheet
fields are not affected. By a redefinition $X_L\to -X_L$, $\psi_L\to 
-\psi_L$, we can leave the vertex
operators unchanged, but this changes the boundary condition on $X$ and
$\psi$ from Neumann to Dirichlet:
 \be \label{efh17}
\bar\p X_L=-\p X_R, \qquad \psi_L=-\psi_R\, .
 \ee
This clearly shows that switching on the deformation \refb{efh13} with
$\alpha={1\over 2}$
corresponds to creating a D8-brane transverse to the circle of radius 
$R=1/\sqrt 2$. Taking the
radius back to infinity leaves the D8-brane unchanged.

The analysis can be generalized in many different ways. First of all,
since the
boundary condition on $X^0,\ldots X^8$ played no role in the analysis, we
can choose them to be anything that we like. This establishes in general
that a kink solution on a D$p$-$\bar{\rm D}p$-brane pair represents a
non-BPS D-$(p-1)$-brane.
The analysis showing that a kink solution on a non-BPS D-$p$-brane
represents a BPS D-$(p-1)$-brane is essentially identical. We begin with
the kink solution on the BPS D$p$-$\bd p$-brane pair in type IIB/IIA 
theory
for $p$ odd/even and mod
this out by $(-1)^{F_L}$. This converts type IIB/IIA theory to type
IIA/IIB
theory, the
D$p$-$\bd p$-brane pair to a non-BPS
D-$p$-brane, and the non-BPS D-$(p-1)$-brane to a BPS
D-$(p-1)$-brane.\footnote{One can show\cite{9810188,9812031} that the 
tachyon state as well as all other open string states on a non-BPS
D-brane carrying Chan-Paton factor $\sigma_1$ are odd under $(-1)^{F_L}$. 
Thus these modes are projected out after
modding out the theory by $(-1)^{F_L}$, and we get a tachyon free BPS
D-brane. \label{fn13}} This shows that the kink solution on a non-BPS 
D-$p$-brane
can be identified as a BPS D-$(p-1)$-brane.
Finally the analysis showing that a
codimension $k$
soliton on a D$p$-$\bar{\rm D}p$-brane pair or a non-BPS D$p$ brane
produces a D-$(p-k)$-brane can be done by compactifying $k$ of the
coordinates tangential to the original D-brane on a torus $T^k$ of
appropriate radii, switching on a marginal deformation that creates the
codimension $k$ soliton, and finally proving that the BCFT obtained at the
end of this marginal deformation is a D-$(p-k)$-brane\cite{0003124}.

\subsection{Analysis of the boundary state} \label{s3.4}

Given a boundary CFT describing a D-brane system, we can associate
with it a boundary state
$|\BB\ra$\cite{callan1,callan2,9604091,9707068,9912275}. This is a {\it
closed string state} of ghost number 3, and has the following
property. Given any closed string state $|V\ra$ and the associated
vertex operator $V$, the BPZ inner product $\la\BB|V\ra$ is given
by the one point function of $V$ inserted at the centre of a unit
disk $D$, the boundary condition on $\p D$ being the one
associated with the particular boundary CFT under consideration:
 \be \label{efg1}
\la\BB|V\ra \propto \la V(0)\ra_D\, .
 \ee
Note that in order to find the contribution to the boundary state at
oscillator level $(N,N)$, we only need to compute the inner product of the
boundary state with closed string states of level $(N,N)$. This in turn
requires computation of one point function on the disk of closed string
vertex operators of level $(N,N)$.

{}From the definition \refb{efg1} it is clear that the boundary state
contains
information about what kind of source for the closed string states is
produced by the D-brane system under consideration. This has been made
more
precise in appendix \ref{appb}. There it has been shown that if the
boundary state $|\BB\ra$ associated with a D-brane in bosonic string 
theory has an expansion of the form:
 \be \label{efg5.5}
|\BB\ra =
\int {d^{26} k\over (2\pi)^{26}} [\wt F(k) +  (\wt A_{\mu\nu}(k)+ \wt
C_{\mu\nu}(k))
\alpha^\mu_{-1}
\bar\alpha^\nu_{-1}
+ \wt B(k) (b_{-1} \bar c_{-1}  + \bar b_{-1} c_{-1}) + \ldots ] (c_0
+\bar
c_0)
c_1
\bar c_1 |k\ra\, ,
 \ee
where $\wt F$, $\wt A_{\mu\nu}=\wt A_{\nu\mu}$, $\wt
C_{\mu\nu}=-\wt C_{\nu\mu}$, $\wt B$ etc. are fixed functions,
$\alpha^\mu_{-n}$, $\bar\alpha^\mu_{-n}$ are oscillators of
$X^\mu$, and $b_{-n}$, $c_{-n}$, $\bar b_{-n}$, $\bar c_{-n}$ are
ghost oscillators, then the energy momentum tensor $T_{\mu\nu}(x)$, 
defined as the source for the graviton field, is given by
 \be \label{efg9a}
T_{\mu\nu}(x) \propto (A_{\mu\nu}(x) + \eta_{\mu\nu} B(x) )\, .
 \ee
We shall choose the normalization of $|\BB\ra$ in such a way that the
above equation takes the form:
 \be \label{efg9}
T_{\mu\nu}(x) = {1\over 2} \, (A_{\mu\nu}(x) + \eta_{\mu\nu} B(x) )\, .
 \ee

Let us apply this to the specific D-brane system studied in
section \ref{s3.1}, namely D-$p$-brane wrapped on a circle of
radius 1. We take the directions transverse to the brane to be
$x^1,\ldots x^{25-p}$, and the spatial directions along the brane
to be $x^{26-p},\ldots x^{25}$. The boundary state of the initial
D-$p$-brane without any perturbation is given
by\cite{callan1,callan2,9604091,9707068,9912275}:
 \be \label{efg10}
|\BB\ra = \TT_p \,  |\BB\ra_{c=1} \otimes |\BB\ra_{c=25} \otimes
|\BB\ra_{ghost}\, ,
 \ee
where $\TT_p$ is the tension of the D$p$-brane as given in \refb{e2.16},
$|\BB\ra_{c=1}$ denotes the boundary state associated with the
$X^{25}\equiv X$ direction, $|\BB\ra_{c=25}$ denotes the boundary state
associated with the other 25 directions $X^0,\ldots X^{24}$, and
$|\BB\ra_{ghost}$ denotes the
boundary state associated with the ghost direction. We have:
 \be \label{efg11}
|\BB\ra_{c=25} =
\int {d^{25-p} k_\perp\over (2\pi)^{25-p}} \,
\exp\left(\sum_{\mu,\nu=0}^{25}\sum_{n=1}^\infty
{1\over
n} \eta_{\mu\nu} \, (-1)^{d_\mu} \, \alpha^\mu_{-n} \bar
\alpha^\nu_{-n} \right)
|k_\parallel=0,
k_\perp\ra\, ,
 \ee
 \be \label{efg12}
|\BB\ra_{ghost} = \exp\left(-\sum_{n=1}^\infty (\bar b_{-n} c_{-n} +
b_{-n}
\bar c_{-n})
\right) (c_0+\bar c_0)c_1\bar c_1 |0\ra\, ,
 \ee
and,
 \be \label{efg13}
|\BB\ra_{c=1} = \sum_{m=-\infty}^\infty \exp\left(-\sum_{n=1}^\infty
{1\over
n} \alpha_{-n} \bar
\alpha_{-n} \right) |k=0, w=m\ra\, ,
 \ee
where $d_\mu=1$ for Neumann directions and 0 for Dirichlet directions,
$k_\parallel$ denotes momentum
along the D-$p$-brane in directions other than $x^{25}$, $k_\perp$ denotes
momentum transverse to the D-$p$-brane,  $\alpha_n$, $\bar \alpha_n$
without any superscript denote the
$X^{25}\equiv X$ oscillators, and $k$ and $w$ denote the momentum and
winding number respectively
along the
circle along $x^{25}$. Let us denote by $x^M$ the coordinates other
than $x^{25}\equiv x$ along the D-$p$-brane world-volume, and by $x^m$
the
coordinates transverse to the D-$p$-brane world-volume. We shall choose
$M$ to run over the values 0 and
$(26-p),\ldots, 24$ and $m$ to run over the values $1,\ldots, (25-p)$.
Expanding the boundary state in powers of the various oscillators, and
comparing this expansion with \refb{efg5.5}, we get the following
non-zero components of $A_{\mu\nu}$ and $B$:
 \be \label{efg14}
A_{xx} = - \TT_p \, \delta(x_\perp) \, , \quad A_{MN} = -
\TT_p \, \eta_{MN} \, \delta(x_\perp) \, , \quad A_{mn}
= \TT_p \,
\delta_{mn} \, \delta(x_\perp) \, ,
\quad B = - \TT_p \, \delta(x_\perp)\, .
 \ee
and hence
 \be \label{efg15}
T_{xx} = - \TT_p \, \delta(x_\perp) \, , \qquad T_{MN} = - \TT_p \,
\eta_{MN} \, \delta(x_\perp)\, , \qquad
T_{mn} = 0.
 \ee
This is the energy-momentum tensor associated with a D-25-brane.

We shall now study the change in the boundary state under the deformation
of the boundary CFT by the marginal operator 
\refb{e3.2}\cite{9402113,9811237}. Using the
boundary condition \refb{eb1} we can rewrite \refb{e3.2} as
 \be \label{egf16}
- \al \int dt \cos (2X_L(t)) \, dt = - \al \int dt J^1_L(t)\, ,
 \ee
Adding such a perturbation at the boundary effectively rotates the
left-moving world-sheet component of the boundary state by an angle
$2\pi\al$ about the 1-axis. In particular for $\al={1\over 2}$, the 
effect of
this perturbation  is a rotation by $\pi$ about the 1-axis which changes
$J^3_L$ to $-J^3_L$, {\it i.e.} $\alpha_n$ to $-\alpha_n$ in the exponent
of \refb{efg13}, and also
converts the winding number $w=m$ to momentum $k=m$ along $x^{25}$.
Thus $|\BB\ra_{c=1}$ is transformed to:
 \be \label{etras}
\sum_{m=-\infty}^\infty \exp\left( \sum_{n=1}^\infty {1\over n} \al_{-n}
\bar\al_{-n}\right) |k=m, w=0\ra\, .
 \ee
The other components of $|\BB\ra$ given in \refb{efg11}, \refb{efg12}
remain
unchanged.
The
result
is precisely the boundary state associated with a D-$(p-1)$-brane, and the
corresponding $T_{\mu\nu}$ computed using \refb{efg5.5}, \refb{efg9}
precisely reproduces the energy-momentum tensor of a D-$(p-1)$-brane
situated
at $x=0$. We shall shortly derive this as a special case of a more general
result.

For a general $\al$ the effect of this rotation on the boundary state is
somewhat complicated but can nevertheless be done by expressing the
boundary state in a suitable basis\cite{9811237}. This analysis
gives\cite{9402113,9811237,0203211}
 \ben \label{efg16.5}
|\BB\ra_{c=1} &=& |k=0\ra + \sum_{n=1}^\infty \sin^n(\pi\al) \Big(
|k=n\ra + |k=-n\ra\Big) - \cos(2\pi\al) \alpha_{-1}
\bar\alpha_{-1} |k=0\ra
\nonumber \\
&& + \sum_{n=1}^\infty \sin^n(\pi\al) \alpha_{-1}
\bar\alpha_{-1}\Big( |k=n\ra + |k=-n\ra\Big) + \ldots \, ,
 \een
where $\ldots$ denote terms with oscillator level higher than
(1,1) and terms with winding modes. A more detailed discussion of
the higher level zero winding number terms will be given in
section \ref{s10} (see eq.\refb{ebs1} and discussion below this equation). 
Combining \refb{efg16.5} with
\refb{efg11}, \refb{efg12}, and using eqs.\refb{efg5.5},
\refb{efg10}, we get
 \ben \label{efg18}
&& \wt B(k) = (2\pi)^{p+1} \, \TT_p \, \left[-\delta(k) -
\sum_{n=1}^\infty \Big(
\delta(k-n) +
\delta(k+n)
\Big)  sin^n(\al\pi)\right] \, \delta(k_\parallel) \, , \nonumber \\
&& \wt A_{xx}(k) = \left[-(2\pi)^{p+1} \, \TT_p \, \Big(1 +
\cos(2\pi\al)\Big) \delta(k) \, \delta(k_\parallel) -
\wt B(k)
\right] \, ,
\nonumber \\
&& \wt A_{xM} = 0, \qquad \wt A_{MN} = \wt B(k) \, \eta_{MN}\, ,
\quad \wt A_{xm} = 0, \quad \wt A_{mn} =-\wt B(k)\, \delta_{mn},
\quad \wt A_{mM}=0 \, . \nonumber \\
 \een
The Fourier transform of these equations give:
 \ben \label{efg20}
&& B = -\TT_p\, f(x)\, \delta(x_\perp)\, , \quad A_{xx} = -\TT_p\,
g(x) \, \delta(x_\perp), \quad A_{MN} =
- \TT_p\, f(x) \, \eta_{MN}\, \delta(x_\perp), \nonumber \\
&&  A_{xM} = 0, \quad A_{xm} = 0, \quad A_{mn} = \TT_p\, f(x) \,
\delta_{mn}\, \delta(x_\perp)\,  , \quad A_{mM} = 0,
 \een
where
 \ben \label{efg21}
f(x) &=& 1 + \sum_{n=1}^\infty \, \sin^n(\al\pi) \, (e^{inx} +
e^{-inx})
= {1\over 1 - e^{ix} \sin(\al\pi)}
+ {1\over 1 - e^{-ix} \sin(\al\pi)} - 1\, , \nonumber \\ g(x) &=& \Big(1 +
\cos(2\pi\al)\Big) - f(x)\, .
 \een
In arriving at the right hand side of eqs.\refb{efg21} we have performed
the sum over $n$, using the fact that it is a convergent sum for
$|\sin(\pi\al)| < 1$. Using eq.\refb{efg9} the energy momentum
tensor
$T_{\mu\nu}$ is now given by:
 \ben \label{efg22}
&& T_{xx} = -  \TT_{p} \cos^2(\pi\al) \, \delta(x_\perp), \quad
T_{MN} = -\TT_{p}
\,    f(x) \,
\eta_{MN} \,\delta(x_\perp) \nonumber \\
&& T_{xM} = 0, \quad T_{xm} = 0, \quad T_{mM}=0, \quad T_{mn}=0\,
\nonumber \\
&& \qquad \qquad \hbox{for} \quad
M,N =0, \, (26-p), \ldots
24, \quad m,n=1,\ldots (25-p)\, .
 \een
This gives the energy momentum tensor associated with the boundary CFT of
section \ref{s3.1} for arbitrary value of $\al$. Note that $T_{xx}$ is $x$ 
independent. This is a consequence of the conservation law $\p_x T_{xx} + 
\eta^{MN}\p_M T_{Nx}=0$.

Using \refb{edchargeapp}
we can also see that the function
$\wt B(k)$ measures the source of the dilaton field $\wt\phi(k)$. This
suggests that we define the dilaton charge density to be
 \be \label{edcharge}
Q(x) = -  B(x) = \TT_{p} \, f(x)\, \delta(x_\perp)\, .
 \ee
The overall normalization of $Q$ is a matter of convention.

An interesting limit is the $\al\to {1\over 2}$ limit. For $x\ne
2n\pi$ with integer $n$, both $f(x)$ and $g(x)$ can be seen to
vanish in this limit. On the other hand for any $\al< {1\over 2}$,
we can compute $\int_{-\pi}^{\pi} f(x) dx$ by a contour integral,
and the answer turns out to be $2\pi$. Thus we would conclude that
as $\al\to {1\over 2}$, $f(x)$ approaches a delta function
concentrated at $x=0$ (and hence also at $2n\pi$). Hence in this
limit,
 \be \label{efg23}
T_{xx} = T_{xM} = T_{xm}=T_{mn}=T_{mM}= 0, \quad T_{MN} = -2\,
\pi\, \TT_{p}\, \delta(x_\perp) \, \, \eta_{MN}
\sum_{n=-\infty}^\infty \delta(x-2n\pi)  \, .
 \ee
This is precisely the energy momentum tensor of a D-$(p-1)$-brane
situated on a circle at $x=0$, since $2\pi \, \TT_p$ is the
D-$(p-1)$-brane tension $\TT_{p-1}$.

Following the same logic as the one given for the $\alpha\to{1\over
2}$ limit, we can see
that for $\al\to -{1\over 2}$ we again get a
D-$(p-1)$-brane situated on a circle, but this time at $x=\pi$ instead of 
at
$x=0$. Thus \refb{efg23} is now replaced by:
 \be \label{ealphamh}
T_{xx} = T_{xM} = T_{xm}=T_{mn}=T_{mM}= 0, \quad T_{MN} = -2\,
\pi\, \TT_{p}\, \delta(x_\perp)  \, \eta_{MN} \,
\sum_{n=-\infty}^\infty \delta\left(x-(2n+1)\pi\right) \, .
 \ee

A similar analysis can be carried out for the superstring theory as 
well\cite{9903123}.
Instead of going through the details of the analysis, we quote here the
final answer\cite{0203265}. For the deformed boundary CFT described in
section
\ref{s3.3}, the energy-momentum tensor is given by:
 \ben \label{efg24}
&& T_{xx} = -  \EE_p\, 
\cos^2(\pi\al) \, \delta(x_\perp), \quad T_{MN} = -\EE_{p}
\,    f(x) \, \delta(x_\perp) \, \eta_{MN}, \nonumber \\ &&
T_{xM} =
T_{xm} =
T_{mM} =
T_{mn} =0,
 \een
where $x^m$ denote directions transverse to the D-brane, $x^M$
denote directions (other than $x$) tangential to the D-brane, $\EE_p$
denotes the tension of the original brane
system ($\wt \TT_p$ for the non-BPS D$p$-brane and $2\TT_p$ for the
brane-antibrane system), and
 \be \label{efg25}
f(x) = {1\over 1 - e^{i\sqrt 2 x} \sin^2(\al\pi)}
+ {1\over 1 - e^{-i\sqrt 2 x} \sin^2(\al\pi)} - 1\, .
 \ee
The dilaton charge density is given by:
 \be \label{esupdil}
Q(x) = \EE_p \, f(x) \, \delta(x_\perp)\, .
 \ee
As $\al\to 0$, we get back the energy-momentum tensor of the original
D9-brane system. On the other hand, as $\al\to {1\over 2}$, $f(x)$
approaches sum of delta functions concentrated at $x=2\pi n/ \sqrt 2$ for 
integer $n$. Since the $x$ coordinate is compactified on a circle of 
radius $1/\sqrt 2$, the
resulting $T_{\mu\nu}$ reduces to that of a D-$(p-1)$-brane situated at
$x=0$.

Finally, a kink solution on a non-BPS D-$p$-brane (but not on a
brane-antibrane pair) also produces a source for the Ramond-Ramond
$p$-form gauge field, given by\cite{0204143}:
 \be \label{erreuclid}
Q^{(p)}_{M_1\ldots M_p} \propto \epsilon_{M_1\ldots M_p} \,
\sin(\alpha\pi) \, \left[ {e^{ix/\sqrt{2}} \over 1 -
\sin^2(\alpha\pi) e^{\sqrt{2} ix}} + {e^{-ix/\sqrt{2}} \over 1 -
\sin^2(\alpha\pi) e^{-\sqrt{2} ix}} \right]\, \delta(x_\perp)\, .
 \ee
This result can be derived from the Ramond-Ramond component of the
boundary state. In the $\alpha\to {1\over 2}$ limit \refb{erreuclid} 
correctly 
reproduces the RR charge of the D-$(p-1)$-brane located at $x=2\pi n/\sqrt 
2$.

Discussion on various other aspects of conformal field theory methods 
reviewed in this section can be found in 
refs.\cite{9902160,9903123,9903139,0001066,
0002061,0003110,
0005114,
0006122,0009090,
0009252,0101211,0102192,
0401003}.

\sectiono{Open String Field Theory} \label{s4}

Although the results on tachyon dynamics on a D-brane were stated in
section \ref{s2} in terms of the effective action obtained by formally
integrating out the heavy fields, in general it is difficult to do this in
practice. Conformal field theory methods described in the last section
provide an indirect way of constructing solutions of the classical
equations of motion without knowing the effective action. But if we want a
more direct construction of the classical solutions, we need to
explicitly take into account the coupling of the
tachyon to infinite number of other fields associated with massive open
string states.
The formalism
that allows us to tackle this problem head on  is string field theory, --
a field theory with
infinite number of fields. 
This will be the topic of discussion of the present section.
We begin by reviewing the formulation of first
quantized open bosonic string theory, followed by a review of bosonic open
string field theory. We then show that this string field theory can be
used to test the various conjectures about the tachyon effective field
theory. At the end we briefly discuss the case of superstring field
theory. An excellent and much more detailed review of string field theory 
with
application to the
problem of tachyon condensation can be found in \cite{0311017}.

\subsection{First quantized open bosonic string theory} \label{s4.0}

For a given space-time background which is a solution of the classical
equations of motion in string theory, we have a two dimensional CFT. This
CFT is a direct sum of two CFT's, the matter CFT of central charge 26, and
the ghost CFT of central charge $-26$. The matter CFT depends on the
choice of the space-time background, but the ghost CFT is universal and is
described by anti-commuting fields $b$, $\bar b$, $c$, $\bar c$ of
conformal dimensions (2,0), (0,2), (1,0) and (0,1) respectively.
Physically this conformal field theory describes the propagation of
closed string in this space-time background.  A
D-brane in this space-time background is in one to one correspondence to a
two dimensional conformal field theory on the upper half plane (or unit
disk) with specific conformally invariant boundary condition on the real
axis (unit circle).  This conformal field theory describes the
propagation of an open string living on the D-brane in this
space-time background. The boundary conditions on the matter fields
depend on
the specific D-brane that we are considering, but those on the ghost
fields
are universal, and take the form:
 \be \label{e4.1}
b=\bar b, \qquad c =
\bar c\, ,
 \ee
on the real axis. This gives rise to the mode expansion:
 \be \label{e4.2}
b = \sum_n b_n z^{-n-2}, \qquad c = \sum_n c_n
z^{-n+1}, \qquad \bar b = \sum_n b_n \bar z^{-n-2}, \qquad \bar c =
\sum_n c_n \bar z^{-n+1}\, ,
 \ee
where $z$ denotes the complex coordinate labelling the upper
half-plane. We shall denote by $t$ the coordinate
labelling the real axis.
The SL(2,R) invariant vacuum $|0\ra$ of
the open string state space satisfies:
 \be \label{e4.3}
b_n|0\ra =0
\quad \hbox{for} \quad n\ge -1, \qquad c_n|0\ra =0 \quad \hbox{for} \quad
n\ge 2\, .
 \ee

Let us denote by $\HH$ the vector space of states in the combined
matter-ghost BCFT,
obtained by acting on $|0\ra$ the ghost oscillators $b_{-n}$ ($n\ge 2$),
$c_{-n}$
($n\ge -1$) and the matter vertex operators. By the usual state-operator
correspondence
in BCFT, for every state $|\phi\ra$ in $\HH$, there is a unique local
{\it boundary}
vertex operator $\phi(t)$ such that
 \be \label{e4.4}
\phi(0)|0\ra = |\phi\ra\, .
 \ee
The states in $\HH$ can be classified by their ghost numbers, defined
through the
following rules:
 \begin{enumerate}
\item $b$, $\bar b$ have ghost number $-1$.
\item $c$, $\bar c$ have ghost number 1.
\item All matter operators have ghost number 0.
\item The SL(2,R) invariant vacuum $|0\ra$ has ghost number 0.
\end{enumerate}
We define by $\HH_n$ the subspace of $\HH$ containing states of ghost number
$n$.

Given a pair of states $|A\ra,|B\ra\in \HH$, and
the associated vertex operators $A(t)$ and $B(t)$, we define the BPZ
inner product
between the states as:
 \be \label{e4.12}
\la A | B\ra = \la I\circ A(0) B(0)\ra_{UHP}\, ,
 \ee
where $\la \cdot\ra_{UHP}$ denotes the correlation function of the BCFT
on the upper
half plane, $f\circ A(t)$ for any function $f(z)$ denotes the
conformal
transform of
$A(t)$ under the map $f$, and
 \be \label{e4.13}
I(z) = - 1/z\, .
 \ee
Thus for example if $A(t)$ is a primary operator of weight $h$ then 
$f\circ A(t) = \left( f'(t)\right)^h A(f(t))$.
It is a well known property of the correlation function of the matter
ghost
BCFT on
the upper half plane that the correlator is non-zero only if the total
ghost number
of all the operators add up to three. Thus $\la A|B\ra$ is non-zero only
if the
ghost
numbers of $|A\ra$ and $|B\ra$ add up to three.

 \begin{figure}[!ht]
 \begin{center}
\leavevmode
\epsfysize=5cm
\epsfbox{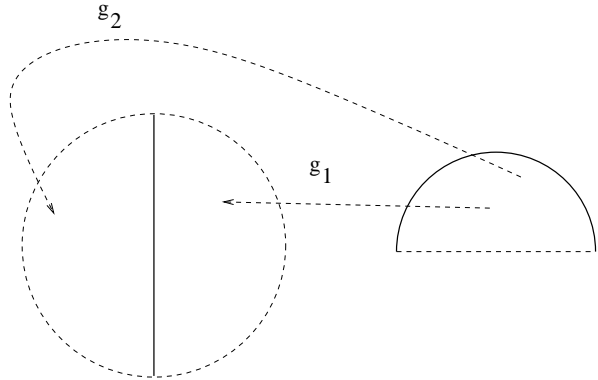}
\end{center}
 \caption{Images of the upper half unit disk under the maps $g_1$ and $g_2$.
The solid line gets mapped to the solid line and the dashed line
gets mapped to the dashed line.} \label{f7}
 \end{figure}
By a conformal transformation $w=(1+iz) / (1-iz)$ that takes the upper
half plane to the unit
disk, we
can reexpress \refb{e4.12} as
 \be \label{e4.14}
\la A | B\ra = \la g_2\circ A(0) g_1\circ B(0)\ra_{D}\, ,
 \ee
where $\la \cdot \ra_D$ denotes correlation function on a unit disk, and,
 \be \label{e4.15}
g_1(z) = {1+iz \over 1-iz}, \qquad g_2(z) = - g_1(z)\, .
 \ee
An intuitive understanding of the maps $g_1$ and $g_2$ may be obtained by
looking at
the image of the upper half unit disk under these maps. This has been
shown in
Fig.\ref{f7}.

Physical open string states on the D-brane are states in
$\HH_1$
satisfying the following criteria:
 \begin{enumerate}
\item The state $|\phi\ra$ must be BRST invariant:
 \be \label{e4.5}
Q_B|\phi\ra = 0\, ,
 \ee
where
 \be \label{e4.6}
Q_B = {1\over 2\pi i} \left[ \ointop c(z) T_m(z) dz + \ointop b(z) c\p
c(z) dz\right]\, ,
 \ee
is the BRST charge carrying ghost number 1 and  satisfying
 \be \label{e4.7}
(Q_B)^2=0\, .
 \ee
Here $T_m(z)=\sum_n L^{(m)}_n z^{-n-2}$ stand for the $zz$ component of
the world-sheet stress
tensor of the matter part of
the BCFT
describing the D-brane. $\ointop$ denotes a contour around the origin.
\item Two states $|\phi\ra$ and $|\phi'\ra$ are considered equivalent
if they differ by a state of the form $Q_B|\Lambda\ra$:
 \be \label{e4.8}
|\phi\ra \equiv |\phi\ra + Q_B|\Lambda\ra\, ,
 \ee
for any state $|\Lambda\ra\in \HH_0$.
\end{enumerate}
Thus physical states are in one to one correspondence with the elements of
BRST
cohomology in $\HH_1$.

In the first quntized formulation there is also a well defined
prescription, known
as Polyakov prescription, for computing tree and loop amplitudes
involving physical
open string states as external lines.

\subsection{Formulation of open bosonic string field theory} \label{s4.1}

The open string field theory describing the dynamics of a D-brane is, by
definition,
a field theory satisfying the following two criteria:
 \begin{enumerate}
\item Gauge inequivalent solutions of the linearized equations of
motion are in
one to one
correspondence with the physical states of the open string.
\item The S-matrix computed using the Feynman rules of the string field
theory
reproduces the S-matrix computed using Polyalov prescription to all
orders in the
perturbation theory.
\end{enumerate}
If we are interested in studying only the classical properties of string
field
theory, we can relax the second constraint a bit by requiring that the
S-matrix
elements agree only at the tree level. Nevertheless for the open bosonic
string
field theory that we shall be describing \cite{osft}, the
agreement has
been
verified to
all
orders in perturbation theory \cite{GIDDMARW}.

The first step in the construction of string field theory will be to
decide what
corresponds to a general off-shell string field configuration. Usually
when one goes
from the first to second quantized formulation, the wave-functions /
states of the
first quantized theory become the field configurations of the second
quantized
theory. However, a generic off-shell field configuration in the second
quantized theory
does not satisfy the physical state condition.\footnote{For example
second quantization of a non-relativistic Schrodinger problem
describing a particle of mass $m$ moving under a potential $V$ in
three
dimensions is described by the action $\int \, dt \, \int \, d^3 x \,
\Psi^*\left( i {\p\Psi\over \p t} +{\hbar^2\over 2m} \vec\nabla^2\Psi
- V\Psi\right)$. A general off-shell field configuration $\Psi(\vec
x,t)$ does not satisfy the Schrodinger equation. Rather, Schrodinger
equation appears as the classical equation of motion for $\Psi$
derived from this action.}  This condition comes as the linearized
equation
of motion of the field theory. In the same spirit we should expect that
the a generic off-shell string field configuration should correspond to a
state in
the BCFT without the restriction of BRST invariance, and the physical state
condition, {\it i.e.} the BRST invariance of the state,
should emerge as the linearized equation of motion of the string field
theory.
This however still does not uniquely fix the space of string field
configurations,
since we can, for example take this to be the whole of $\HH$, or $\HH_1$,
or even a
subspace of $\HH_1$ that contains at least one representative from
each BRST cohomology class. It turn out that the simplest form of open
string field theory
is obtained by taking a general off-shell string field configuration to
be a state
in $\HH_1$, {\it i.e.} a state $|\Phi\ra$ in $\HH$ of ghost number 1
\cite{osft}.

Since we are attempting to construct a string {\it field theory}, one
might wonder
in what sense a state $|\Phi\ra$ in $\HH_1$ describes a field
configuration. To see this
we need to
choose a basis of states $|\chi_{1,\alpha}\ra$ in $\HH_1$.
Then we can expand $|\Phi\ra$ as
 \be \label{e4.9}
|\Phi\ra = \sum_\alpha \phi_\alpha |\chi_{1,\alpha}\ra\, .
 \ee
Specifying $|\Phi\ra$ is equivalent to specifying the coefficients
$\phi_\alpha$. Thus the set of numbers $\{\phi_\alpha\}$ labels a given
string field
configuration.
Let us for example consider
the case of a
D$p$-brane in (25+1) dimensional Minkowski space.
We define Fock
vacuum states
$|k\ra$ labelled by $(p+1)$ dimensional momentum $k$ along the D$p$-brane
as:
 \be \label{e4.40}
|k\ra = e^{ik.X(0)}|0\ra\, .
 \ee
A generic state carrying momentum $k$ will be created by a set of
oscillators acting on the state
\refb{e4.40}.
Thus in this case the index
$\alpha$ in \refb{e4.9}
includes the $p+1$ dimensional continuous momentum index $\{k^\mu\}$
along the
D$p$-brane, and a discrete
index $r$ which originate from various oscillators, and runs over
infinite number of
values. Hence we can write:
 \be \label{e4.10}
\{ \phi_\alpha \} \to \{ \phi_{\{k^\mu\},r} \} \equiv \{ \phi_r(k^0,
\ldots
k^p)\}\, , \qquad \{|\chi_{1,\alpha}\ra\} \to \{|\chi_{1,r}(k)\ra\}, 
\qquad \sum_\alpha\to \sum_r \, \int\, {d^{p+1}k\over
(2\pi)^{p+1}}\, .
 \ee
In other words the string field configuration is labelled by infinite
number of
functions $\{\phi_r(k^0,\ldots k^p)\}$. The Fourier transforms
 \be \label{efoursft}
\wt\phi_r(x^0,\ldots
x^p) \equiv \int \, {d^{p+1}k\over
(2\pi)^{p+1}} \, e^{ik.x} \, \phi_r(k^0,\ldots k^p)
 \ee
give rise
to infinite number of
fields in
$(p+1)$ dimensions. Thus we see that the configuration space of string
field theory
is indeed labelled by infinite number of fields.

For later use we shall now choose some specific normalization convention
for the open string states in this theory.
We normalize $|k\ra$ as
 \be \label{e4.41}
\la k| c_{-1} c_0 c_1 | k\ra = (2\pi)^{p+1} \delta(k+k')\, .
 \ee
This gives
 \be \label{e4.42}
\la 0| c_{-1} c_0 c_1 | 0\ra = (2\pi)^{p+1} \delta(0) = V_{p+1}\, ,
 \ee
where $V_{p+1}$ denotes the total volume of the space-time occupied by
the D-brane.
In arriving at the right hand side of \refb{e4.42} we have used the usual
interpretation of delta function of momentum at zero argument as the
space-time volume. (This can be seen more explicitly by putting the
system in a
periodic box.) This suggests that given two states $|A\ra$ and $|B\ra$,
both
carrying zero momentum, it is useful to define a modified inner product:
 \be \label{e4.43}
\la A | B\ra' = {1\over V_{p+1}} \, \la A | B\ra\, .
 \ee
In this convention
 \be \label{ecnormp}
\la 0| c_{-1} c_0 c_1 |0\ra' = 1\, .
 \ee

Let us now turn to the task of
constructing an action for the string field theory. Given a string field
configuration
$|\Phi\ra$, the action $S(|\Phi\ra)$ should give a number. The action
proposed in
\cite{osft} is:
 \be \label{e4.11}
S = -{1\over g^2} \, \left[ {1\over 2} \la \Phi | Q_B|\Phi\ra + {1\over
3} \la
\Phi|\Phi* \Phi\ra \right]\, .
 \ee
Here $g$ is a constant, known as the open string coupling constant.
This
is to be
distinguished from the closed string coupling constant $g_s$ introduced
earlier in
eq.\refb{e2.16} for example. The precise relation between $g$ and $g_s$
depends on
which D-brane system we are considering and will be discussed later. In
general $g_s\propto g^2$ with the constant of proportionality depending on
the brane system on which we are formulating the open string field theory.
$Q_B$
is the BRST charge defined in \refb{e4.6}. The BPZ inner product $\la
\cdot|\cdot\ra$ has
been defined earler in eqs.\refb{e4.12}, \refb{e4.14}. The only
operation appearing on the right hand side of \refb{e4.11} that has
not been defined so far is the $*$-product
$|A*B\ra$ \cite{osft} for $|A\ra, |B\ra \in \HH$. We shall define this
now.

 \begin{figure}[!ht]
 \begin{center}
\leavevmode
\epsfysize=5cm
\epsfbox{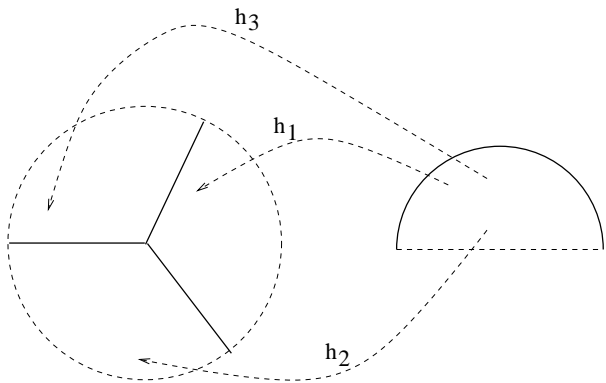}
\end{center}
\caption{The images of the upper half unit disk under the maps $h_1$,
$h_2$ and
$h_3$.
} \label{f8}
\end{figure}
One can show that the BPZ inner
product is non-degenerate. Hence $|A*B\ra$ is completely specified if
we
specify the
inner product $\la C|A*B\ra$ for any state $|C\ra$. This is given as
follows:
 \be \label{e4.16}
\la C|A*B\ra = \la h_1\circ C(0) h_2\circ A(0) h_3\circ B(0)\ra_D\, ,
 \ee
where
 \be \label{e4.17}
h_1(z) = \left({1+iz \over 1-iz}\right)^{2/3}, \qquad
h_2(z) = e^{-2\pi i/3} h_1(z), \qquad h_3(z) = e^{-4\pi i/3} h_1(z)\, .
 \ee
The $\circ$ operation and $\la\cdot|\cdot \ra$ have been defined
around
eqs.\refb{e4.12} - \refb{e4.13}. $\la\cdot\ra_D$ denotes correlation 
function on the unit disk as usual.
The images of the upper half unit disk under the maps $h_1$, $h_2$ and
$h_3$ are
shown in Fig.\ref{f8}.

Using this definition of the $*$-product we can compute the action given in
\refb{e4.11} for any string field configuration $|\Psi\ra$. It turns out
that $Q_B$,
$*$ and $\la\cdot|\cdot\ra$ satisfy some important identities:
 \begin{enumerate}
\item $Q_B$ is nilpotent:
 \be \label{e4.18}
(Q_B)^2 = 0\, .
 \ee
\item $Q_B$ can be `integrated by parts':
 \be \label{e4.19}
\la Q_B A|B\ra = -(-1)^{n_A} \la A |Q_B|B\ra\, ,
 \ee
where $n_A$ denotes the ghost number of the state $|A\ra$.
\item $Q_B$ distributes over the $*$-product:
 \be \label{e4.20}
Q_B|A*B\ra = (Q_B|A\ra)*|B\ra + (-1)^{n_A} |A\ra * (Q_B|B\ra)\, .
 \ee
\item The BPZ inner product is symmetric:
 \be \label{e4.21}
\la A|B\ra = \la B|A\ra\, .
 \ee
\item The quantity $\la A|B*C\ra$ is cyclic:
 \be \label{e4.22}
\la A|B*C\ra = \la C|A*B\ra\, .
 \ee
\item The $*$-product is associative:
 \be \label{e4.23}
(|A\ra * |B\ra)*|C\ra = |A\ra * (|B\ra * |C\ra)\, .
 \ee
\end{enumerate}
These identities can be proved by using the general properties of
the
matter-ghost BCFT, without restricting ourselves to any specific
choice
of the
matter BCFT. We however do need to use the fact that the matter BCFT has
total
central charge
26.

Using these identities one can show that the action \refb{e4.11} is
invariant under
an infinitesimal gauge transformation:
 \be \label{e4.24}
\delta|\Phi\ra = Q_B|\Lambda\ra +|\Phi\ra * |\Lambda - |\Lambda\ra *
|\Phi\ra\, ,
 \ee
where the infinitesimal gauge transformation parameter $|\Lambda\ra$ is
an arbitrary
state in $\HH_0$. More specifically, if $\{|\chi_{0,s}\ra\}$ denote a set
of basis
states in $\HH_0$, and if we expand $|\Lambda\ra$ as
 \be \label{e4.25}
|\Lambda\ra = \sum_s \lambda_s |\chi_{0,s}\ra\, ,
 \ee
then the coefficients of expansion $\lambda_s$ are infinitesimal and
represent the
gauge transformation parameters.\footnote{Again by regarding the sum over
$s$ as a sum over a discrete index and integration over the continuous
momentum index we can regard $\{\lambda_s\}$ as a set of functions of the
momentum along the D-brane, or by taking their Fourier transform, a set
of functions of the coordinates along the D-brane world-volume.} The
variation of the action $S$ under
the transformation \refb{e4.24} vanishes to first order in $\lambda_s$.

The equations of motion obtained by requiring $\delta S=0$ under
arbitrary variation
$\delta |\Phi\ra$ to first order in $\delta|\Phi\ra$, gives
 \be \label{e4.26}
Q_B|\Phi\ra + |\Phi\ra * |\Phi\ra =0\, .
 \ee
Thus at the linearized level the equations of motion take the form:
 \be \label{e4.27}
Q_B|\Phi\ra = 0\, .
 \ee
This agrees with the physical state condition \refb{e4.5}. Furthermore,
at the
linearized level the gauge transformation \refb{e4.24} takes the form:
 \be \label{e4.28}
\delta |\Phi\ra = Q_B|\Lambda\ra\, .
 \ee
Thus equivalence under linearized gauge transformation reproduces the
equivalence
relation \refb{e4.8} of the first quantized theory. This shows that gauge
inequivalent solutions of the linearized equations of motion of string
theory are in
one to one correspondence to the physical states of the first quantized
theory. This
is one of the requirements that the string field theory must satisfy.

It can be shown that the other requirement, that the S-matrix elements
involving
physical external states computed using the Feynman rules of string field
theory
reproduce the S-matrix elements computed using Polyakov prescription, is
also
satisfied by the open string field theory described here
\cite{osft,giddings,GIDDMARW}. The
computation in
string field theory requires a gauge fixing. The most commonly used gauge
is the
Siegel gauge, where we require:
 \be \label{e4.29}
b_0 | \Phi\ra =0\, .
 \ee
We shall make use of this gauge condition later.

\subsection{Reformulation of the tachyon condensation conjectures in 
string field theory} \label{s4.2}

We shall now reformulate the three conjectures about the tachyon
potential on a
bosonic D-brane in the language of string field theory. Since the
conjectures
involve properties of classical solutions of open string field
theory
(translationally invariant vacuum solution and lump solutions) we
shall begin
by reviewing certain properties of classical solutions in open string field
theory. Let $|\Phi_{cl}\ra$ denote a specific solution of the
classical
equations of
motion \refb{e4.26}:
 \be \label{e4.30}
Q_B|\Phi_{cl}\ra + |\Phi_{cl}\ra * |\Phi_{cl}\ra =0\, .
 \ee
If we want to study string field theory around this classical
solution, it is convenient to define shifted field $|\Psi\ra$ as
 \be \label{e4.31}
|\Psi\ra = |\Phi\ra - |\Phi_{cl}\ra\, ,
 \ee
and rewrite the original action as
 \be \label{e4.32}
S(|\Phi\ra) = S(|\Phi_{cl}\ra) + \wt S(|\Psi\ra)\, ,
 \ee
where
 \be \label{e4.33}
\wt S(|\Psi\ra) = -{1\over g^2} \left[ {1\over 2} \la \Psi| \QQ|\Psi\ra
+{1\over 3} \la \Psi|\Psi * \Psi\ra \right]\, ,
 \ee
with
 \be \label{e4.34}
\QQ|A\ra \equiv Q_B|A\ra + |\Phi_{cl}\ra * |A\ra - (-1)^{n_A} |A\ra *
|\Phi_{cl}\ra\, ,
 \ee
for any state $|A\ra\in \HH$. One can show that as long as
$|\Phi_{cl}\ra$
satisfies
the equation of motion \refb{e4.30}, all the identities satisfied by
$Q_B$,
$*$-product and BPZ inner product hold with $Q_B$ replaced by $\QQ$.
This, in turn shows that $\wt S(|\Psi\ra)$ is invariant under a gauge
transformation:
 \be \label{e4.35}
\delta|\Psi\ra = \QQ|\Lambda\ra + |\Psi\ra * |\Lambda\ra - |\Lambda\ra
* |\Psi\ra \, .
 \ee
{}From the structure of the action \refb{e4.33} and the gauge
transformation law
\refb{e4.35} it follows that the spectrum of perturbative physical open
string
states around the solution $|\Phi_{cl}\ra$, obtained by finding the
gauge
inequivalent
solutions of the linearized equations of motion derived from $\wt
S(|\Psi\ra)$, is
in one to one correspondence with the cohomology of $\QQ$. In other words,
they are given by eqs.\refb{e4.5}, \refb{e4.8} with $Q_B$ replaced by
$\QQ$.

We are now in a position to restate the tachyon condensation conjectures
in the
language of string field theory. For simplicity we shall restrict our
discussion to
static D$p$-branes in flat (25+1) dimensional space-time, but many of the
results
hold for D-branes in more general space-time background.
 \begin{enumerate}
\item There is a translationally invariant solution $|\Phi_0\ra$ of the
string field
theory equations of motion:
 \be \label{e4.36}
Q_B|\Phi_0\ra + |\Phi_0\ra * |\Phi_0\ra =0\, ,
 \ee
such that
 \be \label{e4.37}
-{1\over V_{p+1}} \, S(|\Phi_0\ra) + \TT_p = 0\, .
 \ee
Here $V_{p+1}$ is the volume of the D-brane world-volume. Since for a space-time
independent solution the value of the action is given by $-V_{p+1}$
multiplied by
the value of the potential at $|\Phi_0\ra$, \refb{e4.37} is a restatement
of
\refb{e2.17}.
\item Associated with the solution $|\Phi_0\ra$ we have a nilpotent
operator $\QQ$
defined through eq.\refb{e4.34}
 \be \label{e4.38}
\QQ|A\ra \equiv Q_B|A\ra + |\Phi_0\ra * |A\ra - (-1)^{n_A} |A\ra *
|\Phi_0\ra\, .
 \ee
Cohomology of $\QQ$ represents the spectrum of physical open string states 
around
$|\Phi_0\ra$. Since we do not expect any physical open string state around 
the
tachyon vacuum
solution, the cohomology of $\QQ$ must be trivial.
\item $\forall q$ such that $0<q\le p$, there should be a solution
$|\Phi^q\ra$
of the equations of motion which depends on $q$ of the spatial
coordinates and
represents a D-$(p-q)$-brane. The requirement that the energy per unit
$(p-q)$-volume
of this solution agrees with the tension of the D-$(p-q)$-brane gives:
 \be \label{e4.39}
S(|\Phi^q\ra) - S(|\Phi_0\ra) = - V_{p-q+1} \TT_{p-q}\, .
 \ee
Note that in computing the energy of the solution we take the zero of the
energy to
be at the tachyon vacuum solution, since by the first conjecture this
represents the
vacuum without any D-brane. Note also that the solution $|\Phi^0\ra$,
which is
supposed to describe the original D$p$-brane, is to be identified as the
trivial
solution ($|\Phi^0\ra = 0$). Thus $S(|\Phi^0\ra)=0$, and for $q=0$ 
eq.\refb{e4.39} reduces to \refb{e4.37}.

\end{enumerate}

\subsection{Verification of the first conjecture}

In this subsection we shall discuss verification of
eq.\refb{e4.37}.
For this we need to study the component form of the action. Expanding the
string
field $|\Phi\ra$ in a basis as in \refb{e4.9} and substituting
it in the
expression for the action \refb{e4.11}, we get
 \be \label{e4.44}
S(|\Phi\ra) = -{1\over g^2} \left[{1\over 2} \AAA_{\alpha  \beta}
\phi_\alpha \phi_\beta + {1\over 3} \CC_{\alpha\beta\gamma}
\phi_\alpha \phi_\beta \phi_\gamma\right]
  \ee
 \begin{eqnarray} \label{e4.45}
\AAA_{\alpha \beta} &=& \la
\chi_{1,\alpha} | Q_B| \chi_{1,\beta}\ra \nonumber
\\ \cr
\CC_{\alpha\beta\gamma} &=&
\la h_1\circ \chi_{1,\alpha}(0) \, h_2\circ \chi_{1,\beta}(0) \,
h_3\circ \chi_{1,\gamma}(0)\ra_D \, .
\end{eqnarray}
As discussed before, for a D$p$-brane in flat space-time, the
label $\alpha$ can be split into a pair of labels, -- a discrete
label $r$ and a continuous momentum label $\{k^\mu\}$ along
directions tangential to the D$p$-brane. Thus $\{\phi_\alpha\}$
can be regarded as a set of functions $\{\phi_r(k)\}$, and
$\sum_\alpha$ in the action can be replaced by $\sum_r \int
d^{p+1} k$. The basis states $\{|\chi_{1,\alpha}\ra\}$ can be
thought of as the set $\{|\chi_{1,r}(k)\ra\}$ with
$|\chi_{1,r}(k)\ra$ being a state built on the Fock vacuum $|k\ra$
by the action of various oscillators.

Analysis of the tachyon vacuum solution is simplified due to the fact
that the
solution is translationally invariant. In momentum space this allows
us to
write:
 \be \label{e4.46}
\phi_r(k) = \phi_r \, (2\pi)^{p+1} \delta^{(p+1)}(k) \, .
 \ee
Thus the expansion \refb{e4.9} can be rerwitten as
 \be \label{e4.47}
|\Phi\ra = \sum_r \phi_r |\chi_{1,r}\ra\, ,
 \ee
where $\{ |\chi_{1,r}\ra\}\equiv \{|\chi_{1,r}(k=0)\ra\}$ is a
basis of zero momentum states in $\HH_1$. The component form of
the action restricted to this subspace is given by,
 \be \label{e4.48}
S(|\Phi\ra)= -{1\over g^2} \Big[{1\over 2} \AAA_{rs}
\phi_r
\phi_s + {1\over 3} \CC_{rst} \phi_r \phi_s
\phi_t\Big]\, ,
 \ee
where $\AAA_{rs}$ and $\CC_{rst}$ are defined in the same way as in
\refb{e4.45},
with
the state $\chi_{1,\alpha}$ etc. replaced by $\chi_{1,r}$.
Note however that since $|\chi_{1,s}\ra$ carries zero momentum, both
$\AAA_{rs}$ and
$\CC_{rst}$ carry explicit factors of $V_{p+1}$ due to the normalization
condition \refb{e4.42}, and hence it will be more
convenient for our analysis to define new coefficients $A_{rs}$ and
$C_{rst}$ by
removing this volume factor:
 \be \label{e4.49}
\AAA_{rs} = V_{p+1} A_{rs}, \qquad \CC_{rst} = V_{p+1} C_{rst}\, ,
 \ee
 \be \label{e4.49a}
 A_{rs} = \la \chi_{1,r}|Q_B|\chi_{1,s}\ra', \qquad C_{rst} = \la 
\chi_{1,r}|\chi_{1,s}*\chi_{1,t}\ra'\, ,
 \ee
where $\la \cdot\ra'$ has been defined in \refb{e4.43}. 
The action \refb{e4.48} now may be written as
 \be \label{e4.50}
S(|\Phi\ra) = -{1\over g^2} \, V_{p+1} \, \VV(|\Phi\ra)\, ,
 \ee
where
 \ben \label{e4.51}
\VV(|\Phi\ra) &=& {1\over 2} A_{rs} \phi_r \phi_s + {1\over 3} C_{rst}
\phi_r \phi_s
\phi_t \nonumber \\
&=& {1\over 2} \la \Phi| Q_B|\Phi\ra' + {1\over 3}
\la\Phi|\Phi*\Phi\ra'\, .
 \een
Conjecture
1, given in \refb{e4.37} can now be rewritten as
 \be \label{e4.52}
{1\over g^2} \VV(|\Phi_0\ra) + \TT_p = 0\, .
 \ee
We can bring this into a more suggestive form by expressing $\TT_p$ in
terms of
$g^2$. An expression for $\TT_p$ in terms of closed string coupling
constant $g_s$
has been given in \refb{e2.16}, but we would like to express this in
terms of the
open string coupling constant $g$. This can be done in many ways.
One way of doing this is to examine the open string field theory
action carefully to determine the inertial mass per unit volume of
the D-$p$-brane and identify this with $\TT_p$. This analysis yields the
relation\cite{9911116}:
 \be \label{e4.53}
\TT_p = {1\over 2\pi^2 g^2}\, .
 \ee
Substituting this into \refb{e4.52} we get
 \be \label{e4.54}
2\pi^2 \VV(|\Phi_0) + 1 = 0\, .
 \ee

Thus our task is to begin with the $\VV(|\Phi\ra)$ given in \refb{e4.51},
find a
local minimum of this expression by varying the various coefficients
$\phi_r$, and
show that the value of $\VV(|\Phi\ra)$ at the local minimum satisfies
\refb{e4.54}.
Several simplifications occur in this computation which allows us to
restrict
$|\Phi\ra$ to a subspace smaller than the space of all zero momentum
states in
$\HH_1$.
First of all it turns out that the structure of $S(|\Phi\ra)$ allows a
consistent
truncation of the action where we restrict
$|\Phi\ra$ to linear combination of states created from the vacuum
$|0\ra$ by the
action of the ghost oscillators and matter {\it Virasoro generators}
\cite{9911116},
instead of
letting $|\Phi\ra$ be an arbitrary linear combination of matter and ghost
oscillators acting on the vacuum.\footnote{Consistent truncation of the
action means
that if we restrict $|\Phi\ra$ to this subspace, then the equations of
motion
associated with the components of $|\Phi\ra$ outside this subspace are
automatically
satisfied.} We shall call the subspace generated by these
states the universal subspace.\footnote{Once we restrict $|\Phi\ra$ to the
universal subspace, all the conformal field theory correlation functions
which go into the
computation of $\VV(|\Phi\ra)$ are independent of the specific choice of
the conformal field theory used for this
computation. This shows that once we have established \refb{e4.54} for
some unstable D-brane in some closed string background, it proves the
first conjecture for any D-brane in any closed string background in the
bosonic string theory\cite{9911116}. A similar argument also works for
the superstring theory.} Second, the $\VV(|\Phi\ra)$
has a $Z_2$ symmetry known as the twist invarince, under which
 \be \label{e4.55}
|\Phi\ra \to (-1)^{L_0+1} |\Phi\ra\, ,
 \ee
where $L_n$'s denote the total Virasoro generators of the matter-ghost
BCFT. If we
denote by `twist'  the $(-1)^{L_0+1}$ eigenvalue of a state, then this
$Z_2$
symmetry allows us to restrict $|\Phi\ra$ to twist even sector in our
search for the
tachyon vacuum solution. Finally, due to the gauge invariance of the
action, we can
impose a gauge condition.
One can show that in the twist even sector the Siegel gauge
\refb{e4.29} is a good
choice of
gauge around the point $|\Phi\ra=0$ \cite{9912249}. Thus to begin with, we
could
look for a
solution $|\Phi_0\ra$ in the Siegel gauge, and after we have obtained the
solution,
verify that
the Siegel gauge is still a good choice of gauge near the solution
$|\Phi_0\ra$. We shall denote by 
$\wt\HH_1$ the restricted subspace of $\HH_1$ satisfying all
these requirements.\footnote{In the Siegel gauge the action has an
SU(1,1) invariance\cite{siegel-zwie,0010190} which allows us to further
restrict $|\Phi\ra$ to
SU(1,1) singlet
subspace\cite{0010190}, but this has not so far been used effectively in
simplifying the
analysis. Nevertheless, once the solution has been found, one can explicitly
check that the solution is an SU(1,1) singlet\cite{0010190}.}

Let us define the level of a state $|s\ra$ to be the difference
between the $L_0$ eigenvalue $h$ of $|s\ra$ and the $L_0$
eigenvalue of the state $c_1|0\ra$ representing the zero momentum
tachyon. Since the latter state has $L_0$ eigenvalue $-1$, the
level of $|s\ra$ is given by $(1+h)$. Thus, for example,
$c_1|0\ra$ has level 0,  $c_{-1}|0\ra$ and $c_1 L^{(m)}_{-2}|0\ra$
has level 2 etc. Using this definition we can partially order the zero 
momentum
basis states in the order of increasing level. Thus for example in
$\wt\HH_1$, $|\Phi\ra$ can be expanded as:
 \be \label{e4.56}
|\Phi\ra = \phi_0 c_1 |0\ra + \phi_1 c_{-1} |0\ra + \phi_2
L^{(m)}_{-2} c_1 |0\ra + \cdots\, ,
  \ee
where $\cdots$ involves states of level 4 and higher. Let us now
make a drastic approximation where we set all the coefficients
other than that of level zero state $c_1|0\ra$ to zero.
Substituting this into \refb{e4.51} we get
 \be \label{e4.57}
\VV(\phi_0) = {1\over 2} \, \phi_0^2 \, \, \la 0 |c_{-1} Q_B
c_1|0\ra' +{1\over 3} \, \phi_0^3 \, \, \la h_1\circ c(0) h_2\circ
c(0) h_3\circ c(0)\ra'_D \, .
  \ee
The relevant correlation functions can be easily evaluated and
give
 \be \label{ehalf}
  \VV(\phi_0) = -{1\over 2} \, \phi_0^2 + {1\over 3} \,
\left({3\sqrt 3 \over 4}\right)^3 \, \phi_0^3\, .
  \ee
This has a local minimum at $\phi_0=(4/3\sqrt 3)^3$, and at this
minimum,
 \be \label{e4.58}
  2\pi^2 \VV(\phi_0) = -(2\pi^2)
(4/3\sqrt 3)^6 / 6 \simeq -.684 \, .
  \ee
This is about $68\%$ of the conjectured answer $-1$ given in \refb{e4.54}.

This is the beginning of a systematic approximation scheme known as the
level
truncation\cite{kost-sam1,kost-sam2,kost-pot1,9912249,0002237,0211012}. We 
define the
level of a coefficient $\phi_r$ to be
the level of
the state $|\chi_{1,r}\ra$ that it multiplies in the expansion of
$|\Phi\ra$.
We now define a level $(M,N)$ approximation to $\VV(|\Phi\ra)$
as follows:
 \begin{enumerate}
\item Keep all fields $\phi_r$ of level $\le M$.

\item Keep all terms in the action for which the sum of the levels of all
fields in
that term is $\le N$.
Thus for example at level (2,4) we shall include interaction terms of the
form
0-0-0, 0-0-2, 0-2-2 but ignore interaction terms of the form 2-2-2.

\item This gives an expression for $\VV(|\Phi\ra)$ involving finite
number of fields
and finite number of terms.
We find a (local) minimum of this $\VV(|\Phi\ra)$ and evaluate the value
of
$\VV(|\Phi\ra)$ at this minimum.

\end{enumerate}
This defines the level $(M,N)$ approximation to $\VV(|\Phi_0\ra)$.
In order for this to be a sensible approximation scheme, we need to ensure
that the answer converges as we
increase the values of $M,N$.
In actual practice
this method converges quite rapidly. For example the value of
$-2\pi^2\VV(|\Phi_0\ra)$ in level $(L,2L)$ approximation increases
monotonically
towards 1 as we increase the value of $L$ up to $L=10$, reaching the
value .9991 at
$L=10$ \cite{0002237}. However beyond level 12 the value of
$-2\pi^2\VV(|\Phi_0\ra)$
overshoots the
expected value 1, and continiue to increase with $L$ till about level 18
approximation\cite{0211012}. Nevertheless analysis of the tachyon
potential obtained by integrating out all fields other than $\phi_0$
shows that $-2\pi^2\VV(|\Phi_0\ra)$ eventually
turns back
and approaches the expected value 1 from above \cite{0208149,0211012}.

Since the solution is constructed in the Siegel gauge, we need to verify
that
Siegel gauge is a valid gauge choice for this solution. Operationally
what this
amounts to is the following. In arriving at the solution, we have made
sure that the
variation $\delta S$ of the action under a variation $\delta|\Phi\ra$
vanishes
to first order in $\delta|\Phi\ra$ around the solution,
{\it provided $\delta|\Phi\ra$ satisfies the Siegel gauge condition
$b_0\delta|\Phi\ra=0$.} In order to check that the solution
satisfies the
full string field theory equations of motion we need to make sure that
$\delta S$
vanishes to first order in $\delta|\Phi\ra$ even if $\delta|\Phi\ra$ does
not
satisfy the Siegel gauge condition. To check this, we can simply take the
first
order variation of $\VV(|\Phi\ra)$ with respect to components of
$|\Phi\ra$ which
violate Siegel gauge condition and verify that these derivatives vanish
when
evaluated in the background of the solution $|\Phi_0\ra$ found using the
level
truncation scheme. This has been checked explicitly in
refs.\cite{0009105,0211012}.

\subsection{Verification of the second and third conjectures} \label{s4.3}

We shall now briefly discuss the verification of the second and the third
conjectures. Of these  the analysis of the third conjecture,
eq.\refb{e4.39},
proceeds in a way very
similar to that of the  first 
conjecture\cite{0003031,0005036,0008053,0008101}. The main difference is 
that since the
solution depends on $q$ of the spatial coordinates, we can no longer
restrict
$|\Phi\ra$ to be in the zero momentum sector for finding the solution
$|\Phi^q\ra$;
instead we must allow $|\Phi\ra$ to carry momentum along these $q$
directions. The
analysis can be simplified by compactifying the $q$ directions along
which we want
the lump to form. This makes the momenta in these directions discrete,
and as a
result $|\Phi\ra$ can still be expanded in the discrete basis.

The explicit construction of the solution now proceeds via a modified
level
truncation scheme where the level of a state, given by $(L_0+1)$,
includes not only
the oscillator contribution but also the contribution to $L_0$ due to the
momentum
along the
compact directions. The level $(M,N)$ approximation is defined exactly as
before,
and for finite $(M,N)$ we still have a finite number of variables with
respect to
which we need to extremize the action. For $q=1$ the result converges
fast
towards
the expected value as we increase the level of approximation
\cite{0005036}. In particular, the tension of the soliton becomes
independent of the radius of compactification, as is expected of a D-brane
whose transverse direction is compactified. The convergence is also
reasonably good for $q=2$
\cite{0008053,0008101}. Today the best available results for the tension
of the
codimension $q$
lump solution for $q=1$ and $q=2$ differ from their conjectured values
by $1\%$
\cite{0005036} and
$13\%$ \cite{0008053,0008101} respectively.
For larger values of $q$ the number of fields below a given level
increases rapidly,
slowing down the
convergence. The number of fields also increases rapidly as we increase
the radius of
compactification since states carrying different momenta become closely
spaced in
level. Nevertheless the analysis has been done for different radii, and
the tension
of the resulting lump solution has been shown to be independent of the
radius to a
very good approximation as is expected of a D-brane.

All the lump solutions described above have been constructed in the Siegel 
gauge. For the codimension one lump, the validity of the Siegel gauge 
choice has been tested in \cite{0101014}. 

For verifying conjecture 2, we need to check that $\QQ$ defined in
eq.\refb{e4.38} has trivial cohomology. This can be done as follows
\cite{0103085}:
 \begin{enumerate}
\item Take the best available value of
$|\Phi_0\ra$ obtained using the level truncation scheme and construct
$\QQ$ from there using eq.\refb{e4.38}.
\item Construct solutions of $\QQ|A\ra=0$ by taking
$|A\ra$ to be arbitrary linear combinations of states up to a certain
level.
\item Show that for every such $|A\ra$, there is a state $|B\ra$
such that $|A\ra=\QQ|B\ra$.
\end{enumerate}
There is however a further
complication due to the fact that $|\Phi_0\ra$ obtained in the level
truncation scheme is only an approximate solution of the equations of
motion, and as a result $\QQ$ defined in
eq.\refb{e4.38} does not square to zero exactly when we use this
approximate value
of $|\Phi_0\ra$. Hence a state $|A\ra = \QQ\, |B\ra$ does not
satisfy $\QQ|A\ra=0$
exactly. We can circumvent this problem by using an
approximate rather than an exact analysis of the 
$\QQ$-cohomology\cite{0103085}.
Given a state $|A\ra$ satisfying
$\QQ|A\ra=0$, we check if there is a state $|B\ra$ such that the
ratio of the norm of $(|A\ra-\QQ|B\ra)$ to the norm of $|A\ra$ can be made
{\it small}. Of course there is no natural norm in the space of the
string field since the BPZ inner product of ghost number 1 state
with itself vanishes by ghost charge conservation, but we could use
several different artificial norms ({\it e.g.} by explicitly inserting a 
factor of $c_0$ in the BPZ inner product), and
check if the final conclusion is sensitive to the choice of the norm.
It was found in \cite{0103085} that
if we
carry out this analysis in a subspace which includes string states up
to a given
oscillator level
$L$, and carrying momentum $k$ with $|k^2|
\le L$,
then all $\QQ$ closed states are also $\QQ$-exact to within 1\%
accuracy. This gives numerical evidence for the absence of the physical
open string states around the tachyon vacuum.

A somewhat different approach to this problem has been suggested in 
\cite{0309164}.
Various other aspects of tachyon condensation in
bosonic open string field
theory have been discussed in
refs.\cite{0001201,0002117,0003031,0006240,0007235,0007153,
0008033,0010034,
0010247,
0011238,0101162,0103103,
0105024,0105156,0105246,
0105264,
0106068,0107046,
0109182,0201095,
0201159,0202133,0205275,0302182,0304261,
0311115,
0403200,0406023,
0409249}. 

\subsection{Superstring field theory} \label{s4.4}

In this subsection we shall briefly discuss the use of superstring field
theory in verifying the conjectured properties of the tachyon potential in
superstring theory. For this discussion we shall use the
Berkovits version of superstring field 
theory\cite{9503099,9912121,0105230,0109100}. An
alternative approach to this problem based on various cubic 
versions of
superstring field theory\cite{inverse,AREF1,AREF2,pty} has been proposed 
in
\cite{0004112,0011117,0107197},
but we shall not discuss it here.

We shall begin by reviewing the formulation of first quantized
open superstring  theory on a non-BPS
D-brane\cite{9806155,9808141,9809111} in a convention which will
facilitate the formulation of superstring field
theory\cite{9503099,9912121,0002211}. We restrict ourselves to the
Neveu-Schwarz (NS) sector of the theory, since only the bosonic
fields arising in this sector are involved in the construction of
the classical solutions describing the tachyon vacuum and various
lower dimensional D-branes. In the first quantized formulation the
bulk world-sheet theory is given by a $c=15$ superconformal field
theory together with a set of anticommuting ghosts $b$, $c$, $\bar
b$, $\bar c$ and commuting ghosts $\beta$, $\gamma$, $\bar\beta$,
$\bar\gamma$. The fields $\beta,\gamma$ can be replaced by a pair
of fermions $\xi,\eta$, and a scalar $\phi_g$ through the
relations\cite{FMS}
 \be \label{ei1}
\beta=\p\xi e^{-\phi_g}, \qquad \gamma =\eta e^{\phi_g}\, .
 \ee
$\xi$ and $\eta$ have dimensions (0,0) and (1,0) respectively,
whereas $\phi_g$ is a chiral scalar field with background charge
so that $\la e^{q\phi_g(0)}\ra_D$ is non-zero only for $q=-2$.
There are similar relations involving the anti-holomorphic fields.
The fields $\xi$, $\eta$ and $\phi_g$ are normalized such that
 \be \label{eghostnorm}
 \xi(z) \eta(w) \simeq {1\over z-w}\, , \qquad \p \phi_g(z)
 \p\phi_g(w) \simeq -{1\over (z-w)^2}\, ,
 \ee
with a similar relation among the anti-holomorphic components.
Since the open string vertex operators will involve the boundary
values of various fields, and since on the boundary the
holomorphic and the anti-holomorphic ghost fields are set equal,
we shall not need to refer to the anti-holomorphic ghost fields
explicitly. The ghost number ($n_g$) and the picture number
($n_p$) assignments of various fields are defined as follows:
 \ben \label{ei2}
&& b:\quad n_g=-1, n_p=0, \qquad c:\quad n_g=1, n_p=0 \, , \nonumber \\
&& e^{q\phi_g}:\quad n_g=0, n_p=q\, , \nonumber \\
&& \xi:\quad n_g=-1, n_p=1, \qquad \eta:\quad n_g=1, n_p=-1\, . \nonumber
\\
 \een
The matter fields as well as the SL(2,R) invariant vacuum carry zero
ghost and picture number.
The GSO operator is given by:
 \be \label{egso}
(-1)^F (-1)^q\, ,
 \ee
where $F$ denotes the world-sheet fermion number of the matter
fields, and $q$ denotes the $\phi_g$ momentum.

The physical open string states are states of ghost number 1, with
each physical state having different representations in different
picture numbers\cite{FMS}. Furthermore, they are required to satisfy
the following conditions:
 \begin{enumerate}
\item $|\phi\ra$ is BRST invariant:
 \be \label{esuperbrst}
Q_B|\phi\ra = 0\, ,
 \ee
where $Q_B$ is the BRST charge defined as
 \be \label{eqb4}
Q_B = \oint dz \, j_B(z) = \oint dz \Bigl\{  c \bigl( T_m +
T_{\xi\eta} + T_{\phi_g}) + c \partial c b +\eta \,e^{\phi_g} \, G_m
- \eta\p \eta e^{2\phi_g} b \Bigr\}\, ,
 \ee
 \be \label{eb5}
T_{\xi\eta}=\p\xi\,\eta, \quad T_{\phi_g}=-{1\over 2} \p\phi_g \p
\phi_g -\p^2\phi_g \, .
 \ee
$T_m$ is the matter stress tensor and
$G_m$ is the matter
superconformal generator.
\item Two states $|\phi\ra$ and $|\phi'\ra$ are considered to be
equivalent
if they differ by $Q_B|\Lambda\ra$ for some state $|\Lambda\ra$:
 \be \label{esuperequiv}
|\phi\ra \equiv |\phi\ra + Q_B|\Lambda\ra\, .
 \ee
\item $|\phi\ra$ can be either GSO odd or GSO even. The vertex
operators of GSO even states are accompanied by Chan-Paton factors $I$
($I$ being the $2\times 2$ identity matrix), whereas vertex operators
of GSO odd states are accompanied by Chan-Paton factor $\sigma_1$. This 
rule is inherited from the parent brane-antibrane system for which the GSO 
even states carry Chan-Paton facors $I$ and $\sigma_3$ and GSO odd states 
carry Chan-Paton factors $\sigma_1$ and $\sigma_2$. The $(-1)^{F_L}$ 
projection removes the states carrying Chan-Paton factors $\sigma_2$ and 
$\sigma_3$.

\item 
\label{smallp} 
The field $\xi$ appears in the vertex operators for physical
states only through its derivatives. In terms of the state $|\phi\ra$
it means that $|\phi\ra$ is annihilated by $\eta_0$, where
$\eta_n$, $\xi_n$ denote the modes of the fields $\eta$ and $\xi$
defined
through the
expansion
 \be \label{eetaxiexp}
\eta(z) = \sum_n \eta_n z^{-n-1}, \qquad \xi(z) = \sum_n \xi_n
z^{-n}\, .
 \ee
\end{enumerate}

This condition given in item \ref{smallp} above gives what is
known as the small picture representation of the physical 
states\cite{9108021}. For the formulation of
the superstring field theory it is more convenient to use the big
picture\cite{9503099,9912121} where given a state $|\phi\ra$
satisfying the conditions given above, we use the state
$\xi_0|\phi\ra$ to represent the same physical
state.\footnote{Consequently a physical state in the big picture
is annihilated by $Q_B\eta_0$ instead of $Q_B$.} In terms of
vertex operators this corresponds to multiplying the vertex
operator in the small picture representation by $\xi$. Thus for
example a representation of the on-shell tachyon vertex operator
carrying momentum $k$ in the small picture is $c e^{-\phi_g}
e^{ik.X} \otimes \sigma_1$. The same vertex operator in the big
picture will be given by $\xi\, c e^{-\phi_g} e^{ik.X} \otimes
\sigma_1$. According to our convention given in \refb{ei2} the
small picture representation is in picture number $-1$ whereas the
big picture representation is in picture number 0.

Let us now turn to the construction of open superstring field theory.
As in the case of bosonic string theory,
a general open string field configuration $\wh \Phi$ is represented by
a state
in the world-sheet boundary conformal field theory which do not
satisfy all the requirements of a physical state. It turns out that
the choice of a general off-shell field configuration in superstring
field theory is as follows. In the NS sector it contains two
components, a GSO even component $\Phi_+$
accompanied by a Chan-Paton factor $I$ and a GSO odd component $\Phi_-$
accompanied by the Chan-Paton factor $\sigma_1$.
Both $\Phi_+$ and $\Phi_-$ are required to have picture number 0 and
ghost number 0.
Thus we can
write:\footnote{We shall adopt the convention that fields or operators
with internal
CP factors
included are denoted by symbols with a
hat on them, and fields or operators without internal
CP factors included are
denoted by symbols without a hat.}
 \be
\label{efsf}
\hp = \Phi_+ \otimes I  + \Phi_- \otimes \sigma_1\, .
 \ee

The string field theory action to be given below will involve
calculating correlation functions inolving the vertex operators
$\Phi_\pm$. In manipulating these correlation functions, we need
to keep in mind that the string field {\it components} in the NS
sector are always grassman even, whereas the open string vertex
operators which these components multiply in the expression of
$\Phi_\pm$ may be grassman even or grassman odd depending on the
world-sheet fermion number carried by the vertex operator. In
particular the fermion $\psi^\mu$ associated with the matter BCFT,
and $b$, $c$, $\xi$, $\eta$ and $e^{q\phi_g}$ for odd $q$ are
grassman odd fields. Using eqs.\refb{ei2}, \refb{egso} one can
then show that $\Phi_+$ is grassman even whereas $\Phi_-$ is
grassman odd. In particular, the zero momentum tachyon vertex
operator $\xi c e^{-\phi_g}\otimes \sigma_1$ is GSO odd and
grassman odd. Note the extra factor of $\xi$ in the tachyon vertex
operator compared to the conventions used in section \ref{s3}.
This is due to the fact that we are using the big picture.

The BRST charge $\wh Q_B$
acting on the string field and the operator $\wh\eta_0$ is defined by:
 \be \label{ei3}
\wh Q_B = Q_B\otimes \sigma_3, \qquad
\wh \eta_0 = \eta_0\otimes \sigma_3\, .
 \ee
Also, given a set of open string vertex operators $\ha_1,\ldots \ha_n$
on a non-BPS D$p$-brane, we define:
 \be \label{e200}
\lll \wh A_1\ldots \wh A_n \rrr = Tr \Bigl\langle f^{(n)}_1 \circ
\ha_1(0)\cdots
f^{(n)}_n\circ \ha_n(0)\Bigr\rangle_D \, ,
 \ee
where the trace is over the internal CP matrices, and
 \be \label{esk3}
f^{(n)}_k(z) = e^{2\pi i (k-1)\over n} \Big({1+iz\over 1-iz}
\Big)^{2/n}\,\quad  \hbox{for} \quad n\geq 1 .
 \ee
$\la \cdot\ra_D$ denotes correlation function on a unit disk as usual.

In terms of these quantities the open superstring field theory action on a
non-BPS D-brane can be
written as\cite{0001084,0002211},
 \be \label{e00}
S={1\over 4g^2} \lllb (e^{-\hp} \hq e^{\hp})(e^{-\hp}\he e^\hp) -
\int_0^1 ds (e^{-s\hp}\p_s e^{s\hp})\{ (e^{-s\hp}\hq e^{s\hp}),
(e^{-s\hp}\he e^{s\hp})\}\rrrb\, ,
 \ee
where we have divided the overall normalization by a factor
of two in order to compensate for the trace operation on the
internal matrices. $s$ in the second term is just an integration 
parameter. By expanding the various exponentials in a Taylor series 
expansion, and explicitly carrying out the $s$ integral for each term one 
can represent the action as a power series expansion in the string field. 
This is given in eq.\refb{e3a}.

\refb{e00} can be shown to 
be invariant under the infinitesimal gauge transformation\cite{0001084}
 \be \label{egtrsa}
\delta e^{\hp} = (\hq \wh\Omega) e^{\hp} + e^{\hp}(\he\wh\Omega') \, ,
 \ee
where the infinitesimal gauge transformation parameters $\wh\Omega$ and
$\wh\Omega'$ are states
with $(n_g,n_p)$ values $(-1,0)$ and $(-1,1)$
respectively. The internal CP indices carried by the gauge  parameters
are as follows
 \be
\label{ingp}
\wh\Omega = \Omega_+ \otimes \sigma_3  + \Omega_- \otimes i\sigma_2\,,
\qquad \wh\Omega' = \Omega'_+ \otimes \sigma_3  + \Omega'_- \otimes
i\sigma_2\, .
 \ee
The GSO even $\Omega_+$, $\Omega'_+$
are
Grassmann odd, while the GSO odd $\Omega_-$, $\Omega'_-$ are Grassmann
even.
The proof of invariance of \refb{e00} under \refb{egtrsa} can be carried
out by
straightforward algebraic manipulations\cite{0001084,0002211}.

An analysis similar to that in the case of bosonic string theory shows
that with the normalization we have used here, the tension of the non-BPS
D$p$-brane is given by:
 \be \label{ec1}
\TT_p = {1\over 2 \pi^2 g^2}\, .
 \ee
Thus in order to prove the first conjecture about the tachyon potential we
need to show the existence of a solution $|\hp_0\ra$ of the equations of
motion derived from the string field theory action, such that
 \be \label{epreds}
S(|\hp_0\ra) = V_{p+1} \, {1\over 2 \pi^2 g^2}\, ,
 \ee
where $V_{p+1}$ is the volume of the D$p$-brane world-volume.

The calculation proceeds as in the case of bosonic string theory using the
level truncation scheme\cite{0001084,0002211}. For this we need to first
expand the
action \refb{e00} in a power series expansion in the string field. This
gives
 \be \label{e3a}
S = {1\over 2 g^2} \sum_{M,N=0}^\infty {1 \over (M+N+2)!} {M+N \choose N}
(-1)^N \left\la\left\la  (\hq \hp) \hp^M (\he\hp)\hp^N
\right\ra\right\ra\, .
 \ee
As in the case of bosonic string theory, for finding the tachyon
vacuum solution we can restrict the string field to the universal
subspace created from the SL(2,R) invariant vacuum by the action
of the ghost oscillators and matter super-Virasoro generators. The
string field theory action in this universal subspace is invariant
under a $Z_2$ twist symmetry, which, acting on a vertex operator
of conformal weight $h$, has an eigenvalue given by $(-1)^{h+1}$
for even $2h$, and $(-1)^{h+{1\over 2}}$ for odd $2h$. Using this
symmetry, and the fact that the zero momentum tachyon vertex
operator $\xi c e^{-\phi_g}\otimes \sigma_1$ is twist even, we can
restrict the string field to twist even sector for finding the
tachyon vacuum solution.  Finally, we choose the gauge conditions
 \be \label{egauge}
b_0\hp=0, \qquad \xi_0\hp=0\, ,
 \ee
for fixing the gauge symmetries generated by $\wh\Omega$ and $\wh\Omega'$
respectively.

We can now proceed to find the tachyon vacuum solution using the level
truncation scheme as in the case of bosonic string theory. Since the zero
momentum tachyon state has conformal weight $-{1\over 2}$, we define the
level of the state (and the coefficient multiplying it) to be ${1\over
2}+h$ where $h$ is the conformal weight of the state. It turns out that
although the action \refb{e3a} has infinite
number of terms, up to a given level only a finite number of terms
contribute due to various charge conservation\cite{0002211}.
This allows us to express the action to a given level of approximation as
a finite order polynomial in a finite number of string fields.
The resulting action is then extremized with respect to the component
fields to find the vacuum solution $|\hp_0\ra$.

The result for $S(|\hp_0\ra)$ converges rapidly to the expected
value \refb{epreds}. At level (0,0) approximation, where we keep
only the zero momentum tachyon vertex operator $\xi c
e^{-\phi_g}\otimes \sigma_1$, the action contains two terms,
proportional to $t^2$ and $t^4$ respectively, where $t$ is the
coefficient of $\xi c e^{-\phi_g}\otimes \sigma_1$ in the
expansion of $\hp$. By minimizing the action with respect to $t$,
we get $\pi^2/16$ (about 60\%) times the expected
answer\cite{0001084}. At level $(3/2, 3)$ we get about 85\% of the
conjectured answer\cite{0002211} and at level (2,4) we get 89\% of
the conjectured answer\cite{0003220,0004015}.

 \begin{figure}[!ht]
\leavevmode
 \begin{center}
\epsfbox{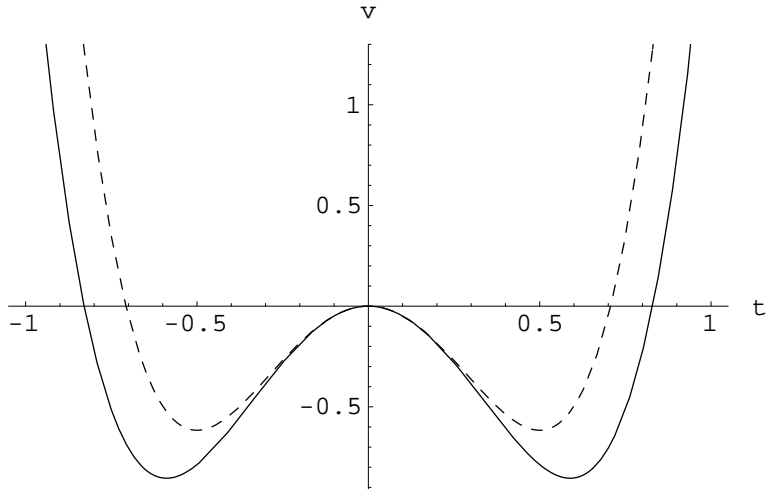}
\end{center}
\caption[]{\small The tachyon potential $v(t)=V(t)/\TT_p$ in level (3/2,
3) approximation (solid line).
For reference we also show
the
zeroeth order potential (dashed line).} \label{fber1}
\end{figure}
In this case we can also try to integrate out all the fields other
than the tachyon $t$ labelling the coefficient of $\xi c
e^{-\phi_g}\otimes \sigma_1$ and obtain a tachyon effective
potential. To level (3/2, 3), this can be done analytically. The
shape of the tachyon potential at level (0,0) and level (3/2, 3)
has been shown in Fig.\ref{fber1}. {}From this we can clearly see
the emergence of the double well shape of the potential.

In principle, the conjecture describing a codimension one D-brane as a
kink solution on
the non-BPS D$p$-brane can be analyzed following a procedure similar to
that in the case of bosonic open string theory. Similarly the verification
of the absence of physical open string states around the tachyon vacuum
can also be carried out in a manner similar to that in the case of bosonic
string theory. However in practice, due to technical complexities
involved in these
analyses, not much work has been done in these directions.

Various other aspects of tachyon condensation in open superstring field
theory have been discussed in 
\cite{0104230,0204155,0209186,0305103,0008127}.

\subsection{Vacuum String Field Theory}

The analysis of section \ref{s4.2} shows that it should be possible to
reformulate string field theory by expanding the action around the tachyon 
vacuum
solution $|\Phi_0\ra$. The resulting string field theory action should
have the
same form as the original string field theory action with cubic
interaction, but
in the quadratic term the BRST operator $Q_B$ is replaced by another
operator $\QQ$:
 \be \label{evv0}
\wt S(|\Psi\ra) = -{1\over g^2} \left[ {1\over 2} \la \Psi| \QQ|\Psi\ra
+{1\over 3} \la \Psi|\Psi * \Psi\ra \right]\, .
 \ee
$|\Psi\ra$ as usual is a state of ghost number 1 of the first
quantized open string living on the original D-brane. For future
reference we shall call the boundary conformal field theory
associated with this D-brane BCFT$_0$. In order to determine the
precise form of $\QQ$, we need to know the analytic form of the
solution $|\Phi_0\ra$ describing the tachyon vacuum. Given that no
analytic expression for $|\Phi_0\ra$ is known at present, we could
ask if it is possible to {\it guess} a form for $\QQ$ which satisfies
all the conditions and conjectures described in section
\ref{s4.2}. A general form of $\QQ$ proposed in
\cite{0012251,0106010,0108150,0111129} is:
 \be \label{evv1}
\QQ = \sum_{n=0}^\infty a_n \left(  c_n +  (-)^n \, c_{-n} \right) \, ,
 \ee
where $a_n$ are some coefficients and $c_n$ are the usual ghost
oscillators. The important point to note is that $\QQ$ is independent of
the matter part of the BCFT. For any choice of the $a_n$'s, this $\QQ$
can be shown to satisfy
the conditions \refb{e4.18} - \refb{e4.20} with $Q_B$ replaced by $\QQ$,
as is required for the gauge
invariance of the theory. Further it has vanishing cohomology in the Fock
space. To see this note that if $a_0\ne 0$, then given any solution
$|\psi\ra$ of the equation $\QQ|\psi\ra=0$ we
have $|\psi\ra = \QQ (a_0)^{-1} b_0 |\psi\ra$. If $a_0=0$, but $a_m\ne 0$
for some $m$, we can write $|\psi\ra =
{1\over 2a_m} \QQ ((-1)^m b_m - b_{-m})|\psi\ra$.
Hence $|\psi\ra$ is $\QQ$
trivial.

It now remains to find the classical solutions predicted by the first and
the third conjecture in this field theory.
Clearly the tachyon vacuum solution corresponds to the
configuration $|\Psi\ra=0$.
Thus the non-trivial solution we need to look for are those describing
various D-branes, including the original D-brane associated with BCFT$_0$.
In order to find them, we
look for classical solutions of the form:
 \be \label{eo3}
\Psi = \Psi_g \otimes \Psi_m\, ,
 \ee
where $\Psi_g$
denotes a state obtained by acting with the ghost
oscillators on the SL(2,R) invariant vacuum of the ghost
sector of BCFT$_0$, and
$\Psi_m$  is a
state obtained by acting with matter vertex operators on the SL(2,R)
invariant
vacuum of the matter sector of BCFT$_0$.
If we
denote by
$*^g$ and $*^m$ the star product in the ghost and matter sector
respectively, the equations of motion
 \be \label{eo2}
\QQ \Psi + \Psi * \Psi = 0\,
 \ee
factorize as
 \be \label{eo4}
\QQ \Psi_g = - K \Psi_g *^g \Psi_g \,,
 \ee
and
 \be \label{eo5}
\Psi_m = K^{-1} \, \Psi_m *^m \Psi_m\, ,
 \ee
where $K$ is an arbitrary constant that can be changed by scaling $\psi_g$ 
and $\psi_m$ in opposite directions.
We further assume that the ghost part $\Psi_g$ is universal for all
D-$p$-brane solutions. Under this assumption the ratio of energies
associated with two different D-brane solutions, with matter parts
$\Psi_m'$ and $\Psi_m$ respectively, is given by:
 \be \label{eo7}
{\EE'\over\EE} = {\wt S(\Psi_g\otimes \Psi_m')\over \wt S(\Psi_g\otimes 
\Psi_m)} = {\langle \Psi_m' | \Psi_m'\rangle_m \over \langle \Psi_m
|
\Psi_m\rangle_m} \, ,
 \ee
with $\langle \cdot| \cdot\rangle_m$ denoting BPZ inner product in
the matter BCFT. Thus the ghost part drops out of this calculation.
The equation \refb{eo5} for the mattter part tells us that $|\Psi_m\ra$ is
a projector under the $*$-product in the matter sector up to a constant of
proportionality.
 \begin{figure}[!ht]
\leavevmode
 \begin{center}
\epsfxsize 14cm
\epsfbox{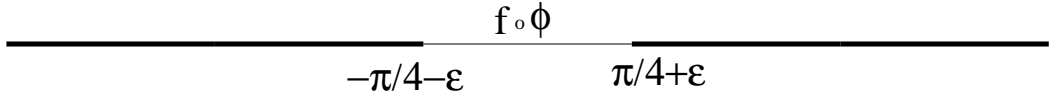}
\end{center}
\caption[]{\small The geometry used for defining the right hand side of
eq.\refb{eo8}. The thin part of the
real line has boundary conditions / interactions relevant to BCFT$_0$,
and the thick part of the real line has boundary conditions /
interactions relevant to BCFT.}
\label{fvsft}
\end{figure}

General methods for constructing such projectors have been
developed in
\cite{0008252,0102112,0105058,0105059,0105168,0106036,0202151}.
Let BCFT denote some boundary conformal field theory with the same
bulk world-sheet action as BCFT$_0$, {\it i.e.} both BCFT and
BCFT$_0$ represent D-branes in the same space-time background.
Consider now a state $|\Psi_m^{BCFT}\ra$ in the matter part of
BCFT$_0$, defined through the relation:\footnote{Note that once we have 
chosen the reference BCFT$_0$, the string field $|\Psi\ra$ is always a 
state in  BCFT$_0$ even if it describes a D-brane 
associated with another BCFT.}
 \be \label{eo8}
\la \Psi_m^{BCFT} | \phi\ra = \la f\circ \phi(0)\ra'_{BCFT}
 \ee
for any state $|\phi\ra$ in the matter part of BCFT$_0$.
Here $f(\xi)$ is the conformal map
 \be \label{eo9}
f(\xi) = \tan^{-1}\xi\, ,
 \ee
and $\la\cdot\ra'_{BCFT}$ denotes correlation function of the matter 
theory on the
upper half plane, with the boundary condition associated with
BCFT$_0$ in the range $-{\pi\over 4} -\epsilon \le x \le{\pi\over
4}+\epsilon$ and the boundary condition / interaction associated
with some other boundary conformal field theory BCFT in the range
${\pi\over 4}+\epsilon<x<\infty$ and $-\infty< x < -{\pi\over
4}-\epsilon$. Here $\epsilon$ is a small positive number which
should be taken to zero at the end. The geometry has been shown in
Fig.\ref{fvsft} with the thin part of the real line having
boundary conditions / interactions relevant to BCFT$_0$, and the
thick part of the real line having boundary conditions /
interactions relevant to BCFT. Note that since $f(0)=0$, the
vertex operator $\phi$ is inserted at the origin where we have
boundary condition associated with BCFT$_0$. This is consistent
with the fact that $|\phi\ra$ is a state in BCFT$_0$. It was shown
in \cite{0105168} that $|\Psi_m\ra$ defined in \refb{eo8} satisfy
the projector equation \refb{eo5}, with a constant $K$ that is
independent of the choice of BCFT. Furthermore, if we denote by
$|\Psi_m^{BCFT'}\ra$ a state defined through eq.\refb{eo8} with
BCFT replaced by another boundary conformal field theory BCFT$'$
(with the same bulk CFT as BCFT or BCFT$_0$), one can
show that the ratio
 \be \label{eo11}
{\langle \Psi_m^{BCFT} | \Psi_m^{BCFT}\rangle \over \langle
\Psi_m^{BCFT'} | \Psi_m^{BCFT'}\rangle} \, ,
  \ee
is equal to the ratio of tensions of the D-branes associated with
BCFT and BCFT$'$ respectively\cite{0105168}. \refb{eo7} then
suggests that we identify
 \be \label{eo12}
|\Psi_g\ra \times |\Psi_m^{BCFT}\ra 
 \ee
as the classical solution in vacuum string field theory describing
the D-brane associated with BCFT. In particular the D-brane
associated with BCFT$_0$ is described by the state
$|\Psi_g\ra \times |\Psi_m^{BCFT_0}\ra$ where in the computation of 
$|\Psi_m^{BCFT_0}\ra$ the correlation
function on the right hand side of \refb{eo8} is calculated with
the boundary condition associated with BCFT$_0$ along the whole
real axis.

The conformal map $f$ defined in \refb{eo9} has the property
that it maps the point $i$
to $\infty$, and the points $\pm 1$ to $\pm{\pi\over 4}$. If we
think of the open string to be situated along the unit semi-circle
on the upper half plane, then the point $i$ is the mid-point of
the open string, and the map $f$ sends the mid-point to $\infty$
which is a point on the boundary of the world-sheet. 
It turns out that this is the important property that makes the state 
$|\Psi_m^{BCFT}\ra$ into a projector\cite{0202151}.
One can
construct other conformal transformations which map the mid-point of the 
open string 
to the boundary, and these have
also been used to construct projectors of the $*$-algebra using
formula similar to \refb{eo8} \cite{0202151}. It is generally
believed that different projectors associated with different
conformal maps but same BCFT give gauge equivalent solutions, so
that we have one inequivalent solution for a given BCFT.

Note that in the discussion so far we did not have to know anything about
the coefficients $a_n$ in \refb{evv1}. It turns out that various
consistency conditions leads to a unique choice of $\QQ$ up to a constant
of proportionality:
 \be \label{evv5}
\QQ = \gamma c(i) \, ,
 \ee
where $\gamma$ is a constant. The coefficient $\gamma$ can be fixed by
requiring that the classical solutions described above not only reproduces
the ratios of various D-brane tensions, but also give the overall
normalization of the tension correctly. Unfortunately this leads to a
singular coefficient $\gamma$\cite{0111129}. This indicates that vacuum
string
field theory described by the kinetic term \refb{evv5} must be related to
the original open string field theory expanded around
the tachyon vacuum by a singular field redefinition\cite{0111129}. A
complete
understanding of this regularization procedure remains a challenge as of
today, although there has been quite a lot of
progress\cite{0204012,0310264}.

Various other aspects of vacuum string field theory have been discussed in 
\cite{0105184,0110124,0110136,
0111034,0111069,0111087,0111092,
0111153,0111281,
0112169,0112202,
0112214,
0112231,0201015,0201060,
0201136,0201149,
0201177,
0201197,0201229,
0202087,0202139,0203071,0203188,0203227,
0204031,0204138,0204172,0204233,
0206208,0207001,0207044,
0208009,0208067,0209186,0211286,
0212055,0212335,
0301079,
0302151,0304270,0305010,
0311198,0403031,0403283,
0404154,0409063}.

\sectiono{Boundary String Field Theory} \label{s7}

So far we have discussed two different approaches to studying the
problem of tachyon condensation, -- the approach based
on the correspondence between two dimensional conformal field theories and
classical solutions of string field theory equations of motion, and the 
direct
approach based on the analysis of classical equations of motion of open
string field theory. Although it is generally believed that these two
approaches are equivalent, this equivalence is not manifest. In
particular there is no known procedure for finding an explicit
solution of the string field theory equations of motion associated with a
given two dimensional conformal field theory (or vice versa). In this
section we shall discuss a
different version of string field theory which makes the relationship
between these two approaches more
explicit\cite{9208027,9210065,9303067,9303143,9311177}. This version of
string field
theory has been given the name boundary string field theory (BSFT). We
shall restrict our discussion to bosonic string theory only; for boundary
string field theory associated with superstring theory see
\cite{0010108,0103089,0106231}.

As in the case of the cubic string field theory, the open string field in
boundary string field theory is a state $|\Phi\ra$ of ghost number 1 of
the first quantized open string. We
can associate with it a boundary vertex operator $\Phi$ of ghost number 1.
Let $\{\chi_{1,\alpha}\}$ denote a complete set of
vertex operators of ghost number 1, so that we can expand $\Phi$ as
 \be \label{eeb2}
\Phi = \sum_\alpha \vp_\alpha \chi_{1,\alpha}\, .
 \ee
$\{\vp_\alpha\}$ are the dynamical variables of the string field
theory.\footnote{Presumably these $\vp_\alpha$'s are related to the 
$\phi_\alpha$'s
appearing in eq.\refb{e4.9} by a complicated field redefinition.} We also 
define
 \be \label{eeb0a}
|V\ra = b_{-1} |\Phi\ra\, ,
 \ee
and let $V$ be the vertex operator associated with $|V\ra$. Clearly $V$
has ghost number 0. The boundary string field theory action
$S_B$ is a function of the variables $\{\vp_\alpha\}$ given by the 
solution to the equation\cite{9208027}
 \be \label{eeb3}
{\p S_B\over \p \vp_\alpha} = - {\TT_p\over 2} \, \int
{d\theta\over
2\pi} \, \int {d\theta'\over 2\pi} \, \la \chi_{1,\alpha}(\theta)\, \{
Q_B,
\Phi(\theta')\}\ra_V\, .
 \ee
Here
$\TT_p$ is the tension
of the
D-$p$-brane,
$Q_B$ is the usual BRST charge defined as the contour integral of
the BRST currents $j_B(z)$, $\bar j_B(\bar z)$,
and
$\la \cdot \ra_V$ denotes the correlation function in the two
dimensional
field theory on a unit disk,
described by the world-sheet action:
 \be \label{eeb4}
s_{Bulk} + \int_0^{2\pi} \, {d\theta\over 2 \pi} \, V(\theta)\, .
 \ee
Here the angle $\theta$ parametrizes the boundary of the disk, and
$s_{Bulk}$ denotes the bulk world-sheet action involving matter and ghost
fields. In computing $\la \cdot \ra_V$ we must use unnormalized
correlation functions with the convention that in the absence of any
boundary deformation,
 \be \label{eghno}
-{1\over 2} \, \la c(P) \p c(P) \p^2 c(P)\ra_{V=0}  = V_{p+1}\, ,
 \ee
where $P$ is any point on the disk and $V_{p+1}$ is the volume of 
space-time along the D-$p$-brane whose
dynamics we are describing.
Note that although in general the conformal invariance is broken by the
boundary interaction term for the action
\refb{eeb4}, the theory in the bulk is still conformally invariant, and
hence $j_B(z)$, $\bar j_B(\bar z)$ in the bulk are well defined. This can 
be used to define
$Q_B$ appearing in \refb{eeb3}. The overall normalization constant in
\refb{eeb3} is fixed by
comparing the open string amplitudes computed from the action
\refb{eeb3} with those computed using the world-sheet
theory\cite{9210065} or in cubic open string field
theory\cite{0009191}.

This gives the BSFT action for a given string field configuration
$|\Phi\ra$. The expression for the action simplifies for special class of 
$|\Phi\ra$ of the
form\footnote{At this stage we cannot claim that \refb{espp1} corresponds 
to a consistent truncation. However the string field configuration 
\refb{een14} 
that will be of interest to us does correspond to a consistent 
truncation.}
 \be \label{espp1}
\Phi = c V_m\, ,
 \ee
where $V_m$ is a linear combination of primary operators in the matter
BCFT. Let
$\{V_{mj}\}$ denote a complete set of primary vertex operators in the
matter part
of
the BCFT, so that we can expand $V_m$ as 
 \be \label{evmpara}
V_m = \sum_j \lambda_j
V_{mj}\, .
 \ee  
If $\Delta_j$ denotes the conformal weight of $V_{mj}$ then
 \be \label{exxxx1}
\{Q_B, \Phi(\theta)\} =- \sum_j (\Delta_j-1)
\lambda_j
V_{mj}(\theta) c\partial c(\theta)\, .
 \ee
\refb{eeb3} now gives:
 \be \label{espp1aa}
{\p S_B \over \p \lambda_i} =-{\TT_p\over 2} \, \sum _j (\Delta_j-1) \,
\lambda_j
\, \int
{d\theta\over
2\pi} \, \int {d\theta'\over 2\pi} \, \la\, c(\theta) \,
V_{mi}(\theta)\,
c(\theta')\, \p c(\theta')\, V_{mj}(\theta')\, \ra_V \, .
 \ee
This equation however is not covariant under a change in the coordinate
system $\lambda_i$ labelling the two dimensional quantum field
theories\footnote{Such a change in
the coordinate system may be induced by a
change in the
renormalization scheme in this quantum field theory.} 
and holds only in
a
special coordinate system in which the
relation \refb{exxxx1} continues to hold even in the deformed theory 
described by the action \refb{eeb4}.
A covariant version of this equation which is independent of the way we 
choose the parameters $\vec\lambda$ labelling $V_m$ takes the 
form\cite{9303143}
 \be \label{espp2}
{\p S_B \over \p \lambda_i} \propto \sum_j\, \beta^j(\vec\lambda)
G_{ij}(\vec
\lambda)\, ,
 \ee
where $\beta^j$ is the beta-function determining the renormalization group
flow of $\lambda^j$ and $G_{ij}$ is the Zamolodchikov metric in the space
of two dimensional theories deformed by boundary vertex operators in
the matter sector. For the parametrization of $V_m$ given in 
\refb{evmpara} and a suitable renormalization scheme we have
 \be \label{echeck}
\beta^j(\vec\lambda)\propto (\Delta_j-1)\lambda_j, \qquad
\la c(\theta) V_{mi}(\theta) c(\theta') \p c(\theta') V_{mj}(\theta')\ra_V 
\propto G_{ij}(\vec\lambda)\, ,
 \ee 
and 
\refb{espp2}
agrees with \refb{espp1aa}. Eq.\refb{espp2} makes it clear that a
conformally
invariant
two dimensional field theory in the matter sector, for which
$\beta^j$'s
vanish, corresponds to a solution of the BSFT equations of motion.

There is however a difficulty in
defining the action using \refb{eeb3} for string field configurations
associated with vertex operators of dimension $>1$ since they give rise to
non-renormalizable world-sheet field theory.
For this reason a systematic quantization procedure for this field
theory has not yet been developed.
Fortunately in our analysis
of tachyon condensation using classical BSFT we shall not need to deal
with such
operators. In particular in order to study classical solutions in BSFT
involving the tachyon field, we shall focus on the string field
configurations $|\Phi\ra$ of the form
 \be \label{eeb6}
\Phi = c T(X)
 \ee
where $T(x)$ is some function which has the interpretation of
being the tachyon profile in BSFT. In this case $V=T(X)$ and
 \be \label{elapl}
\{Q_B, \Phi\} = \{ Q_B, c T(X)\} = c\, \p c \, (\square_X+1)T(X)\, ,
 \ee
where
 \be \label{edefbox}
\square_x = \eta^{\mu\nu} {\p\over \p x^\mu} {\p\over \p x^\nu}\, .
 \ee

First consider the case of constant tachyon $T=a$. In this case
$\Phi(\theta)=a\, c(\theta)$
and $V(\theta) = a$ (multiplied by the
the identity operator $I$ in the world-sheet theory).
This gives
 \be \label{eeb7}
\int_0^{2\pi} \, {d\theta\over 2 \pi} \, V(\theta) = a\, .
 \ee
Thus the world-sheet correlation functions are
the same as the ones in the original theory except for a factor of
$e^{-a}$. This in particular shows that the right hand side of \refb{eeb3}
vanishes for $a\to\infty$ and hence in this limit we have a solution of
the equations of motion. In order to find the value of the action as
$a\to\infty$ we note that, using $\{Q_B, c(z)\} = c \p c(z)$, \refb{eeb3} 
gives
 \be \label{eeb8}
{\p S_B \over \p a} =  -{\TT_p\over 2} \, a \, e^{-a} \, \int
{d\theta\over
2\pi} \, \int {d\theta'\over 2\pi} \, \la c(\theta)\, c\p
c(\theta')\ra\, ,
 \ee
where the correlation function on the right hand side now has to be
calculated in the original theory.
Using
 \be \label{eeb9}
\la c(\theta)\, c\p
c(\theta')\ra = - V_{p+1} \, |e^{i\theta} - e^{i\theta'}|^2 = - 4
V_{p+1} \, \sin^2{\theta -
\theta'\over 2} \, ,
 \ee
we get
 \be \label{eeb10}
{\p S_B \over \p a} = \TT_p\, V_{p+1} \, a \, e^{-a}\, .
 \ee
This gives\cite{0009103,0009148}
 \be \label{eeb11}
S_B(a) = - V_{p+1} \, \TT_p \, (a+1) \, e^{-a}\, ,
 \ee
up to an additive constant.
Since $-S_B / V_{p+1}$ can be identified as the potential, we see that the
tachyon potential as a function of $a$ is given by
 \be \label{eeb12}
\VV(a) = \TT_p \, (a+1) \, e^{-a}\, .
 \ee
Thus
the difference between the values of the potential at the maximum $a=0$
and the local minimum at $a=\infty$ is given by $\TT_p$,
in accordance with the conjecture 1.

In order to see how a D-$q$-brane may be obtained as a
solution of the BSFT equations of motion on a D-$p$-brane,
we consider a
more general tachyon profile\cite{0009148}
 \be \label{eeb13}
T = a + \sum_{i=q+1}^p u_i (x^i)^2 \, .
 \ee
This corresponds to a boundary deformation of the form \refb{eeb4} with
 \be \label{een14}
V = a + \sum_{i=q+1}^p u_i
(X^i)^2\, .
 \ee
Since this term is quadratic in the fields, the world-sheet theory is
still exactly solvable. In particular on the unit
disk the
normalized two point function of the $X^i$'s in the presence
of this boundary deformation is
given by\cite{9210065}:
 \ben \label{e2pta}
\GG_{ij}\left(\theta, \theta'\right)  &\equiv& \la X^i(
\theta) X^j(\theta') c\p c\p^2
c(P)\ra_V /
\la c\p c\p^2 c(P) \ra_V \nonumber \\
&=& \left[-2\ln\Big|1 -
e^{i(\theta'-\theta)}\Big| +
{1\over 2
u_i} - 4
u_i \sum_{k=1}^\infty {1\over k (k+2u_i)} \,
\cos\left(k(\theta-\theta')\right)\right]\, \delta_{ij}\, . \nonumber \\
 \een
Here $P$ is any point on the disk. The only role of the $c\p c\p^2c$
factor in this
expression is to soak up the ghost zero modes.
Defining
 \be \label{enorma}
:(X^i(\theta))^2: = \lim_{\theta'\to \theta} \left(
X^i(\theta')
X^i(\theta) +
2\, \ln
\Big| 1 -
e^{i(\theta'-\theta)}\Big |\right)
 \ee
we get
 \be \label{enormvev-}
\la :(X^i(\theta))^2: c\p c\p^2 c(P)\ra_V / \la c\p c\p^2 c(P) \ra_V
= {1\over 2
u_i} - 4
u_i \sum_{k=1}^\infty {1\over k (k+2u_i)} \, .
 \ee
This may be rewritten as\cite{9210065}
 \be \label{enormvev}
\la :(X^i(\theta))^2: c\p c\p^2 c(P)\ra_V / \la c\p c\p^2 c(P) \ra_V
= -
{d\over du_i} \ln
Z_1(2u_i)\, ,
 \ee
where
 \be \label{eeb15a}
Z_1(v) = \sqrt{v} e^{\gamma v} \Gamma(v)  \, .
 \ee
$\gamma$ is the Euler number.
The
partition
function $\la  c\p c\p^2 c(P)\ra_V$ on the disk in the presence of
the boundary
deformation $\int{d\theta\over 2\pi} V(\theta)$ with $V(\theta)$
given by \refb{een14} is now obtained by solving the equation:
 \be \label{epdiff}
{\p \la  c\p c\p^2 c(P) \ra_V \over \p u_i} = -\int {d\theta\over 2\pi}
\, \la :(X^i(\theta))^2:  c\p
c\p^2 c(P)\ra_V = \la  c\p c\p^2 c(P) \ra_V \, {d\over du_i} \ln Z_1(2u_i)
\, .
 \ee
A solution to this equation
is given by
 \be \label{epart}
-{1\over 2} \, \la  c\p c\p^2 c(P)\ra_V = e^{-a} \, V_{q+1} \,
(2\pi)^{(p-q)/2}  \,
\prod_{i=q+1}^p Z_1(2u_i)\, ,
 \ee
where
$V_{q+1}$ denotes the volume of the $(q+1)$ dimensional space-time
spanned by $x^0$, $x^1$, $\ldots$ $x^q$.
Note that the differential equation \refb{epdiff} determines
only the $u_i$ dependence of the right hand side of
\refb{epart}, and does not determine its dependence on $a$ or
the overall normalization. The $a$ dependence is fixed trivially by
noting that its presence
gives rise to an additive constant $a$ in the world-sheet action and
hence a multiplicative factor of $e^{-a}$ in any unnormalized
correlation function.
The constant part of the normalization factor is determined by
requiring that the answer reduces to $V_{p+1}$ for $a=u_i=0$ in
accordance with \refb{eghno}. This is however somewhat subtle since the 
right hand side of \refb{epart} contains a factor of $V_{q+1}$ instead of 
$V_{p+1}$. The reason for this
is
that for non-zero $u_i$'s the integration over the zero modes $\int d x^i$ 
for $i=q+1,\ldots p$
is replaced by:
 \be \label{eeb16}
\int d x^i e^{-u_i (x^i)^2} = \sqrt{\pi/u_i}\, .
 \ee
Thus in the $u_i\to 0$ limit, we should make the replacement
 \be \label{ereplacem}
V_{q+1}
\prod_{i=q+1}^p \sqrt{\pi/u_i}\quad\longrightarrow\quad V_{p+1}\, .
 \ee
 With this
prescription
the right hand side of \refb{epart} reduces to $V_{p+1}$ for $a\to 
0$, $u_i\to 0$ in
accordance
with \refb{eghno}.

In order to calculate
$S_B(a,\vec u)$ using \refb{eeb3} for the field configuration
\refb{eeb13} we need to compute $\{Q_B\, , \,
c(\theta')(X^i(\theta'))^2\}$. This is
done using \refb{elapl}. Since
$\square_X (X^i)^2=2$, we get
 \be \label{ethird}
\{Q_B\, , \,
c(\theta')(X^i(\theta'))^2\} =  c\p c(\theta') \left(2 +
(X^i(\theta'))^2\right)\, .
 \ee
Thus \refb{eeb3} gives
 \be \label{ebnew1}
{\p S_B \over \p u_i} = -{\TT_p\over 2} \, \int
{d\theta\over
2\pi} \, \int {d\theta'\over 2\pi} \, \left \la c(\theta)
(X^i(\theta))^2 \, c\p
c(\theta') \left(
a + \sum_{i=q+1}^p \, u_i \left(2+(X^i(\theta'))^2\right)\right)
\right\ra_V \, .
 \ee
Also
 \be \label{ebnew2}
{\p S_B \over \p a} = -{\TT_p\over 2} \, \int
{d\theta\over
2\pi} \, \int {d\theta'\over 2\pi} \, \left\la c(\theta) \, c\p
c(\theta')  \left(
a + \sum_{i=q+1}^p \, u_i \left(2+(X^i(\theta'))^2\right)\right)
\right\ra_V\, .
 \ee
The right hand side of \refb{ebnew2} can be easily evaluated using
eqs.\refb{enormvev}, \refb{epart}. For example, we have
 \be \label{eextran1}
\la c(\theta) c\p c(\theta')\ra_V = 2\, \sin^2\left({\theta - \theta'\over
2}\right) \, \la c \p c \p^2 c(P)\ra_V\, ,
 \ee
 \be \label{eextran2}
\left\la c(\theta) c\p c(\theta') \left( X^i(\theta')\right)^2\right\ra_V
= 2\,
\sin^2\left({\theta - \theta'\over
2}\right) \, \la c \p c \p^2 c(P)\ra_V \left( -{d\over d u_i}\ln Z_1(2
u_i)\right)\, .
 \ee
The correlation function appearing in \refb{ebnew1} can also be evaluated
with the help of Wick's theorem
using the propagator \refb{e2pta} for the $X^i$-fields. The
result for $S_B$, obtained after integrating eqs.\refb{ebnew1},
\refb{ebnew2} is given by\cite{9210065}
 \be \label{eeb15}
S_B(a, \vec u) = -\TT_p \, V_{q+1} \, (2\pi)^{(p-q)/2}\, \left( a + 1 +
2\sum_i u_i - \sum_i
u_i
{\p\over
\p
u_i}\right) e^{-a} \prod_i Z_1(2 u_i)
\, .
 \ee

In order to find a solution of the equations of motion with non-vanishing
$\{u_i\}$, we first solve for $\p S_B/\p a=0$. This gives:
 \be \label{eeb17}
a = \left(-2 \sum u_i + \sum u_i {\p\over \p u_i}\right) \ln Z_1(2
u_i) \,
.
 \ee
Substituting this back into \refb{eeb15} gives
 \be \label{eeb18}
S_B = -\TT_p \, V_{q+1} \, (2\pi)^{(p-q)/2} \, \prod_i \, \exp\left[ 2 u_i
- u_i {\p\over \p
u_i} \ln Z_1(2 u_i) + \ln Z_1(2 u_i) \right] \, .
 \ee
Using the definition of $Z_1(v)$ given in \refb{eeb15a} one can show
that $-S_B$ is a monotone decreasing function of $u_i$ and
hence has minima at $u_i=\infty$. Using Stirling's formula one gets, for
large $u$,
 \be \label{estir}
\ln Z_1(2u) \simeq 2u \ln(2u) - 2u + 2\gamma u + \ln\sqrt{2\pi}\, .
 \ee
This gives, at $u_i=\infty$,
 \be \label{eeb19}
S_B = -\TT_p \, V_{q+1} \, (2\pi)^{p-q} \, .
 \ee
This describes a $q$-dimensional brane along $x^1,\ldots x^q$ with tension
 \be \label{eeb20}
\TT_p \,(2\pi)^{p-q} \, .
 \ee
This is precisely the tension of a D-$q$-brane. Thus we see that BSFT on a
D-$p$-brane
contains as a classical solution D-$q$-branes with $q\le p$, in accordance
with the third conjecture.

Various other aspects of open string tachyon condensation in boundary 
string field theory have been discussed in 
\cite{0010021,0010218,
0011009,0011002,0011033,0012089,
0012198,0012210,0101002,
0101087,0102063,
0103021,0103056,0103079,
0104099,
0104143,0104164,
0105076,0105098,
0105227,0105238,
0105245,0107098,0109032,
0109187,0203185,
0206025,0206102,
0211180,
0301179,
0302125,0308123,
0404237,0408157}.

\sectiono{Non-commutative Solitons} \label{snc}

In closed string theory constant anti-symmetric gauge field configuration
$B_{\mu\nu}$ describes a pure gauge configuration. However this has
non-trivial effect on the dynamics of a D-brane situated in such a
background. For simplicity let us consider the case where only the spatial
components of the $B_{\mu\nu}$ field are non-zero, and the closed
string metric is the flat Minkowski metric. It was shown in
\cite{9711165,9903205,9908142,0006071} that if $S(\{\phi_r\})$ denotes the 
complete tree
level open string field theory action on a D-brane in the absence of
anti-symmetric tensor field background, with $\phi_r$
denoting various
components of string fields, then the open string field theory action in
the presence of the anti-symmetric tensor field background is obtained by
replacing
in $S(\{\phi_r\})$ the original closed string metric $\eta_{\mu\nu}$
by a new `open string metric' $G_{o\mu\nu}$, and
all ordinary products between two field combinations $A$ and $B$ by
the star product:\footnote{In order to make sense of this prescription
we need to know the precise order in which we should arrange the open
string fields before replacing the ordinary product by star product.
The correct prescription is to begin with a system of $N$ coincident
D-branes so that each field is replaced by an $N\times N$ matrix
valued field and ordinary products of fields get replaced by trace
over product of these matrix valued fields in a given cyclic order. In
order to
represent the effect of a background anti-symmetric field
configuration in terms of non-commutative field theory, we arrange the
fields in the same cyclic order even for $N=1$ and then replace the
ordinary product by star product.}
 \be \label{estar1}
A * B \equiv \exp\left( {i\over 2} \theta^{\mu\nu} \p_{x^\mu}
\p_{y^\nu}\right)
A(x)
B(y)\bigg|_{x=y}\, .
 \ee
The anti-symmetric tensor $\theta^{\mu\nu}$
and the symmetric open string metric
$G_o^{\mu\nu}$ along the D-brane
world-volume are
given by the equation:
 \be \label{estar2}
G_o^{\mu\nu} + {1\over 2\pi} \, \theta^{\mu\nu} = \left((\eta +
B)^{-1}\right)_{\mu\nu} \, .
 \ee
Here $B_{\mu\nu}$ on the right hand side is the pull-back of the
anti-symmetric tensor field on the D-brane world-volume. There is also
a change in the effective string coupling constant\cite{9908142}
causing a change in the overall normalization of the action.
We shall determine this indirectly through eq.\refb{eceq} below.

A simplification occurs in the limit of large $B_{\mu\nu}$.
For
simplicity let us consider the case where only the $1-2$ component of
$B_{\mu\nu}$ is non-zero, with $x^1$ and $x^2$ being directions tangential
to the D-brane, and we take $B_{12}\equiv B$ to
be large. In this limit:
 \be \label{estar3}
G_o^{11} = G_o^{22} \simeq {1\over B^2}, \qquad G_o^{12} =
G_o^{21}=0,
\qquad
\theta^{12} = -\theta^{21} \simeq -{2\pi\over B}\, .
 \ee
Note that in the $B\to\infty$ limit $G_o^{ij}/\theta^{12}$ vanishes for
$i,j=1,2$. Since after replacing $\eta^{\mu\nu}$ by $G_o^{\mu\nu}$
in $S(\{\phi_r\})$
all explicit derivatives with respect to the $x^1$
and $x^2$
coordinates in
$S(\{\phi_r\})$
are contracted with the metric $G_o^{ij}$, we see that such derivative
terms may be neglected in the large $B$ limit. If we focus on classical
solutions of the effective action which depend only on these two
coordinates, then we can simply drop all explicit derivative terms
from the effective action, keeping only the implicit derivatives coming
through the definition of the star-product \refb{estar1}. In other words
if $-\int d^{p+1} x \, V(\{\phi_r\})$ denotes the part of the original
action
$S(\{\phi_r\})$ without any
derivative
terms,
then in order to look for solutions which depend only on $x^1$ and
$x^2$ in the presence of a strong anti-symmetric tensor field in the
1-2 plane, we can work with
the string field theory action:
 \be \label{estar4}
-C\int d^{p+1} x \, V_*(\{\phi_r\})\, ,
 \ee
where the subscript $*$ in $V$ denotes that all the products of fields
inside $V$ are to be interpreted as star product, and the constant
$C$ takes into account the change in the overall normalization of the
action due to the $\sqrt{-\det G_o}$ term that will multiply the
Lagrangian density, and the overall change in the normalization of the
action due to the change in the string coupling constant. The
equations of motion now are:\footnote{In order to determine how the
fields are arranged inside $(\p_sV)_*$, we begin with a term in 
$V_*(\{\phi_r\})$ and, using the cyclicity of the star product under the 
integral, bring the $\phi_s$ factor with respect to which we are 
differentiating to the extreme left. The derivative of this term with 
respect to $\phi_s$ is then given by removing the $\phi_s$ factor from the 
string, leaving the rest of the terms in the same order.}
 \be \label{estar5}
(\p_s V)_*(\{\phi_r\}) = 0\, .
 \ee
For later convenience we shall choose the string field $\{\phi_r\}$
such
that $\{\phi_r=0\}$
denotes the tachyon vacuum, and $V(\{\phi_r\})$ vanishes at
$\{\phi_r=0\}$. This convention is different from the one used in
section \ref{s4} for example, where $\{\phi_r=0\}$ describes the
original D-brane on which the string field theory is formulated.

A general method for constructing soliton solutions to
eq.\refb{estar5} was developed in \cite{0003160}. Suppose
$\{\phi^{(0)}_r\}$ denotes a translationally invariant
configuration corresponding to a local extremum of the tachyon
potential:
 \be \label{estar6}
\p_s V(\{\phi^{(0)}_r\}) = 0\, .
 \ee
In particular we can choose $\{\phi^{(0)}_r\}$ to be the configuration
describing the original D-brane on which open string field theory is
defined.
Now suppose $f(x^1,x^2)$ denotes a function such that
 \be \label{estar7}
f * f = f\, .
 \ee
Then for
 \be \label{estar8}
\phi_r(x) = \phi^{(0)}_r \, f(x^1,x^2)\, ,
 \ee
we have
 \be \label{epsvstar}
(\p_sV)_*(\{\phi_r\}) = \p_s V(\phi^{(0)}_r) \, f(x^1, x^2) = 0\, .
 \ee
Thus \refb{estar8}
is a solution of \refb{estar5}. For this solution
 \be \label{estar}
V_*(\{\phi_r(x)\}) = V(\{\phi^{(0)}_r\}) \, f(x^1,x^2)\, .
 \ee
Thus from eq.\refb{estar4} we see that the energy per unit
$(p-2)$-volume associated
with this solution, obtained by integrating the energy density
$CV_*(\{\phi_r(x)\})$ over $x^1$ and $x^2$, is:
 \be \label{estar9}
C \, V(\{\phi^{(0)}_r\}) \, \int \, d
x^1\, dx^2 \, f(x^1, x^2) \, .
 \ee

This allows us to construct space dependent solutions by starting with a
translationally invariant solution. We shall now show that if we take
$\phi^{(0)}_r$ to be the translationally invariant solution
describing
the original D-brane, then for a suitable choice of $f$
satisfying \refb{estar7}, \refb{estar8}
represents a soliton solution describing a codimension 2
D-brane\cite{0005006,0005031,0006071}.
In this case $V(\{\phi^{(0)}_r\})$
denotes the tension
$\TT_p$ of
the
original D-brane on which we have formulated the string field theory
in the absence of any background anti-symmetric tensor field. Also due
to eq.\refb{estar4}
$CV(\{\phi^{(0)}_r\})$ should describe the tension of the D-brane in
the presence of the background $B$-field. Since in the presence of the
$B$-field the tension of a D-brane gets multipled by a factor of
$\sqrt{-\det(\eta+B)}$ which in the present example takes the value
$\sqrt{1+B^2}\simeq B$ for large $B$, we have
 \be \label{eceq}
C\, V(\{\phi^{(0)}_r\}) = B\, \TT_p \, .
 \ee
Thus \refb{estar9}
takes the form:
 \be \label{estar10}
B\, \TT_{p} \, \int \, d
x^1\, dx^2 \, f(x^1, x^2) \, .
 \ee
This is
the energy per unit $(p-2)$-volume of the solution \refb{estar8}.

An $f$ satisfying \refb{estar7} is given by\cite{0003160}:
 \be \label{estar11}
f(x^1, x^2) = 2\exp\left(-{B\over 2\pi} \left((x^1)^2 + (x^2)^2\right)
\right)\, .
 \ee
This 
can be 
checked by rewriting \refb{estar7} in
terms of Fourier transform of $f$, and noting that if $\wt A(k)$, $\wt 
B(k)$ and $\wt{A*B}(k)$ denote the Fourier transforms of $A(x)$, $B(x)$ 
and 
$A*B(x)$ respectively, then eq.\refb{estar1} takes the form
 \be \label{estarfourier}
 \wt{A*B}(k) = \int {d^{p+1}q\over (2\pi)^{p+1}} \, \exp\left( -{i\over 
2} \, \Theta^{\mu\nu} q_\mu (k-q)_\nu\right) \wt A(q) \wt B(k-q)\, .
 \ee
In 
this case
 \be \label{etar12}
\int d x^1 dx^2 f(x^1, x^2) = {4\pi^2\over B}\, ,
 \ee
and hence the tension of the codimension 2 solution, as given by
\refb{estar10}, is
 \be \label{estar13}
4\pi^2 \, \TT_p \, .
 \ee
This is precisely the tension of a D-$(p-2)$-brane. This shows
that we can identify the non-commutative soliton given in
eqs.\refb{estar8}, \refb{estar11} with a codimension two D-brane,
in accordance with conjecture 3.\footnote{Note that this analysis
does not provide a verification of conjecture 1, but given
conjecture 1, it provides a verification of conjecture 3.}

Eq.\refb{estar7} for $f$ is the requirement that $f$ is a projection
operator under the star product. The $f$ given in \refb{estar11}
describes a rank one projector\cite{0003160}. In general, if we take
$f$ to be a rank $n$ projector, it describes a configuration of $n$
D-$(p-2)$-branes. By making appropriate choice of $f$, one can
construct multi-soliton
solutions describing multiple D-branes located at arbitrary points in the
transverse space\cite{0003160,0103256}. We shall not discuss this
construction,
and refer
the interested reader to the original literature.

Various other aspects of non-commutative tachyon condensation have been 
discussed in 
\cite{9912274,0007226,0008013,0008023,0008214,
0009038,0009142,0010016,
0010028,0010058,0010060,0010101,0011094,
0012081,0007078,0007217,
0008064,0009002,
0009030,0009101,
0011079,0011090,
0011223,0012217,
0101001,
0101125,
0101145,0101199,
0104090,0104176,
0104263,0105115,
0106142,0301119,
0407229}.

\sectiono{Time Dependent Solutions} \label{s10}

In this section we shall begin by outlining a general procedure for
constructing time dependent solutions in string theory describing time
evolution of a field (or a set of fields) from a given `initial
condition', and then apply this to the construction of time dependent
solutions describing the rolling of the tachyon on an unstable D-brane
system away from the maximum of the potential. Our discussion will closely
follow that in \cite{0207105}. Throughout this analysis we shall restrict 
our attention to time independent closed string background for which the 
matter part of the bulk CFT is given by the direct sum of the theory of a 
free 
scalar (super-)field $X^0$ representing the time coordinate and a 
unitary conformal field theory of central charge $c=25$ ($\hat c=9$).

\subsection{General procedure}

Let us begin with some general unstable D-brane system with a tachyonic
field $\phi$ of mass$^2=-m^2$. This could either describe the tachyons of
the kind we discussed earlier, or more general tachyon, {\it e.g.} the
tachyon on a D2-D0-brane system coming from open strings stretched from
the D0 to the D2-brane. Since $\phi$ is tachyonic, its potential $V(\phi)$
has a
maximum at $\phi=0$ with $V''(0)=-m^2$.
If $\phi$ had been described by the
action of a standard scalar
field theory with two derivative kinetic term plus a potential term, then
the motion of $\phi$ away from the maximum will be characterized by two
parameters, the initial value of $\phi$ and its first time
derivative.\footnote{For simplicity we are considering only spatially
homogeneous field configurations here.}$^,$\footnote{One of these
parameters can be
fixed using the time translation invariance of the system; we simply
choose the origin of time where either the field or its first time
derivative vanishes.}
However, given that string field theory action has inifinite number of
time (and space) derivatives, it is not {\it a priori} clear if such a
set of solutions can be constructed in string theory as well. We shall now
argue that it is indeed possible to construct a similar set of solutions
in string theory; and give an algorithm for constructing these solutions.

The construction will be carried out by using the well-known
correspondence between the solutions of classical equations of
motion in string theory and two dimensional conformal field
theories. Boundary conformal field theories associated with time dependent 
open string field configurations involve boundary interaction terms which 
depend on the time coordinate field $X^0$ in a non-trivial manner. 
Since such conformal field theories are difficult to analyze directly,
we shall first construct a solution that
depends non-trivially on a space-like coordinate $x$ and then
replace $x$ by $i x^0$. The new
configuration will represent a time dependent solution of the
equations of motion. All we need to ensure is that the solution
obtained this way is real.

Whereas this gives a general procedure for constructing time
dependent solutions in string theory, we are looking for a
specific kind of time dependent solution, -- that which describes
the rolling of $\phi$ away from the maximum of $V(\phi)$. So the
next question is: which particular euclidean solution should we
begin with in order to generate such time dependent solutions? The
clue to this answer comes from looking at the solution of the
linearized equation of motion near the maximum of the potential.
Since the higher derivative terms in string field theory are all
in the interaction term, they do not affect the linerarized
equation of motion for $\phi$, which takes the standard form:
 \be \label{eff1}
(\p_0^2 + m^2)\phi \simeq 0\, .
 \ee
Using time translation invariance we can choose the boundary condition on
$\phi$ to be either
 \be \label{eff2}
\phi=\lambda, \quad \p_0 \phi=0, \quad \hbox{at} \quad x^0=0,
 \ee
or
 \be \label{eff3}
\phi=0, \quad \p_0 \phi=m\lambda, \quad \hbox{at} \quad x^0=0\, .
 \ee
For a conventional scalar field
\refb{eff2} holds when the total energy density of the system is less than
$V(0)$
so that the field comes to rest at a point away from 0 and \refb{eff3}
holds when the total energy density of the system is larger than $V(0)$ so
that
the field $\phi$ passes 0 with non-zero velocity during its motion. We 
shall see that the same interpretation holds for open string field theory 
as well. For
the boundary condition \refb{eff2} the solution  to \refb{eff1}
is given by:
 \be \label{eff4}
\phi(x^0) \simeq \lambda \cosh(m x^0)\, ,
 \ee
whereas for the boundary condition \refb{eff3} the solution
is:
 \be \label{eff5}
\phi(x^0) \simeq \lambda \sinh(m x^0)\, .
 \ee
Both solutions are valid for small $\lambda$ and finite $x^0$ {\it i.e.}
as long as $\phi$ is small. Thus the one parameter family of solutions of
the full string field theory equations of motion that we are looking for
must have the property that for small $\lambda$ it reduces to the form
\refb{eff4} or \refb{eff5}.

For definiteness we shall from now on concentrate on the class of
solutions
with total energy density less than $V(0)$, but the analysis can be easily
generalized to the other case.\footnote{In fact the solution of type
\refb{eff5}
can be obtained from those of type \refb{eff4} by the formal replacement
$x^0\to x^0 - i\pi/2m$, $\lambda\to i\lambda$.} We now note that
\refb{eff4} can be obtained from the Euclidean
solution:
 \be \label{eff6}
\phi(x) 
\simeq \lambda \cos(mx)\, ,
 \ee
under the replacement $x\to ix^0$. Thus we need to search for a one
parameter family of euclidean solutions which for small value of the
parameter $\lambda$, reduces to \refb{eff6}. Given such a one parameter
family of solutions, we can construct one parameter family of time
dependent solutions by the replacement $x\to ix^0$.\footnote{The original
idea of constructing time dependent solution in open string theory by Wick 
rotating Euclidean
solution is due to ref.\cite{0202210}.}

To proceed further, we shall, for definiteness, concentrate on the bosonic
string theory, although the analysis can be easily generalized to the
superstring theory. Since the tachyon has mass$^2=-m^2$,
the zero momentum tachyon vertex
operator $V_\phi$ must have dimension $(1-m^2)$ so that $V_\phi e^{ik.X}$
has dimension 1 for $k^2=m^2$. For small $\lambda$,
switching on
the background \refb{eff6} corresponds to deforming the original boundary
CFT by a boundary perturbation of the form:
 \be \label{eff7}
\lambda \, \int dt \, V_\phi(t) \, \cos(m X(t))\, ,
 \ee
where $t$ denotes a parameter labelling the boundary of the world-sheet.
Since $V_\phi \cos(mX)$
has conformal dimension 1, the deformed theory represents a conformal
field theory to first order in $\lambda$. In order to construct a one
parameter family of solutions of the string field theory equations of
motion, we need to construct a family of conformal field theories labelled
by $\lambda$, which to first order in $\lambda$ agrees with the deformed
theory
\refb{eff7}.

In special cases, $V_\phi\cos(mX)$ may represent an exactly
marginal operator in string field theory, in which case \refb{eff7}
represents a conformal field theory even for finite $\lambda$, and our
task
is over. However even in this case, for finite $\lambda$ the parameter
$\lambda$ appearing
in \refb{eff7} may not agree exactly with the value of $T$ at $x^0=0$
(which in turn depends on the precise definition of the tachyon field, and
varies between different formulations of string field theory for example.)
Thus it is more appropriate to label the perturbation as
 \be \label{eff7a}
\tl \, \int dt \, V_\phi(t) \, \cos(m X(t))\, ,
 \ee
with $\tl=\lambda+\OO(\lambda^2)$.

In the
generic case $V_\phi\cos(mX)$ is not an
exactly marginal operator since its $\beta$-function $\beta_{\tl}$ will
have
contribution
of order $\tl^3$ and higher.\footnote{Note that due to $X\to X+\pi/m$,
$\tl\to-\tl$ symmetry of the deformation,  $\beta_{\tl}$ receives
contribution only to odd orders in $\tl$.} In this case we proceed as
follows. Instead of considering the deformation \refb{eff7a}, we consider
deformation by the operator:
 \be \label{eff8}
\wt \lambda \, \int dt \, V_\phi(t) \, \cos(\omega X(t))\, ,
 \ee
where $\omega$ is a constant that will be fixed shortly. Since the
perturbing operator has dimension $(\omega^2-m^2+1)$, $\beta_{\tl}$ now
receives contribution linear in $\tl$, and the full $\beta$-function is
given by:
 \be \label{eff9}
 \beta_{\tl} = (\omega^2 - m^2) \tl + g(\omega, \tl)\, ,
 \ee
where $g(\omega, \tl)$ denotes higher order contribution to the
$\beta$-function.
We now adjust $\omega$ so that the right hand side of \refb{eff9} 
vanishes.
Since $g(\omega, \tl)\sim \tl^3$, we can get a solution of the form:
 \be \label{eff10}
\omega=m + \OO(\tl^2)\, .
 \ee

This gives a way to generate a one parameter family of boundary CFT's
labelled by $\tl$, which for small $\tl$ corresponds to the solution
\refb{eff6} with $\lambda\simeq \tl$. Given this conformal field theory,
we can calculate the
energy momentum tensor and sources of other massless fields like the
dilaton, anti-symmetric tensor field etc. from the boundary state
associated with this BCFT. (If the deformed boundary CFT is solvable then
we can find an exact expression for the boundary state as in section
\ref{s3.4}; otherwise we may need
to compute it as a perturbation series in $\tl$.)

Note that in the above analysis we have not included the
$\beta$-functions for any operator other than the original perturbing
operator. In order to get a conformal field theory we need to ensure
that the $\beta$-function for every other operator also vanishes. What we
have done here is to implicitly assume that the other operators have been
`integrated out' (in a space-time sense) so that we can talk in terms of
the `effective' $\beta$-function of this single operator
$V_\phi\cos(\omega X)$. To see explicily what this means
let $\{\OO_i\}$ denote a complete set of boundary operators in the theory
other than $V_\phi\cos(\omega X)$,
and let
$h_i$ be the conformal weight of $\OO_i$. We add to the action the
boundary term
 \be \label{eff11}
\sum_i \, g_i\,  \int dt \, \OO_i(t)\, ,
 \ee
besides \refb{eff8}. Then the $\beta$-function $\beta_i$ of $g_i$ and
$\beta_{\tl}$ have the form:
 \be \label{eff12}
\beta_i = (h_i - 1) g_i + F_i(\omega, \vec g, \tl)\, , \qquad
\beta_{\tl} = (\omega^2 - m^2) \tl + F(\omega, \vec g, \tl)\, .
  \ee
$F_i$ and $F$ contains terms quadratic and higher order in $g_i$ and
$\tl$. {}From
this we see that as long as $h_i \ne 1$, we can solve for the $g_i$'s and
$(\omega-m)$ in
power series in $\tl$ beginning at quadratic or higher order. In 
particular we can first solve the $\beta_i=0$ equations to express each 
$g_i$ as 
a function of $\tl$, and substitute this back into the expression for 
$\beta_{\tl}$ to get an equation of the form \refb{eff9}. This procedure 
of `integrating out' the other operators
breaks down if there are operators of dimension $\simeq 1$ other than
$V_\phi\cos(\omega X)$ since in this case we can no longer solve the 
$\beta_i=0$ equations to express $g_i$ as a power series in $\tl$.
This difficulty is of course a
reflection of the well known difficulty in integrating out the massless
fields.

In particular we can consider operators $\p X$ or $V_\phi \sin(\omega X)$
both of
which will have dimension $\simeq 1$. Fortunately both these operators are
odd under $X\to -X$, and hence are not generated in the operator product
of
$V_\phi \cos(\omega X)$ with itself, since the latter operator is even
under $X\to -X$. In the generic case we do not expect any other dimension
$\simeq 1$ operator to appear in the operator product of $V_\phi
\cos(\omega X)$ with itself, but there may be special cases where such
operators do appear.\footnote{In fact the general method outlined here
also holds
for generating time dependent solutions in closed string theory, but
the dimension (1,1) operators associated with the zero momentum
graviton/dilaton vertex operators cause obstruction to this procedure. It
may be possible to avoid this problem in
case of localized tachyons\cite{0108075}, since in this case the rolling
tachyon
generates a localized source for the bulk graviton/dilaton field, and
hence we can solve the equations of motion of these massless fields in the
presence of these localized sources in order to get a consistent conformal
field theory.}

In order to generate the time dependent solution, we now
make the replacement
$X\to iX^0$. This corresponds to deforming the original boundary CFT by
the operator
 \be \label{eff13}
\tl\, \int dt \, V_\phi(t) \, \cosh(\omega X^0(t))\, .
 \ee
$\omega$ is determined in terms of $\tl$ by the same equation as in the
euclidean case.\footnote{If we are using the picture in which the higher
modes have not been integrated out,  then we
also need
to add perturbations generated by the Wick rotated version of
\refb{eff11}. These pictures differ from each other by a choice of
renormalization scheme for the world-sheet field theory.} The
corresponding energy-momentum tensor and sources
of other massless fields are also obtained by the making the replacement
$x\to ix^0$ in the corresponding expressions in the euclidean theory.

This illustrates the general method for constructing a one parameter
family of solutions labelled by $\tl$ which reduce to \refb{eff4} for
small $\tl$. In order to generate another one parameter family of BCFT
labelled by $\tl$ which reduce to \refb{eff5} for small $\tl$, we simply
make a replacement $X^0\to X^0-i{\pi\over 2\omega}$, $\tl\to i\tl$ in the
solution derived above. This corresponds to deforming
the original BCFT by the operator:
 \be \label{eff14}
\tl\, \int dt \, V_\phi(t) \, \sinh(\omega X^0(t))\, ,
 \ee
and hence corresponds to a solution of the form \refb{eff5} for small 
$\tl$ since $\omega\to m$ as $\tl\to 0$.

It is clear that the same method can be used to construct time
dependent solutions describing rolling of tachyons on unstable D-brane
systems in superstring theory as well.
The method can also be generalized
to describe simultaneous rolling of multiple tachyons\cite{0207105}.

\subsection{Specific applications}

We shall now apply the general method discussed in the last section to
describe the rolling of spatially homogeneous tachyon field configuration
on a D$p$-brane of bosonic string theory lying along $x^0,x^{26-p},\ldots 
,
x^{25}$\cite{0203211,0203265}. In this case the vertex operator
$V_\phi$ of the zero momentum tachyon is just the identity operator, and
the tachyon has mass$^2=-1$. As a
result, to lowest order in $\tl$, the analog of the perturbation
\refb{eff7a}
in the euclidean theory is given by:
 \be \label{ehx1}
\wt \lambda \, \int dt \, \cos(X(t))\, .
 \ee
This is identical to the perturbation \refb{e3.2} with $\tl$ identified to
$-\al$ and represents an
exactly marginal deformation of the original BCFT.\footnote{In the
analysis of the perturbation \refb{e3.2}
we took $X$ to be compact whereas here $X$, being related by Wick
rotation to the time coordinate $X^0$, is non-compact. However the
marginality
of the operator does not depend on whether $X$ is compact or not.}
As a result, the deformation \refb{ehx1} represents a BCFT for finite
$\tl$ as well. Put another way, in this case eq.\refb{eff10} is replaced
by
 \be \label{ehx2}
\omega = m=1\, ,
 \ee
for all $\tl$.

Making the replacement $X\to i X^0$ we see that the rolling
tachyon solution is given by perturbing the original BCFT by the operator:
 \be \label{ehx3}
\wt \lambda \, \int dt \, \cosh(X^0(t))\, .
 \ee
The energy-momentum tensor $T_{\mu\nu}$ and the dilaton charge density $Q$
associated with this solution can be obtained
by making the replacement $X\to iX^0$, $X^0\to -i X^{25}$, $\al\to
-\tl$ in
\refb{efg21}, \refb{efg22}, \refb{edcharge}. This gives:
 \ben \label{ehx4}
&& Q = \TT_{p} \, \wt f(x^0) \, \delta(x_\perp)\, ,
\qquad T_{00} =  \TT_{p} \,
\cos^2(\pi\tl) \, \delta(x_\perp), \quad
\nonumber \\
&& T_{MN}
= -\TT_{p}
\,   \wt f(x^0) \, \delta_{MN} \, \delta(x_\perp) \quad \hbox{for}
\quad (26-p)\le M,N\le 25\, ,
 \een
with all other components of $T_{\mu\nu}$ being zero.
Here
 \be \label{ehx5}
\wt f(x^0) = f(i x^0)\big|_{\alpha=-\tl} = {1\over 1 + e^{x^0}
\sin(\tl\pi)} + {1\over 1 + e^{-x^0} \sin(\tl\pi)} - 1\, .
 \ee
This reproduces \refb{e2.18}-\refb{e2.20} and \refb{ebosdil} after taking 
into account the fact that the results in section \ref{s2} were quoted for 
a D$p$-brane lying along $x^0, x^1, \ldots x^p$, whereas the results 
obtained here are for D$p$-branes lying along $x^0$, $x^{26-p}$, $\ldots$ 
$x^{25}$. 

As discussed in the previous subsection, the other one parameter family of
solutions, valid when the total energy carried by the tachyon field
configuration is larger than the tension of the D-brane, is obtained by
making the replacement $x^0\to x^0 - i\pi/2$, $\tl\to i\tl$. Making these
replacements in \refb{ehx4}, \refb{ehx5} gives the non-zero
components of the energy momentum tensor and the dilaton charge
density $Q(x)$
to be
 \ben \label{ehx4a}
&& Q = \TT_{p} \, \wt f(x^0) \, \delta(x_\perp)\, ,
\qquad T_{00} = \TT_{p} \, 
\cosh^2(\pi\tl) \, \delta(x_\perp),
\nonumber \\
&& T_{MN} = -\TT_{p}
\,   \wt f(x^0) \, \delta_{MN} \, \delta(x_\perp) \quad \hbox{for}
\quad (26-p)\le M,N\le 25\, ,
 \een
where
 \be \label{ehx6}
\wt f(x^0) = {1\over 1 + e^{x^0} \sinh(\tl\pi)}
+ {1\over 1 - e^{-x^0} \sinh(\tl\pi)} - 1\, .
 \ee
This reproduces \refb{e2.18}, \refb{e2.22}, \refb{e2.23},
\refb{ebosdil}.

The analysis in the case of D$p$-$\bar{\rm D}p$-brane pair in
superstring theory proceeds in an identical manner. If the total
energy per unit $p$-volume is less than $\EE_p$ where $\EE_p$ is
the tension of the original brane system, we need to switch on a
tachyon background\footnote{The $\sqrt 2$ multiplying $\tl$ is
part of the normalization convention for $\tl$.}
 \be \label{etacb}
T=\sqrt 2 \, {\tl}\, \cosh(X^0/\sqrt 2)\, .
 \ee
This is related to the background \refb{efh5} for $R=1/\sqrt 2$ by
the replacement
 \be \label{ehx10}
 x\to i x^0+ \pi/\sqrt 2\, , \qquad \alpha\to\tl\, ,
 \ee
and hence represents an exactly marginal deformation of the
theory. Making these replacements in \refb{efg24}-\refb{esupdil},  we
get the non-zero components of $T_{\mu\nu}$ and the dilaton charge density 
$Q$ to be:
 \ben \label{ehx11}
T_{00} &=& \EE_p\, 
\cos^2(\pi\tl) \, \delta(x_\perp), \nonumber \\
T_{MN}
&=& -\EE_p
\,    \wt f(x^0) \,
\delta_{MN} \, \delta(x_\perp) \quad \hbox{for} \quad (10-p)\le M,N
\le 9\, , \nonumber \\
Q(x) &=& \EE_p
\,    \wt f(x^0) \,
 \delta(x_\perp) \, ,
 \een
 \be \label{ehx12}
\wt f(x^0) = {1\over 1 + e^{\sqrt 2 x^0} \sin^2(\tl\pi)}
+ {1\over 1 + e^{-\sqrt 2 x^0} \sin^2(\tl\pi)} - 1\, .
 \ee
This reproduces \refb{e2.11}-\refb{e2.13} and \refb{edilcharge} after 
taking 
into account the fact that the results of section \ref{s2} were quoted 
for a brane-antibrane system along $x^0,\ldots x^p$ whereas here the 
original brane-antibrane system is taken to be along $x^0$, 
$x^{10-p},\ldots 
x^9$. 

For total energy $>\EE_p$ we need to switch on a tachyon background
 \be \label{eepge1}
 T = \sqrt 2 \, {\tl}\, \sinh(X^0/\sqrt 2)\, .
 \ee
This is related to \refb{etacb} by the replacement $x^0\to
x^0-i\pi/\sqrt 2$, $\tl\to i\tl$. Making these replacements in
\refb{ehx11}, \refb{ehx12} gives
 \ben \label{ehx13}
T_{00} &=& \EE_p \, 
\cosh^2(\pi\tl) \, \delta(x_\perp), \nonumber \\
T_{MN} &=& -\EE_p
\,    \wt f(x^0) \,
\delta_{MN} \, \delta(x_\perp) \quad \hbox{for} \quad (10-p)\le M,N
\le 9\, , \nonumber \\
Q(x) &=& \EE_p
\,    \wt f(x^0) \,
 \delta(x_\perp) \, ,
 \een
with
 \be \label{ehx14}
\wt f(x^0) = {1\over 1 + e^{\sqrt 2 x^0} \sinh^2(\tl\pi)}
+ {1\over 1 + e^{-\sqrt 2 x^0} \sinh^2(\tl\pi)} - 1\, .
 \ee
All other components of $T_{\mu\nu}$ vanish.
This reproduces \refb{e2.14} - \refb{edilcharge}.

Similar method can be used for getting the corresponding results
for the non-BPS D-brane. In fact since a non-BPS D-brane is
obtained by modding out a brane-antibrane system by $(-1)^{F_L}$,
and since $T_{\mu\nu}$ is invariant under this transformation, the
results for the sources for NS-NS sector fields produced by
non-BPS D-brane are identical to those on a D-$\bd$ pair except
for a change in the overall normalization. Information
about the sources for the RR fields associated with the rolling
tachyon background on a non-BPS D-brane, as given in
eqs.\refb{errcharge}, can be obtained from eq.\refb{erreuclid} by
the replacement \refb{ehx10}. \refb{errchargep} may be obtained
from \refb{errcharge} by the replacement $x^0\to x^0-i\pi/\sqrt
2$, $\tl\to i\tl$.

The boundary state can also be used to determine the source terms
for massive closed string states. For this we need to find the
complete boundary state in the euclidean theory and replace $X$ by
$iX^0$. We shall illustrate this in the context of bosonic string
theory, but in principle the same procedure works for superstring
theory. The $|\BB\ra_{c=25}$ and $|\BB\ra_{ghost}$ parts of the
boundary state are given by eqs.\refb{efg11} and \refb{efg12}
respectively. Thus the non-trivial part is $|\BB\ra_{c=1}$. The
complete expression for $|\BB\ra_{c=1}$ for a general deformation
parameter $\tl$ can be obtained by following the procedure
outlined in section \ref{s3.4}\cite{9402113,9811237}.\footnote{The
analysis of section \ref{s3.4} was carried out for compact $X$.
The result for non-compact $X$ is obtained by simply dropping from
the result of section \ref{s3.4} all terms in the boundary state
carrying non-zero winding number.} In order to describe the
results we need to review a few facts about the 
spectrum
of the $c=1$ conformal field theory described by a single
non-compact scalar field $X$. This theory contains a set of
Virasoro primary states labelled by SU(2) quantum numbers $(j,m)$
and have the form\cite{DVV}\footnote{The underlying SU(2) group is 
inherited from
the theory with $X$-coordinate compactified on a circle of radius
1. Although SU(2) is not a symmetry of the theory when $X$ is
non-compact, it is still useful to classify states with integer
$X$-momentum. In fact the theory with compact $X$ actually has an
SU(2)$_L\times$SU(2)$_R$ symmetry and the theory with non-compact
$X$ contains a more general set of primary states carrying left
and right SU(2) quantum numbers $(j,m)$ and $(j',m)$ respectively.
The states with $j\ne j'$ will not be important for our
discussion.}
 \be \label{epriform}
|j,m\ra = \wh \PP_{j,m} \, e^{2 \, i \, m \, X(0)} |0\ra\, ,
 \ee
where $\wh \PP_{j,m}$ is some combination of the $X$ oscillators
of level $(j^2 - m^2, j^2 - m^2)$. Thus $|j,m\ra$ carries
$X$-momentum $2m$ and $(L_0,\bar L_0)$ eigenvalue
$(j^2,j^2)$. For any given primary state $|j,m\ra$ we have an
associated Virasoro Ishibashi state\cite{ishibashi} $|j,m\ra\ra$, defined 
as the 
unique 
linear combination of the primary $|j,m\ra$ and all Virasoro 
descendants of $|j,m\ra$ such that $L_n - \bar L_{-n}$ 
annihilate the
state for every $n$. We shall normalize the Ishibashi states so
that when we express $\exp\left(\sum_{n=1}^\infty \, {1\over n}
\alpha_{-n}\bar\alpha_{-n}\right)e^{2im X(0)}|0\ra$ as a linear
combination of the Ishibashi states built on various primaries,
the state $|j,m\ra\ra$ 
appears with coefficient 1 for every $j\ge |m|$:\footnote{In this 
convention,
the usual $\delta$-function normalized primary state $|j,m\ra$ appears in 
$|j,m\ra\ra$ with unit coefficient\cite{0408064}.}
 \be \label{edefjm}
\exp\left(\sum_{n=1}^\infty \, {1\over
n}
\alpha_{-n}\bar\alpha_{-n}\right)e^{2im X(0)}|0\ra = \sum_{j\ge |m|}
|j,m\ra\ra\, .
 \ee

For the deformation \refb{ehx1}
$|\BB\ra_{c=1}$
may be expressed in terms of these Ishibashi states 
as\cite{9402113,9811237,0402157}
 \be \label{ebs1}
|\BB\ra_{c=1} = \exp\left(\sum_{n=1}^\infty \, {1\over n}
\alpha_{-n}\bar\alpha_{-n}\right)\, f(X(0)) \, |0\ra +
|\wt\BB\ra_{c=1} \, ,
 \ee
where
 \be \label{edeffagain}
f(x) = {1\over 1 + \sin(\pi\tl) e^{ix}} + {1\over 1 +
\sin(\pi\tl) e^{-ix}} - 1\, ,
 \ee
 \be \label{esecondform}
|\wt\BB\ra_{c=1} = \sum_{j\ge 1} \, \sum_{m=-j+1}^{j-1} \,
f_{j,m}(\tl) \, |j,m\ra\ra\, ,
 \ee
 \be \label{efjm}
f_{j,m}(\tl) = D^j_{m,-m}(2\pi\tl) \, {(-1)^{2m} \over
D^j_{m,-m}(\pi)} -
(-1)^{2m} \, \sin^{2|m|}(\pi\tl)\, .
 \ee
$D^j_{m,m'}(\theta)$ are the representation matrices of the
SU(2)
group element $e^{i\theta\sigma_1/2}$ in the
spin
$j$
representation.

If $|\wh\BB\ra_{c=1}$ denotes the continuation of $|\wt\BB\ra_{c=1}$
to
the Minkowski space, then the complete boundary state $|\BB\ra$ in the
Minkowski
space may be expressed as:
 \be \label{ebs2}
|\BB\ra = |\BB_1\ra + |\BB_2\ra\, ,
 \ee
where
 \ben \label{ebs3}
|\BB_1\ra &=& \TT_p\,  \exp\left(-\sum_{n=1}^\infty \, {1\over n}
\alpha^0_{-n}\bar\alpha^0_{-n}\right)\, \wt f(X^0(0))\, |0\ra \nonumber \\
&& \otimes
\int {d^{25-p} k_\perp \over (2\pi)^{25-p}}\,
\exp\left(\sum_{n=1}^\infty
\sum_{s=1}^{25} (-1)^{d_s} \, {1\over
n} \,  \alpha^s_{-n} \bar
\alpha^s_{-n} \right) |\vec k_\parallel=0,
\vec k_\perp\ra\, \nonumber \\
&& \otimes
\exp\left(-\sum_{n=1}^\infty (\bar b_{-n} c_{-n} +
b_{-n}
\bar c_{-n})
\right) (c_0+\bar c_0)c_1\bar c_1 |0\ra\, ,
 \een
and
 \ben \label{ebs4}
|\BB_2\ra &=& \TT_p\, |\wh\BB\ra_{c=1} \, \otimes
\int {d^{25-p} k_\perp \over (2\pi)^{25-p}} \,
\exp\left(\sum_{n=1}^\infty
\sum_{s=1}^{25} (-1)^{d_s} \,{1\over
n} \,  \alpha^s_{-n} \bar
\alpha^s_{-n} \right) |\vec k_\parallel=0,
\vec k_\perp\ra\, \nonumber \\
&& \otimes
\exp\left(-\sum_{n=1}^\infty (\bar b_{-n} c_{-n} +
b_{-n}
\bar c_{-n})
\right) (c_0+\bar c_0)c_1\bar c_1 |0\ra\, .
 \een
$d_s$ is an integer which is zero for Dirichlet directions
and one for Neumann directions. $\tf(x^0)=f(ix^0)$ is given in 
\refb{ehx5}.

$|\BB_1\ra$ produces source terms proportional to $\wt f(x^0)$ for
various closed string fields, and hence these sources fall off to
zero as $x^0\to\pm \infty$. On the other hand since
$|\wt\BB\ra_{c=1}$ is a linear combination of Virasoro descendants
over higher level primaries, and since these primaries occur at
discrete values of momenta, these terms cannot be reorganized by
summing over momenta as in eq.\refb{efg21}. Thus its Minkowski
version $|\wh\BB\ra_{c=1}$ will involve linear combinations of
Virasoro descendants of $\exp\left(\pm n X^0(0)\right)|0\ra$ for
integer $n$ and $|\BB_2\ra$, which contains
$|\wh\BB\ra_{c=1}$, will have source terms for various higher
level closed string fields which grow exponentially
\cite{0208196,0305177,0308172}. Thus $|\BB_2\ra$ has the form:
 \be \label{ebb2}
|\BB_2\ra = \TT_p\, \sum_{n=-\infty}^\infty \, \sum_{N=1}^\infty
\, \int {d^{25-p} k_\perp \over (2\pi)^{25-p}} \, \wh\OO^{(n)}_N
\, (c_0+\bar c_0) \, c_1 \, \bar c_1 \, e^{n X^0(0)} \, |k^0=0,
\vec k_\parallel=0,\vec k_\perp\ra\, ,
 \ee
where $\wh\OO^{(n)}_N$ is some fixed combination of negative moded
oscillators of total level $(N,N)$. We shall return to a
discussion of these terms in sections \ref{s9} and \ref{s11}.

An interesting limit to consider is the $\tl\to {1\over 2}$ limit.
As pointed out in eq.\refb{elhalf}, in this limit $\wt f(x^0)$
vanishes, and hence $|\BB_1\ra$ vanishes. It is also easy to see
using \refb{efjm} that $f_{j,m}(1/2)=0$. Thus in the
$\tl\to{1\over 2}$ limit $|\wh\BB\ra_{c=1}$, its analytic
continuation $|\wt\BB\ra_{c=1}$ and hence $|\BB_2\ra$ vanishes.
This shows that at $\tl={1\over 2}$ not only the sources for 
the
massless closed string fields vanish, but the sources for all the
massive closed string fields also vanish. This is consistent with
the identification of the $\tl={1\over 2}$ point as the vacuum
without any D-brane.

It is easy to verify that $|\BB_1\ra$ is BRST invariant, {\it i.e.}
 \be \label{ebrinv}
(Q_B+\bar Q_B) |\BB_1\ra = 0\, .
 \ee
Indeed we have the stronger relation
 \ben \label{estrong}
&& (Q_B+\bar Q_B)\, \bigg[
\exp\left(-\sum_{n=1}^\infty {1\over n}
\alpha^0_{-n}\bar\alpha^0_{-n}\right)|k^0\ra \otimes
\exp\left(\sum_{n=1}^\infty
\sum_{s=1}^{25} (-1)^{d_s} \, {1\over
n} \, \alpha^s_{-n} \bar
\alpha^s_{-n} \right) |\vec k_\parallel=0,
\vec k_\perp\ra\, \nonumber \\
&& \qquad \qquad \otimes
\exp\left(-\sum_{n=1}^\infty (\bar b_{-n} c_{-n} +
b_{-n}
\bar c_{-n})
\right) (c_0+\bar c_0)c_1\bar c_1 |0\ra\,
\bigg]\nonumber \\
&& = 0\, ,
 \een
for any $k^0$ and $\vec k_\perp$. Thus $(Q_B+\bar Q_B)|\BB_1\ra$,
which may be expressed as
a linear combination of the states appearing on the left hand side of
\refb{estrong}, also vanishes. Since $|\BB\ra=|\BB_1\ra + |\BB_2\ra$ is
BRST invariant, this shows that $|\BB_2\ra$ is also BRST invariant:
 \be \label{eb2gi}
(Q_B+\bar Q_B)|\BB_2\ra=0\, .
 \ee

{}From \refb{estrong} it follows that $|\BB_1\ra$  given in \refb{ebs3} is 
BRST invariant for any choice of the function $\tf(x^0)$. Thus
the BRST invariance of $|\BB_1\ra$
does not impose any condition on the time dependence $\wt f(x^0)$
of this boundary state. In contrast the time dependence of the boundary 
state
$|\BB_2\ra$ is fixed by the requirement of BRST
invariance\cite{0402157,0408064} since in
the $c=1$ conformal field theory the primary states $|j,m\ra$ for
$|m|<j$ exist only for integer $X$-momentum $2m$. This suggests
that the coefficients of $|j,m\ra\ra$ appearing in $|\BB_2\ra$ can
be thought of as conserved charges\cite{0402157,0309074}. We shall 
discuss this point in
detail in the context of two dimensional string theory in section
\ref{s11} where we shall also identify these charges with
appropriate conserved charges in the matrix model description of
the theory.

Before we conclude this section we note that in principle we should be
able to study the time dependent solutions described here in string field
theory by starting with the euclidean solution describing the lump or a
kink on a
circle of appropriate radius (1 for bosonic string theory and $1/\sqrt 2$
for superstring) and then making an inverse Wick
rotation\cite{0207107,0304163,0301137}. So far however
this has
not yielded any useful insight into the structure of these solutions.
Various other approaches to studying these time dependent solutions in 
string field theory have
been discussed in 
\cite{0208028,0209090,0302146,0306026,0308205,0406199,0409179}

Other aspects of time dependent classical solutions describing 
rolling of the tachyon away from the maximum of the potential have been 
discussed in refs.\cite{0205085,0303172,
0310253,
0306026,0308205,0311179}.

\sectiono{Effective Action Around the Tachyon Vacuum}
\label{s5.3}

A question that arises naturally out of the studies in the previous
sections is: Is it possible to describe the physics around the tachyon
vacuum by a {\it low energy effective action}? Given that the tachyon
field near the top of the potential has a mass$^2$ of the order of the
string scale, one wouldn't naively expect any such action to exist.
Furthermore, since around the tachyon vacuum we do not expect to get any
physical open string states, there is no S-matrix with which we could
compare the predictions of the effective action. Thus it would seem that
not only is it unlikely that we have a low energy effective action
describing
the physics around the tachyon vacuum, but that the very question does not
make sense due to the absence of physical states around such a vacuum.

We should keep in mind however that in string theory there is another way
of checking the correctness of the classical effective action, -- namely
by demanding that the solutions of the classical equations of
motion derived from this effective
action should
correspond to appropriate conformal field theories. In particular for open
string theory the classical solutions should be in one to one
correspondence with conformally invariant boundary deformations of the
world-sheet theory. As we
saw in section \ref{s10}, there are families of known time dependent
solutions around the tachyon vacuum labelled by the parameter $\tl$,
and we could ask if it is possible to
construct an effective action that reproduces these solutions.  Also the
effective action must have the property that perturbative quantization of
the theory based on this action should fail to give rise to any physical
particle like states. We shall show that it is indeed possible to
construct an effective action satisfying these criteria at a qualitative
level. However
we shall not be able to derive this effective action from first principles
{\it e.g.} by comparison with any S-matrix elements, or make any
definitive statement about the region of validity of this effective
action. Some attempts to partially justify the validity of this effective 
action has been made in 
\cite{0304045,0401066}.

\subsection{Effective action involving the tachyon}

We begin with the purely tachyonic part of the action, ignoring
the massless fields on the D-brane world-volume. In this case the
proposed action around the tachyon vacuum is given
by\cite{0003122,0003211,0003221,0204143,0209068,0303239,0304145,0307197,
0310138,0004106}:
 \ben \label{ez1}
S &=& \int d^{p+1} x \, \LL\, , \nonumber \\
\LL &=& -  V(T) \, \sqrt{1 + \eta^{\mu\nu}\p_\mu T
\p_\nu T} \, = \, - V(T) \, \sqrt{-\det A}
\, ,
 \een
where
 \be \label{ey2}
A_{\mu\nu} = \eta_{\mu\nu} + \p_\mu T \p_\nu T\, .
 \ee
The potential $V(T)$ has a maximum at $T=0$ and has the asymptotic form
 \be \label{ezz1}
V(T) \simeq e^{-\alpha T/2}  \quad \hbox{for large $T$}\, ,
 \ee
with
 \ben \label{e2}
\alpha &=& 1 \qquad \hbox{for bosonic string theory} \nonumber \\
&=& \sqrt 2 \qquad \hbox{for superstring theory}\, .
 \een
In this parametrization the potential has a minimum at infinity.
The energy momentum tensor can be computed
from the action \refb{ez1} by first minimally coupling it to a background
metric, and then calculating the functional derivative of the action with
respect to the background metric. The result is given
by\footnote{In writing down the expression for the energy momentum
tensor, it will be understood that these are localized on the plane of the
brane by a position space delta function in the transverse coordinates.
Also only the components of the energy-momentum tensor along the
world-volume of the brane are non-zero.}
 \be \label{etens}
T_{\mu\nu} = {V(T) \, \p_\mu T \p_\nu T \over \sqrt{1 + \eta^{\rho\sigma}
\p_\sigma T \p_\rho T}} - V(T)\,
\eta_{\mu\nu}\, \sqrt{1 + \eta^{\rho\sigma}
\p_\rho
T \p_\sigma T}\, .
 \ee

We shall first verify that the action \refb{ez1} produces the correct
large
$x^0$ behaviour of the pressure for
spatially homogeneous, time dependent field configurations.
For such configurations the conserved energy
density is
given
by
 \be \label{e2b}
T_{00} = V(T) (1-(\p_0T)^2)^{-1/2}\, .
 \ee
Since $T_{00}$ is conserved, and $V(T)\to 0$ for large $T$, we see that
for any given $T_{00}$,
as $T\to\infty$, $\p_0 T\to 1$ and hence $T\to x^0~+$~constant. In 
particular using \refb{ezz1} we can
show that for large $x^0$ the
solution has the
form
 \be \label{e2a}
T = x^0 +  C e^{-\alpha x^0} + \OO(e^{-2\alpha x^0}) \, ,
 \ee
after shifting the origin of $x^0$ so as to remove the additive constant 
that might otherwise appear in the expression for $T$.
One way to check that \refb{e2a} gives the correct form of the solution is
to note that the leading contribution to $T_{00}$
computed
from this configuration
remains constant in time:
 \be \label{e3}
T_{00} \simeq {1\over
\sqrt{2\alpha C}}\, .
 \ee
The pressure associated with this configuration is given
by:
 \be \label{e4}
p= T_{11} = - V(T) (1-(\p_0T)^2)^{1/2} \simeq -\sqrt{2\alpha C}
e^{-\alpha x^0}\, .
 \ee
Using the choice of $\alpha$ given in eq.\refb{e2} one can see that
\refb{e4}
is in precise agreement with the asymptotic forms of
\refb{e2.20} and \refb{e2.23} for the bosonic string or
\refb{e2.13} and \refb{e2.15} for the superstring for large $x^0$.
In particular the pressure vanishes asymptotically.

Given the success of the effective action in reproducing the
asymptotic form of $T_{\mu\nu}$, it is natural to ask if it can
also reproduce the sources for the dilaton and the RR fields
associated with the rolling tachyon solution on an unstable
D-brane. For this we need to know how the dilaton $\Phi_D$ and the
RR $p$-form fields $C^{(p)}$ couple to this effective field
theory. For a non-BPS D-brane of type II string theory, following
coupling to $\Phi_D$, $C^{(p)}$ and the string metric $G_{\mu\nu}$
seems to reproduce qualitatively the dilaton and RR source terms
and $T_{\mu\nu}$ associated with a rolling tachyon solution:
 \be \label{efullcoup}
S = - \int d^{p+1} x \,  e^{-\Phi_D} \, V(T) \, \sqrt{-\det{\bf
A}} + \int W(T)\, dT \wedge C^{(p)}\, ,
 \ee
 \be \label{efull1}
{\bf A}_{\mu\nu} = G_{\mu\nu} + \p_\mu T \p_\nu T\, ,
 \ee
where $W(T)$ is some even function of $T$ which goes to zero
asymptotically as $e^{-T/\sqrt 2}$\cite{9810188,9812135,9904207,0204143}. 
The sources for $G_{\mu\nu}$,
$\Phi_D$ and $C^{(p)}$ can be calculated by varying this action
with respect to these closed string fields and then setting
$G_{\mu\nu}=\eta_{\mu\nu}$, $\Phi_D=0$, $C^{(p)}=0$. The source
for $G_{\mu\nu}$ gives us back the $T_{\mu\nu}$ given in
\refb{etens} whereas the sources for $\Phi_D$ and $C^{(p)}$,
evaluated for the asymptotic solution \refb{e2a} reproduces the
asymptotic form of the exact stringy answers \refb{edilcharge} -
\refb{errchargep}. For a D-$p$-brane in bosonic string theory and
a D-$\bd$ system of type II string theory the coupling to the
Ramond-Ramond field $C^{(p)}$ is absent from \refb{efullcoup}, but the 
dilaton coupling has the same form.

\subsection{Classical solutions around the tachyon vacuum}

Next we shall demonstrate the absence of perturbative states upon
quantization
of the theory around the tachyon vacuum. Since {\it a priori} it is not
clear how to quantize a
non-linear theory of this type, we shall use a pragmatic definition of
the absence of perturbative states. Since in conventional field theory
perturbative states are associated with plane wave solutions, we shall
assume that absence of perturbative quantum states implies absence of
plane-wave solutions (which are not pure gauge) and vice versa. Thus we
need to show the absence of plane-wave solutions around the tachyon vacuum
in this theory.

This leads us to the analysis of classical solutions in this theory. Since
around the tachyon vacuum $V(T)=0$ and hence the action \refb{ez1}
vanishes, it is more convenient to work in the Hamiltonian
formalism\cite{0009061,9901159,9704051,9807064,9910159,
0010240,0101213,
0204143,0209034}. 
Defining
the momentum conjugate to $T$ as:
 \be \label{ex1}
\Pi(x) = {\delta S \over \delta (\p_0 T(x))} = {V(T) \p_0 T\over \sqrt{1
- (\p_0 T)^2 + \dt}}\, ,
 \ee
we can construct the Hamiltonian $H$:
 \be \label{ex2}
H = \int d^p x \, (\Pi \p_0 T - \LL) \equiv \int d^p x \, \HH, \qquad
\HH = T_{00} = \sqrt{\Pi^2 + (V(T))^2 } \, \sqrt{1 + \dt}
\, .
 \ee
The equations of motion derived from this hamiltonian take the form:
 \be \label{ex4p}
\p_0 \Pi(x)= -{\delta H\over \delta T(x)} = \p_j \bigg( \sqrt{\Pi^2 +
V^2}
\, { \p_j
T
\over
\sqrt{1 + \dt} }\bigg) - {V(T) V'(T) \over \sqrt{\Pi^2 + V^2}} \,
\sqrt{1 + \dt}\, ,
 \ee
 \be \label{ex5p}
\p_0 T(x) = {\delta H \over \delta \Pi(x)} = {\Pi\over
\sqrt{\Pi^2 + V^2}}\, \sqrt{1 + \dt} \, .
 \ee
In the limit of large $T$ ({\it i.e.}
near the tachyon vacuum) at fixed $\Pi$,
we can
ignore the
$V^2\simeq e^{-\alpha T}$ term, and the Hamiltonian and the equations
of motion take the form:
 \be \label{eham}
H = \int d^p x \, |\Pi| \sqrt{1 + \dt}\, ,
 \ee
 \be \label{ex4}
\p_0 \Pi(x)= \p_j \bigg( |\Pi|\, { \p_j
T
\over
\sqrt{1 + \dt} }\bigg)\, ,
 \ee
 \be \label{ex5}
\p_0 T(x) = {\Pi\over
|\Pi|}\, \sqrt{1 + \dt} \, .
 \ee
{}From \refb{ex5}, we see that
in this limit we have $(\p_0T)^2 - \dt= 1$.

These equations can be rewritten in a suggestive form by
defining\cite{0204143}
 \be \label{eadd1}
u_\mu \equiv -\p_\mu T, \qquad \epsilon(x) \equiv |\Pi(x)|/
\sqrt{1 + \dt}\, .
 \ee
Eqs.\refb{ex4}, \refb{ex5} then take the form:
 \be \label{eadd2}
\eta^{\mu\nu} u_\mu u_\nu = -1, \qquad \p_\mu (\epsilon(x) u^\mu) = 0\, .
 \ee
Expressed in terms of these new variables, $T_{\mu\nu}$ given in
\refb{etens} take the form:
 \be \label{eadd3}
T_{\mu\nu} = \epsilon(x) u_\mu u_\nu\, ,
 \ee
where we have used eq.\refb{ex5p} and the small $V(T)$ approximation. We 
now note that \refb{eadd2}, \refb{eadd3}
are precisely the equations governing the motion of non-rotating,
non-interacting dust,
with $u_\mu$ interpreted as the local $(p+1)$-velocity
vector\cite{0204143}, and
$\epsilon(x)$ interpreted as the local rest mass density.
Conversely, any configuration describing flow of non-rotating,
non-interacting dust can
be interpreted as a solution of the equations of motion \refb{ex4},
\refb{ex5}.

It is now clear that there are no plane wave solutions in this classical
theory. For example if we begin with an initial
static configuration with an inhomogeneous distribution of energy, this
disturbance does not propagate. On the other hand a plane wave solution
always propagates. Thus the particular field theory described here does
not have any plane wave solution, and is not expected to have any
perturbative physical state upon quantization.

Given this large class of classical solutions in the effective field
theory, we can now ask if there are boundary conformal field theories
corresponding to these solutions. Existence of 
such boundary
conformal field theories is a necessary condition for the validity of
the effective action description due to the correspondence between
classical solutions of open string field theory and two dimensional
boundary conformal field theories. As we have seen, a classical
solution of the effective field theory at late time 
corresponds to
a configuration of non-rotating, non-interacting dust. The latter on
the other hand can be thought of as a configuration of massive particles
moving around freely in space. As there is no lower bound to the density
of the dust in the effective field theory, there is no lower bound
to the masses of these particles. Are there boundary CFT's
corresponding to such
configurations? It turns out that the answer is in the affirmative.
To see this consider a non-BPS D0-brane (in bosonic or type
IIB string theory) or a D0-$\bd$0 pair (in type IIA string theory) and set
up the rolling tachyon solution on this. This allows us to construct a
configuration of arbitrary energy by adjusting the parameter $\tl$ in
eq.\refb{e2.12} or \refb{e2.19} for $p=0$. With the help of Lorentz
transformation, we can now construct a
configuration where this 0-brane system is moving with arbitrary
velocity. Since at open string tree
level different D-branes do not interact with each other, we can also
construct a configuration by superposing an arbitrary number of such
0-brane systems with arbitrary mass and velocity distribution. Such
configurations precisely describe a configuration of non-interacting dust,
{\it i.e.} classical solutions of the field theory described by the
action \refb{ez1}
at late time.

\subsection{Inclusion of other massless bosonic fields} \label{s8.3}

On a non-BPS D$p$-brane world volume we have, besides the tachyonic field
and infinite tower of massive fields, a U(1) gauge field $A_\mu$
($0\le\mu\le p$), and a
set of scalar fields $Y^I$, one for each
direction $y^I$ transverse to the D-brane ($(p+1)\le I\le D$, $D$ being 9 for
superstring theory and 25 for bosonic string theory).
One could try to generalize \refb{ez1} by including these massless fields
in the action.
The proposed form of the action is\cite{0003122}
 \be \label{exz1}
S = \int d^{p+1} x \, \LL\, , \qquad \qquad
\LL = -  V(T) \, \sqrt{-\det A} \, ,
 \ee
where
 \be \label{exy2}
A_{\mu\nu} = \eta_{\mu\nu} + \p_\mu T \p_\nu T + F_{\mu\nu} + \p_\mu Y^I
 \p_\nu Y^I\, ,
 \ee
 \be \label{esst4}
F_{\mu\nu} = \p_\mu A_\nu - \p_\nu A_\mu\, ,
 \ee
and $V(T)$ has the same form as in \refb{ezz1} for large $T$.
The action described in \refb{exz1}, \refb{exy2} satisfies the
requirement that
for $T=0$ where
$V(T)$ has a maximum it reduces to the usual
Dirac-Born-Infeld form. Furthermore, this action obeys
various restrictions involving the
universality of the tachyon potential\cite{9909062,9911116} and T-duality
invariance\cite{0003221}. As we shall see later, this form of the action
can also
be supersymmetrized easily.

Coupling of this action to background string metric $G_{\mu\nu}$,
anti-symmetric tensor field $B_{\mu\nu}$ and the dilaton $\Phi_D$
is given by:
 \be \label{exz1c}
S = -\int d^{p+1} x \, e^{-\Phi_D} \, V(T) \, \sqrt{-\det {\bf A}}
\, ,
 \ee
where
 \ben \label{exy2c}
{\bf A}_{\mu\nu} &=& G_{\mu\nu} +B_{\mu\nu}  + \p_\mu T \p_\nu T +
F_{\mu\nu} +
\p_\mu Y^I
 \p_\nu Y^I\, \nonumber \\
&& +(G_{IJ}+B_{IJ})\p_\mu Y^I
\p_\nu Y^J + (G_{\mu I}+B_{\mu I}) \p_\nu Y^I + (G_{I\nu}+B_{I\nu})
\p_\mu Y^I\, .
 \een
Using this action we can compute the source terms for various
closed string fields produced by the brane. For the non-BPS
D-brane of type II string theory there is also a coupling to the
RR fields that generalizes \refb{efullcoup}, but we shall not
describe it here.

The dynamics
of this brane is again best described in the Hamiltonian formalism. If we
denote by $p^I$ the momentum conjugate to $Y^I$, by $\Pi$ the momentum
conjugate to $T$ and by $\Pi^i$ the momentum conjugate to $A_i$
($1\le i\le p$), and
consider the limit of large
$T$ where we can ignore the $V(T)$ term, the
Hamiltonian of the system described by the action \refb{exz1} - 
\refb{esst4} is given by\cite{0009061},
 \be \label{esst1}
H = \int d^p x \, \HH\, ,
 \ee
 \be \label{esst2}
\HH = \sqrt{\Pi^i \Pi^i + p_I p_I + \Pi^2 + \Pi^i \p_i Y^I \Pi^j \p_j Y^I
+ \Pi^i \p_i T \Pi^j \p_j T + b_i b_i}\, ,
 \ee
where
 \be \label{esst3}
b_i \equiv F_{ij} \, \Pi^j + p_I \, \p_i Y^I + \Pi\, \p_i T \, .
 \ee
Furthermore the $\Pi^i$'s satisfy a constraint:
 \be \label{esst5}
\p_i \Pi^i = 0\, .
 \ee

A particular classical solution of the equations of motion derived from
this Hamiltonian is given by:
 \be \label{esst6}
\Pi^1 = f(\vec x_\perp), \qquad \Pi^i = 0 \quad \hbox{for} \quad i\ge 2,
\qquad \Pi = 0, \qquad A_1 = x^0\, ,
 \ee
with all other fields set to zero. Here $\vec x_\perp$ denotes
coordinates along the D$p$-brane world-volume transverse to $x^1$,
and $f(\vec x_\perp)$ denotes an arbitrary function which is
everywhere positive. In particular if we choose $f(\vec x_\perp) =
\delta(\vec x_\perp)$ we get a string like object along the $x^1$
direction localized at $\vec x_\perp=0$.\footnote{Since $f(\vec
x_\perp)$ needs to be positive everywhere, it is more appropriate
to regard this as the limit of a gaussian.} Using \refb{exz1c},
\refb{exy2c} it can be shown that this string acts as a source for
the anti-symmetric tensor field $B_{\mu\nu}$ like a fundamental
string, and has the same mass to charge ratio as that of the
fundamental string\cite{0002223,0009061}. Furthermore the
classical dynamics of this string-like solution that follows from
the Hamiltonian given in \refb{esst2} is described exactly by the
Nambu-Goto action\cite{0005031,0009061,0010181,0010240} . In
particular, even if we begin with a D$p$-brane that breaks the
full Lorentz invariance of the string theory, the dynamics of this
string-like solution has the full Lorentz
invariance\cite{0010240}.

This tends to suggest that the effective field theory described by
the action \refb{exz1}, \refb{exy2}, which was proposed for
describing the tree level open string dynamics on an unstable
D-brane system, contains closed strings as classical solutions.
This interpretation, however, cannot be quite correct due to
various reasons, one of them being that the same effective action
contains solutions describing continuous distribution of electric
flux as given in \refb{esst6}, while the fundamental strings must
carry quantized flux. An interpretation, proposed in
\cite{0305011,0306137,0402027} is that the classical solution of
the open string effective action should be trusted only for energy
densities of order $1/g_s$. In the $g_s\to 0$ limit this
corresponds to a very high density of flux and energy, and the
effective field theory represents average properties of a dense
system of closed strings. This interpretation is consistent with
the general open string completeness conjecture to be discussed in
section \ref{sopenclosed}.

\subsection{Supersymmetrization of the effective action}

We now turn to the issue of supersymmetrization of the action
\refb{exz1} in the case of type IIA or type IIB string theory in (9+1)
dimensions. For this we shall first rewrite the action \refb{exz1} in a
slightly different way. We introduce a set of 10 scalar fields $X^M$
instead of $(9-p)$ scalar fields $Y^I$ and take the action:
 \be \label{esst9}
S = \int d^{p+1} x \, \LL\, , \qquad \qquad
\LL = -  V(T) \, \sqrt{-\det A} \, ,
 \ee
where
 \be \label{esst10}
A_{\mu\nu} = \p_\mu T \p_\nu T + F_{\mu\nu} + \eta_{MN} \p_\mu X^M
 \p_\nu X^N \, .
 \ee
$\eta_{MN}$ is the (9+1)-dimensional Minkowski metric.
This action is invariant under an arbitrary reparametrization of the
world-volume coordinates $\{x^\mu\}$. We can fix this reparametrization
invariance by choosing the gauge condition:
 \be \label{esst11}
X^\mu = x^\mu \quad \hbox{for} \quad 0\le \mu \le p\, .
 \ee
If we furthermore call the coordinates $X^{I}$ to be $Y^I$ for $(p+1)\le
I\le 9$, we recover the action \refb{exz1}.

We shall now write down the supersymmetric generalization of the action
\refb{esst9}, \refb{esst10} \cite{9909062,9912255}. To do this we first 
need
to know the
spectrum of massless fermions on the world-volume of a non-BPS D$p$-brane.
For
type IIA string theory the massless fermionic fields on the world volume
theory
can
be thought of as a single non-chiral Majorana spinor $\theta$ of the
(9+1)-dimensional Lorentz group, whereas for type IIB string theory we
have a pair of chiral spinors $\theta_A$ ($A=1,2$) of the
(9+1)-dimensional Lorentz group. We shall concentrate on the type IIA
theory first. Let us define:
 \be \label{esp1}
\Pi^M_\mu = \p_\mu X^M -\bar\theta \Gamma^M \p_\mu\theta\, ,
 \ee
 \be \label{esp2}
\GG_{\mu\nu}=\eta_{MN}\Pi_\mu^M\Pi_\nu^N\, ,
 \ee
and
 \be \label{esp3}
\FF_{\mu\nu}=F_{\mu\nu} -[\bar\theta \Gamma_{11}\Gamma_M\p_\mu\theta
(\p_\nu X^M-{1\over 2}\bar\theta\Gamma^M\p_\nu\theta)
-(\mu\leftrightarrow\nu)]\, ,
 \ee
where $\Gamma^M$ denote the ten dimensional gamma matrices and
$\Gamma_{11}$
is the product of all the gamma matrices.
The supersymmetric world-volume action on the non-BPS D-brane is given by:
 \be \label{esp4}
S = \int d^{p+1} x \, \LL\, , \qquad \qquad
\LL = -  V(T) \, \sqrt{-\det \AAA} \, ,
 \ee
where
 \be \label{esp5}
\AAA_{\mu\nu} = \p_\mu T \p_\nu T + \FF_{\mu\nu} + \GG_{\mu\nu}\, .
 \ee
In fact, both $\GG_{\mu\nu}$ and $\FF_{\mu\nu}$ and hence the
action $S$ given in \refb{esp4} can be shown to be invariant under
the supersymmetry transformation\cite{9612080}:
 \ben \label{esp6}
&& \delta_\eps\theta=\eps, \qquad \delta_\eps X^M=\bar\eps \Gamma^M\theta,
\qquad \delta_\eps T = 0\, , \nonumber \\
&& \delta_\eps A_\mu=\bar\eps\Gamma_{11}\Gamma_M\theta \, \p_\mu X^M
-{1\over 6}(\bar\eps\Gamma_{11}\Gamma_M\theta\,
\bar\theta\Gamma^M\p_\mu\theta
+\bar\eps\Gamma_M\theta\, \bar\theta\Gamma_{11}\Gamma^M\p_\mu\theta)\,
,
 \een
where the supersymmetry transformation parameter $\eps$ is a Majorana
spinor of SO(9,1) Lorentz group.

Since the gauge conditions \refb{esst11} are not invariant under
supersymmetry transformation, once we fix this gauge, a supersymmetry
transformation must be accompanied by a compensating world-volume
reparametrization. For any world-volume field $\Phi$, this corresponds to
a modified supersymmetery transformation law:
 \be \label{esp7}
\wh\delta_\eps\Phi = \delta_\eps\Phi - \bar\eps \Gamma^\mu\theta \,
\p_\mu\Phi\, .
 \ee

Let us now turn to non-BPS D$p$-branes in type IIB string theory.
As mentioned earlier, in this case the world-volume theory
contains a pair of Majorana-Weyl spinors $\theta_1$ and $\theta_2$
of SO(9,1) Lorentz group. For definiteness we shall take these spinors to 
be right-handed. Let us define
 \be \label{esp18}
\theta=\pmatrix{\theta_1\cr\theta_2}\, ,
 \ee
and let $\tau_3$ denote the matrix $\pmatrix{I & \cr & -I}$ acting on
$\theta$, where $I$ denotes the identity matrix acting on $\theta_1$ and
$\theta_2$. We also define
 \be \label{edefga}
\wh\Gamma^M = \pmatrix{\Gamma^M & 0\cr 0 & \Gamma^M}\, ,
 \ee
 \be \label{esp1a}
\Pi^M_\mu = \p_\mu X^M -\bar\theta \wh\Gamma^M \p_\mu\theta\, ,
 \ee
 \be \label{esp2a}
\GG_{\mu\nu}=\eta_{MN}\Pi_\mu^M\Pi_\nu^N\, ,
 \ee
and
 \be \label{esp3a}
\FF_{\mu\nu}=F_{\mu\nu} -[\bar\theta \tau_3\wh\Gamma_M\p_\mu\theta
(\p_\nu X^M-{1\over 2}\bar\theta\wh\Gamma^M\p_\nu\theta)
-(\mu\leftrightarrow\nu)]\, .
 \ee
The supersymmetric world-volume action is then given by:
 \be \label{espsp4}
S = \int d^{p+1} x \, \LL\, , \qquad \qquad
\LL = -  V(T) \, \sqrt{-\det \AAA} \, ,
 \ee
where
 \be \label{espsp5}
\AAA_{\mu\nu} = \p_\mu T \p_\nu T + \FF_{\mu\nu} + \GG_{\mu\nu}\, .
 \ee
Both $\GG_{\mu\nu}$ and $\FF_{\mu\nu}$ and hence the action is invariant
under the supersymmetry transformation:
 \ben \label{esp6a}
&& \delta_\eps\theta=\eps, \qquad \delta_\eps X^M=\bar\eps
\wh\Gamma^M\theta,
\nonumber \\
&& \delta_\eps A_\mu=\bar\eps\tau_3\wh\Gamma_M\theta\p_\mu X^M
-{1\over
6}(\bar\eps\tau_3\wh\Gamma_M\theta\, \bar\theta\wh\Gamma^M\p_\mu\theta
+\bar\eps\wh\Gamma_M\theta\, \bar\theta\tau_3\wh\Gamma^M\p_\mu\theta)\, ,
 \een
where the supersymmetry transformation parameter $\eps$ is given by
$\pmatrix{\eps_1\cr \eps_2}$, with $\eps_1$ and $\eps_2$ both right-handed
Majorana
spinor of SO(9,1) Lorentz group.

As before, if we choose to work with a gauge fixed action, then the
supersymmetry transformations have to be modified by including a
compensating gauge transformation as in \refb{esp7}.

Just as \refb{efullcoup}, \refb{efull1} describe the coupling of the
massless closed string fields to the open string tachyon, we can
generalize the supersymmetric action described in this section to
include coupling to the complete set of massless closed string fields. 
This can be done by following the general procedure developed in 
\cite{9611173} (see 
also \cite{9512062,9610148,9610249,9611159,9612080}), 
but we shall not describe this
construction here.

\subsection{Kink solutions of the effective field theory}

The effective field theory described by the action \refb{ez1} provides
a good description of the rolling tachyon solution at late time when $|T|$
becomes large and $|\dot T|$ approaches 1.
Since for a non-BPS D-brane or a D-$\bd$ system in superstring theory a
kink
solution interpolates between the vacua at $T=\pm \infty$, and hence $T$
must pass through 0, there is no reason to expect this effective field
theory to
provide a good description of the kink solution. Nevertheless, we shall
now see that the effective field theory provides a good description of the
kink solution as well.
Our discussion
will follow the analysis of
\cite{0208217,0303057,0004106,0012222}.

To begin with we shall not commit ourselves to any specific form of $V(T)$
except
that it is symmetric under $T\to -T$ and falls off to 0 asymptotically
with the behaviour given in \refb{ezz1}. We look for a solution that
depends on one spatial direction $x\equiv x^p$ and is time independent.
For such a system:
 \ben \label{ezsol1}
&& T_{xx} = -V(T) / \sqrt{1+(\p_x T)^2}\, , \qquad T_{\mu x} = 0,
\nonumber \\
&& T_{\mu\nu} = - V(T) \, \sqrt{1+(\p_x T)^2} \, \eta_{\mu\nu}\, ,
\quad
\hbox{for} \quad 0\le\mu,\nu\le (p-1)\, .
 \een
The energy-momentum conservation gives,
 \be \label{ezsol2}
\p_x T_{xx} = 0\, .
 \ee
Thus $T_{xx}$ is independent of $x$. Since for a kink solution
$T\to\pm\infty$ as $x\to\pm\infty$, and $V(T)\to 0$ in this limit, we see
that $T_{xx}$ must vanish for all $x$. This, in turn, shows that we must
have\cite{0004106}
 \be \label{ezsol3}
T=\pm\infty \quad \hbox{or} \quad \p_x T= \infty \quad \hbox{(or
both)} \quad \hbox{for all $x$}\, .
 \ee
Clearly the solution looks singular. We shall now show that despite this
singularity, the solution has finite energy density which is independent
of the way we regularize the singularity, and for which the energy
density is localized on a
codimension 1 subspace, just as is expected of a D$(p-1)$-brane. For this
let us consider the following field configuration:
 \be \label{ezsol4}
T(x) = f(ax)\, ,
 \ee
where $f(x)$ is an odd, monotone increasing function of $x$ that
approaches $\pm\infty$ as $x\to\pm\infty$ but is otherwise arbitrary, and
$a$ is a constant that we shall take to $\infty$ at the end. Clearly in
this limit we have $T=\infty$ for $x>0$ and $T=-\infty$ for $x<0$, therby
producing a singular kink.

Let us compute the energy momentum tensor associated with the
configuration \refb{ezsol4}. From \refb{ezsol1} we see that the non-zero
components are:
 \ben \label{ezsol5}
T_{xx} &=& -V(f(ax))/ \sqrt{1 + a^2 (f'(ax))^2}\, , \nonumber \\
T_{\mu\nu} &=& - V(f(ax))\, \sqrt{1 + a^2 (f'(ax))^2}\,
\eta_{\mu\nu}\,
, \quad 0\le \mu,\nu\le (p-1).
 \een
Clearly in the $a\to\infty$ limit, $T_{xx}$ vanishes everywhere since
the numerator vanishes and the denominator blows up. Hence
the conservation law \refb{ezsol2} is automatically satisfied. This in
turn shows that this configuration is a solution of the equations of
motion in this limit.

{}From \refb{ezsol5} we see furthermore that in the $a\to\infty$ limit, we
can write $T_{\mu\nu}$ as:
 \be \label{ezsol6}
T_{\mu\nu} = -a\, \eta_{\mu\nu}\,  V(f(ax))\, f'(ax)\, .
 \ee
Thus the integrated $T_{\mu\nu}$, associated with the codimension 1
soliton, is given by:
 \be \label{ezsol7}
T^{tot}_{\mu\nu} = -a\, \eta_{\mu\nu} \, \int_{-\infty}^\infty\, d x \,
V(f(ax))\, f'(ax) = -\eta_{\mu\nu} \, \int_{-\infty}^\infty \, d y \,
V(y)\, ,
 \ee
where $y=f(ax)$. This shows that
the
final answer depends only on the form of $V(y)$ and not on the shape of
the function $f(x)$ used to describe the soliton. It is also clear from
the exponential fall off in $V(y)$ for large $y$ that most of the
contribution to $T^{tot}_{\mu\nu}$
is contained within a finite range of $y$.
The relation $y=f(ax)$ then implies that
the contribution comes from a region of $x$ integral of
width $1/a$ around $x=0$. In the $a\to\infty$ limit such a distribution
approaches a $\delta$-function. Thus the energy density associated with
this solution is given by:
 \be \label{ezsol8}
T_{\mu\nu} = - \eta_{\mu\nu} \, \delta(x) \, \int_{-\infty}^\infty \, d y
\,
V(y)\, .
 \ee
This is precisely what is expected of a D-$(p-1)$-brane, provided the
integral $\int_{-\infty}^\infty \, d y \, V(y)$ equals the tension of the
D-$(p-1)$-brane. For comparison, we recall that $V(0)$ denotes the tension
of a D$p$-brane.

One can check that the solution constructed this way satisfies the
complete set of equations of motion of the effective field theory.
Furthermore one can construct the world-volume effective action on this
kink solution, and this turns out to be exactly the Dirac-Born-Infeld
action, as is expected if the kink has to describe a BPS
D-$(p-1)$-brane\cite{0303057,0104218,0012080}. We shall not 
discuss this
construction in detail,
and refer the reader to the original papers.

So far in our discussion we have not committed ourselves to a specific
form of $V(T)$. It turns out that we can get a lot more quantitative
agreement with string theory results if we
choose\cite{0301076,0303035,0303139}:
 \be \label{evform}
V(T) = \wt \TT_p / \cosh(T/\sqrt 2)\, .
 \ee
Here for definiteness we have considered the case of a non-BPS
D-$p$-brane. The overall normalization of $V$ has been adjusted so that
$V(0)$ reproduces the tension of this brane. We now note that:
 \begin{enumerate}
\item If we expand the action around $T=0$ and keep terms up to quadratic
order in $T$, we find that the field $T$ describes a particle of
mass$^2=-{1\over 2}$. This agrees with the mass$^2$ of the tachyon on an
unstable D-brane in superstring theory.

\item The tension of the kink solution in the effective field theory is
now given by:
 \be \label{evform2}
\int_{-\infty}^\infty dy \, V(y) = \sqrt 2 \pi \wt\TT_p \, .
 \ee
This gives the correct tension of a BPS D-$(p-1)$-brane.

\item $V(T)$ given in \refb{evform} has the behaviour given in
\refb{ezz1}, \refb{e2} for large $T$.
\end{enumerate}
Finally, it was shown in \cite{0303139} that this
effective field theory admits a one parameters family of solutions of the
form
 \be \label{evform3}
T = \sqrt 2 \sinh^{-1} \left( {\lambda} \sin \left({x\over
\sqrt 2}\right) \right)\, ,
 \ee
where $x$ denotes any of the spatial coordinates on the D-$p$-brane.
These are precisely the analogs of the solutions \refb{efh5}
for $R=1/\sqrt 2$. Thus the effective
field theory has solutions in one to one correspondence with the family of
BCFT's associated with marginal
deformations of the original BCFT. The moduli space of these solutions in 
the effective field theory closely resembles the moduli space of the 
corresponding BCFT's but they are not identical\cite{0312003}. This shows 
that the effective field theory does not reproduce quantitatively all the 
features of the classical solutions in the full open string theory.

Various other aspects of the tachyon effective action described in this
section have been discussed in
refs.\cite{9905195,
0102174,0107087,
0202079,0204203,0206212,
0207235,0208019,0208094,
0209142,0210108,
0210221,
0211090,
0301101,
0303204,0304108,0304180,0304197,0305092,0305229,0305249,
0306294,0306295,
0307184,0308069,
0310066,
0310079,0310253,
0312086,0312149,
0401163,
0401195,
0401236,0403073,
0403124,0403217,
0404163,0404253,
0405058,0406120,
0408073,0409030,
0409151,0410030}

\sectiono{Toy Models for Tachyon Condensation} \label{s5}

In section \ref{s5.3} we have discussed a specific form of effective field
theory that reproduces many of the features of the tachyon dynamics on
an
unstable D-brane. In this section we describe two other types of field
theory models which share some properties of the tachyon dynamics in the
full string theory, namely absence of physical states around the tachyon
vacuum, and lower dimensional branes as solitons. The first of these
models will be based on
singular potential but regular kinetic term, while the second type of
model will be based on smooth potential but
non-local kinetic term. Both these classes of models mimick many of the
properties of the time independent solutions of the tachyon effective
action in open string theory. However unlike the model described in
section \ref{s5.3} these models do not seem to reproduce the features of
the time dependent solutions involving the open string tachyon. For this
reason we shall restrict our analysis to time independent solutions
only.
For some analysis of time dependent solutions in these models, see
\cite{0207107,0209197,math-ph/0306018}.

\subsection{Singular potential model} \label{s5.1}

Let us consider a field theory of a scalar field $\phi$ in $(p+1)$
dimensions, described by the
action\cite{0008231}:
 \be \label{esin1}
S = -\int d^{p+1}x \left[ {1\over 2} \eta^{\mu\nu} \p_\mu \phi \p_\nu\phi
+
V(\phi)\right]
 \ee
where
 \be \label{esin2}
V(\phi) = -{1\over 4} \phi^2 \ln \phi^2\, .
 \ee
The potential has a local maximum at $\phi = e^{-1/2}$. At this point
 \be \label{esin3}
V(e^{-1/2}) = {1\over 4 e}, \qquad V''(e^{-1/2})=-1\, .
 \ee
This shows that the scalar field excitation around the maximum describes
a
tachyonic mode with mass$^2=-1$. The potential also has minimum at
$\phi=0$ where it vanishes. The second derivative of the potential is
infinite at $\phi=0$, showing that there is no finite mass scalar particle
obtained by perturbative quantization of this theory around this minimum.
This is consistent with the second conjecture. The difference between the
values of the potential at the minimum and the local maximum can be
thought of as the tension of the $p$-brane that the $\phi=e^{-1/2}$
solution describes.

We now examine the classical soliton solutions in this field theory. As
already mentioned, it has a translationally invariant vacuum solution
$\phi=0$ and a translationally invariant solution $\phi=e^{-1/2}$
corresponding to a local maximum of the potential. But the theory also has
codimension $(p-q)$ lump solutions, given by\cite{0008231}:
 \be \label{esin4}
\phi = F\left(\sqrt{(x^{q+1})^2 + \cdots + (x^p)^2}\right)\, ,
 \ee
where
 \be \label{esin5}
F(\rho) = \exp\left(-{1\over 4} \rho^2 + {1\over 2} (p-q-1)\right)\, .
 \ee
The tension of this $q$-brane solution is given by:
 \be \label{esin6}
\TT_q = {1\over 4} e^{p-q-1} (2\pi)^{(p-q)/2}\, .
 \ee
Thus
 \be \label{esin7}
\TT_q / \TT_{q+1} = \sqrt{2\pi} e\, .
 \ee
This is independent of $q$. This reproduces the string theoretic
feature
that the ratio of the tension of a D-$q$-brane and D-$(q-1)$-brane is
independent of $q$.

Given the lump solution \refb{esin4}, \refb{esin5}, we can anlyze the
spectrum of small fluctuations around the solution. For a codimension
$(p-q)$ lump the spectrum coincides with that of $(p-q)$ dimensional
harmonic oscillator\cite{0008231} up to an overall additive constant. To 
be more specific,
the spectrum of excitations on a codimension $(p-q)$ lump is labelled by
$(p-q)$
integers $(n_1, \cdots n_{p-q})$, and the mass$^2$ of the states
associated
with these excitations are given by:
 \be \label{esin8}
m_{n_1, \cdots n_{p-q}}^2 = (n_1 +\cdots n_{p-q} - 1)\, .
 \ee
This shows that on each of these D-$q$-branes the lowest excitation
mode, corresponding to $n_1=\ldots = n_{p-q}=0$, is tachyonic and has
mass$^2=-1$. This is again in accordance with the results in the
bosonic string theory that the mass$^2$ of the tachyonic mode on any
D-$q$-brane has the same value $-1$ independent of the value of $q$.

Various generalizations of these models as well as other field theory 
models can be found in 
refs.\cite{0004131,0008227,
0009246,0011226,0104229,0106103,
0107070,0107075,0112088,
0203108,0211127,0405125,
0407081}.

\subsection{$p$-adic string theory} \label{s5.2}

The $p$-adic open string theory is obtained from ordinary bosonic open 
string
theory on a D-brane
by replacing, in the Koba-Nielson amplitude, the integral over the
real world-sheet coordinates by $p$-adic integral associated with a prime
number
$p$\cite{Freund1,Freund2,Frampton1,padic}. We
shall not
review this construction here. For our purpose it will be
sufficient to know
that in this case there is an exact expression for the tachyon effective
action which reproduces correctly all the tree level amplitudes involving
the tachyon. This effective action for the tachyon on a Dirichlet
$(d-1)$-brane is given by\cite{padic,Frampton2}:
 \ben \label{epa1}
S &=& \int d^d x \, \LL \nonumber \\
&=& {1\over g^2} {p^2 \over p-1} \int d^d x \left[ -{1\over 2}
\phi \, p^{-\half\Box}\, \phi + {1\over p+1} \, \phi^{p+1} \right]\, ,
 \een
where $\Box$ denotes the $d$ dimensional Laplacian, $\phi$ is the
tachyon field (after a rescaling and a shift), and $g$ is the open
string coupling constant.\footnote{Although the $p$-adic string
theory is defined only for $p$ prime, once the action \refb{epa1}
is written down, we can analyze its properties for any integer
$p$. In the $p\to 1$ limit the action \refb{epa1} reduces to the
action given in \refb{esin1}, \refb{esin2}\cite{0009103}.} The
potential of the model, defined as $-\LL$ evaluated for spatially
homogeneous field configurations, is given by:
 \be \label{epa22}
V(\phi) = {1\over g^2} {p^2 \over p-1} \left[ {1\over 2} \phi^2 -
{1\over p+1} \phi^{p+1} \right]\, .
 \ee
This has a local minimum at $\phi=0$ and maxima at $\phi^{p-1}=1$.

The classical equation of motion derived from the action \refb{epa1} is
 \be \label{epa30}
p^{-\half\Box} \phi = \phi^p\, .
 \ee
Different known solutions of this equation
are as follows\cite{padic}:
 \begin{itemize}
\item The configuration $\phi=0$ is a local minimum of
$V(\phi)$ with $V(0)=0$.
We shall identify this solution
with the tachyon vacuum configuration. By definition we have taken
the energy density of this vacuum to be zero.

To analyze the spectrum of perturbative physical excitations around 
$\phi=0$
we examine the linearized equation of motion around this point:
 \be \label{epa1d}
p^{-\half\Box} \phi = 0\, .
 \ee
If we look for a plane-wave solution of this equation of the form:
 \be \label{epa1f}
\phi = \phi_0 e^{ik.x}\, ,
 \ee
then eq.\refb{epa1d} gives
 \be \label{epa1e}
p^{\half k^2} = 0\, .
 \ee
This has no solution for finite $k^2$. Thus there are no perturbative
physical excitations around the configuration $\phi=0$. This is in 
accordance with
conjecture 2 if we identify the $\phi=0$ configuration as the vacuum 
without any D-brane.

\item The configuration $\phi=1$, being the maximum of $V(\phi)$,
represents the original D-brane
configuration around
which we quantized the string.\footnote{For odd $p$, there is also
an equivalent solution corresponding to $\phi=-1$. Since the action
is symmetric under $\phi\to -\phi$, we shall restrict our analysis to
solutions with positive $\phi$.} We shall call this the
D-$(d-1)$-brane solution. The energy density associated with this
configuration, which can be identified as the tension $\TT_{d-1}$ of
the D-$(d-1)$-brane according to the first conjecture on open string 
tachyon dynamics, is given by
 \be \label{epa1a}
\TT_{d-1} = - \LL(\phi=1) = {1\over 2 g^2} {p^2 \over p+1}\, .
 \ee
Unfortunately in $p$-adic string theory there is no independent method of 
calculating the tension of the D-$(d-1)$-brane and compare this with 
\refb{epa1a} to verify the first conjecture.

The linearized equation of motion around the $\phi=1$ solution is found 
by
defining $\chi\equiv (\phi-1)$ and expanding \refb{epa30} to first order
in $\chi$. This gives:
 \be \label{epa1b}
\left( p^{-\half\Box} -p\right) \chi = 0\, .
 \ee
This has plane wave solutions of the form
 \be \label{epa1c}
\chi = \chi_0 e^{ik.x}\, ,
 \ee
provided $k^2 = 2$. Thus the excitation around the point $\phi=1$
describes a tachyonic mode with mass$^2=-2$.

\item The configuration:
 \be \label{epa2}
\phi(x) = f(x^{q+1}) f (x^{q+2}) \cdots f(x^{d-1})\equiv
F^{(d-q-1)}(x^{q+1},\ldots,x^{d-1})\, ,
 \ee
with
 \be \label{epa3}
f(\eta) \equiv p^{1\over 2(p-1)} \exp\left(-{1\over 2}\,
{p-1\over p \ln p} \, \eta^2\right) \, ,
 \ee
describes a soliton solution with energy density localized
around the hyperplane $x^{q+1} = \cdots = x^{d-1}=0$. 
Indeed, by using
the identity
 \be \label{epas1}
p^{-\half \p_\eta^2} f(\eta) = \left(f(\eta)\right)^p
 \ee
one can show
that \refb{epa2}, \refb{epa3} solves the classical equations of
motion \refb{epa30}. 
\refb{epas1} in turn can be proven easily by working in the Fourier
transformed space.

We shall call \refb{epa2}
the solitonic $q$-brane solution. Let us denote by
$x_\perp=(x^{q+1},\ldots, x^{d-1})$ the coordinates transverse to the
brane and by $x_\parallel=(x^0,\ldots,x^q)$ those tangential to it.
The energy density per unit $q$-volume of this brane, which can
be identified as its tension $\TT_q$, is given by
 \be \label{epa4}
\TT_q = - \int d^{d-q-1} x_\perp\; \LL(\phi = F^{(d-q-1)}(x_\perp))
= {1\over 2g_q^2} \, {p^2 \over p+1}\, ,
 \ee
where,
 \be \label{epa9}
g_q = g \left[{p^2 - 1\over
2 \pi\, p^{2p/(2p-1)}\ln p}\right]^{(d-q-1)/4}\, .
 \ee
\end{itemize}
{}From eqs.\refb{epa1a},\refb{epa4} and \refb{epa9} we see that the
ratio
of the tension of a $q$-brane and a $(q-1)$-brane is
 \be \label{epa5}
{\TT_q\over \TT_{q-1}} = \left[ { 2\pi\, p^{2p\over p-1}\ln p
\over p^2 -1}\right]^{-{1\over 2}}  \, .
 \ee
This is independent of $q$. Since
this is also a feature of the D-branes in ordinary
bosonic string theory, it suggests that the solitonic $q$-branes of
$p$-adic string theory should have interpretation as
D-branes. Unfortunately so far we do not have an independent way of
calculating the tension of a D-$q$-brane in $p$-adic string theory for
arbitrary $q$ and verify \refb{epa5} explicitly. However, as we shall show 
now, the spectrum of fluctuations around a solitonic $q$-brane does match 
the spectrum of open strings on a $p$-adic D-$q$-brane. 

We can study fluctuations around a solitonic $p$-brane solution by taking
the following ansatz for the field $\phi$:
 \be \label{epa6}
\phi(x) = F^{(d-q-1)}(x_\perp) \psi(x_\parallel) \, ,
 \ee
with $F^{(d-q-1)}(x_\perp)$ as defined in \refb{epa2}, \refb{epa3}.
For $\psi=1$ this describes the solitonic $q$-brane. Fluctuations of
$\psi$ around 1 denote fluctuations of $\phi$ localized on the
soliton; thus $\psi(x_\parallel)$ can be regarded as one
of the fields on its world-volume. We shall call this
the tachyon field on the solitonic $q$-brane world-volume. Substituting
\refb{epa6} into \refb{epa30} and using \refb{epas1} we get
 \be \label{epa7}
p^{-\half\Box_\parallel} \psi = \psi^p\, ,
 \ee
where $\Box_\parallel$ denotes the $(q+1)$ dimensional Laplacian
involving the world-volume coordinates $x_\parallel$ of the $q$-brane. Any 
solution of eq.\refb{epa7}, after being substituted into eq.\refb{epa6}, 
gives a solution to the original equation of motion \refb{epa30}. The 
action involving 
$\psi$ can be obtained by substituting \refb{epa6} into \refb{epa1}:
 \ben
S_q(\psi) &=& S\left(\phi=F^{(d-q-1)}(x_\perp)\psi(x_\parallel)\right)
\nonumber\\
&=& {1\over g_q^2} {p^2 \over p-1} \int d^{q+1} x_\parallel
\left[ -{1\over 2} \psi p^{-\half \Box_\parallel}
\psi + {1\over p+1} \psi^{p+1} \right]\, ,\label{epa8}
 \een
where $g_q$ has been defined in eq.\refb{epa9}.
\refb{epa7}, \refb{epa8} are precisely the tachyon equation
of motion and tachyon effective action (up to an overall normalization) 
that we would have
gotten by quantizing the open $p$-adic string on a D-$q$-brane directly.
This correspondence continues to hold for other massless and massive
excitations on the
solitonic brane as well\cite{0003278,0102071}.
This is a strong indication that these solitonic $q$-brane solutions on
the world-volume of the D-$(d-1)$-brane actually describe lower
dimensional D-$q$-brane, in accordance with conjecture 3.

Various other aspects of tachyon condensation in $p$-adic string theory 
have been studied in \cite{0105312,0209197,
0304213,0406120,
0406259,0409311}.

\sectiono{Closed String Emission from `Decaying' D-branes}
\label{s9}

So far we have carried out our analysis in tree level open string theory.
Although we have used the coupling of closed strings to D-branes to
determine the sources for various closed string fields and 
construct the boundary state associated with a D-brane, we have not
treated the closed strings as dynamical objects and studied what kind of
closed string background the D-brane produces. In this section we shall
address this problem in the context of time dependent solutions associated
with the rolling tachyon configuration on an unstable D-brane
\cite{0303139,0304192,0402157}.
For
simplicity we shall restrict our analysis to bosonic string theory only,
but the results can be generalized to superstring theories as
well.

A brief review of some aspects
of closed string field theory has been 
given in appendix \ref{appb}.
We begin with the closed string field equation in the presence of a
D-brane, as given in eq.\refb{efg4app}:
 \be \label{efg4}
2\, (Q_B+\bar Q_B)\, |\Psi_c\ra = K\, g_s^2 \, |\BB\ra\, .
 \ee
As described in appendix A, 
the closed string field $|\Psi_c\ra$ is represented by a ghost number 
two state in the CFT on the full complex plane (which is conformally 
equivalent to a cylinder) satisfying the constraints \refb{efg1.5}.
$g_s$ is the closed string coupling constant, $K$ is
a numerical constant determined in eq.\refb{ezopen}, and $|\BB\ra$ is
the boundary state associated with the D-brane. Noting that $|\BB\ra$
is BRST invariant,
and that $\{Q_B+\bar Q_B, b_0+\bar b_0\} = (L_0 + \bar L_0)$, we can write
down a
solution to equation \refb{efg4} as:
 \be \label{es9.1}
|\Psi_c\ra = K\, g_s^2 \, [2(L_0 + \bar L_0)]^{-1} \, (b_0+\bar b_0) \,
|\BB\ra\, .
 \ee
This solution satisfies the Siegel gauge condition and Siegel
gauge equations of motion:
 \be \label{eseqom}
(b_0+\bar
b_0)|\Psi_c\ra=0\, , \qquad 2(L_0+\bar L_0)|\Psi_c\ra = K g_s^2 (b_0+\bar
b_0)|\BB\ra\, .
 \ee
We can of course construct other solutions which
are
gauge equivalent to this one by adding to $|\Psi_c\ra$ terms of the
form
$(Q_B+\bar Q_B)|\Lambda\ra$. However even within Siegel gauge, the
right
hand side of \refb{es9.1} is not defined
unambiguously due to the presence of the zero eigenvalues of the
operator $(L_0+\bar L_0)$.\footnote{Since acting on a level $(N,N)$
state the operator
$(L_0+\bar L_0)$ takes the form of a differential operator $-{1\over
2}
\square + 2(N-1)$,
free closed string field theory in Minkowski space has infinite
number of
plane
wave solutions of the equations
$(L_0+\bar L_0)|\Psi_c\ra
= 0$, $(b_0+\bar b_0)|\Psi_c\ra
= 0$.}
Thus we need to carefully choose a
prescription for defining the right hand side of \refb{es9.1}.
A natural prescription (known as the
Hartle-Hawking prescription) is to begin with the
solution of the associated equations of motion in the Euclidean theory
where there is a unique solution to eq.\refb{es9.1} (which therefore
satisfies the full equation \refb{efg4}) and then analytically
continue
the
result to the Minkowski space along the branch passing through the
origin $x^0=0$\cite{0303139,0304192,0402124}. This is the prescription
we
shall
follow.

We shall now describe the results obtained using this formalism.
This will be carried out in two steps:
 \begin{enumerate}
\item First we describe the
closed string
background produced by the $|\BB_1\ra$
component of the boundary state
as defined in \refb{ebs3}.
 \item We then discuss the computation of the closed string background
produced by the $|\BB_2\ra$ component of the boundary state  associated
with the rolling tachyon solution.
\end{enumerate}

\subsection{Closed string radiation produced by 
$|\BB_1\ra$} \label{eccl}

Let us denote by $|\Psi_c^{(1)}\ra$ and $|\Psi_c^{(2)}\ra$ the
closed string field configurations produced by $|\BB_1\ra$ and
$|\BB_2\ra$ as given in \refb{ebs3} and \refb{ebs4} respectively.
We begin with the analysis of $|\Psi_c^{(1)}\ra$. Let us define
$\wh A_N$ to be an operator of level $(N,N)$ acting on closed string 
states, composed of negative
moded oscillators of $X^0$, $X^s$, $b$, $c$, $\bar b$ and $\bar c$
such that
 \be \label{eexpan}
\exp\left[\sum_{n=1}^\infty \, \left( -{1\over n}
\alpha^0_{-n}\bar\alpha^0_{-n} +
\sum_{s=1}^{25} (-1)^{d_s} \, {1\over
n} \,  \alpha^s_{-n} \bar
\alpha^s_{-n} - (\bar b_{-n} c_{-n} +
b_{-n}
\bar c_{-n}) \right)
\right] = \sum_{N=0}^\infty \wh A_N\, .
 \ee
Here $\wh A_0=1$. Then $|\BB_1\ra$ given in \refb{ebs3} can be
expressed as
 \be \label{ebb1exp}
|\BB_1\ra = \TT_p \, \int {d^{25-p} k_\perp \over (2\pi)^{25-p}} \,
\sum_{N=0}^\infty \, \wh A_N
\,
(c_0+\bar c_0) \, c_1 \, \bar c_1 \, \wt f\left(X^0(0)\right) \,
|k^0=0, \vec k_\parallel=0, \vec k_\perp\ra\, .
 \ee
In terms of the operators $\wh A_N$, the result for
$|\Psi_c^{(1)}\ra$ is given by:
 \be \label{esa0}
|\Psi_c^{(1)}\ra = 2\, K\, g_s^2 \, \TT_p \, \int {d^{25-p}
k_\perp\over
(2\pi)^{25-p}} \,
\sum_{N\ge 0} \, \wh A_N \, h^{(N)}_{\vec k_\perp}(X^0(0)) \, c_1\,
\bar c_1\, |k^0=0,k_\parallel=0, \vec k_\perp\ra\, ,
 \ee
where $h^{(N)}_{\vec k_\perp}(x^0)$ satisfies:
 \be \label{eheq}
\left( \p_0^2 +\vec k_\perp^2 + 4(N-1)\right) \, h^{(N)}_{\vec
k_\perp}(x^0) =\wt f(x^0)\, .
 \ee
It is easy to see that \refb{esa0} satisfies eq.\refb{eseqom} since
acting on a level $(N,N)$ state the operator
$(L_0+\bar L_0)$ takes the form of a differential operator
$-{1\over 2} \square + 2(N-1)$. It can also be shown explicitly
that \refb{esa0} actually satisfies the full set of equations
\refb{efg4}  with $|\BB\ra$ replaced by $|\BB_1\ra$\cite{0402157}.

The result for $h^{(N)}_{\vec k_\perp}(x^0)$ following the
Hartle-Hawking prescription is\cite{0303139,0304192,0402157}
 \be \label{esa-1}
h^{(N)}_{\vec k_\perp}(x^0)
= {i \over 2\onk} \, \left[\int_C \, e^{-i \onk
(x^0 -
x^{\prime 0})} \wt f(x^{\prime
0}) d x^{\prime 0} - \int_{C'} e^{ i \onk (x^0 - x^{\prime 0})}
\wt f(x^{\prime 0}) d x^{\prime 0}\right]
\, ,
 \ee
with
 \be \label{defonk}
\onk= \sqrt{\vec k_\perp^2
+ 4(N-1)}\, .
 \ee
Here the contour $C$ runs from $i\infty$ to the origin along the
imaginary $x^{\prime 0}$ axis, and then to $x^0$ along the real
$x^{\prime
0}$ axis, and the contour $C'$ runs from $-i\infty$ to the origin
along
the
imaginary $x^{\prime 0}$ axis, and then to $x^0$ along the real
$x^{\prime
0}$ axis. These are known as the Hartle-Hawking contours. It is easy to 
see that $h^{(N)}_{\vec k_\perp}$ given in \refb{esa-1} satisfies 
\refb{eheq}.

Since we shall be
interested in
the asymptotic form of the closed string fields in the $x^0\to\infty$
limit, we
can
take the contours $C$ and $C'$ to run all the way to $+\infty$ along
the
real $x^{\prime 0}$ axis. By closing the contours in the first and the
fourth quadrangles
respectively, we can easily show that\cite{0301038,0303139} as
$x^0\to\infty$,
 \ben \label{es9.16}
\int_C \, e^{i \onk
x^{\prime 0}} \wt f(x^{\prime
0}) d x^{\prime 0} \to -{i\pi\over \sinh (\pi\onk)} e^{-i\onk \ln
\sin(\pi\tl)} \, , \nonumber \\
\int_{C'} \, e^{-i \onk
x^{\prime 0}} \wt f(x^{\prime
0}) d x^{\prime 0} \to {i\pi\over \sinh (\pi\onk)} e^{i\onk \ln
\sin(\pi\tl)} \, .
 \een
Thus in this limit
 \be \label{es9.17aa}
h^{(N)}_{\vec k_\perp}(x^0) \to
{\pi\over
\sinh(\pi\onk)} \, {1\over 2\onk} \, \left[ e^{-i\onk (x^0 + \ln
\sin(\pi\tl)) } +  e^{i\onk (x^0 + \ln
\sin(\pi\tl)) } \right] \, .
 \ee
Substituting this into \refb{esa0} we get
 \ben \label{es9.17bb}
|\Psi_c^{(1)}\ra &\to& 2\, K\, g_s^2 \, \TT_p \, \int {d^{25-p}
k_\perp\over (2\pi)^{25-p}} \, \sum_{N\ge 0} \, {\pi\over
\sinh(\pi\onk)} \, {1\over 2\onk} \, \wh A_N\,  c_1\, \bar c_1\,
\,
\nonumber \\
&& \left[ e^{-i\onk \, \ln
\sin(\pi\tl)} |k^0=\onk, \vec k_\parallel=0, \vec k_\perp\ra +
e^{i\onk
\, \ln
\sin(\pi\tl)} |k^0=-\onk, \vec k_\parallel=0, \vec k_\perp\ra\right]
\, .
\nonumber \\
 \een
Since $\wt f(x^0)$ vanishes as $x^0\to\infty$, in this limit we
should be left with pure closed string background satisfying free
field
equations of motion $(Q_B+\bar Q_B)|\Psi_c^{(1)}\ra=0$, {\it i.e.}
on-shell
closed string field
configuration\cite{0304192}. This can be verified
explicitly\cite{0304192,0402157}.

One amusing point to note is that \refb{es9.17bb} does not vanish even in
the $\tl\to {1\over 2}$ limit, although the boundary state
$|\BB_1\ra$
vanishes in this limit. This is because in the euclidean theory the
boundary state $|\BB_1\ra$ for $\tl={1\over 2}$ represents an array of
D-branes with Dirichlet
boundary condition on $X=i X^0$, located at $x=(2n+1)\pi$. This produces
a non-trivial background in the euclidean theory, which, upon
the replacement $x\to i x^0$, produces a source free closed string
background in the Minkowski
theory\cite{0304192}.

\refb{es9.17bb} gives the on-shell closed string radiation produced by the
rolling tachyon background. In appendix \ref{appa} we have computed
the energy $\EE$ per unit
$p$-volume carried by this radiation. The answer is
 \be \label{eedensityr}
\EE = \sum_N \EE_N = 4\, K \, (g_s \TT_p)^2 \, \sum_N \, s_N \,
\int {d^{25-p} k_\perp\over (2\pi)^{25-p}} \, {\pi^2\over
\sinh^2(\pi\onk)} \, ,
 \ee
where $s_N$ is defined through the generating function:
 \ben \label{esn3r}
\sum_N s_N \, q^{2N} &=&  {1\over 2} \, \la 0 |
\exp\left[\sum_{m=1}^\infty \sum_{s=1}^{24} (-1)^{d_s} \, {1\over
m} \,  \alpha^s_{m} \bar \alpha^s_{m}\right]  q^{L_0^{matter}+\bar
L_0^{matter}} \nonumber \\  && \exp\left[\sum_{n=1}^\infty
\sum_{r=1}^{24} (-1)^{d_r} \, {1\over n} \,  \alpha^r_{-n} \bar
\alpha^r_{-n}\right] \, | 0\ra''_{matter}\, .
 \een
Here $\la \cdot|\cdot\ra''_{matter}$ denotes the BPZ inner product in
the
matter sector with the normalization convention $\la
0|0\ra''_{matter}=1$. $d_s$ is an integer which can take values 0 or 1
but the final answer is independent of $d_s$. The $N$-th term $\EE_N$ in
the
sum in eq.\refb{eedensityr} gives the total energy carried by all the
closed string modes at level $(N,N)$.

Since $g_s \TT_p\sim 1$, $K$ is a numerical constant determined
using eq.\refb{ezopen}, and $s_N$ is a dimensionless
number we see that for a fixed $N$, $\EE_N$
is of order 1. Thus the total energy per unit $p$-volume carried by closed
string states of a
given mass$^2$ level is of order unity. Since for small $g_s$ this is much
smaller that
$\TT_p\sim (g_s)^{-1}$ -- the tension of the D-$p$-brane --
we see that the amount of energy per unit $p$-volume carried by the closed
string modes at a
given level is much smaller than the tension of the brane. However the
question that we need to address is whether the
total energy density $\EE$ carried by all the modes of the closed string
is also small
compared to $\TT_p$. For this we need to estimate the large $N$
behaviour
of $s_N$ and also
need to carry out the momentum integral in \refb{eedensityr}. Let
us first do the momentum integral. Since for large $N$,
$m_N\equiv \sqrt{4(N-1)}$ is large, \refb{defonk} reduces to
 \be \label{es9.21}
\onk \simeq m_N + {\vec k_\perp^2 \over 2 m_N} \, .
 \ee
This gives
 \be \label{es9.21a}
\sinh(\pi\onk) \simeq {1\over 2} \exp\left(\pi m_N + \pi \, {\vec
k_\perp^2
\over 2 m_N} \right) \, .
 \ee
The integration over $\vec k_\perp$ then becomes gaussian integral and
gives:
 \be \label{es9.22}
\int {d^{25-p} k_\perp \over (2\pi)^{25-p}} \, \, \left({\pi \over
\sinh(\pi\onk)}\right)^2 \simeq 4\pi^2 \, {1 \over (2\pi)^{25-p}}
\exp(-4\pi \sqrt{N}) \, (4N)^{(25-p)/4} \, .
 \ee
The dominant contribution comes from $|k_\perp| \sim (m_N)^{1/2}$.

To find the large $N$ behaviour of $s_N$, we note from \refb{esn3r}
that $2s_N$ simply
counts the number of closed string states at oscillator level $(N,N)$ 
which
are created by identical combination of $X^s$ oscillators
in the left and the
right sector. Alternatively this can be regarded as the number of states
at oscillator level $N$ created by the 24 right moving oscillators
$\alpha^s_{-n}$ acting on the Fock vacuum. For large $N$ this is given
by\cite{0303139}:
 \be \label{es9.19}
2 s_N \simeq {1\over \sqrt 2} \, N^{-27/4} \, \exp(4\pi\sqrt{N})
\, .
 \ee
Substituting \refb{es9.22} and \refb{es9.19} into \refb{eedensityr} we
get:
 \be \label{es9.23}
\EE_N \simeq  \, K \, (g_s \TT_p)^2 \, \pi^2 \, {2^{(30-p)/2}
\over (2\pi)^{25-p}} \, N^{-1/2-p/4}\, ,
 \ee
for large $N$. Thus $\EE=\sum_N\EE_N$
is divergent for $p\le 2$. This shows that the total amount of
energy per unit $p$-volume carried by {\it all} the closed string
modes during the rolling of
the tachyon is infinite\cite{0303139,0304192}!

{}From now on we shall focus on the $p=0$ case. For this the results
obtained so far may be summarized as follows:
 \begin{enumerate}
\item Total amount of energy in closed string modes below any given mass
level is finite. More precisely \refb{es9.23} for $p=0$ shows that the
total energy carried by the closed string modes of mass less than
some fixed value $M$ is proportional to $\sum_{N\le M^2/4} N^{-1/2}
\sim M$.

\item The total amount of energy in all the closed string modes is
infinite since the sum over $N$ diverges.

\item The contribution to the energy of a closed string mode of mass
$m_N$ comes predominantly from modes with momentum $|\vec k_\perp|\sim
(m_N)^{1/2}$.

\end{enumerate}

Of course since the D0-brane has a finite energy, the total energy
carried by the closed string fields cannot
really be infinite. We should expect that once the backreaction of the
closed string emission process on the rolling of the tachyon is taken into
account there will be a natural upper cut-off on the sum over $N$ so that
we get a finite answer.
In particular since
the original D0-brane has energy of order $1/g_s$, it
suggests that the backreaction of the closed string emission on the
rolling tachyon solution will put a natural cut-off of order $1/g_s$ on
the emitted closed string modes.
In
that case the results of the calculation may be reinterpreted as follows:

 \begin{enumerate}

\item All the energy of the D0-brane is converted into closed string
radiation.

\item Most of the energy is stored in the closed string modes of mass
$\sim g_s^{-1}$. This follows from the fact that the total energy
carried by all closed string modes of mass$\le M$ is of order $M$, and
for $M<< g_s^{-1}$ this energy is small compared to $g_s^{-1}$.

\item Typical momentum of these closed string modes is of order
$g_s^{-1/2}$.

\end{enumerate}

If we have a D-$p$-brane with all its tangential directions compactified
on a torus, then it is related to the D0-brane via T-duality, and
hence we expect that similar results will hold for this system as well. In
particular since under a T-duality transformation momentum along a circle
gets mapped to the winding charge along the dual circle, we expect the
following results:

 \begin{enumerate}

\item All the energy of the D$p$-brane wrapped on a torus is converted
into closed string
radiation.

\item Most of the energy is stored in the closed string modes of mass
$\sim g_s^{-1}$.

\item For these closed string modes, the typical momentum along
directions
transverse to the brane and typical winding along directions tangential to
the brane are of order
$g_s^{-1/2}$.

\end{enumerate}

These results suggest that the effect of closed string emission from a 
D-brane produces a large backreaction and invalidates the classical open 
string analysis. However a different interpretation based on the open 
string completeness conjecture 
will be discussed in section \ref{sph}.

\subsection{Closed string fields produced by $|\BB_2\ra$} \label{ecc2}

We now turn to the analysis of closed string fields generated by
$|\BB_2\ra$ given in \refb{ebb2}.
Since the state $ \wh\OO^{(n)}_N
\, (c_0+\bar c_0) \, c_1 \, \bar c_1 \, e^{n X^0(0)} \, |\vec
k_\perp\ra$ appearing in \refb{ebb2} is an eigenstate of
$2(L_0+\bar L_0)$ with eigenvalue $(4(N-1)+n^2 + \vec k_\perp^2)$,
the natural choice of the closed string field produced by
$|\BB_2\ra$, obtained by replacing $|\BB\ra$ by 
$|\BB_2\ra$ in eq.\refb{es9.1}, is
 \ben \label{ebb2cl}
|\Psi_c^{(2)}\ra &=& 2 \,  K\, g_s^2 \, \TT_p \, \sum_{n\in Z} \,
\sum_{N=1}^\infty
\,  \int \,{ d^{25-p}
k_\perp \over (2\pi)^{25-p}} \,
\left(4(N-1)+n^2 + \vec
k_\perp^2\right)^{-1} \nonumber \\
&& \qquad \qquad \qquad \qquad \qquad \, \wh \OO^{(n)}_N
\, c_1 \, \bar c_1 \, e^{n X^0(0)} \, |k^0=0, \vec k_\parallel=0,\vec
k_\perp\ra\, .
 \een
Clearly, this is the result that we shall get if we begin with the
closed
string background produced by the boundary state in the euclidean
theory
and then analytically continue it to the Minkowski space.

Since the source for the closed string fields produced by $|\BB_2\ra$
is localized at $\vec x_\perp=0$, $|\Psi^{(2)}_c\ra$ should satisfy
source
free closed
string
field equations away from the origin. It is easy to see that this is
indeed the case\cite{0402157}.
The space-time
interpretation of this state for a given value of $n$ is that it
represents a field which grows as $e^{n x^0}$. For positive $n$ this
diverges as $x^0\to\infty$ and for negative $n$ this diverges as
$x^0\to-\infty$.
On the other hand in the transverse spatial directions the solution
falls
off as
$G(\vec x_\perp, \sqrt{4(N-1) +n^2})$ where $G(\vec x_\perp, m)$
denotes
the Euclidean Green's function of a scalar field of mass $m$ in
$(25-p)$
dimensions. Since $G(\vec x_\perp, m) \sim e^{-m |\vec x_\perp|}/
|\vec
x_\perp|^{(24-p)/2}$ for large $|\vec x_\perp|$, we
see
that in position space representation the closed 
string field associated with the
state $\wh \OO^{(n)}_N c_1\bar c_1|k\ra$ behaves as
 \be \label{eexpon}
\exp\left(n x^0 - \sqrt{4(N-1) +n^2} \, |\vec x_\perp|\right)/ |\vec
x_\perp|^{(24-p)/2}\,
 \ee
for large $|\vec x_\perp|$.
Thus at any given time $x^0$, the field associated with $\wh
\OO^{(n)}_N
c_1\bar c_1|k\ra$ is small for $|\vec x_\perp|>> n x^0/\sqrt{4(N-1)
+n^2}$
and large for $|\vec x_\perp| <<  n x^0/\sqrt{4(N-1) +n^2}$. We can
view such a field configuration as a disturbance
propagating outward in
the transverse directions from
$\vec x_\perp=0$ at a speed of $ n/\sqrt{4(N-1) +n^2}$. For $N>
1$
this
is less than the speed of light but approaches the speed of light for
fields for which $N-1<< n^2$.

\refb{eexpon} shows that for any $\vec x_\perp$, the closed string
field configuration eventually grows to a value much larger than
1, and hence the linearized closed string field equation which we
have used for this computation is no longer valid. This also suggests that 
the classical open string analysis of the rolling tachyon solution will 
suffer a large backreaction due to these exponentially growing closed 
string fields. However a different 
interpretation of this phenomenon will be discussed in section
\ref{sph} in the context of open string completeness conjecture.

We have already seen earlier that
$|\wh\BB\ra_{c=1}$ and hence
$|\BB_2\ra$
vanishes for $\tl={1\over 2}$. As a result the operators
$\wh\OO^{(n)}_N$
defined
through \refb{ebb2} vanish, and hence $|\Psi_c^{(2)}\ra$ given in
\refb{ebb2cl} also vanishes. Thus in the $\tl\to {1\over 2}$ limit the
$|\Psi_c^{(1)}\ra$ given in \refb{es9.17bb} is the only contribution
to
the closed string background. This of course is manifestly finite in
the
$x^0\to\infty$ limit (although, as we have seen, it carries infinite
energy).

There is an alternative treatment\cite{0301038} of the boundary state
associated with the rolling tachyon solution in which the
exponentially growing contributions to $|\BB_2\ra$ are absent
altogether. This follows a different analytic continuation
prescription in which instead of beginning with the Euclidean $c=1$
theory of a scalar field $X$, we begin with a theory with $c>1$
by giving the scalar field $X$ a small amount of background charge. We
then analytically continue the results to the Minkowski space and then
take the $c\to 1$ limit. In the context of the 26 dimensional critical
string theory that we have been discussing, we do not have any
independent way of deciding which of
the analytic continuation procedure is correct.
We shall however
see in section \ref{s11} that at least in the context of two
dimensional string theory the
exponentially growing terms in $|\BB_2\ra$ do carry some physical
information about
the system.

Other aspects of closed string field produced by 
unstable D-branes and other time dependent configurations have been 
discussed in 
\cite{0005242,
0204071,0204144,
0204191,0205198,0207004,
0207089,0209222,
0212150,0301095,
0303035,0305055,
0305191,
0306096,0307034,
0307078,
math-ph/0308034,
0309017,0403050,
0403147,0403156,0404039,
0407147,
0409019,
0409044}.

\sectiono{D0-brane 
Decay in Two Dimensional String Theory} \label{s11}

In the last few sections we have discussed various aspects of the 
dynamics of 
unstable
D-branes in critical string theory. Due to the complexity of the
problem our analysis has been restricted mostly to the
level of disk amplitudes. In this section we shall study the process of 
D-brane decay
in two dimensional string
theory\cite{0304224,0305159,0305194,0307083,0307195}. One of the
reasons for doing this is that in this theory we can carry out the
analysis in two different ways: 1) by regarding this as an
ordinary string theory\cite{KPZ,DAVID,DISKAW} and applying the
techniques developed in the earlier sections for studying the
dynamics of unstable D-branes, and 2) by using an exact description
of the theory in terms of matrix model\cite{GROMIL,BKZ,GINZIN}. We
shall see that the matrix model results agree with the open string
tree level analysis in the appropriate limit, and allows us to
extend the results beyond tree level. In section \ref{sopenclosed} we
shall see that these all order results lend support to the open
string completeness conjecture that will be formulated in that
section.

\subsection{Two dimensional string theory} \label{s2d}

We begin by reviewing the bulk conformal field theory associated with
the
two dimensional string theory. The world-sheet action of this
CFT is given by the sum of three separate components:
 \be \label{e11.1}
s = s_L + s_{X^0} + s_{ghost}\, ,
 \ee
where $s_L$ denotes the Liouville field theory with central charge 25,
$s_{X^0}$ denotes the conformal field theory of a single scalar field
$X^0$ describing the time coordinate
and
$s_{ghost}$ denotes the usual
ghost action involving the fields $b$, $c$, $\bar b$ and $\bar c$. Of
these $s_{X^0}$ and $s_{ghost}$ are familiar objects. The Liouville
action
$s_L$ on a flat world-sheet is given by:
 \be \label{e11.2}
s_L = \int d^2 z \left ({1\over 2\pi}\, \p_z\vp \p_{\bar z} \vp + \mu
e^{2\vp}\right)
 \ee
where $\vp$ is a world-sheet scalar field and $\mu$ is a constant
parametrizing the theory. We shall set $\mu=1$ by shifting $\vp$
by ${1\over 2} \ln\mu$. The scalar field $\vp$ carries a
background charge $Q=2$ which is not visible in the flat
world-sheet action \refb{e11.2} but controls the coupling of $\vp$
to the scalar curvature on a curved world-sheet. This is
equivalent to switching on a background dilaton field
 \be \label{ebdil}
\Phi_D=Q\vp=2\vp\, .
 \ee
The resulting theory has a central 
charge
 \be \label{e11.3}
c = 1 + 6 Q^2 = 25\, .
 \ee
Note that for large negative $\vp$ the potential term in \refb{e11.2}
becomes small and $\vp$ behaves like a free scalar field with background
charge. Also in this region the 
string coupling constant $e^{\Phi_D}$ is small.

For our analysis we shall not use the explicit world-sheet action
\refb{e11.2},
but
only use the abstract properties of the Liouville field theory
described
in \cite{9206053,9403141,9506136,0101152,0104158,0207041,0303150,0311202}. 
In particular
the
property of
the bulk
conformal
field
theory that we shall be using is that it has a one parameter ($P$)
family
of primary
vertex operators, denoted by $V_{Q + i P}$, of conformal weight:
 \be \label{e11.4}
\left({1\over 4} (Q^2 + P^2), {1\over 4} (Q^2 + P^2)\right) =
\left(1+{1\over 4} P^2, 1+{1\over 4} P^2\right)\, .
 \ee
A generic $\delta$-function normalizable state of the bulk Liouville
field theory
is
given by a
linear combination of the secondary states built over the primary
 \be \label{defpl}
|P\ra= V_{Q+iP}(0)|0\ra\, , \quad P\quad \hbox{real}\, ,
 \ee
where $|0\ra$ denotes the
SL(2,C) invariant vacuum in the Liouville field theory.

For large negative $\vp$ the world-sheet scalar field $\vp$ behaves like a
free field,
and hence one might expect that the primary vertex operators in this
region take the form $e^{iP\vp}$. There are however two subtleties.
First, due to the linear dilaton background, the delta-function 
normalizable
vertex operators are not of the form $e^{iP\vp}$ but of the form
$e^{(Q+iP)\vp}$. Also due to the presence of the exponentially growing
potential for large positive $\vp$ we effectively have a wall that
reflects any incoming wave into an outgoing wave of equal and opposite
$\vp$-momentum. Thus a primary vertex operator should be an appropriate
linear superposition of $e^{(Q+iP)\vp}$ and $e^{(Q-iP)\vp}$ for large 
negative $\vp$. $V_{Q+iP}$
represents precisely this  vertex operator. In particular for $Q=2$,
$V_{2+iP}$ has the asymptotic form
 \be \label{evasymp}
V_{2+iP} \simeq e^{(2+iP)\vp} - \left({\Gamma(iP)\over
\Gamma(-iP)}\right)^2  e^{(2-iP)\vp}\, .
 \ee
With this choice of normalization the primary states $|P\ra$ satisfy
 \be \label{enormp}
\la P|P'\ra_{liouville} = 2\pi\left( \delta(P+P') - \left({\Gamma(iP)\over
\Gamma(-iP)}\right)^2 \delta(P-P')\right)\, ,
 \ee
where $\la\cdot|\cdot\ra_{liouville}$ denotes BPZ inner product in the
Liouville sector.
{}From this analysis we see that the $V_{2+iP}$ and $V_{2-iP}$ should not
be regarded as independent vertex operators. Instead there is an
identification\cite{9403141,9506136}
 \be \label{evidenti}
V_{2+iP} \equiv -  \left({\Gamma(iP)\over
\Gamma(-iP)}\right)^2 \, V_{2-iP}\, .
 \ee

The closed string field $|\Psi_c\ra$ in this two dimensional string
theory is a ghost
number 2 state satisfying \refb{efg1.5} in the combined state space
of the
ghost, Liouville and $X^0$ field theory. We can expand $|\Psi_c\ra$ as
 \be \label{eexppsi}
|\Psi_c\ra = \int_0^\infty {d P\over 2\pi} \, \int_{-\infty}^\infty\, {d 
E\over 2\pi} \,
\wt\phi(P,
E) \,
c_1 \bar c_1 e^{-i E
X^0(0)} |P\ra + \cdots \, ,
 \ee
where $\cdots$ denote higher level terms. Note that due to the reflection
symmetry \refb{evidenti} we have restricted the range of $P$ integration
to be from 0 to $\infty$. $\wt\phi$ may be regarded as the Fourier
transform of a scalar field $\phi(\vp,x^0)$:
 \be \label{edefftphi}
\phi(\vp, x^0) = \int_0^\infty {dP\over 2\pi} \int_{-\infty}^\infty \,
{dE\over 2\pi} \, e^{-iEx^0} \, \left(e^{i P \vp} -\left({\Gamma(iP)\over
\Gamma(-iP)}\right)^2 \,
e^{-i P\vp}\right)\,
\wt\phi(P, E)
 \ee
for large negative $\vp$. $\phi(\vp,x^0)$ is known as the closed
string tachyon field.\footnote{Throughout this section we shall
use the same symbol  to denote a field and its Fourier transform
with respect to $x^0$. However for the liouville coordinate $\vp$,
a field in the momentum space representation will carry a `tilde'
whereas the corresponding field in the position space
representation will be denoted by the same symbol without a
`tilde'. \label{fn45}} Despite its name, it actually describes a massless
particle in this $(1+1)$ dimensional string theory, since the
condition that the state $c_1 \bar c_1 e^{-i E X^0(0)} |P\ra$ is
on-shell is $E^2 - P^2=0$. This is the only physical closed string
field in this theory. The condition that $\phi(\vp, x^0)$ is real
translates to the following condition on $\wt\phi(P,E)$:
 \be \label{ereal}
 \wt\phi(P,E) = - \left( {\Gamma(-iP)\over \Gamma(iP)}\right)^2 \,
 \wt\phi^*(P,-E) \, .
 \ee

We shall normalize $|\Psi_c\ra$ so that its kinetic term is given by:
 \be \label{ekinlio}
-\la \Psi_c|c_0^- (Q_B+\bar Q_B) |\Psi_c\ra\, .
 \ee
Substituting \refb{eexppsi} into \refb{ekinlio} and using \refb{enormp}
we see that the
kinetic
term for $\wt\phi$ is given by:
 \be \label{ekinphi}
{1\over 2} \, \int_0^\infty {dP\over 2\pi} \, 
\int_{-\infty}^\infty {dE\over 2\pi} \, \wt\phi(P,
-E)
(P^2
-
E^2) \wt\phi(P,E)\, \left({\Gamma(iP)\over
\Gamma(-iP)}\right)^2 \, .
 \ee
Using \refb{edefftphi} this may be expressed as
 \be \label{eactionphi}
-{1\over 2} \int dx^0\, \int d\vp \, (\p_0\phi\p_0\phi - \p_\vp\phi
\p_\vp\phi) \, ,
 \ee
in the large negative $\vp$ region.
Thus $\phi$ is a scalar field with conventional normalization.

If we define a new field
 \be \label{edefpsi}
\wt\psi(P,E) = \left({\Gamma(iP)\over
\Gamma(-iP)}\right) \, \wt\phi(P,E)\, ,
 \ee
then \refb{ekinphi} may be written as
 \be \label{enewac}
{1\over 2} \, \int_0^\infty {dP\over 2\pi} \, 
\int_{-\infty}^\infty \, {dE\over 2\pi} \, \wt\psi(P,
-E)
(P^2
-
E^2) \wt\psi(P,E)\, .
 \ee
If we consider the Fourier transform $\psi(\vp,x^0)$ of $\wt\psi$, defined
through
 \be \label{edefftpsi}
\psi(\vp, x^0) = \int_0^\infty {dP\over 2\pi} \int_{-\infty}^\infty \,
{dE\over 2\pi} \, e^{-iEx^0} \, \left(e^{i P \vp} - e^{-i P\vp}\right)\,
\wt\psi(P, E)
 \ee
for large negative $\vp$,
then the action in terms of $\psi$ takes the form:
 \be \label{eactionpsi}
-{1\over 2} \int dx^0\, \int d\vp \, (\p_0\psi\p_0\psi - \p_\vp\psi
\p_\vp\psi) \, .
 \ee
Thus $\psi$ is also a scalar field with conventional
normalization. Also the reality condition \refb{ereal} guarantees
that the field $\psi(\vp, x^0)$ defined in \refb{edefftpsi} is
real. Although in the momentum space $\wt\psi$ and $\wt\phi$ are
related to each other by multiplication by a phase factor, this
translates to a non-local relation between the two fields in the
position space.

\subsection{D0-brane and its boundary state in two dimensional string
theory} \label{sdb2d}

The two dimensional string theory also has an unstable D0-brane
obtained by
putting an appropriate boundary condition on the world-sheet field $\vp$,
and the
usual Neumann boundary condition on $X^0$ and the ghost
fields\cite{0305159,0305194}.
Since
$\vp$
is an interacting field in the world-sheet theory, it is more
appropriate to describe the
corresponding
boundary CFT associated with the Liouville field
by specifying its abstract properties. The relevant properties are as
follows:
 \begin{enumerate}

\item The open string
spectrum in this boundary CFT is described by a single Virasoro module
built
over the SL(2,R) invariant vacuum state.

\item In the Liouville theory the one point function on the disk of the
closed string vertex
operator $V_{Q +
i P}$ is given
by\cite{0101152,0305159}:
 \be \label{e11.5}
\la V_{Q +
i P} \ra_D = {2\ C\over \sqrt \pi} \, i\, \sinh(\pi P) \,
{\Gamma(i P)\over \Gamma(-iP)}\, ,
 \ee
where $C$ is a normalization constant to be determined in
eq.\refb{e11.5cc}.
\end{enumerate}
Since $V_{Q +
i P}$ for any real $P$ gives the complete set of primary states in the
theory, we get the boundary state associated with the D0-brane to
be:\footnote{The normalization factor of $1/2$ has been included for
convenience so as to compensate for the factor of 2 in the ghost
correlator
$\la
0|c_{-1}\bar c_{-1} c_0^- c_0^+ c_1\bar c_1|0\ra$. In the end the overall
normalization of $|\BB\ra$,
encoded in
the constant $C$, will be determined from eq.\refb{eformb} by requiring
that the
classical action reproduces correctly the one loop string partition
function.}
 \ben \label{eboulio}
|\BB\ra &=& {1\over2} \, \exp\left(\sum_{n=1}^\infty
{1\over
n} \alpha^0_{-n} \bar
\alpha^0_{-n} \right)|0\ra \, \otimes \, \int_0^\infty {d P\over 2\pi} \,
\la
V_{Q
-
i P} \ra_D |P\ra\ra \nonumber \\
&& \otimes \, \exp\left(-\sum_{n=1}^\infty
(\bar b_{-n}
c_{-n} +
b_{-n}
\bar c_{-n})
\right) (c_0+\bar c_0)c_1\bar c_1 |0\ra\, ,
 \een
where $|P\ra\ra$ denotes the
Ishibashi state in the Liouville theory, built on the primary
$|P\ra\equiv V_{Q+iP}(0)|0\ra$.\footnote{Note that $|P=0\ra$ is not the 
SL(2,R) invariant vacuum $|0\ra$ in the Liouville sector.}
The normalization constant $C$ is
determined
by requiring that if we eliminate $|\Psi_c\ra$ from the combined
action
 \be \label{ecombined}
-\la \Psi_c|c_0^- (Q_B+\bar Q_B) |\Psi_c\ra + \la \Psi_c|c_0^-
|\BB\ra\, ,
 \ee
using its equation of motion, then the resulting value of the action
reproduces the one loop open string partition function $Z_{open}$ on
the
D0-brane. In fact, in this case the contribution from the higher
closed string modes cancel (with the contribution from $b$ and $c$ 
oscillators cancelling the contribution from the $X^0$ oscillator and the 
Liouville Virasoro generators). Thus we can restrict $|\Psi_c\ra$ 
to only the closed string tachyon mode given in eq.\refb{eexppsi}. 
Substituting \refb{eexppsi} into
\refb{ecombined}, and expressing the result in terms of the field 
$\wt\psi(P,E)$ defined
in \refb{edefpsi}, we get
 \be \label{eforma}
-{1\over 2} \, \int_0^\infty {d P\over 2\pi} \, 
\int_{-\infty}^\infty \, {dE\over 2\pi} \,
\wt\psi(P,-E) (P^2 - E^2) \, \wt\psi(P,E) + \int_0^\infty {dP\over 2\pi}
\la
V_{Q+iP}\ra_D \,  {\Gamma(-iP)\over \Gamma(iP)} \,
\wt\psi(P,0)\, .
 \ee
Eliminating $\wt\psi$ using its equation of motion and requiring that the
answer is equal to
$Z_{open}$ gives
 \ben \label{eformb}
Z_{open} &=& -{1\over 2}\, T \, \int_0^\infty {dP\over 2\pi} \, \left(\la
V_{Q+iP}\ra_D\right)^2 \, {1\over P^2}\, \left( {\Gamma(-iP)\over
\Gamma(iP)}\right)^2 \nonumber \\
&=& {2 C^2\over \pi} \, T\, \int_0^\infty {dP\over 2\pi} \, \sinh^2(\pi P)
\, {1\over P^2}\, ,
 \een
where $T=2\pi\delta(E=0)=\int dx^0$ denotes the total length of the
time interval.
We can rewrite this as
 \be \label{erew1}
Z_{open} = {2 C^2\over \pi} \, T\, \int_0^\infty ds \int_0^\infty {dP\over
2\pi} \, e^{-s P^2} \, {1\over 4} \, (e^{2\pi P} + e^{-2\pi P} -2)\, .
 \ee
After doing the $P$ integral and making a change of variable $t=\pi/2s$ we
get
 \ben \label{erew2}
Z_{open} &=& {T\over 4\sqrt 2 \pi} \, C^2 \, \int_0^\infty {dt\over
t^{3/2}}
\left(e^{2\pi t} - 1\right) \nonumber \\
&=& T C^2 \int_0^\infty \, {dt\over 2t} \, \int_{-\infty}^\infty \,
{dE\over 2\pi} \, e^{-2\pi t E^2} \, \left(e^{2\pi t} -1\right)\, .
 \een
This can be regarded as the one loop open string amplitude with the open
string spectrum consisting of a single tachyonic mode of mass$^2=-1$ and a
single gauge field,\footnote{In 0+1 dimension a gauge field produces a
constraint and removes one degree of freedom. This is the origin of the
$-1$ in \refb{erew2}.} provided we choose\cite{0305159}
 \be
\label{e11.5cc} C=1\, .
  \ee
Thus this choice of normalization reproduces the result that the open 
string spectrum in the Liouville sector consists of a single Virasoro 
module built over the SL(2,R) invariant vacuum . The open string tachyon 
is associated with the state $c_1|0\ra$ and the gauge field is associated 
with the state $\alpha^0_{-1} c_1 |0\ra$. 

We can now add a boundary interaction term $\tl\, \int dt \cosh(X^0(t))$ 
to deform the free field theory involving the coordinate $X^0$ to
the
rolling tachyon boundary CFT, and leave the Liouville and the ghost parts 
unchanged. This gives a rolling tachyon solution on the
D0-brane in
two dimensional string theory.
As in the critical string theory, we divide the boundary state into
two
parts, $|\BB_1\ra$ and $|\BB_2\ra$, with
 \ben \label{ebb1l}
|\BB_1\ra &=& {1\over 2} \, \exp\left(-\sum_{n=1}^\infty
{1\over
n} \alpha^0_{-n} \bar
\alpha^0_{-n} \right) \wt f(X^0(0)) |0\ra \, \otimes \,
\int_0^\infty {dP\over
2\pi}
\,
\la
V_{Q-iP}\ra_D \, |P\ra\ra \nonumber \\
&& \otimes
\exp\left(-\sum_{n=1}^\infty (\bar
b_{-n} c_{-n} +
b_{-n}
\bar c_{-n})
\right) (c_0+\bar c_0)c_1\bar c_1 |0\ra\nonumber \\
&\equiv& {1\over 2} \, \int_0^\infty {dP\over 2\pi} \, \la
V_{Q-iP}\ra_D \, \sum_N \, \wc A_N(P) \, \wt f(X^0(0))
(c_0+\bar c_0) c_1 \bar c_1 |P\ra\, ,
 \een
and
 \ben \label{ebb2l}
|\BB_2\ra &=& {1\over 2} \, |\wh\BB\ra_{c=1} \, \otimes \,
\int_0^\infty {dP\over
2\pi} \,
\la
V_{Q-iP}\ra_D \, |P\ra\ra \nonumber \\
&& \otimes 
\exp\left(-\sum_{n=1}^\infty
(\bar
b_{-n} c_{-n} +
b_{-n}
\bar c_{-n})
\right) (c_0+\bar c_0)c_1\bar c_1 |0\ra\, \nonumber \\
&\equiv& {1\over 2} \, \sum_{n\in Z} \, \sum_{N=1}^\infty \, \int_0^\infty
{d P
\over 2\pi} \, \la V_{Q-iP}  \ra_D
\, \wc \OO^{(n)}_N(P)
\,
(c_0+\bar c_0) \, c_1 \, \bar c_1 \, e^{n X^0(0)} \,
|P\ra\, .
 \een
Here $|\wh\BB\ra_{c=1}$ is the inverse Wick rotated version of $|\wt
B\ra_{c=1}$ as defined in eq.\refb{esecondform}, and $\wc A_N(P)$ and
$\wc\OO^{(n)}_N(P)$ are operators of
level $(N,N)$,
consisting of non-zero mode ghost and $X^0$ oscillators, and the
Virasoro
generators
of the Liouville theory. The $P$ dependence of $\wc A_N$ and $\wc 
O^{(n)}_N$ originates from the fact that the Virasoro Ishibashi state in 
the Liouville sector, when expressed as a linear combination of Liouville 
Virasoro generators acting on the primary $|P\ra$, has $P$ dependent 
coefficients.

As in the case of critical string theory, it is easy to show that
$|\BB_1\ra$ and $\BB_2\ra$ are separately BRST invariant.

\refb{ebb1l} and \refb{ebb2l} shows that the sources for the various
closed string fields in the momentum space are proportional to $ \la
V_{Q-iP}  \ra_D$. It is instructive to see what they correspond to in
the
position space labelled by the Liouville coordinate $\vp$. We
concentrate
on the large negative $\vp$ region.
In this region $V_{Q+iP}$ takes the form \refb{evasymp}
and
 \be \label{eprr}
|P\ra\ra \sim \wh \OO_L(P) \, \left( e^{2\vp(0) +
i P\vp(0)} |0\ra - \left( {\Gamma(iP)\over \Gamma(-iP)}\right)^2
e^{2\vp(0) -
i P\vp(0)} \right)|0\ra\, ,
 \ee
where $\wh\OO_L(P)$ is an appropriate operator in the Liouville
field theory which is even under $P\to -P$.
Thus the source terms are of the form
 \ben \label{esource}
&&\int_0^\infty {d
P\over 2\pi} \, \left( e^{2 \vp + i
P \,
\vp}- \left( {\Gamma(iP)\over \Gamma(-iP)}\right)^2
e^{2 \vp - i
P \,
\vp} \right) \la V_{Q-iP}\ra_D \, \wh O_L(P)|0\ra \otimes |s\ra_{X^0,g}
\nonumber 
\\
&\propto&
\int_{-\infty}^\infty {d
P\over 2\pi} \, e^{2 \vp + i
P \,
\vp}  \, \, \sinh(\pi P) \,
{\Gamma(-i P)\over \Gamma(iP)} \, \wh O_L(P)|0\ra \otimes |s\ra_{X^0,g}\, 
,
 \een
where $|s\ra_{X^0,g}$ denotes some state in the $X^0$ and ghost CFT.
As it stands the integral is not well defined since $\sinh(\pi P)$
blows
up for large $|P|$. For negative $\vp$, we shall define this integral
by
closing the contour in the lower half plane, and picking up the
contribution from all the poles enclosed by the contour. Since the
poles
of
$\Gamma(-iP)$ at $P=-in$ are cancelled by the zeroes of $\sinh(\pi P)$
we
see that the integrand has no pole in the lower half plane and hence
the
integral
vanishes.\footnote{In arriving at this conclusion we have to ignore poles 
in $\wh O_L(P)$ in the complex $P$ plane. The residues at these poles 
correspond to null states in the Liouville theory and are set to zero in 
our analysis\cite{0408064}.} Thus the boundary state $|\BB_1\ra$ 
and $|\BB_2\ra$ given in \refb{ebb1l} and \refb{ebb2l} do not produce any
source term for large negative $\vp$. This in turn leads to the
identification
of this
system as a D0-brane that is localized in the Liouville
direction\cite{0305159}.

Following arguments similar to those for the critical string theory
given at the end of section \ref{s10}, one can show that the time
dependence of various terms in the $|\BB_2\ra$ component of the
boundary state is fixed by the requirement of BRST invariance. This
indicates
that
$|\BB_2\ra$ encodes information about conserved charges. To see
explicitly
what these conserved charges correspond to, we first need to
express $|\BB_2\ra$ in a more suggestive form. Making the replacement 
$x\to i x^0$ in eq.\refb{epriform} we get
 \be \label{eneweq}
|j,m\ra = \wh \PP_{j,m} \, e^{-2mX^0(0)}|0\ra\, ,
 \ee
where it is understood that in $\wh \PP_{j,m}$ we replace the $\alpha_n$'s 
by $i\alpha^0_n$'s. Let us define  $\wh
\NN_{j,m}$ to be
an operator made of $\alpha^0_{-n}$,
$\bar\alpha^0_{-n}$ for $n>0$
such that the Ishibashi state $|j,m\ra\ra$ in the Minkowski $c=1$
theory is
given by:
 \be \label{ediff3}
|j,m\ra\ra = \wh \NN_{j,m} \, e^{-2m X^0(0)} \, |0\ra\, .
 \ee
We also define
$\wh\RR_{j,m}(P)$ as
 \be \label{ecrit1.5}
\wh\RR_{j,m}(P) = \wh \NN_{j,m} \,
\wh\OO_L(P)\, \exp\left(-\sum_{n=1}^\infty (\bar b_{-n} c_{-n}
+
b_{-n}
\bar c_{-n})
\right)\, ,
 \ee
where $\wh \OO_L(P)$ has been defined in eq.\refb{eprr}.
{}From eqs.\refb{ediff3}, \refb{ecrit1.5} it is clear that $\wh
\RR_{j,m}(P)$
does not have
any explicit $\tl$ dependence.
We now use
eqs.\refb{ebb2l}, \refb{eprr} and \refb{esecondform} to express
$|\BB_2\ra$ as
 \be \label{ecrit1}
|\BB_2\ra = {1\over 2} \, \sum_{j=1}^\infty \, \sum_{m=-(j-1)}^{j-1}
\,
\int_0^\infty {d P\over
2\pi} \, \la V_{Q-iP}\ra_D\, f_{j,m} (\tl) \, \wh\RR_{j,m}(P) \,
(c_0+\bar c_0)c_1\bar c_1 e^{-2 m X^0(0)} \, |k^0=0, P\ra\, ,
 \ee
where
$f_{j,m}(\tl)$ has been defined in
eq.\refb{efjm}.

If we now generalize
the
source term so that the boundary state has the
same operator structure but
arbitrary time dependence:
 \be \label{ecrit2}
|\BB_2\ra' ={1\over 2} \, \sum_{j=1}^\infty \, \sum_{m=-(j-1)}^{j-1}
\,
\int_0^\infty {d P\over
2\pi} \, \la V_{Q-iP}\ra_D \, \wh\RR_{j,m}(P) \,
(c_0+\bar c_0)c_1\bar c_1 \, g_{j,m}(X^0(0)) \, |k^0=0, P\ra\, ,
 \ee
then requiring $(Q_B+\bar Q_B)|\BB_2\ra'=0$ gives:
 \be \label{ecrit3}
\p_0 \left(e^{2m x^0}  g_{j,m}(x^0)\right) = 0\, .
 \ee
Thus $e^{2m x^0}
g_{j,m}(x^0)$
can be thought of as a conserved charge which takes value $f_{j,m}
(\tl)$
for $|\BB_2\ra$ given in \refb{ecrit1}.
In particular the total energy carried by the system is proportional
to the conserved charge $g_{1,0}$\cite{0408064} and is given by
 \be \label{eennc}
E = {1\over g_s} \, \cos^2(\pi\tl)\, .
 \ee
This is the analog of eq.\refb{ehx4} 
for the critical string theory. The
$\tl =0$ configuration represents the original D0-brane which has mass
$1/g_s$ in our convention for $g_s$.

Note that this procedure for identifying the conserved charges requires
{\it a priori} knowledge of the boundary state and hence does not provide
us with a systematic method for computing these charges for an
arbitrary boundary state. Such a systematic procedure was developed in
\cite{0408064} where it was shown how given any boundary state in the two
dimensional string theory one can construct infinite number of conserved
charges. Applying this method to the present problem one finds that the
conserved
charges are proportional to the combinations $f_{j,m}(\tl) -
f_{j-1,m}(\tl)$ for $|m|\le j-2$, and $f_{j,m}(\tl)$ for $|m|=j-1$. By
taking
appropriate linear combinations of these charges one can construct
conserved charges whose values are directly given by $f_{j,m}(\tl)$.

\subsection{Closed string background produced by $|\BB_1\ra$}
\label{s2closed}

We now calculate the closed string field produced by this time
dependent
boundary state\cite{0305159,0402157}.
The contribution
from
the $|\BB_1\ra$ part of the boundary state can be easily computed as
in
the case of critical string theory, and in the $x^0\to\infty$ limit
takes the form:
 \be \label{eb1cont}
|\Psi_c^{(1)}\ra \to \int_0^\infty {dP\over 2\pi} \la
V_{Q-iP}\ra_D\, \sum_N \wc A_N(P) \, h^{(N+1)}_{
P}(X^0(0))
c_1 \bar c_1 |P\ra\, ,
 \ee
where $h^{(N)}_{\vec
k_\perp}(x^0)$ has been defined in \refb{esa-1}. The $(N+1)$ in the
superscript of $h$ in \refb{eb1cont} can be traced to the fact that in
this theory a level $(N,N)$ state has mass$^2=4N$ whereas in the
critical
string theory a level $(N,N)$ state had mass$^2=4(N-1)$.

Since the source terms represented by $|\BB_1\ra$ vanish as 
$x^0\to\infty$, in this limit
$|\Psi_c^{(1)}\ra$ is on-shell, {\it i.e.} it is annihilated by the BRST
charge
$(Q_B+\bar Q_B)$. Since the only physical states in the theory come
from
the
closed string `tachyon' field, it must be possible to remove all the
other
components of $|\Psi_c^{(1)}\ra$ by an on shell gauge transformation
of
the
form
$\delta |\Psi_c\ra = (Q_B+\bar Q_B)|\Lambda\ra$ by suitably choosing
$|\Lambda\ra$. Thus the non-trivial contribution to $|\Psi_c^{(1)}\ra$
is encoded completely in the field $\wt\phi(P,E)$ defined through 
eq.\refb{eexppsi}.
Furthermore, since the action of $Q_B$ and $\bar Q_B$
does
not
mix states of different levels, the gauge transformation
that removes the
$N>0$ components of $|\Psi_c^{(1)}\ra$ does not modify the $N=0$
component.
Using \refb{eb1cont}, the expression for $h^{(1)}_{P}$ from
\refb{es9.17aa}, \refb{defonk}, expression for $\la V_{Q-iP}\ra_D$ from 
\refb{e11.5}, the
definition of $\wt\phi(P,E)$ from \refb{eexppsi}, and the fact that $\wc
A_{N=0}=1$,
we get the following expression for the closed string tachyon field
$\wt\phi(P,x^0)$ in the $x^0\to\infty$ limit\cite{0305159}
 \ben \label{e11.6}
\wt\phi(P, x^0\to\infty) &=& - {\pi\over
\sinh(\pi\omega_P)} \, {1\over 2\omega_P} \, {2\over \sqrt \pi} \, i\,
\sinh(\pi
P) \,
{\Gamma(-i P)\over \Gamma(iP)} \nonumber \\
&& \qquad \, \left[ e^{-i\omega_P (x^0 + \ln
\sin(\pi\tl)) } +  e^{i\omega_P (x^0 + \ln
\sin(\pi\tl)) } \right] \, ,
 \een
where $\omega_P = |P|$. As stated in footnote \ref{fn45}, $\wt\phi(P, 
x^0)$ denotes the Fourier transform of $\wt\phi(P,E)$ in the variable $E$.
\refb{e11.6} can be simplified as
 \be \label{e11.7}
\wt\phi(P, x^0\to \infty) = - i\, {\sqrt\pi\over
P} \,
{\Gamma(-i P)\over \Gamma(iP)}
\, \left[ e^{-i P (x^0 + \ln
\sin(\pi\tl)) } +  e^{i P (x^0 + \ln
\sin(\pi\tl)) } \right] \, .
 \ee
The position space representation $\phi(\vp, x^0)$ of this field,
as defined through \refb{edefftphi}, is somewhat complicated. 
However the expression simplifies if we use the field 
$\psi(\vp, x^0)$ to represent this configuration.
Eq. \refb{edefpsi} and \refb{e11.7} gives
 \be \label{e11.7a}
\wt\psi(P, x^0\to \infty) = - i\, {\sqrt\pi\over
P} \,
\left[ e^{-i P (x^0 + \ln
\sin(\pi\tl)) } +  e^{i P (x^0 + \ln
\sin(\pi\tl)) } \right] \, .
 \ee
Its Fourier transform, defined in \refb{edefftpsi}, is given by
 \be \label{epsiposition}
\psi(\vp, x^0) = -\sqrt{\pi} [H(x^0+ \ln
\sin(\pi\tl) - \vp) -
H(x^0+ \ln
\sin(\pi\tl)+\vp)] \, ,
 \ee
where $H$ denotes the step
function:
 \be \label{es9.7}
H(u) = \cases{1 \quad \hbox{for} \quad u>0\cr
0  \quad \hbox{for} \quad u<0} \, .
 \ee

Eq.\refb{epsiposition} is valid only in the $x^0\to\infty$ limit.
Since $\vp<0$, in this limit the first term goes to a constant and
we get
 \be \label{epsinew}
\psi(\vp, x^0) = \sqrt{\pi} \, H(x^0 + \vp + \ln
\sin(\pi\tl)) + \hbox{constant}\, .
 \ee
Thus we see that the $\psi$ field background produced by the rolling
tachyon configuration takes the form of a single step function of height
$\sqrt{\pi}$ in the $x^0\to\infty$ limit. Since $\psi(\vp, x^0)$ is a 
scalar field with conventional kinetic term for large negative
$\vp$, \refb{epsinew} carries infinite energy as in the case
of critical string theory due to the infinite spatial gradient at 
$\vp=-x^0-\ln \sin(\pi\tl)$\cite{0305159}. Thus 
one would again be tempted to conclude that all the energy of the D0-brane 
is 
converted to closed string radiation, thereby invalidating the tree level 
open string analysis. In this case however there is 
a simple interpretation of this infinity as will be discussed in sections
\ref{s11.2} and \ref{sopenclosed}.

This finishes our discussion of closed string field configuration
produced by the
$|\BB_1\ra$ component of the boundary state.

\subsection{Closed string background produced by $|\BB_2\ra$}
\label{s12.4}

We shall now discuss the
closed string background produced by $|\BB_2\ra$\cite{0402157}.
This can be analyzed
in the same way as in the case of critical string theory. We begin
with the expression \refb{ebb2l} of
$|\BB_2\ra$.
Since $ \wc\OO^{(n)}_N(P)
\,
(c_0+\bar c_0) \, c_1 \, \bar c_1 \, e^{n X^0(0)} \,
|P\ra$
in this expression is
an eigenstate of $2(L_0+\bar L_0)$ with eigenvalue $(4N+n^2 +P^2)$, we
can
choose the closed string field produced by $|\BB_2\ra$ to be:
 \be \label{es12.4b}
|\Psi_c^{(2)}\ra = \sum_{n\in Z} \, \sum_{N=2}^\infty \, \int_0^\infty \,{
dP
\over 2\pi} \, \la V_{Q-iP}  \ra_D \,
\left(4N+n^2 + P^2\right)^{-1}
\, \wc\OO^{(n)}_N(P)
\, c_1 \, \bar c_1 \, e^{n X^0(0)} \, |P\ra\, .
 \ee
This corresponds to closed string field configurations which grow as
$e^{nx^0}$ for large $x^0$.

A special class of operators among the
$\wc O^{(n)}_N$'s are those which involve only excitations involving
the
$\alpha^0$,
$\bar\alpha^0$ oscillators and correspond to higher level {\it
primaries}
of the
$c=1$
conformal field theory. As described before, these primaries are
characterized by SU(2) quantum numbers
$(j,m)$ with $j\ge 1$, $-j< m < j$, and has dimension $(j^2,
j^2)$.  From \refb{esecondform}, \refb{epriform} and
\refb{ebb2l} we see
that the
contribution to $|\BB_2\ra$ from these primary states has the form:
 \be \label{eb2new}
{1\over 2} \, \sum_{j\ge 1} \, \sum_{m=-j+1}^{j-1} \, \int_0^\infty \,{ dP
\over 2\pi} \, \la V_{Q-iP}  \ra_D \,
f_{j,m}(\tl) \, \wh
\PP_{j,m} \, (c_0+\bar c_0) \, c_1
\, \bar c_1 \, e^{-2m X^0(0)} \, |P\ra\, .
 \ee
The level of the operators $\wh\PP_{j,m}$ is
 \be \label{espace1}
N = (j^2 - m^2) \, .
 \ee
Thus the $|\Psi^{(2)}_c\ra$ produced by this part of $|\BB_2\ra$ takes
the
form:
 \be \label{espace2}
|\bar\Psi_c^{(2)}\ra = \sum_{j,m} \, f_{j,m}(\tl) \, \int_0^\infty \,{ dP
\over 2\pi} \, \la V_{Q-iP}  \ra_D
\, (4 j^2 + P^2)^{-1}
\, \wh\PP_{j,m}
\, c_1 \, \bar c_1 \, e^{-2m X^0(0)} \, |P\ra\, .
 \ee

As in the case of critical string theory, it is instructive to
study the behaviour of $|\Psi^{(2)}_c\ra$ in the position space
characterized by the Liouville coordinate $\vp$ instead of the
momentum space expression given in \refb{es12.4b}. We concentrate
on the large negative $\vp$ region as usual. Let us first focus on the 
$|\bar\Psi^{(2)}_c\ra$ part
of $|\Psi^{(2)}_c\ra$ as given in \refb{espace2}. If $|\Psi\ra$
contains a term
 \be \label{epjm1}
 \int_0^\infty {dP\over 2\pi}   \, \wh\PP_{j,m} \, c_1\bar c_1\, \wt
 \psi_{j,m}(P,X^0(0)) \,
 |P\ra \, ,
 \ee
then the corresponding position space representation
$\psi_{j,m}(\vp, x^0)$ of this field is defined to be
 \be \label{epjm2}
 \psi_{j,m}(\vp, x^0) = \int_0^\infty {dP\over 2\pi} 
\, \wt \psi_{j,m}(P,x^0)  \, \left[ e^{2\vp + i
 P \vp} - \left( \Gamma(iP)\over \Gamma(-iP)\right)^2 \, e^{2\vp -
 iP\vp}\right]\, .
 \ee
Comparing \refb{epjm1} with \refb{espace2}, and using \refb{epjm2}
and the expression for $\la V_{Q-iP}\ra$ given in \refb{e11.5},
we get
 \ben \label{espace4}
\psi_{j,m}(\vp,x^0) &=& f_{j,m}(\tl) \, e^{-2 m x^0} \, 
\int_0^\infty {d
P\over
2\pi}
\, (4 j^2 + P^2)^{-1} \, \la V_{Q-iP}  \ra_D \nonumber \\
&& \qquad \left[ e^{2 \vp + i
P \,
\vp} -\left({\Gamma(iP)\over \Gamma(-iP)}\right)^2 \, e^{2\vp - iP\vp}
\right]\, \nonumber \\
&=& -{2\over \sqrt \pi} \, i\, f_{j,m}(\tl) \, e^{-2
m
x^0} \, \int_{-\infty}^\infty {d
P\over 2\pi} \, e^{2 \vp + i
P \,
\vp} \, (4 j^2 + P^2)^{-1} \, \sinh(\pi P) \,
{\Gamma(-i P)\over \Gamma(iP)}\, . \nonumber \\
 \een
This integral is not well defined since $\sinh(\pi P)$ blows
up for large $|P|$. As in the analysis of \refb{esource}, for negative
$\vp$ we shall define this integral by
closing the contour in the lower half plane, and picking up the
contribution from all the poles. Since the poles of
$\Gamma(-iP)$ at $P=-in$ are cancelled by the zeroes of $\sinh(\pi
P)$,
the only pole that the integral has in the lower half plane is at
$P=-2ij$. Evaluating the residue at this pole, we get
 \be \label{espace6}
\psi_{j,m}(\vp,x^0) = {1
\over \left((2j)!\right)^2} \, 
\sqrt \pi \, f_{j,m}(\tl) \,  e^{-2m x^0 +
2 (1+j) \vp} \, .
 \ee
In the language of string field theory, this corresponds to
 \be \label{espace7}
|\bar\Psi_c^{(2)}\ra = \sum_{j,m} \, {\sqrt \pi\over
\left((2j)!\right)^2}
\,  f_{j,m}(\tl) \, \wh \PP_{j,m} \,
e^{-2 m X^0(0)}|0\ra_{X^0} \otimes e^{2 (1+j)\vp(0)}|0\ra_L \otimes
c_1 \, \bar c_1 |0\ra_{ghost}\, .
 \ee
The states appearing in \refb{espace7} are precisely the discrete
states
of two dimensional string theory\cite{LIAN,9108004} (after the
replacement $X^0 \to -iX$).

Contribution to $|\Psi_c^{(2)}\ra$ from the terms in
$|\BB_2\ra$ involving excitations by Liouville Virasoro generators have 
been analyzed in \cite{0408064}. Since we shall not need
these results for later analysis we refer the reader to the original paper
for details. 

\subsection{Matrix model description of the two dimensional string theory}
\label{s11.2}

\begin{figure}[!ht]
 \begin{center}
\leavevmode
\epsfysize=5cm
\epsfbox{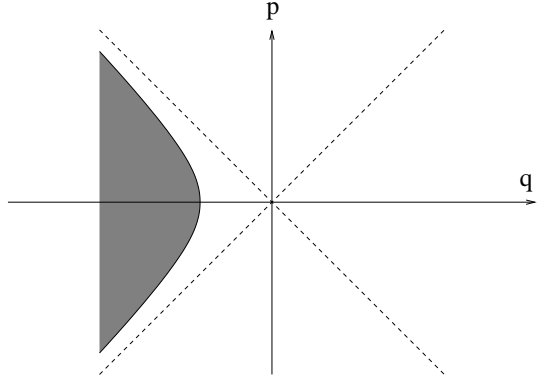}
\end{center}
\caption{Semi-classical representation of the vacuum state in the
matrix
model.
} \label{f12.1}
\end{figure}

The two dimensional string theory described above also has an
alternative
description to all orders in perturbation theory as a matrix
model\cite{GROMIL,BKZ,GINZIN}. This matrix description, in turn, can
be
shown to be equivalent to a theory of infinite number of
non-interacting
fermions,
each moving in an
inverted harmonic oscillator potential with hamiltonian
 \be \label{e12.1}
h(q,p) = {1\over 2} (p^2 - q^2) + {1\over g_s}\, ,
 \ee
where $(q,p)$ denotes a canonically conjugate pair of variables.
The coordinate variable $q$ is related to the eigenvalue of an
infinite dimensional matrix, but this information will not be
necessary for our discussion. Clearly $h(q,p)$ has a continuous
energy spectrum spanning the range $(-\infty, \infty)$. The vacuum
of the theory corresponds to all states with negative $h$
eigenvalue being filled and all states with positive $h$
eigenvalue being empty. Thus the fermi surface is the surface of
zero energy. Since we shall not go beyond perturbation theory, we
shall ignore the effect of tunneling from one side of the barrier
to the other side and work on only one side of the barrier. For
definiteness we shall choose this to be the negative $q$ side. In
the semi-classical limit, in which we represent a quantum state by
an area element of size $\hbar$ in the phase space spanned by $p$
and $q$, we can restrict ourselves to the negative $q$ region, and
represent the vacuum by having the region $(p^2 - q^2) \le
-{2\over g_s}$ filled, and rest of the region
empty\cite{POLCH,9212027} (see Fig.\ref{f12.1}). In this picture
the fermi surface in the phase space corresponds to the curve:
 \be \label{e12.2}
{1\over 2} (p^2 - q^2) + {1\over g_s} = 0\, .
 \ee

If $\Psi(q,t)$ denotes the second quantized fermion field describing
the
above non-relativistic system, then
the massless `tachyon' field in the closed string sector is identified
with the scalar field obtained by the bosonization of the fermion
field
$\Psi$\cite{DASJEV,SENWAD,GROSSKLEB}. The precise correspondence goes
as
follows.
The classical equation of motion satisfied by the field $\Psi(q,x^0)$
has
the form:
 \be \label{e12.1b}
i {\p\Psi\over \p x^0} + {1\over 2} \, {\p^2 \Psi\over \p q^2} +
{1\over
2} \, q^2 \Psi - {1\over g_s} \, \Psi=
0\, .
 \ee
We now define the
`time of flight' variable $\tau$ that is related to $q$ via the
relation:
 \be \label{e12.1a}
q = -\sqrt{2\over g_s} \, \cosh\tau\, , \qquad \tau < 0\, .
 \ee
$|\tau|$ measures the time taken by a zero energy classical particle
moving
under the Hamiltonian \refb{e12.1} to travel from $-\sqrt{2\over g_s}$
to $q$. We also define
 \be \label{e12.1c}
v(q) = -\sqrt{q^2-{2\over g_s}} = \sqrt{2\over g_s} \, \sinh \tau\, .
 \ee
$|v(q)|$ gives the magnitude of the
classical velocity of a zero energy particle when
it is at position
$q$.
Using these variables, it is easy to see that for large negative
$\tau$
the
solution to eq.\refb{e12.1b} takes the form:
 \be \label{e12.1d}
\Psi(q,x^0) = {1\over \sqrt{-2 v(q)}} \, \left[ e^{-i\int^q v(q') dq'
+
i\pi/4} \, \Psi_R(\tau, x^0) + e^{i\int^q v(q') dq' -
i\pi/4} \, \Psi_L(\tau, x^0)\right]\, ,
 \ee
where $\Psi_L$ and $\Psi_R$ satisfy the field equations:
 \be \label{e12.1e}
(\p_0 - \p_\tau) \, \Psi_L(\tau, x^0)=0, \qquad (\p_0 + \p_\tau) \,
\Psi_R(\tau, x^0)=0 \, .
 \ee
Thus at large negative $\tau$ we can regard the system as a theory of
a
pair of
chiral fermions, one left-moving and the other right-moving. Of course
there is an effective boundary condition at $\tau=0$ which relates the
two
fermion fields, since a particle coming in from $\tau=-\infty$ will be
reflected from $\tau=0$ and will go back to $\tau=-\infty$. Since
$\tau$
ranges from 0 to $-\infty$, we can interprete $\Psi_R$ as the incoming
wave
and $\Psi_L$ as the outgoing wave.

Eq.\refb{e12.1e} shows that $\Psi_L$ and $\Psi_R$ represent a pair
of relativistic fermions for large negative $\tau$. 
Thus we can bosonize them into a pair of
chiral bosons $\chi_L$ and $\chi_R$. This pair of chiral bosons
may in turn be combined into a full scalar field $\chi(\tau, x^0)$
which satisfy the free field equation of motion for large negative
$\tau$ and satisfies an appropriate boundary condition at
$\tau=0$. If $\chi$ is defined with the standard normalization,
then for large negative $\tau$ a single right moving fermion is
represented by the configuration\cite{COLEMAN,MANDELSTAM}
 \be \label{ebo1}
\chi=\sqrt{\pi}\, H(x^0-\tau)\, ,
 \ee
and
a
single
left-moving fermion is represented by the configuration
 \be \label{ebo2}
\chi=\sqrt{\pi}\, H(x^0+\tau)\, ,
 \ee
where $H(u)$ denotes the step function defined in \refb{es9.7}.

The field $\chi(\tau,x^0)$ is related to the tachyon field $\phi(\vp,
x^0)$ in the continuum description of string theory by a non-local
field
redefinition\cite{MOORESEI}. In momentum space the relation
is\cite{poly1,9108019,9109005,9402156}:
 \be \label{e12.1f}
\wt\chi(P, E) = {\Gamma(iP) \over \Gamma(-iP)} \,
\wt\phi(P, E)\, .
 \ee
Using \refb{edefpsi} this gives
 \be \label{echps}
\wt\chi(P, E) = \psi(P,E)\, ,
 \ee
and hence
 \be \label{echpspo}
\chi(\tau, x^0) = \psi(\tau, x^0)\, .
 \ee
Using \refb{epsinew} and \refb{echpspo} we see that
the background $\chi$ associated with a rolling tachyon
solution is given by
 \be \label{e12.1i}
\chi(\tau, x^0) = \sqrt{\pi} \, H(x^0 + \tau + \ln
\sin(\pi\tl))
 \ee
for large positive $x^0$ and large negative $\vp$.
According to \refb{ebo2} this precisely represents a single
left-moving
(outgoing)
fermion. This shows that
the
non-BPS D0-brane of the two dimensional string theory can be
identified as
a state of the matrix theory where a single fermion is excited from
the
fermi level to some energy
$>0$\cite{0304224,0305194,0305159}.

The classical
configuration \refb{e12.1i} 
has infinite energy in the
scalar field theory. In the fermionic description this infinite
energy is the result of infinite quantum uncertainty in momentum
for a sharply localized particle in the position space. Thus the
classical limit of the fermionic theory does not have this
infinite energy. This is the origin of the apparent discrepancy
between the classical open string calculation of the D0-brane
energy which gives a finite answer \refb{eennc} and the classical closed 
string
calculation which gives infinite answer\cite{0305159}. We hope that
a similar interpretation can be given for the infinite energy carried
by the closed string configuration produced by a `decaying' D-brane
in the critical string theory as discussed in section \ref{eccl}.

 \begin{figure}[!ht]
\leavevmode
 \begin{center}
\epsfysize=5cm
\epsfbox{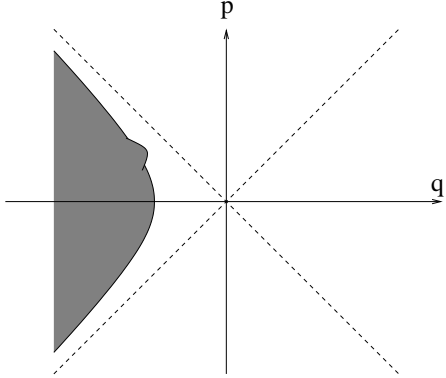}
\end{center}
\caption{Semiclassical representation of a closed string tachyon field
configuration in the matrix model.} \label{f12.2}
\end{figure}

Note that strictly speaking the above analysis, leading to the
identification of the D0-brane with the single fermion excitation,
holds only close to the fermi level, {\it i.e.} near $\tl=1/2$.
{}From the continuum viewpoint this requirement comes from the
fact that the effect of $|\BB_2\ra$ which has not been taken into
account so far, can be ignored only in the $\tl\to {1\over 2}$
limit. {}From the matrix model side this requirement comes due to
the fact that the bosonization of the fermion system in terms of a
single scalar field holds only for excitations close to the fermi
level\cite{POLCH}. Nevertheless it is natural to assume that the
correspondence between a D0-brane and single fermion excitations
continues to hold for general $\tl$.  In this description the
D0-brane with the tachyon field sitting at the maximum of the
potential corresponds to the configuration $p=0$, $q=0$. The mass
of the D0-brane is then given by $h(0,0)=1/g_s$. On the other hand
the rolling tachyon solution in open string theory, characterized
by the parameter $\tl$, corresponds to the phase space trajectory
 \be \label{etrajec}
q = -\sqrt{2\over g_s} \, \sin(\pi\tl) \, \cosh x^0,
\qquad p = -\sqrt{2\over g_s} \, \sin(\pi\tl) \, \sinh x^0
 \ee
as can be seen by comparing the energies of the rolling tachyon
system (eq.\refb{eennc}) and the system described by the
Hamiltonian \refb{e12.1}. The $\tl\to{1\over 2}$
limit corresponds to a trajectory at the fermi level.

We can now use this
correspondence to find an interpretation for the exponentially
growing component $|\Psi^{(2)}_c\ra$ of the string field produced
by $|\BB_2\ra$. Since $|\BB_2\ra$ contains information about the
conserved charges, what we need is the identification of these
charges in the matrix model. An infinite set of conserved charges
of this type do indeed exist in the quantum theory of a single
fermion described by \refb{e12.1}. These are of the
form\cite{SENWAD,MOORESEI,UTTG-16-91,9108004,9110021,9201056,9209036,9210105,
9302106,9507041}:
 \be \label{econs1old}
e^{(k-l) x^0} (p+q)^l (q-p)^k\, ,
 \ee
where $k$ and $l$ are integers.
Requiring that the canonical transformations generated by these
charges
preserve the fermi level $h(q,p)=0$ \cite{9108004} gives us a more
restricted class of
charges:
 \be \label{econs1}
h(q,p) \, e^{(k-l) x^0} (p+q)^l (q-p)^k = \left( {1\over 2}
(p^2-q^2) + {1\over g_s}\right) \, e^{(k-l) x^0} (p+q)^l
(q-p)^k\, .
 \ee
Thus it is natural to identify these with linear combinations of
the charges $e^{2m
x^0} g_{j,m}(x^0)$ in the continuum theory. In order to find the
precise
relation between these charges we can first compare the
explicit $x^0$ dependence of the two sets of charges. This gives:
 \be \label{econs2}
k-l = 2m\, .
 \ee
Thus the conserved charge $e^{2m
x^0} g_{j,m}(x^0)$ should correspond to some specific linear
combination
of the charges given in \refb{econs1} subject to the condition
\refb{econs2}:
 \be \label{econs3}
g_{j,m}(x^0) \leftrightarrow g_s\, \left( {1\over 2} (p^2-q^2) +
{1\over g_s}\right) \, \sum_{k\in Z \atop k\ge 0, 2m}\, \left({2\over
g_s}\right)^{m-k}
\,  a^{(j,m)}_k \, (q-p)^k
(q+p)^{k-2m}
\, .
 \ee
Here $a^{(j,m)}_k$ are constants and the various $g_s$
dependent
normalization factors have been introduced for later convenience.
In order to find the precise form of the coefficients $a_k^{(j,m)}$ we
compare the
$\tl$ dependence of the two sides for the classical trajectory
\refb{etrajec}. Since for this trajectory
 \be \label{econs4}
q\pm p = - \sqrt{2\over g_s} \, \sin(\pi\tl) \, e^{\pm x^0}\, ,
 \ee
and $g_{j,m}(x^0) = e^{-2m x^0} \, f_{j,m}(\tl)$, we have:
 \be \label{econs5}
f_{j,m}(\tl) = (-1)^{2m} \, \left(1-\sin^2(\pi\tl)\right) \,
\sum_{k\in Z \atop k\ge 0,
2m}\,
a^{(j,m)}_k \,  \sin^{2k -
2m}(\pi\tl) \, .
 \ee
Thus by expanding $f_{j,m}(\tl)$ given in \refb{efjm} in powers of
$\sin(\pi\tl)$ we can determine the coefficients $a^{(j,m)}_k$.
One consistency check for this procedure is that on the right hand
side the expansion in powers of $\sin(\pi\tl)$ starts at order
$\sin^{2|m|}(\pi\tl)$. It can be verified that the expansion of
$f_{j,m}(\tl)$ also starts at the same order. The other
consistency check is that the right hand side of \refb{econs5}
vanishes at $\tl={1\over 2}$, which is also the case for
$f_{j,m}(\tl)$. One can also show \cite{0402157,0408064} that these
relations are invertible, {\it i.e.} the charges $h(q,p)\,
(q-p)^{2m+l}\, (q+p)^l$ may be expressed as linear combinations
of $g_{j,m}(x^0)$ for $|m|+1\le j\le m+l+1$. This shows that the
conserved charges $g_{j,m}(x^0)$ in the continuum theory contains
information about the complete set of symmetry generators in the
matrix model description of the D0-brane.

Given that the boundary state $|\BB_2\ra$ carries information
about the conserved charges, the closed string field produced by
$|\BB_2\ra$ must also carry the same information. These can be
regarded as the analog of the long range electric or gravitational
field produced by a particle carrying charge or mass.
Ref.\cite{0408064} gives a systematic procedure for relating the
conserved charges to the asymptotic closed string field
configuration at large negative $\vp$ associated with the discrete
states.

 \begin{figure}[!ht]
\leavevmode
 \begin{center}
\epsfysize=5cm
\epsfbox{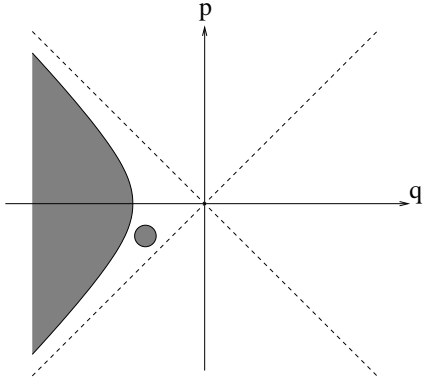}
\end{center}
\caption{Semi-classical representation of a state of the D0-brane in
the
matrix model.} \label{f12.3}
\end{figure}

In the context we note that by carefully examining the bosonization
rules
for a fermion moving under the influence of inverted harmonic
oscillator
potential, ref.\cite{0401067} has argued that in order to
describe the
motion of a single fermion in the language of closed string theory, we
need to switch on infinite number of closed string fields besides the
tachyon. The essential point is that whereas a closed string tachyon
mode describes a deformation of the fermi surface of the form shown in
Fig.\ref{f12.2}, the D0-brane is represented by a blob in the phase
space disconnected from the Fermi surface as in Fig.\ref{f12.3}.
We believe that the presence of the additional closed
string
background \refb{espace7} associated with the discrete
states is a reflection of this effect. As a consistency check we note
that
at $\tl=1/2$ the additional background \refb{espace7} vanish. This is
expected to be true in the matrix model description as
well since in this limit the blob merges with the fermi sea\cite{0401067}.

To summarize, we see that a D0-brane in the continuum two
dimensional string theory is described in the matrix model as a single
fermion excited from the fermi level to some positive energy state.
This provides a satisfactory picture of the D0-brane. But this
correspondence also raises a puzzle. In the matrix model there is
another natural class of states, namely the single hole states, which
correspond to a single fermion exited from some state below
the fermi level to the fermi level. One might expect that just like
the single fermion states, the single hole states should also have
some natural description
in the continuum string theory. So far however we do not have a
completely satisfactory description of these states in the matrix
model. Some proposal for what these states might correspond to have
been made in \cite{0307195,0307221}. According to this proposal the
boundary state
describing a hole is obtained by analytically continuing the boundary
state of a rolling tachyon configuration to $\tl={1\over 2}+i\alpha$ for a
real parameter $\alpha$ and then changing the overall sign of the boundary
state. Although this produces the right values for the conserved charges
carried by the hole, neither the space-time nor the world-sheet
interpretation of these states is very clear\cite{0408064}. An alternative
suggestion, according to which the hole states correspond to ordinary
D0-branes moving under the influence of the linear dilaton
background has been put forward in \cite{0408064}. We hope that this
question will be resolved in the near future.

\subsection{D0-brane decay in type 0B string theory} \label{s0b}

Although two dimensional bosonic string theory provides a useful
arena for studying the decay of D0-branes beyond leading order in
perturbation theory, it suffers from one problem; the theory,
although well defined to all orders in string perturbation
theory, is not non-perturbatively stable. In the continuum
description of the D0-brane this is due to the fact that the
tachyon potential on the D0-brane, while having a local minimum at
some positive value of the tachyon field, is expected to be
unbounded from below on the negative side. The problem associated 
with this can be seen for
example in the analysis of the time dependent solutions where the
tachyon rolling on the wrong side of the potential causes a
divergence in the  dilaton charge at a finite time. In the matrix
model description the instability is
related to the fact that the fermi
level is filled on only one side of the potential leaving the
other side empty. Such a vacuum, while perturbatively stable, is
non-perturbatively unstable. Hence while the matrix
model - two dimensional string theory correspondence illustrates
many aspects of D0-brane decay in perturbation theory, it does not
allow us to go beyond perturbation theory.

On the matrix model side there is an easy way to solve the
problem, -- we just fill the fermi level on both sides. This will
have the feature that the spectrum of closed strings, which
represent excitations around the fermi level, will be doubled
since now we can excite fermion hole pairs on either side of the
potential. In particular bosonizing the fermion field on the two
sides of the potential we should either find two different
asymptotic regions with a massless scalar field living in each of
these regions, or find one asymptotic region with
two massless scalar fields living in this region. This is clearly 
not the
two dimensional bosonic string theory
that we have discussed so far which has only one asymptotic region
($\vp\to -\infty$) with one massless scalar
field. Is there some other two dimensional string theory that
corresponds to this particular matrix model? It turns out that the
answer is yes: it is two dimensional type 0B string
theory\cite{0307083,0307195,0402196}.

The local world sheet dynamics of type 0 string theories is
identical to that of type II string theories, {\it i.e.} it has
both left and right moving world-sheet supersymmetry. Thus the
critical dimension for this theory is 10, which is equivalent to
saying the matter part of the theory must have $\hat c=10$. The
difference with type II  string theories comes in the GSO
projection rules. Whereas the type II string theories have
separate GSO projection on the left and the right sector, type 0
theories have one combined GSO projection\cite{polbook}. Due to
modular invariance this then requires that the theory contains
NS-NS and RR sector states, but no NS-R or R-NS sector states. In
other words, the closed string sector does not contain fermions!
As in the case of type II string theories, we can construct two
types of type 0 string theories, 0A and 0B, which differ from each
other in the sign of the GSO projection operator in the RR sector.

As in the case of bosonic string theory where
we can construct a
two dimensional string theory by replacing 25 of the scalar fields
by a Liouville theory with total central charge 25,  we can also
construct two dimensional type 0A and 0B string theories by
replacing nine of the space-like world-sheet superfields by a 
super-Liouville theory. This theory contains a single superfield with 
exponential potential and background charge such that it describes a 
super-conformal field theory 
with $\hat 
c=9$. This gives rise to the so called two dimensional type 0 string
theories. In particular type 0B theory has two independent
massless scalar fields in the closed string spectrum, one coming
from the RR sector and the other coming from the NS-NS sector.
Thus it is natural to conjecture that this two dimensional string
theory is equivalent to the matrix model described earlier, with
the two scalars being identified with appropriate linear
combinations of the excitations on the fermi level on two sides of
the potential\cite{0307083,0307195}. This is verified by computing the 
S-matrix
involving these fields in the type 0B theory and comparing them
with the predictions of the matrix model\cite{9109005,0309148}.
The only
subtle point to keep in mind is that the matrix model Hamiltonian
\refb{e12.1} seems to correspond to type 0B theory with
$\alpha'=1/2$ rather than $\alpha'=1$. We shall work in this unit
in the rest of this section.

It turns out that the type 0B string theory has an unstable
D0-brane that corresponds to Dirichlet
boundary condition on the Liouville coordinate, and the usual
Neumann boundary condition on the time coordinate. The open string
spectrum on this brane has a tachyonic mode. In the $\alpha'=1$
unit the tachyon has mass$^2=-{1\over 2}$ as in the case of
unstable D-branes of superstring theory, but in the
$\alpha'={1\over 2}$ unit that we are using the tachyon has
mass$^2=-1$. This agrees with the tachyon mass$^2$ obtained by
quantizing the inverted harmonic oscillator Hamiltonian \refb{e12.1}
around $q=0$.
We can also construct the rolling tachyon
solution by switching on tachyon background proportional to $\cosh
x^0$ or $\sinh x^0$. The world-sheet analysis is identical to that
in the case of unstable D-branes in superstring theory except for
the scaling of $\alpha'$. In particular we can construct the
boundary state describing this D0-brane following the procedure
discussed earlier and use this to study closed string radiation
from the D0-brane. The result gives a kink configuration of the
type given in eq.\refb{e12.1i}
showing that the D0-brane is naturally identified with single
fermion excitations in the theory\cite{0307083,0307195}.

Thus we see that we now have an example of a completely consistent
two dimensional string theory and its matrix model description.
This allows us to study aspects of D0-brane dynamics in this
theory not only to all orders in perturbation theory but also
non-perturbatively. We hope that this can be used to derive useful
insight into the dynamics of unstable D-branes in critical string
theories, particularly in the context of the open string
completeness conjecture to be discussed in section
\ref{sopenclosed}.

Various other aspects of D0-brane decay in two dimensional string theory
have been discussed in \cite{0308047,
0310106,0407270,0408049}.

\sectiono{Open String Completeness Conjecture} \label{sopenclosed}

The straightforward analysis of closed string emission from unstable
D0-branes (or D$p$-branes wrapped on $T^p$) tell us that all the energy of
the D0-brane is radiated away into closed strings, both in the critical
string theory and in the two dimensional string theory. Naively this would
suggest that the backreaction
due to the closed string emission process invalidates the classical open
string results. However we shall argue in this section that results
obtained from the tree level
open string theory actually give dual description of the closed
string emission process.

\subsection{Open string completeness in the critical string theory}
\label{sph}

We begin our discussion by analyzing the results for D-brane decay
in critical string theory. In order to illustrate the proposed
duality between open string and closed string description, let us
compare the properties of the emitted closed strings from an
unstable D$p$-brane wrapped on $T^p$ with those infered from the
tree level open string
analysis\cite{0304192,0305011,0306132,0306137}. First of all, tree
level open string analysis tells us that the final system has:
 \be \label{est1}
Q / T_{00} = 0\, ,
 \ee
where $Q$ and $T_{00}$ denote the dilaton charge density and energy
density of the system respectively. On the other hand by examining
the closed string world-sheet action in the background  {\it
string metric} $G_{\mu\nu}$, the anti-symmetric tensor field
$B_{\mu\nu}$ and the dilaton $\Phi_D$ at zero momentum,
 \be \label{est2}
S_{world-sheet} = {1\over 2\pi} \, \int d^2 z (G_{\mu\nu}(X) +
B_{\mu\nu}(X)) \p_z X^\mu \p_{\bar z} X^\nu\, ,
 \ee
we see that the closed string world-sheet does not
couple to the zero momentum dilaton.
This shows the final state closed strings carry zero total dilaton
charge. Hence the dilaton charge of the final state closed
strings
agrees with that computed in the open string description.

Next we note that the tree level open string analysis tells us that the
final system
has:
 \be \label{est3}
p / T_{00} = 0\, ,
 \ee
where $p$ denotes the pressure of the system.
On the other hand, closed string analysis tells us that the final
closed
strings have mass $m$ of order $1/g_s$, momentum $k_\perp$
transverse to the
D-brane of order $1/\sqrt{g_s}$ and
winding $w_\parallel$ tangential to the D-brane of order $1/\sqrt{g_s}$.
For such a system the ratio of transverse pressure to the energy density
is of order $(k_\perp/m)^2\sim g_s$ and the ratio of tangential pressure
to
the energy density is of order $-(w_\parallel/m)^2\sim -g_s$.
Since both these ratios
vanish in the $g_s\to 0$ limit, we again see that the pressure of the
final state closed strings match the
result
computed in the open string description.

Such agreements between open and closed string results also hold
for
more general cases, {\it e.g.} in the decay of unstable branes in the
presence of electric field. Consider, for example, the decay of a
D$p$-brane along $x^1,\ldots x^p$ plane, with
an
electric field $e$ along the $x^1$ axis. In this case the final state is
characterized by its energy-momentum tensor
$T_{\mu\nu}$, source $S_{\mu\nu}$ for anti-symmetric tensor field
$B_{\mu\nu}$ and the
dilaton charge density $Q$. One can show that in the $x^0\to\infty$
limit\cite{0208142,0301049,0303133}:
 \be \label{est4}
T^{00}=|\Pi|\, e^{-1} \, \delta(\vec x_\perp) \, , \quad T^{11} =
- |\Pi| \, e\, \delta(\vec x_\perp) \,   , \quad S^{01} = \Pi \,
\delta(\vec x_\perp) \, ,
 \ee
where $\Pi$ is a parameter labelling the solution. All other components of
$T_{\mu\nu}$ and $S_{\mu\nu}$, as well as the dilaton charge vanishes in
this limit. It can be shown that these tree level open string results
again agree exactly with
the
properties of the final state closed strings into which the D-brane
decays\cite{0306137,0409050}.

Since in all these cases the tree level open string results for
various properties of the final state agree with the properties of
the closed strings produced in the decay of the brane, we are led
to conjecture that the tree level open string theory provides a
description of the rolling tachyon system which is {\it dual} to
the description in terms of closed  string
emission\cite{0305011,0306137}.\footnote{This correspondence has
been checked only for the space averaged values of various
quantities, and not for example, for the local distribution of the
various charges like the stress tensor, dilaton charge and
anti-symmetric tensor field charge.  This is due to the fact that
we can easily give a gauge invariant definition of the
space-averaged quantities since they are measured by coupling to
on-shell (zero momentum) closed string states, but it is more
difficult to give a gauge invariant definition of local
distribution of these charges\cite{0308068}.}$^,$\footnote{At present
it is not clear how exactly the open string theory encodes 
information about closed strings. A hint of how this might happen can
be seen in the analysis of the effective field 
theory\cite{9901159,9909062,0002223,0005031,0009061,0010181,0010240,
0305011,0402027}. This has been reviewed briefly in section \ref{s8.3}.}
This is different
from the usual open closed duality where one loop open string
theory contains information about closed strings. In order to put
this conjecture on a firmer footing, one must show that it arises
from a more complete conjecture involving full quantum open string
theory on an unstable D-brane. The full
conjecture, suggested in
\cite{0308068,0312153}, takes the following
form:

{\it There is a quantum open string field theory
(OSFT) that describes the full dynamics of an unstable Dp-brane without an
explicit coupling to closed strings. Furthermore, Ehrenfest theorem holds
in the weakly coupled OSFT; the classical results correctly
describe the time evolution of the quantum expectation values.}

Since the conjecture in essence says that quantum open string theory is 
fully capable of describing the complete dynamics of an unstable D-brane, 
we call this the {\it open string completeness conjecture}. 
Stated this way, this conjecture also embodies the usual perturbative 
open-closed string duality for a stable D-brane where the open string loop
amplitudes contain information about closed string exchange
processes. In
this case the quantum open string field theory
is fully capable of reproducing the open string scattering
amplitudes at least to all orders in perturbation
theory\cite{GIDDMARW,THORN}.
For example for this system quantum open 
string 
theory gives rise to the cylinder contribution to the string partition 
function at one loop order. Coupling closed strings to this system 
as in eqs.\refb{efg2}, 
\refb{efg3} will ensure that by eliminating the closed string fields by 
their 
classical equations of motion we get the cylinder amplitude again
as in eq.\refb{ezopen}.
This gives a 
clear indication that including closed strings in open string field theory 
amounts to double counting.\footnote{It is of course possible to contruct 
open-closed string field theories where part of the contribution to a 
given process comes from the closed string sector and part of it comes 
from the open string sector\cite{9705241}, but we are not discussing these 
theories here.}

Note that this open string completeness conjecture {\it does
not} imply that the quantum open string theory on a given system
of unstable (or stable) D-branes gives a complete description of
the {\it full string theory}.\footnote{In this sense it is
different from the cases discussed in \cite{9811131} where open
string theory on a set of D-branes is completely equivalent to
closed string theory on a certain background.} It only states that
this open string field theory describes a quantum mechanically
consistent subsector of the full string theory, and is fully
capable of describing the quantum dynamics of the 
D-brane.\footnote{For a similar phenomenon involving stable D-branes in 
two 
dimensional string theory see ref.\cite{0312196}.} 
One of the
consequences of this correspondence is that the notion of
naturalness of a solution may differ dramatically in the open and
the closed string description. A solution describing the decay of
a single (or a few) unstable D-branes may look highly contrived in
the closed string description, since a generic deformation of this
background in closed string theory may not be describable by the
dynamics of a single D-brane, and may require a large (or even
infinite) number of D-branes for its description in open string
theory.

In
our discussion of the properties of the closed string states
emitted from a `decaying' D-brane
we have so far
only included closed string 
radiation produced by the $|\BB_1\ra$ component of the
boundary state.
Hence our analysis is valid only in the
$\tl\to{1\over 2}$ limit when the closed string background produced by
$|\BB_2\ra$ vanishes.
As we have seen in section \ref{ecc2}, for $\tl\ne {1\over 2}$
the closed string fields
produced by the $|\BB_2\ra$ component of the boundary state grow
exponentially with time. Naively,  this exponential growth of the
closed string field configurations again indicates the breakdown of
classical open string description at late time. However in the
spirit of the open string completeness conjecture proposed here
it is more natural to  seek an alternative
interpretation. What this may be indicating is an inadequacy of
the weakly coupled 
closed string description rather than an inadequacy of
the open string
description. As an analogy we can cite the example of closed
string field configurations produced by static stable D-branes.
Often the field configuration is singular near the core of the
brane. However we do not take this as an indication of the
breakdown of the open string description. Instead it is a
reflection of the inadequacy of the closed string description.

While in critical string theory the open string completeness
remains a conjecture, we shall see
in the next subsection that this is bourne out quite clearly
in the two dimensional string theory.

\subsection{Open string completeness in two dimensional string
theory} \label{sdual2d}

The analysis of section \ref{s2closed} shows that a rolling tachyon 
configuration on a D0-brane in two dimensional string theory produces an 
infinitely sharp kink of the closed string tachyon field. Since this 
carries infinite energy, we might naively conclude that all the energy of 
the D0-brane is converted to closed string radiation and hence the results 
of tree level open string analysis cannot be trusted. The analysis of 
section \ref{s12.4} shows that for $\tl\ne {1\over 2}$ the closed string 
field configuration produced by the $|\BB_2\ra$ component of the boundary 
state grows exponentially with time. This is again a potential source for 
large backreaction on the tree level open string results. 

On the other 
hand
we have seen in section \ref{s11.2} that a D0-brane in two
dimensional string theory can be identified in the matrix model
description as a single fermion excitation from the fermi level to
some energy level above 0. Since the fermions are non-interacting,
the states with a single excited fermion do not mix with any other
states in the theory (say with states where two or more fermions
are excited above the fermi level or hole states where a fermion is
excited from below the fermi level to the fermi level). As a
result, the quantum states of a D0-brane are in one to one
correspondence with the quantum states of the single particle
Hamiltonian
 \be \label{ehagain}
 h(q,p) = {1\over 2} (p^2 - q^2) + {1\over g_s}\, ,
 \ee
with one additional constraint, -- the spectrum is cut off sharply
for energy below zero due to Pauli exclusion principle. 
Since by definition quantum open string field theory on a D-brane is a 
field theory that describes the dynamics of that D-brane, we see that 
in
the matrix model description the `quantum open string field
theory' for a single D0-brane is described by the inverted
harmonic oscillator hamiltonian \refb{ehagain} with all the
negative energy states removed by hand\cite{0308068}. The
classical limit of this quantum Hamiltonian is described by the
classical Hamiltonian \refb{ehagain}, with a sharp cut-off on the
phase space variables:\footnote{This Hamiltonian is related to the
D0-brane effective Hamiltonian given in \refb{ex2}, \refb{evform}
by a canonical transformation\cite{0308068} that maps the curve
$h(q,p)=0$ to  $\infty$ in the $\Pi-T$ plane.}
 \be \label{e12.4}
  {1\over 2} (p^2 - q^2) + {1\over g_s} \ge 0\, .
 \ee
This is the matrix model description of `classical open string
field theory' describing the dynamics of a D0-brane.\footnote{It
will be very interesting to understand the precise connection
between this system and the cubic open string field theory that
describes the dynamics of the D0-brane in the continuum
description. It will be even more interesting to study similar
relation between this system and the open string field theory
describing the dynamics of the D0-brane in type 0B string theory,
since there the system is non-perturbatively stable. Since on the
matrix model side we have a free system, the approach of
\cite{0308184,0402063,0409233} 
might provide a useful starting point for
establishing this correspondence.} 

Clearly the quantum system described above provides us with a
complete description of the dynamics of a single D0-brane.  In
particular there is no need to couple this system explicitly to
closed strings, although closed strings could provide an
alternative description of the D0-brane as a kink solution in the
closed string `tachyon' field (as
given in \refb{e12.1i}). This is in
accordance with the open string completeness conjecture
discussed in section \ref{sph}. {}From this it is clear that it is
a wrong notion to think in terms of {\it backreaction of closed
string fields on the open string dynamics}. Instead we should
regard the closed string background produced by the D-brane as a
way of characterizing the open string background (although the
open string theory itself is sufficient for this purpose). For
example, in the present context, we can think of the closed string
tachyon field $\chi(\tau, x^0)$ at late time as the expectation
value of the operator\footnote{$\p_\tau\hat\chi$ is the
representation of the usual density operator of free fermions in
the Hilbert space of first quantized theory of a single fermion.}
 \be \label{eexpec}
\hat \chi(\tau, x^0) \equiv \sqrt{\pi} \, H \left(-\hat q(x^0) -
\sqrt{2\over g_s}
\cosh\tau\right)  \,
 \ee
in the quantum open string theory on a single D0-brane, as described
by
\refb{ehagain}, \refb{e12.4}.
In \refb{eexpec} $\hat q$ denotes the position operator in the quantum
open string theory. When we calculate the expectation value of $\hat
\chi$
in the
quantum state whose classical limit is described by the trajectory
\refb{etrajec}, we can replace $\wh q$ by its classical value
$q=-\sqrt{2\over g_s} \sin(\pi\tl) \cosh(x^0)$. This gives
 \be \label{eexpec2}
\la \hat \chi(\tau, x^0) \ra = \sqrt{\pi} \, H \left(\sqrt{2\over
g_s} \, \sin(\pi\tl) \, \cosh(x^0) - \sqrt{2\over g_s}
\cosh\tau\right) \simeq \sqrt{\pi} \, H(x^0 + \tau + \ln
\sin(\pi\tl))\, ,
 \ee
for large $x^0$ and negative $\tau$. This reproduces
\refb{e12.1i}.

In a similar spirit we note that while the naive analysis of the
closed string field configuration produced by the $|\BB_2\ra$
component of the boundary state indicates that the exponentially
growing closed string fields produce large backreaction and hence
invalidates the analysis based on 
open string theory, the results of section \ref{s11.2} clearly
point to a different direction. According to these results the
exponentially growing terms in $|\BB_2\ra$ are simply consequences
of the conserved charges a D0-brane carries, which in turn may be
calculated completely within the framework of the open string (field)
theory. Thus these exponentially growing terms do not in any way
point to an inadequacy of the open string description of D0-brane
dynamics.

Various other aspects of the open string completeness
conjecture in the
context of two dimensional string theory have been discussed in
\cite{0312135,0312163,0312192}.

\subsection{Generalized holographic principle} \label{sholo}

We have seen in the previous two subsections that the analysis of
unstable D-brane `decay' in critical string theory as well as in
the two dimensional string theory points to the conjecture that
the quantum open string field theory on a given D-brane
describes complete quantum
dynamics of the D-brane. However generically
a given D-brane system does not have the ability to describe an
arbitrary state in string theory, -- it carries only partial 
information about the full theory.\footnote{The possibility of
using infinite number of unstable D-instantons or D0-branes or finite 
number of
space-filling D-branes to give a complete description of string
theory has been discussed in 
refs.\cite{9812135,9904207,0009189,0108085,0212188,0305006}.
Vacuum string field theory even attempts to give a complete description of 
the theory in terms of open string field theory on a single 
D-brane\cite{0105058,0105059,0105168}.}
This conjecture makes it 
clear that open string field theory on a single or
a finite number of D-branes must be encoding information about the
full string theory in a highly non-local manner. For example when
a D-brane `decays' into closed strings we expect that at least
some of the closed strings will eventually 
disperse to infinity. The open string
field theory, defined in terms of variables localized near the
original D-brane, must be able to describe the final closed string
state produced in the decay, although it may not be able to
describe states of the {\it individual closed strings} into which
the D-brane decays. The situation can be described by drawing
analogy to a hologram. If we regard the full string theory as the
complete image produced by a hologram, then a system of
finite number of D-branes can be regarded as a part of the
hologram. This encodes partial information about the full image.
However the information contained in any given piece of the
hologram does not correspond to a given part of the complete
image; instead it has partial information about all parts of the
image. Thus the open string completeness conjecture proposed here
can be thought of as a generalization of the holographic
principle\cite{9802109,9409089,9711200,9802109,9802150,9805114}. As in 
these papers in our proposal
the
relation between the open and closed string description is
non-local, but the space in which the open string degress of
freedom live is not necessarily the boundary of the space in which
the closed string degrees of freedom live.

\medskip

{\bf Acknowledgement:} I would like to thank my collaborators
N.~Berkovits,
D.~Gaiotto, D.~Ghoshal, J.~Majumder, N.~Moeller, P.~Mukhopadhyay,
L.~Rastelli and B.~Zwiebach for collaboration on various aspects
of open string tachyon dynamics.

\appendix

\sectiono{Energy-Momentum Tensor from Boundary State}
\label{appb}

As discussed in section \ref{s3.4},
given a boundary CFT describing a D-brane system, we define the
corresponding boundary state $|\BB\ra$ such that given any closed
string state
$|V\ra$ and the associated vertex operator $V$,
 \be \label{efg1app}
\la\BB|V\ra \propto \la V(0)\ra_D\, ,
 \ee
where  $\la V(0)\ra_D$ is the one point function of $V$ inserted at
the
centre of a unit disk $D$, the boundary condition / interaction
on $\p D$ being the one
associated with the particular boundary CFT under consideration.
{}From this definition it is clear that the boundary state
contains
information about what kind of source for the closed string states is
produced by the D-brane system under consideration.
In this appendix we shall make this more
precise by working with (linearized) closed string field
theory\cite{9705241}.\footnote{We shall not attempt to give a detailed
description of closed string field theory. Section \ref{s4} contains a
self-contained discussion of open string field theory. Closed string
field theory is formulated on similar principles.}

For simplicity we
shall focus on the bosonic string theory.
The closed string field corresponds to a state $|\Psi_c\ra$ of ghost
number 2 in the Hilbert space of matter ghost conformal field theory in
the full complex plane, satisfying the constraint\cite{9705241}
 \be \label{efg1.5}
b_0^-|\Psi_c\ra = 0\, , \qquad L_0^-|\Psi_c\ra = 0\, ,
 \ee
where
 \be \label{efg2.5}
c_0^\pm = (c_0 \pm \bar c_0), \qquad b_0^\pm = (b_0 \pm \bar b_0)\, ,
\qquad
L_0^\pm=(L_0 \pm \bar L_0)\, .
 \ee
$c_n$, $\bar c_n$, $b_n$, $\bar b_n$ are the usual ghost
oscillators and $L_n$, $\bar L_n$ are the total Virasoro generators.
The quadratic part of the closed string field theory action can be
taken to be:
 \be \label{efg2}
-{1\over K g_s^2}\la \Psi_c|c_0^- (Q_B+\bar Q_B) |\Psi_c\ra\, ,
 \ee
where $Q_B$ and $\bar Q_B$ are the holomorphic and
anti-holomorphic components of the BRST charge, $K$ is a
normalization constant to be determined in eq.\refb{ezopen}, and
$g_s$ is the appropriately normalized closed string coupling
constant so that \refb{e2.1} holds. In the presence of the D-brane
we need to add an extra source term to the action:
 \be \label{efg3}
\la \Psi_c| c_0^- |\BB\ra\, .
 \ee
The equation of motion of
$|\Psi_c\ra$ is then
 \be \label{efg4app}
2\, (Q_B+\bar Q_B)\, |\Psi_c\ra = K\, g_s^2 \, |\BB\ra\, .
 \ee
Clearly by a rescaling of $|\Psi_c\ra$ by $\lambda$ we can change
$K$ and $|\BB\ra$ to $K/\lambda^2$ and $\lambda |\BB\ra$
respectively. However once the normalization of $|\BB\ra$ is fixed
in a convenient manner, the normalization constant $K$ can be
determined by requiring that the classical action obtained after
eliminating $|\Psi_c\ra$ using its equation of motion \refb{efg4app}
reproduces the one loop partition function $Z_{open}$ of the open
string theory on the D-brane. Choosing the solution of \refb{efg4app}
to be\footnote{Some subtleties involved in obtaining the solution
in the Minkowski space have been discussed after eq.\refb{es9.1}.}
 \be \label{es9.1prev}
|\Psi_c\ra = {1\over 2} \, K\, g_s^2 \, b_0^+ \, (L_0^+)^{-1} \,
|\BB\ra \, ,
 \ee
we get
 \be \label{ezopen}
Z_{open} = -{1\over 4} \,  K\, g_s^2 \, \la \BB| b_0^+ \, c_0^- \,
(L_0^+)^{-1} \, |\BB\ra\, .
 \ee
We have chosen a convenient normalization of $|\BB\ra$ in
eq.\refb{efg10}. This determines $K$ from
eq.\refb{ezopen}. Since $Z_{open}$ is independent of $g_s$, and
$|\BB\ra$ is inversely
proportional to $g_s$ due to the $\TT_p$ factor in \refb{efg10},
we
see
that $K$ is a purely numerical constant.

Eqs.\refb{efg3}, \refb{efg4app} clearly shows that $|\BB\ra$ represents
the
source for the closed string fields in the presence of a D-brane. We shall
now make it more explicit by expanding $|\Psi_c\ra$ and $|\BB\ra$ in the
oscillator basis.
The expansion of $|\Psi_c\ra$ in the oscillator basis of closed string
states has, as coefficients, various closed string fields. For example the
first few terms in the expansion are:
 \ben \label{efg5}
|\Psi_c\ra &=& \int {d^{26}k \over (2\pi)^{26}}\, \bigg[ \wt T(k) c_1
\bar c_1 +
\left(\wt h_{\mu\nu}(k) + \wt b_{\mu\nu}(k)\right) 
\alpha^\mu_{-1}\bar\alpha^\nu_{-1}
c_1
\bar c_1 \nonumber \\ &&
+ \left(\wt \phi(k) + {1\over 2} \eta^{\mu\nu} \wt
h_{\mu\nu}(k)\right) (c_1
c_{-1} - \bar c_1 \bar c_{-1}) + \ldots\bigg] |k\ra
 \een
where $\alpha^\mu_n$, $\bar\alpha^\mu_n$ are the usual oscillators
associated with the $X^\mu$ fields. In this expansion
$\wt T(k)$ has the interpretation of the Fourier transform of the
closed string tachyon
field, $\wt h_{\mu\nu}=\wt h_{\nu\mu}$ and $\wt\phi(k)$ are
Fourier transforms of the graviton (associated with the string metric) and
the dilaton fields respectively, $\wt
b_{\mu\nu}=-\wt b_{\nu\mu}$ is the Fourier transform of the anti-symmetric
tensor field etc.\footnote{The identification of these fields can be found
by substituting \refb{efg5} into the quadratic action \refb{efg2} and
comparing the resulting action with the quadratic part of the known
effective
action involving the graviton, dilaton and anti-symmetric tensor field.}
On the other hand the boundary state $|\BB\ra$ has an expansion of the
form:
 \ben \label{efg5.5app}
|\BB\ra &=&
\int {d^{26} k\over (2\pi)^{26}} \left[\wt F(k) +  
\left(\wt A_{\mu\nu}(k)+ \wt
C_{\mu\nu}(k)\right)
\alpha^\mu_{-1}
\bar\alpha^\nu_{-1}
+ \wt B(k) (b_{-1} \bar c_{-1}  + \bar b_{-1} c_{-1}) + 
\ldots \right] \nonumber \\
&& \qquad \qquad  \qquad \qquad \qquad \qquad (c_0
+\bar
c_0)
c_1
\bar c_1 |k\ra\, ,
 \een
where $\wt F$, $\wt A_{\mu\nu}=\wt A_{\nu\mu}$, $\wt
C_{\mu\nu}=-\wt C_{\nu\mu}$, $\wt B$ etc. are fixed functions
which can be read out of the boundary state $|\BB\ra$.
Substituting \refb{efg5} and \refb{efg5.5app} into
eq.\refb{efg3} we get terms in the action proportional to
\be \label{elinear}
\int \, {d^{26}k\over (2\pi)^{26}} \, \left[ \wt h_{\mu\nu}(-k) 
\left(\wt
A^{\mu\nu}(k) + \wt B(k) \, \eta^{\mu\nu}\right) 
+ 2\wt\phi(-k) \, \wt B(k) +
\cdots \right] \, ,
\ee
where $\cdots$ denote source terms for other fields. Since the
Fourier transform $\wt T^{\mu\nu}$ of the energy momentum tensor couples
to $\wt h_{\mu\nu}$, we see that $\wt T_{\mu\nu}$ is
proportional to $\wt A_{\mu\nu}(k) + \eta_{\mu\nu}\,  \wt
B(k)$.\footnote{Note however that this
definition of the various source terms is not unique, since we
could redefine the off-shell closed string field, and this will in
general modify the source terms for various fields. A simple
example of this kind is $\wt h_{\mu\nu}(k) \to f(k^2) \wt
h_{\mu\nu}(k)$ with $f(0)=1$. This does not change the definition
of the graviton field on-shell, but modifies it off-shell.
Consequently the energy-momentum tensor $\wt T_{\mu\nu}$ to which
the field couples will get modified as $\wt T_{\mu\nu}(k) \to
(f(k^2))^{-1} \, \wt T_{\mu\nu}(k)$. In our analysis we shall
adopt the particularly simple definition of energy momentum tensor
as follows from the action \refb{efg2}, \refb{efg3}. In this convention 
the quadratic terms in the closed string field theory action have at most 
two 
powers of the 
momentum $k^\mu$. In position space this translates to these quadratic 
terms having at most two derivatives. However one should keep in mind that 
other choices are possible.}

We can also use an alternative but equivalent method for
expressing the energy momentum tensor $T_{\mu\nu}$ in terms of
$A_{\mu\nu}$ and $B$ by using the
conservation law of
$T_{\mu\nu}$. \refb{efg4app} together with
nilpotence of $(Q_B+\bar Q_B)$ gives:
 \be \label{efg6}
(Q_B+\bar Q_B)|\BB\ra = 0\, .
 \ee
Substituting the expansion \refb{efg5.5app} 
into this equation gives, besides
other equations,
 \be \label{efg7}
k^\nu \wt
A_{\mu\nu}(k)
+
k_\mu \wt B(k) = 0\, .
 \ee
If $A_{\mu\nu}(x)$ and $B(x)$ denote the Fourier transforms of $\wt
A_{\mu\nu}$ and $\wt B$ respectively,
 \be \label{efourab}
A_{\mu\nu}(x) = \int {d^{26}k\over (2\pi)^{26}} \, e^{ik\cdot x} \wt
A_{\mu\nu}(k)\, , \qquad
B(x) = \int {d^{26}k\over (2\pi)^{26}} \, e^{ik\cdot x} \wt
B(k)\, ,
 \ee
then \refb{efg7} gives us:
 \be \label{efg8}
\p^\nu (A_{\mu\nu}(x) + \eta_{\mu\nu} B(x) ) = 0\, .
 \ee
This shows that we should identify the conserved energy-momentum tensor as
 \be \label{efg9aapp}
T_{\mu\nu}(x) \propto (A_{\mu\nu}(x) + \eta_{\mu\nu} B(x) )\, .
 \ee

{}From \refb{elinear} we also see that the function $\wt B(k)$
measures the source of the dilaton field $\wt\phi(k)$. This
suggests that in position space we define the source of the
dilaton to be
 \be \label{edchargeapp}
Q(x) \propto  B(x) \, .
 \ee

\sectiono{Computation of the Energy of Closed String Radiation from
Unstable D-brane} \label{appa}

\refb{es9.17bb} gives the on-shell closed string field configuration
produced by the
rolling tachyon background. In this appendix we shall compute the
energy per unit
$p$-volume carried by this background.
For this we
express $(Q_B+\bar Q_B)$, acting on a state carrying momentum
$\{k^\mu\}$,
as
 \be \label{eqbbreak}
Q_B+\bar Q_B = (c_0 L_0 + \bar c_0 \bar L_0) + \wh Q_1(\vec k) k^0 +
\wh
Q_2(\vec k) \, ,
 \ee
where the operators $\wh Q_1(\vec k)$ and $\wh
Q_2(\vec k)$ do not have any explicit $k^0$ dependence
but can depend on the spatial components of the momentum
$\vec k$. The reason
that the right hand side of \refb{eqbbreak} does not contain higher
powers of $k^0$ is that the $k^\mu$ dependence of $(Q_B+\bar Q_B)$
comes through the $k^\mu$ dependence of the matter Virasoro generators
$L^{(m)}_n$ and $\bar L^{(m)}_n$, and for $n\ne 0$ $L^{(m)}_n$ and
$\bar L^{(m)}_n$ are linear in $k^\mu$. Contribution from the $L_0$
and $\bar L_0$ part of $Q_B+\bar Q_B$, which are quadratic in $k^\mu$,
has been written separately in the right hand side of \refb{eqbbreak}.
Let us denote
by $\wh L$ the part of $(L_0+\bar L_0)$ that involves oscillator
contribution and contribution from spatial momenta, so that
 \be \label{el0bl0}
L_0+\bar L_0 = -{1\over 2} (k^0)^2 + \wh L\, .
 \ee
We also
express a general
closed string field configuration $|\Psi_c\ra$ as
 \be \label{eenergy}
|\Psi_c\ra = \int {d k^0\over 2\pi} \, \wt\psi_A(k^0) \, |k^0, A\ra\, ,
 \ee
where the label $A$ runs over discrete oscillator labels as well as
continuous spatial momenta. The free closed string field theory
action \refb{efg2},
expressed in terms of the component fields $\wt\psi_A$, takes the form
 \be \label{eener1}
S = - {1\over K g_s^2} \, \int \, {d k^0\over 2\pi} \, \wt \psi_A(-k^0) \,
\left[ -{1\over 2}
(k^0)^2 N_{AB} + M_{AB} + k^0 \,  M^{(1)}_{AB} + M^{(2)}_{AB}
\right] \wt \psi_B(k^0)\, ,
 \ee
where $N_{AB}$, $M_{AB}$, $M^{(1)}_{AB}$ and $M^{(2)}_{AB}$ are defined
through the relations:
 \ben \label{eener2}
{1\over 2} \la k^0, A| c_0^- c_0^+ | k^{\prime 0}, B\ra &=& N_{AB} \, 2\pi
\, \delta(
k^0 +  k^{\prime 0}) \, , \nonumber \\
{1\over 2} \, \la k^0, A| c_0^- c_0^+ \wh L | k^{\prime 0}, B\ra &=&
M_{AB}\, 2\pi
\, \delta(
k^0 +  k^{\prime 0}) \, , \nonumber \\
\la k^0, A| c_0^- \wh Q_1 | k^{\prime 0}, B\ra &=& M^{(1)}_{AB}\, 2\pi
\, \delta(
k^0 +  k^{\prime 0}) \, , \nonumber \\
\la k^0, A| c_0^- \wh Q_2 | k^{\prime 0}, B\ra &=& M^{(2)}_{AB}\, 2\pi
\, \delta(
k^0 +  k^{\prime 0}) \, . \nonumber \\
 \een
If we denote by
 \be \label{eener3}
\psi_A(x^0) = \int {d k^0\over 2\pi} \, e^{-i k^0 x^0} \, \wt
\psi_A(k^0)\, ,
 \ee
then the action \refb{eener1} may be reexpressed as
 \be \label{eener4}
S = \int d x^0 L\, ,
 \ee
where
 \be \label{eener5}
L = {1\over K g_s^2} \, \left[{1\over 2} \p_0 \, \psi_A N_{AB} \,
\p_0 \psi_B - i\, M^{(1)}_{AB} \, \psi_A \, \p_0 \psi_B -
\left(M_{AB} + M^{(2)}_{AB}\right) \, \psi_A \, \psi_B\right]\, .
 \ee
{}From this we can write down an expression for the conserved energy:
 \be \label{eener6}
E = \p_0 \psi_A {\p L \over \p (\p_0\psi_A)} -
L
= {1\over K g_s^2} \, \left[{1\over 2} \p_0 \,
\psi_A N_{AB} \, \p_0 \psi_B +  \left(M_{AB} + M^{(2)}_{AB}\right) \,
\psi_A \,
\psi_B\right]\, .
 \ee

We shall now evaluate \refb{eener6} for the on-shell closed string
background \refb{es9.17bb}. For this we express $|\Psi_c^{(1)}\ra$ given
in
\refb{es9.17bb} as
 \be \label{eener7}
|\Psi_c^{(1)}\ra = \int {d k^0\over 2\pi} \, \sum_N \, \int {d^{25-p}
k_\perp\over
(2\pi)^{25-p}} \, \wt\psi^{(1)}_N(k^0, \vec k_\perp)
|N,\vec k_\perp, k^0\ra\, ,
 \ee
where
 \be \label{eener8}
|N,\vec k_\perp, k^0\ra =
\wh A_N\,  c_1\, \bar c_1\, |k^0, \vec k_\parallel=0, \vec k_\perp\ra \,
 \ee
and
 \ben \label{eener9}
\wt\psi_N^{(1)}(k^0, \vec k_\perp) &=& 2\, K\, g_s^2 \, \TT_p \,
{\pi\over
\sinh(\pi\onk)} \, {1\over 2\onk} \nonumber \\
&& \left[e^{-i\onk \, \ln
\sin(\pi\tl)} \delta(k^0 -
\onk)
 + e^{i\onk
\, \ln
\sin(\pi\tl)} \delta(k^0 + \onk) \right]\, .
 \een
This gives
 \ben \label{eener10}
\psi_N^{(1)} (x^0,\vec k_\perp) &\equiv& \int {d k^0\over 2\pi} \,
e^{-i k^0 x^0} \,
\wt\psi_N(k^0, \vec k_\perp)   \nonumber \\
&=& 2\, K\, g_s^2 \, \TT_p \, {\pi\over \sinh(\pi\onk)} \, {1\over
2\onk} \, \left[e^{-i\onk \, (x^0+\ln \sin(\pi\tl))}
 + e^{i\onk
\, (x^0+ \ln
\sin(\pi\tl))}
\right]\, . \nonumber \\
 \een
Using the definitions \refb{eenergy} and \refb{eener3} of $\psi_A(x^0)$,
definition \refb{eener2} of $N_{AB}$, $M_{AB}$, $M^{(1)}_{AB}$ and
$M^{(2)}_{AB}$, and eq.\refb{eener6}, we can now write down an expression
for the total energy associated with the configuration $|\Psi_c^{(1)}\ra$
as:
 \ben \label{eener11}
E^{(1)} &=& {1\over K g_s^2}  \, \int {d^{25-p}
k_\perp\over
(2\pi)^{25-p}} \, \sum_N \Big[ S_N \, \p_0 \psi_N^{(1)}(x^0, \vec
k_\perp) \, \p_0
\psi_N^{(1)}(x^0, -\vec k_\perp)  \nonumber \\
&& \qquad \qquad + R_N \, \psi_N^{(1)}(x^0,
\vec k_\perp) \,
\psi_N^{(1)}(x^0,
-\vec k_\perp)
\Big] \, ,
 \een
where $S_N$ and $R_N$ are defined through
 \be \label{eener12-}
{1\over 4} \la N, \vec k_\perp, k^0| c_0^- c_0^+ | N, \vec k_\perp',
k^{\prime 0}\ra
= (2\pi)^{26-p} \, \delta(k^0 + k^{\prime 0}) \, \delta^{(25-p)}(\vec
k_\perp +
\vec k'_\perp) \,
S_N\, ,
 \ee
 \be \label{eener12}
\la N, \vec k_\perp, k^0| c_0^- ({1\over 2} \, c_0^+ \wh L + \wh Q_2) |
N, \vec
k_\perp',
k^{\prime 0}\ra
=(2\pi)^{26-p} \, \delta(k^0 + k^{\prime 0}) \, \delta^{(25-p)}(\vec
k_\perp +
\vec k'_\perp) \, R_N\, .
 \ee
{}From the definition of $|N, \vec k_\perp, k^0\ra$ and $\wh Q_2$ it
follows
that neither of them contains a $c_0$ or a $\bar c_0$ zero mode. As a
result the matrix element of $c_0^- \wh Q_2$ appearing in \refb{eener12}
vanishes. On the other hand $|N, \vec k_\perp\ra$ is an eigenstate of $\wh
L$ with eigenvalue $2(N-1) + {1\over 2} \vec k_\perp^2 = {1\over 2}
(\onk)^2$. Thus comparison of \refb{eener12-} and \refb{eener12}
gives
 \be \label{eener13}
R_N = (\onk)^2 S_N\, .
 \ee
Substituting this into \refb{eener11}, and using \refb{eener10} we get
 \be \label{eener14}
E^{(1)} = 4\, K \, (g_s \TT_p)^2 \, \sum_N \, \int {d^{25-p}
k_\perp\over (2\pi)^{25-p}} \, S_N \, {\pi^2\over
\sinh^2(\pi\onk)} \, .
 \ee

In order
to compute $S_N$ we express the left hand side of
\refb{eener12-} as
 \ben \label{eener15}
&& {1\over 4} \, \la  k^0, \vec k_\parallel=0, \vec k_\perp| c_{-1} \,
\bar
c_{-1} \, (\wh A_N)^c \, c_0^- \, c_0^+\, \wh A_N \, c_1 \, \bar c_1 |
k^{\prime 0}, \vec k_\parallel=0, \vec k_\perp'\ra \nonumber \\
&=& {1\over
2} \,
V_\parallel\, (2\pi)^{26-p} \, \delta(k^0 + k^{\prime 0}) \,
\delta^{(25-p)}(\vec
k_\perp +
\vec k'_\perp) \, \la 0 |  c_{-1} \, \bar
c_{-1} \, (\wh A_N)^c \, c_0 \, \bar c_0\, \wh A_N \, c_1 \, \bar c_1 |
0\ra''\, , \nonumber \\
 \een
where $(\wh A_N)^c$ is the BPZ conjugate of $\wh A_N$, $V_\parallel$
denotes the spatial volume tangential to the D-brane,
coming from the factor of $(2\pi)^p\delta^{(p)}(\vec k_\parallel)$
evaluated at
$\vec k_\parallel=0$, and
$\la\cdot|\cdot\ra''$ denotes a renormalized
BPZ inner product in the zero momentum sector such that\footnote{This
is the analog of the renormalized inner product $\la \cdot|\cdot\ra'$ for 
open string states
as defined in eq.\refb{e4.43}.}
 \be \label{ereninn}
\la 0 |  c_{-1} \, \bar
c_{-1} \,  c_0 \, \bar c_0\, \, c_1 \, \bar c_1 |
0\ra'' = 1\, .
 \ee
This gives
 \be \label{esntosn}
S_N = V_\parallel s_N\, ,
 \ee
where
 \be \label{esn1}
s_N = {1\over
2} \,
\la 0 |  c_{-1} \, \bar
c_{-1} \, (\wh A_N)^c \, c_0 \, \bar c_0\, \wh A_N \, c_1 \, \bar c_1 |
0\ra''\, .
 \ee
Substituting this into \refb{eener14} we get
 \be \label{eener14aa}
E^{(1)} = V_\parallel \EE
 \ee
where the energy per unit $p$-volume $\EE$ is given by:
 \be \label{eedensity}
\EE = \sum_N \EE_N = 4\, K \, (g_s \TT_p)^2 \, \sum_N \, s_N \,
\int
{d^{25-p} k_\perp\over (2\pi)^{25-p}} \, {\pi^2\over
\sinh^2(\pi\onk)} \, .
 \ee

In order to compute $s_N$ using \refb{esn1} we can use the
generating
functional
 \be \label{esn1a}
\sum_N s_N \, q^{2(N-1)} =  {1\over
2} \,
\sum_N \sum_M \la 0 |  c_{-1} \, \bar
c_{-1} \, (\wh A_N)^c \,  q^{L_0+\bar L_0} \, \wh A_M \, c_1 \, \bar c_1 |
0\ra''\, .
 \ee
Note that only the $M=N$ terms contribute to \refb{esn1a}
due to $(L_0+\bar L_0)$ conservation. We can now
replace $\sum_N \wh A_N$ and $\sum_M\wh A_M$ by the left hand side of
\refb{eexpan}. The result is essentially the cylinder amplitude, with
$\ln q$ denoting the ratio of the height to the circumference of
the cylinder. It is however well known that in this computation
the contribution from the ghost sector cancels the contribution
from two of the matter sector fields, leaving behind the
contribution from 24 matter sector fields. Thus in computing the
right hand side of \refb{esn1a} we can replace $\sum_N \wh A_N$ by
 \be \label{esn2}
\exp\left[\sum_{n=1}^\infty
\sum_{s=1}^{24} (-1)^{d_s} \, {1\over
n} \,  \alpha^s_{-n} \bar
\alpha^s_{-n}\right] \, .
 \ee
We can for example take the sum over $s$ to run over all the spatial
directions
tangential to the brane and all the transverse directions except one. The
final formula is insensitive to $d_s$ (since $d_s$ can be changed by a
redefinition $\alpha_s\to -\alpha_s$ without changing $\bar \alpha_s$),
and so it does not matter whether we drop a Neumann or Dirichlet
direction in the sum over $s$. Using the replacement \refb{esn2} in
\refb{esn1a}
the ghost term factorises giving a contribution of
$q^{-2}$, and we get
 \ben \label{esn3}
\sum_N s_N \, q^{2N} &=&  {1\over 2} \, \la 0 |
\exp\left[\sum_{m=1}^\infty \sum_{s=1}^{24} (-1)^{d_s} \, {1\over
m} \,  \alpha^s_{m} \bar \alpha^s_{m}\right]  q^{L_0^{matter}+\bar
L_0^{matter}} \nonumber \\  && \exp\left[\sum_{n=1}^\infty
\sum_{r=1}^{24} (-1)^{d_r} \, {1\over n} \,  \alpha^r_{-n} \bar
\alpha^r_{-n}\right] \, | 0\ra''_{matter}\, ,
 \een
where $\la \cdot|\cdot\ra''_{matter}$ denotes the BPZ inner product in the
matter sector with the normalization convention $\la 0|0\ra''_{matter}=1$.


\begin{thebibliography}{99}

\bibitem{callan1}
A.~Abouelsaood, C.~G.~.~Callan, C.~R.~Nappi and S.~A.~Yost,
Nucl.\ Phys.\ B {\bf 280}, 599 (1987).

\bibitem{0104263}
C.~Acatrinei and C.~Sochichiu,
Phys.\ Rev.\ D {\bf 67}, 125017 (2003)
[arXiv:hep-th/0104263].

\bibitem{0108075}
A.~Adams, J.~Polchinski and E.~Silverstein,
JHEP {\bf 0110}, 029 (2001) [arXiv:hep-th/0108075].

\bibitem{9610249}
M.~Aganagic, C.~Popescu and J.~H.~Schwarz,
Phys.\ Lett.\ B {\bf 393}, 311 (1997)
[arXiv:hep-th/9610249].

\bibitem{9612080}
M.~Aganagic, C.~Popescu and J.~H.~Schwarz,
Nucl.\ Phys.\ B {\bf 495}, 99 (1997)
[arXiv:hep-th/9612080].

\bibitem{0009142}
M.~Aganagic, R.~Gopakumar, S.~Minwalla and A.~Strominger,
JHEP {\bf 0104}, 001 (2001)
[arXiv:hep-th/0009142].

\bibitem{0404253}
G.~L.~Alberghi, R.~Casadio and A.~Tronconi,
JHEP {\bf 0406}, 040 (2004)
[arXiv:hep-th/0404253].

\bibitem{0012222}
M.~Alishahiha, H.~Ita and Y.~Oz,
Phys.\ Lett.\ B {\bf 503}, 181 (2001) [arXiv:hep-th/0012222].

\bibitem{0104164}
M.~Alishahiha,
Phys.\ Lett.\ B {\bf 510}, 285 (2001)
[arXiv:hep-th/0104164].

\bibitem{0203185}
J.~Ambjorn and R.~A.~Janik,
Phys.\ Lett.\ B {\bf 538}, 189 (2002)
[arXiv:hep-th/0203185].

\bibitem{0312163}
J.~Ambjorn and R.~A.~Janik,
Phys.\ Lett.\ B {\bf 584}, 155 (2004) [arXiv:hep-th/0312163].

\bibitem{0407270}
J.~Ambjorn and R.~A.~Janik,
arXiv:hep-th/0407270.

\bibitem{0010218}
O.~Andreev,
Nucl.\ Phys.\ B {\bf 598}, 151 (2001)
[arXiv:hep-th/0010218].

\bibitem{0109187}
O.~Andreev and T.~Ott,
Nucl.\ Phys.\ B {\bf 627}, 330 (2002)
[arXiv:hep-th/0109187].

\bibitem{0112088}
O.~Andreev,
Phys.\ Lett.\ B {\bf 534}, 163 (2002)
[arXiv:hep-th/0112088].

\bibitem{0308123}
O.~Andreev,
Nucl.\ Phys.\ B {\bf 680}, 3 (2004)
[arXiv:hep-th/0308123].

\bibitem{AREF1}
I.~Y.~Arefeva, P.~B.~Medvedev and A.~P.~Zubarev,
Phys.\ Lett.\  {\bf B240}, 356 (1990).

\bibitem{AREF2}
I.~Y.~Arefeva, P.~B.~Medvedev and A.~P.~Zubarev,
Nucl.\ Phys.\  {\bf B341}, 464 (1990).


\bibitem{0011117}
I.~Y.~Aref'eva, A.~S.~Koshelev, D.~M.~Belov and P.~B.~Medvedev,
Nucl.\ Phys.\ B {\bf 638}, 3 (2002) [arXiv:hep-th/0011117].

\bibitem{0107197}
I.~Y.~Arefeva, D.~M.~Belov, A.~S.~Koshelev and P.~B.~Medvedev,
Nucl.\ Phys.\ B {\bf 638}, 21 (2002) [arXiv:hep-th/0107197].

\bibitem{0112214}
I.~Y.~Arefeva, A.~A.~Giryavets and P.~B.~Medvedev,
Phys.\ Lett.\ B {\bf 532}, 291 (2002)
[arXiv:hep-th/0112214].

\bibitem{0201197}
I.~Y.~Arefeva, D.~M.~Belov and A.~A.~Giryavets,
JHEP {\bf 0209}, 050 (2002)
[arXiv:hep-th/0201197].

\bibitem{0203227}
I.~Y.~Arefeva, A.~A.~Giryavets and A.~S.~Koshelev,
Phys.\ Lett.\ B {\bf 536}, 138 (2002)
[arXiv:hep-th/0203227].

\bibitem{0301137}
I.~Y.~Aref'eva, L.~V.~Joukovskaya and A.~S.~Koshelev,
arXiv:hep-th/0301137.

\bibitem{0012080}
G.~Arutyunov, S.~Frolov, S.~Theisen and A.~A.~Tseytlin,
JHEP {\bf 0102}, 002 (2001) [arXiv:hep-th/0012080].

\bibitem{0105238}
G.~Arutyunov, A.~Pankiewicz and B.~.~J.~Stefanski,
JHEP {\bf 0106}, 049 (2001)
[arXiv:hep-th/0105238].

\bibitem{0108085}
T.~Asakawa, S.~Sugimoto and S.~Terashima,
JHEP {\bf 0203}, 034 (2002)
[arXiv:hep-th/0108085].

\bibitem{0212188}
T.~Asakawa, S.~Sugimoto and S.~Terashima,
JHEP {\bf 0302}, 011 (2003)
[arXiv:hep-th/0212188].

\bibitem{0305006}
T.~Asakawa, S.~Sugimoto and S.~Terashima,
Prog.\ Theor.\ Phys.\ Suppl.\  {\bf 152}, 93 (2004)
[arXiv:hep-th/0305006].

\bibitem{0309074}
T.~Asakawa, S.~Kobayashi and S.~Matsuura,
JHEP {\bf 0310}, 023 (2003) [arXiv:hep-th/0309074].

\bibitem{9209036}
J.~Avan and A.~Jevicki,
Nucl.\ Phys.\ B {\bf 397}, 672 (1993)
[arXiv:hep-th/9209036].

\bibitem{0404039}
V.~Balasubramanian, E.~Keski-Vakkuri, P.~Kraus and A.~Naqvi,
arXiv:hep-th/0404039.

\bibitem{0407229}
R.~Banerjee, Y.~Kim and O.~K.~Kwon,
arXiv:hep-th/0407229.

\bibitem{9511194}
T.~Banks and L.~Susskind,
arXiv:hep-th/9511194.

\bibitem{bar1}
K.~Bardakci,
Nucl.\ Phys.\ B {\bf 68}, 331 (1974).

\bibitem{bar2}
K.~Bardakci and M.~B.~Halpern,
Nucl.\ Phys.\ B {\bf 73}, 295 (1974).

\bibitem{bar3}
K.~Bardakci,
Nucl.\ Phys.\ B {\bf 133}, 297 (1978).

\bibitem{0105098}
K.~Bardakci and A.~Konechny,
arXiv:hep-th/0105098.

\bibitem{0406120}
N.~Barnaby,
JHEP {\bf 0407}, 025 (2004)
[arXiv:hep-th/0406120].

\bibitem{0410030}
N.~Barnaby and J.~M.~Cline,
arXiv:hep-th/0410030.

\bibitem{0010101}
I.~Bars, H.~Kajiura, Y.~Matsuo and T.~Takayanagi,
Phys.\ Rev.\ D {\bf 63}, 086001 (2001)
[arXiv:hep-th/0010101].

\bibitem{0302151}
I.~Bars, I.~Kishimoto and Y.~Matsuo,
Phys.\ Rev.\ D {\bf 67}, 126007 (2003)
[arXiv:hep-th/0302151].

\bibitem{0401195}
D.~Bazeia, R.~Menezes and J.~G.~Ramos,
arXiv:hep-th/0401195.

\bibitem{9806155}
O.~Bergman and M.~R.~Gaberdiel,
Phys.\ Lett.\ B {\bf 441}, 133 (1998) [arXiv:hep-th/9806155].

\bibitem{9901014}
O.~Bergman and M.~R.~Gaberdiel,
JHEP {\bf 9903}, 013 (1999)
[arXiv:hep-th/9901014].

\bibitem{9902160}
O.~Bergman, E.~G.~Gimon and P.~Horava,
JHEP {\bf 9904}, 010 (1999)
[arXiv:hep-th/9902160].

\bibitem{9908126}
O.~Bergman and M.~R.~Gaberdiel,
Class.\ Quant.\ Grav.\  {\bf 17}, 961 (2000)
[arXiv:hep-th/9908126].

\bibitem{0009252}
O.~Bergman,
JHEP {\bf 0011}, 015 (2000)
[arXiv:hep-th/0009252].

\bibitem{0002223}
O.~Bergman, K.~Hori and P.~Yi,
Nucl.\ Phys.\ B {\bf 580}, 289 (2000) [arXiv:hep-th/0002223].

\bibitem{0402124}
O.~Bergman, S.S~Razamat,
arXiv:hep-th/0402124.

\bibitem{9611173}
E.~Bergshoeff and P.~K.~Townsend,
Nucl.\ Phys.\ B {\bf 490}, 145 (1997)
[arXiv:hep-th/9611173].

\bibitem{0003221}
E.~A.~Bergshoeff, M.~de Roo, T.~C.~de Wit, E.~Eyras and S.~Panda,
JHEP {\bf 0005}, 009 (2000) [arXiv:hep-th/0003221].

\bibitem{9108021}
N.~Berkovits, M.~T.~Hatsuda and W.~Siegel,
Nucl.\ Phys.\ B {\bf 371}, 434 (1992)
[arXiv:hep-th/9108021].

\bibitem{9503099}
N.~Berkovits,
Nucl.\ Phys.\ B {\bf 450}, 90 (1995) [Erratum-ibid.\ B {\bf 459},
439 (1996)] [arXiv:hep-th/9503099].

\bibitem{9912121}
N.~Berkovits,
Fortsch.\ Phys.\  {\bf 48}, 31 (2000) [arXiv:hep-th/9912121].

\bibitem{0001084}
N.~Berkovits,
JHEP {\bf 0004}, 022 (2000) [arXiv:hep-th/0001084].

\bibitem{0002211}
N.~Berkovits, A.~Sen and B.~Zwiebach,
Nucl.\ Phys.\ B {\bf 587}, 147 (2000) [arXiv:hep-th/0002211].

\bibitem{0105230}
N.~Berkovits,
arXiv:hep-th/0105230.

\bibitem{0109100}
N.~Berkovits,
JHEP {\bf 0111}, 047 (2001)
[arXiv:hep-th/0109100].

\bibitem{0308069}
S.~Bhattacharya, S.~Mukherji and S.~Roy,
Phys.\ Lett.\ B {\bf 584}, 163 (2004)
[arXiv:hep-th/0308069].

\bibitem{0201060}
L.~Bonora, D.~Mamone and M.~Salizzoni,
Nucl.\ Phys.\ B {\bf 630}, 163 (2002)
[arXiv:hep-th/0201060].

\bibitem{0203188}
L.~Bonora, D.~Mamone and M.~Salizzoni,
JHEP {\bf 0204}, 020 (2002)
[arXiv:hep-th/0203188].

\bibitem{0207044}
L.~Bonora, D.~Mamone and M.~Salizzoni,
JHEP {\bf 0301}, 013 (2003)
[arXiv:hep-th/0207044].

\bibitem{0304270}
L.~Bonora, C.~Maccaferri, D.~Mamone and M.~Salizzoni,
arXiv:hep-th/0304270.

\bibitem{0311198}
L.~Bonora, C.~Maccaferri and P.~Prester,
JHEP {\bf 0401}, 038 (2004)
[arXiv:hep-th/0311198].

\bibitem{0404154}
L.~Bonora, C.~Maccaferri and P.~Prester,
arXiv:hep-th/0404154.

\bibitem{0409063}
L.~Bonora, C.~Maccaferri, R.~J.~S.~Santos and D.~D.~Tolla,
arXiv:hep-th/0409063.

\bibitem{0002023}
P.~Bouwknegt and V.~Mathai,
JHEP {\bf 0003}, 007 (2000)
[arXiv:hep-th/0002023].

\bibitem{0005242}
P.~Brax, G.~Mandal and Y.~Oz,
Phys.\ Rev.\ D {\bf 63}, 064008 (2001)
[arXiv:hep-th/0005242].

\bibitem{0304197}
P.~Brax, J.~Mourad and D.~A.~Steer,
arXiv:hep-th/0304197.

\bibitem{0310079}
P.~Brax, J.~Mourad and D.~A.~Steer,
arXiv:hep-th/0310079.

\bibitem{padic}
L.\ Brekke, P.G.O.\ Freund, M.\ Olson and E.\ Witten,
Nucl.\ Phys.\  {\bf B302} (1988) 365.

\bibitem{BKZ}
E.~Brezin, V.~A.~Kazakov and A.~B.~Zamolodchikov,
Nucl.\ Phys.\ B {\bf 338}, 673 (1990).

\bibitem{0207235}
A.~Buchel, P.~Langfelder and J.~Walcher,
Annals Phys.\  {\bf 302}, 78 (2002) [arXiv:hep-th/0207235].

\bibitem{0212150}
A.~Buchel and J.~Walcher,
Fortsch.\ Phys.\  {\bf 51}, 885 (2003)
[arXiv:hep-th/0212150].

\bibitem{0305055}
A.~Buchel and J.~Walcher,
JHEP {\bf 0305}, 069 (2003)
[arXiv:hep-th/0305055].

\bibitem{0105204}
C.~P.~Burgess, M.~Majumdar, D.~Nolte, F.~Quevedo, G.~Rajesh and 
R.~J.~Zhang,
JHEP {\bf 0107}, 047 (2001)
[arXiv:hep-th/0105204].

\bibitem{callan2}
C.~G.~.~Callan, C.~Lovelace, C.~R.~Nappi and S.~A.~Yost,
Nucl.\ Phys.\ B {\bf 293}, 83 (1987).

\bibitem{9402113}
C.~G.~.~Callan, I.~R.~Klebanov, A.~W.~W.~Ludwig and
J.~M.~Maldacena,
Nucl.\ Phys.\ B {\bf 422}, 417 (1994) [arXiv:hep-th/9402113].

\bibitem{9610148}
M.~Cederwall, A.~von Gussich, B.~E.~W.~Nilsson and 
A.~Westerberg,
Nucl.\ Phys.\ B {\bf 490}, 163 (1997)
[arXiv:hep-th/9610148].

\bibitem{9611159}
M.~Cederwall, A.~von Gussich, B.~E.~W.~Nilsson, 
P.~Sundell and A.~Westerberg,
Nucl.\ Phys.\ B {\bf 490}, 179 (1997)
[arXiv:hep-th/9611159].

\bibitem{0103056}
G.~Chalmers,
JHEP {\bf 0106}, 012 (2001)
[arXiv:hep-th/0103056].

\bibitem{0204233}
B.~Chen and F.~L.~Lin,
Phys.\ Rev.\ D {\bf 66}, 126001 (2002)
[arXiv:hep-th/0204233].

\bibitem{0209222}
B.~Chen, M.~Li and F.~L.~Lin,
JHEP {\bf 0211}, 050 (2002)
[arXiv:hep-th/0209222].

\bibitem{0204071}
C.~M.~Chen, D.~V.~Gal'tsov and M.~Gutperle,
Phys.\ Rev.\ D {\bf 66}, 024043 (2002)
[arXiv:hep-th/0204071].

\bibitem{0301119}
T.~Chen, J.~Frohlich and J.~Walcher,
Commun.\ Math.\ Phys.\  {\bf 237}, 243 (2003)
[arXiv:hep-th/0301119].

\bibitem{0311179}
X.~g.~Chen,
arXiv:hep-th/0311179.

\bibitem{0209142}
T.~E.~Clark, M.~Nitta and T.~ter Veldhuis,
Phys.\ Rev.\ D {\bf 69}, 047701 (2004)
[arXiv:hep-th/0209142].

\bibitem{0401163}
T.~E.~Clark, M.~Nitta and T.~ter Veldhuis,
arXiv:hep-th/0401163.

\bibitem{0409030}
T.~E.~Clark, M.~Nitta and T.~t.~Veldhuis,
arXiv:hep-th/0409030.

\bibitem{0409151}
T.~E.~Clark, M.~Nitta and T.~ter Veldhuis,
arXiv:hep-th/0409151.

\bibitem{0301101}
J.~M.~Cline and H.~Firouzjahi,
Phys.\ Lett.\ B {\bf 564}, 255 (2003)
[arXiv:hep-th/0301101].

\bibitem{COLEMAN}
S.~R.~Coleman,
Phys.\ Rev.\ D {\bf 11}, 2088 (1975).

\bibitem{0305177}
N.~R.~Constable and F.~Larsen,
arXiv:hep-th/0305177.

\bibitem{0306294}
E.~J.~Copeland, P.~M.~Saffin and D.~A.~Steer,
Phys.\ Rev.\ D {\bf 68}, 065013 (2003) [arXiv:hep-th/0306294].

\bibitem{0012217}
S.~Corley and S.~Ramgoolam,
JHEP {\bf 0103}, 037 (2001)
[arXiv:hep-th/0012217].

\bibitem{0010021}
L.~Cornalba,
Phys.\ Lett.\ B {\bf 504}, 55 (2001)
[arXiv:hep-th/0010021].

\bibitem{0105227}
B.~Craps, P.~Kraus and F.~Larsen,
JHEP {\bf 0106}, 062 (2001)
[arXiv:hep-th/0105227].

\bibitem{0403217}
B.~C.~da Cunha,
arXiv:hep-th/0403217.

\bibitem{dai}
J. Dai, R. Leigh and J. Polchinski, Mod. Phys. Lett. {\bf A4}
(1989) 2073.

\bibitem{DASJEV}
S.~R.~Das and A.~Jevicki,
Mod.\ Phys.\ Lett.\ A {\bf 5}, 1639 (1990).

\bibitem{9110021}
S.~R.~Das, A.~Dhar, G.~Mandal and S.~R.~Wadia,
Int.\ J.\ Mod.\ Phys.\ A {\bf 7}, 5165 (1992)
[arXiv:hep-th/9110021].

\bibitem{0401067}
S.~R.~Das,
arXiv:hep-th/0401067.

\bibitem{0005006}
K.~Dasgupta, S.~Mukhi and G.~Rajesh,
JHEP {\bf 0006}, 022 (2000)
[arXiv:hep-th/0005006].

\bibitem{0010247}
S.~Dasgupta and T.~Dasgupta,
JHEP {\bf 0106}, 007 (2001)
[arXiv:hep-th/0010247].

\bibitem{0310106}
S.~Dasgupta and T.~Dasgupta,
arXiv:hep-th/0310106.

\bibitem{DAVID}
F.~David,
Mod.\ Phys.\ Lett.\ A {\bf 3}, 1651 (1988).

\bibitem{0012089}
J.~R.~David,
JHEP {\bf 0107}, 009 (2001)
[arXiv:hep-th/0012089].

\bibitem{0105184}
J.~R.~David,
JHEP {\bf 0107}, 024 (2001)
[arXiv:hep-th/0105184].

\bibitem{0007235}
J.~R.~David,
JHEP {\bf 0010}, 004 (2000)
[arXiv:hep-th/0007235].

\bibitem{0011223}
S.~P.~de Alwis and A.~T.~Flournoy,
Phys.\ Rev.\ D {\bf 63}, 106001 (2001)
[arXiv:hep-th/0011223].

\bibitem{0312135}
J.~de Boer, A.~Sinkovics, E.~Verlinde and J.~T.~Yee,
JHEP {\bf 0403}, 023 (2004) [arXiv:hep-th/0312135].

\bibitem{0003031}
R.~de Mello Koch, A.\ Jevicki, M.\ Mihailescu and R.\ Tatar,
Phys.\ Lett.\  {\bf B482}, 249 (2000)
[hep-th/0003031].

\bibitem{0008053}
R.~de Mello Koch and J.~P.~Rodrigues,
Phys.\ Lett.\ B {\bf 495}, 237 (2000) [arXiv:hep-th/0008053].

\bibitem{0107070}
P.~J.~De Smet and J.~Raeymaekers,
JHEP {\bf 0107}, 032 (2001)
[arXiv:hep-th/0107070].

\bibitem{0109182}
P.~J.~De Smet,
arXiv:hep-th/0109182.

\bibitem{0003220}
P.~J.~De Smet and J.~Raeymaekers,
JHEP {\bf 0005}, 051 (2000) [arXiv:hep-th/0003220].

\bibitem{0004112}
P.~De Smet and J.~Raeymaekers,
JHEP {\bf 0008}, 020 (2000)
[hep-th/0004112].

\bibitem{0305191}
Y.~Demasure and R.~A.~Janik,
Phys.\ Lett.\ B {\bf 578}, 195 (2004)
[arXiv:hep-th/0305191].

\bibitem{0309148}
O.~DeWolfe, R.~Roiban, M.~Spradlin, A.~Volovich and 
J.~Walcher,
JHEP {\bf 0311}, 012 (2003)
[arXiv:hep-th/0309148].

\bibitem{9212027}
A.~Dhar, G.~Mandal and S.~R.~Wadia,
Int.\ J.\ Mod.\ Phys.\ A {\bf 8}, 3811 (1993)
[arXiv:hep-th/9212027].

\bibitem{9507041}
A.~Dhar, G.~Mandal and S.~R.~Wadia,
Nucl.\ Phys.\ B {\bf 454}, 541 (1995)
[arXiv:hep-th/9507041].

\bibitem{9109005}
P.~Di Francesco and D.~Kutasov,
Nucl.\ Phys.\ B {\bf 375}, 119 (1992)
[arXiv:hep-th/9109005].

\bibitem{DVV}
R.~Dijkgraaf, E.~Verlinde and H.~Verlinde,
Commun.\ Math.\ Phys.\  {\bf 115}, 649 (1988).

\bibitem{DISKAW}
J.~Distler and H.~Kawai,
Nucl.\ Phys.\ B {\bf 321}, 509 (1989).

\bibitem{9707068}
P.~Di Vecchia, M.~Frau, I.~Pesando, S.~Sciuto, A.~Lerda and
R.~Russo,
Nucl.\ Phys.\ B {\bf 507}, 259 (1997)
[arXiv:hep-th/9707068].

\bibitem{9912275}
P.~Di Vecchia and A.~Liccardo,
arXiv:hep-th/9912275.

\bibitem{0011090}
E.~E.~Donets, A.~P.~Isaev, C.~Sochichiu and M.~Tsulaia,
JHEP {\bf 0012}, 022 (2000)
[arXiv:hep-th/0011090].

\bibitem{9206053}
H.~Dorn and H.~J.~Otto,
Phys.\ Lett.\ B {\bf 291}, 39 (1992)
[arXiv:hep-th/9206053].

\bibitem{9403141}
H.~Dorn and H.~J.~Otto,
Nucl.\ Phys.\ B {\bf 429}, 375 (1994)
[arXiv:hep-th/9403141].

\bibitem{9711165}
M.~R.~Douglas and C.~M.~Hull,
JHEP {\bf 9802}, 008 (1998)
[arXiv:hep-th/9711165].

\bibitem{0202087}
M.~R.~Douglas, H.~Liu, G.~W.~Moore and B.~Zwiebach,
JHEP {\bf 0204}, 022 (2002)
[arXiv:hep-th/0202087].

\bibitem{0307195}
M.~R.~Douglas, I.~R.~Klebanov, D.~Kutasov, J.~Maldacena, E.~Martinec
and
N.~Seiberg,
arXiv:hep-th/0307195.

\bibitem{0004131}
N.~Drukker, D.~J.~Gross and N.~Itzhaki,
Phys.\ Rev.\ D {\bf 62}, 086007 (2000)
[arXiv:hep-th/0004131].

\bibitem{0301079}
N.~Drukker,
JHEP {\bf 0308}, 017 (2003)
[arXiv:hep-th/0301079].

\bibitem{9812483}
G.~R.~Dvali and S.~H.~H.~Tye,
Phys.\ Lett.\ B {\bf 450}, 72 (1999)
[arXiv:hep-ph/9812483].

\bibitem{0105203}
G.~R.~Dvali, Q.~Shafi and S.~Solganik,
arXiv:hep-th/0105203.

\bibitem{0103085}
I.~Ellwood and W.~Taylor,
Phys.\ Lett.\ B {\bf 512}, 181 (2001) [arXiv:hep-th/0103085].

\bibitem{0105024}
I.~Ellwood, B.~Feng, Y.~H.~He and N.~Moeller,
JHEP {\bf 0107}, 016 (2001) [arXiv:hep-th/0105024].

\bibitem{0105156}
I.~Ellwood and W.~Taylor,
arXiv:hep-th/0105156.

\bibitem{0406199}
T.~G.~Erler and D.~J.~Gross,
arXiv:hep-th/0406199.

\bibitem{0409179}
T.~G.~Erler,
arXiv:hep-th/0409179.

\bibitem{9908121}
E.~Eyras,
JHEP {\bf 9910}, 005 (1999)
[arXiv:hep-th/9908121].

\bibitem{0208019}
G.~N.~Felder, L.~Kofman and A.~Starobinsky,
JHEP {\bf 0209}, 026 (2002) [arXiv:hep-th/0208019].

\bibitem{0403073}
G.~N.~Felder and L.~Kofman,
Phys.\ Rev.\ D {\bf 70}, 046004 (2004)
[arXiv:hep-th/0403073].

\bibitem{9406125}
P.~Fendley, H.~Saleur and N.~P.~Warner,
Nucl.\ Phys.\ B {\bf 430}, 577 (1994) [arXiv:hep-th/9406125].

\bibitem{0103103}
B.~Feng, Y.~H.~He and N.~Moeller,
vacuum,''
arXiv:hep-th/0103103.

\bibitem{0310253}
A.~Fotopoulos and A.~A.~Tseytlin,
JHEP {\bf 0312}, 025 (2003)
[arXiv:hep-th/0310253].

\bibitem{Frampton1}
P.~H.~Frampton and Y.~Okada,
Phys.\ Rev.\ Lett.\  {\bf 60}, 484 (1988).

\bibitem{Frampton2}
P.~H.~Frampton and Y.~Okada,
Phys.\ Rev.\ D {\bf 37}, 3077 (1988).

\bibitem{9903123}
M.~Frau, L.~Gallot, A.~Lerda and P.~Strigazzi,
Nucl.\ Phys.\ B {\bf 564}, 60 (2000)
[arXiv:hep-th/9903123].

\bibitem{0012164}
S.~Fredenhagen and V.~Schomerus,
JHEP {\bf 0104}, 007 (2001)
[arXiv:hep-th/0012164].

\bibitem{0308205}
S.~Fredenhagen and V.~Schomerus,
arXiv:hep-th/0308205.

\bibitem{Freund1}
P.~G.~O.~Freund and M.~Olson,
Phys.\ Lett.\ B {\bf 199}, 186 (1987).

\bibitem{Freund2}
P.~G.~O.~Freund and E.~Witten,
Phys.\ Lett.\ B {\bf 199}, 191 (1987).

\bibitem{THORN}
D.~Z.~Freedman, S.~B.~Giddings, J.~A.~Shapiro and C.~B.~Thorn,
Nucl.\ Phys.\ B {\bf 298}, 253 (1988).

\bibitem{FMS1}
D.~Friedan, S.~H.~Shenker and E.~J.~Martinec,
Phys.\ Lett.\ B {\bf 160}, 55 (1985).

\bibitem{FMS}
D.~Friedan, E.~J.~Martinec and S.~H.~Shenker,
Nucl.\ Phys.\ B {\bf 271}, 93 (1986).

\bibitem{0207001}
E.~Fuchs, M.~Kroyter and A.~Marcus,
JHEP {\bf 0209}, 022 (2002)
[arXiv:hep-th/0207001].

\bibitem{0304163}
M.~Fujita and H.~Hata,
arXiv:hep-th/0304163.

\bibitem{0403031}
M.~Fujita and H.~Hata,
arXiv:hep-th/0403031.

\bibitem{9910109}
M.~R.~Gaberdiel and B.~J.~Stefanski,
Nucl.\ Phys.\ B {\bf 578}, 58 (2000)
[arXiv:hep-th/9910109].

\bibitem{0005029}
M.~R.~Gaberdiel,
Class.\ Quant.\ Grav.\  {\bf 17}, 3483 (2000)
[arXiv:hep-th/0005029].

\bibitem{0108102}
M.~R.~Gaberdiel, A.~Recknagel and G.~M.~T.~Watts,
Nucl.\ Phys.\ B {\bf 626}, 344 (2002)
[arXiv:hep-th/0108102].

\bibitem{0108238}
M.~R.~Gaberdiel and A.~Recknagel,
JHEP {\bf 0111}, 016 (2001) [arXiv:hep-th/0108238].

\bibitem{0111129}
D.~Gaiotto, L.~Rastelli, A.~Sen and B.~Zwiebach,
Adv.\ Theor.\ Math.\ Phys.\  {\bf 6}, 403 (2003)
[arXiv:hep-th/0111129].

\bibitem{0201159}
D.~Gaiotto, L.~Rastelli, A.~Sen and B.~Zwiebach,
JHEP {\bf 0203}, 003 (2002)
[arXiv:hep-th/0201159].

\bibitem{0202151}
D.~Gaiotto, L.~Rastelli, A.~Sen and B.~Zwiebach,
JHEP {\bf 0204}, 060 (2002)
[arXiv:hep-th/0202151].

\bibitem{0211012}
D.~Gaiotto and L.~Rastelli,
JHEP {\bf 0308}, 048 (2003) [arXiv:hep-th/0211012].

\bibitem{0304192}
D.~Gaiotto, N.~Itzhaki and L.~Rastelli,
arXiv:hep-th/0304192.

\bibitem{0307221}
D.~Gaiotto, N.~Itzhaki and L.~Rastelli,
Phys.\ Lett.\ B {\bf 575}, 111 (2003)
[arXiv:hep-th/0307221].

\bibitem{0312196}
D.~Gaiotto and L.~Rastelli,
arXiv:hep-th/0312196.

\bibitem{9812226}
H.~Garcia-Compean,
Nucl.\ Phys.\ B {\bf 557}, 480 (1999)
[arXiv:hep-th/9812226].

\bibitem{0003122}
M.~R.~Garousi,
Nucl.\ Phys.\ B {\bf 584}, 284 (2000) [arXiv:hep-th/0003122].

\bibitem{0209068}
M.~R.~Garousi,
Nucl.\ Phys.\ B {\bf 647}, 117 (2002)
[arXiv:hep-th/0209068].

\bibitem{0303239}
M.~R.~Garousi,
JHEP {\bf 0304}, 027 (2003)
[arXiv:hep-th/0303239].

\bibitem{0304145}
M.~R.~Garousi,
Nucl.\ Phys.\ B {\bf 584}, 284 (2000) [arXiv:hep-th/0003122], JHEP
{\bf 0305}, 058 (2003)
[arXiv:hep-th/0304145].

\bibitem{0307197}
M.~R.~Garousi,
JHEP {\bf 0312}, 036 (2003)
[arXiv:hep-th/0307197].

\bibitem{9704006}
E.~Gava, K.~S.~Narain and M.~H.~Sarmadi,
Nucl.\ Phys.\ B {\bf 504}, 214 (1997)
[arXiv:hep-th/9704006].

\bibitem{0009103}
A.~A.~Gerasimov and S.~L.~Shatashvili,
JHEP {\bf 0010}, 034 (2000) [arXiv:hep-th/0009103].

\bibitem{0011009}
A.~A.~Gerasimov and S.~L.~Shatashvili,
JHEP {\bf 0101}, 019 (2001)
[arXiv:hep-th/0011009].

\bibitem{0105245}
A.~A.~Gerasimov and S.~L.~Shatashvili,
JHEP {\bf 0106}, 066 (2001)
[arXiv:hep-th/0105245].

\bibitem{0003278}
D.~Ghoshal and A.~Sen,
Nucl.\ Phys.\ B {\bf 584}, 300 (2000) [arXiv:hep-th/0003278].

\bibitem{0009191}
D.~Ghoshal and A.~Sen,
JHEP {\bf 0011}, 021 (2000)
[arXiv:hep-th/0009191].

\bibitem{0106231}
D.~Ghoshal,
arXiv:hep-th/0106231.

\bibitem{0406259}
D.~Ghoshal,
arXiv:hep-th/0406259.

\bibitem{0409311}
D.~Ghoshal and T.~Kawano,
arXiv:hep-th/0409311.

\bibitem{0009061}
G.~W.~Gibbons, K.~Hori and P.~Yi,
Nucl.\ Phys.\ B {\bf 596}, 136 (2001) [arXiv:hep-th/0009061].

\bibitem{0209034}
G.~Gibbons, K.~Hashimoto and P.~Yi,
JHEP {\bf 0209}, 061 (2002) [arXiv:hep-th/0209034].

\bibitem{0204008}
G.~W.~Gibbons,
Phys.\ Lett.\ B {\bf 537}, 1 (2002)
[arXiv:hep-th/0204008].

\bibitem{giddings}
S.~B.~Giddings,
Nucl.\ Phys.\ B {\bf 278}, 242 (1986).

\bibitem{GIDDMARW}
S.~B.~Giddings, E.~J.~Martinec and E.~Witten,
Phys.\ Lett.\ B {\bf 176}, 362 (1986).

\bibitem{GINZIN}
P.~Ginsparg and J.~Zinn-Justin,
Phys.\ Lett.\ B {\bf 240}, 333 (1990).

\bibitem{0309164}
S.~Giusto and C.~Imbimbo,
Nucl.\ Phys.\ B {\bf 677}, 52 (2004)
[arXiv:hep-th/0309164].

\bibitem{0106195}
C.~Gomez and P.~Resco,
arXiv:hep-th/0106195.

\bibitem{0111169}
C.~Gomez and J.~J.~Manjarin,
arXiv:hep-th/0111169.

\bibitem{9811131}
R.~Gopakumar and C.~Vafa,
Adv.\ Theor.\ Math.\ Phys.\  {\bf 3}, 1415 (1999)
[arXiv:hep-th/9811131].

\bibitem{0003160}
R.~Gopakumar, S.~Minwalla and A.~Strominger,
JHEP {\bf 0005}, 020 (2000)
[arXiv:hep-th/0003160].

\bibitem{0007226}
R.~Gopakumar, S.~Minwalla and A.~Strominger,
hep-th/0007226.


\bibitem{0103256}
R.~Gopakumar, M.~Headrick and M.~Spradlin,
Commun.\ Math.\ Phys.\  {\bf 233}, 355 (2003)
[arXiv:hep-th/0103256].

\bibitem{0308184}
R.~Gopakumar,
Phys.\ Rev.\ D {\bf 70}, 025009 (2004)
[arXiv:hep-th/0308184].

\bibitem{0402063}
R.~Gopakumar,
Phys.\ Rev.\ D {\bf 70}, 025010 (2004)
[arXiv:hep-th/0402063].

\bibitem{0409233}
R.~Gopakumar,
arXiv:hep-th/0409233.

\bibitem{9403040}
M.~B.~Green,
Phys.\ Lett.\ B {\bf 329}, 435 (1994)
[arXiv:hep-th/9403040].

\bibitem{9604091}
M.~B.~Green and M.~Gutperle,
Nucl.\ Phys.\ B {\bf 476}, 484 (1996) [arXiv:hep-th/9604091].

\bibitem{0206025}
G.~Grignani, M.~Laidlaw, M.~Orselli and G.~W.~Semenoff,
Phys.\ Lett.\ B {\bf 543}, 127 (2002)
[arXiv:hep-th/0206025].

\bibitem{GROMIL}
D.~J.~Gross and N.~Miljkovic,
Phys.\ Lett.\ B {\bf 238}, 217 (1990).

\bibitem{GROSSKLEB}
D.~J.~Gross and I.~R.~Klebanov,
Nucl.\ Phys.\ B {\bf 352}, 671 (1991).

\bibitem{0105059}
D.~J.~Gross and W.~Taylor,
JHEP {\bf 0108}, 009 (2001)
[arXiv:hep-th/0105059].

\bibitem{0106036}
D.~J.~Gross and W.~Taylor,
JHEP {\bf 0108}, 010 (2001)
[arXiv:hep-th/0106036].

\bibitem{9802109}
S.~S.~Gubser, I.~R.~Klebanov and A.~M.~Polyakov,
Phys.\ Lett.\ B {\bf 428}, 105 (1998)
[arXiv:hep-th/9802109].

\bibitem{9901042}
S.~Gukov,
Commun.\ Math.\ Phys.\  {\bf 210}, 621 (2000)
[arXiv:hep-th/9901042].

\bibitem{0101001}
Z.~Guralnik and S.~Ramgoolam,
JHEP {\bf 0102}, 032 (2001)
[arXiv:hep-th/0101001].

\bibitem{9807064}
H.~Gustafsson and U.~Lindstrom,
Phys.\ Lett.\  {\bf B440}, 43 (1998)
[hep-th/9807064].

\bibitem{0202210}
M.~Gutperle and A.~Strominger,
JHEP {\bf 0204}, 018 (2002)
[arXiv:hep-th/0202210].

\bibitem{0301038}
M.~Gutperle and A.~Strominger,
Phys.\ Rev.\ D {\bf 67}, 126002 (2003)
[arXiv:hep-th/0301038].

\bibitem{0308047}
M.~Gutperle and P.~Kraus,
Phys.\ Rev.\ D {\bf 69}, 066005 (2004)
[arXiv:hep-th/0308047].

\bibitem{0407147}
M.~Gutperle and W.~Sabra,
arXiv:hep-th/0407147.

\bibitem{0409050}
M.~Gutperle and P.~Yi,
arXiv:hep-th/0409050.

\bibitem{0001143}
J.~A.~Harvey, P.~Horava and P.~Kraus,
JHEP {\bf 0003}, 021 (2000)
[arXiv:hep-th/0001143].

\bibitem{0002117}
J.A.\ Harvey and P.\ Kraus, 
JHEP {\bf 0004}, 012 (2000) [hep-th/0002117].

\bibitem{0003101}
J.~A.~Harvey, D.~Kutasov and E.~J.~Martinec,
arXiv:hep-th/0003101.

\bibitem{0005031}
J.~A.~Harvey, P.~Kraus, F.~Larsen and E.~J.~Martinec,
JHEP {\bf 0007}, 042 (2000) [arXiv:hep-th/0005031].

\bibitem{0008064}
J.~A.~Harvey, P.~Kraus and F.~Larsen,
Phys.\ Rev.\ D {\bf 63}, 026002 (2001)
[arXiv:hep-th/0008064].

\bibitem{0009030}
J.~A.~Harvey and G.~W.~Moore,
J.\ Math.\ Phys.\  {\bf 42}, 2765 (2001)
[arXiv:hep-th/0009030].

\bibitem{0010060}
J.~A.~Harvey, P.~Kraus and F.~Larsen,
hep-th/0010060.

\bibitem{0102076}
J.~A.~Harvey,
arXiv:hep-th/0102076.

\bibitem{9703217}
A.~Hashimoto and W.~I.~Taylor,
Nucl.\ Phys.\ B {\bf 503}, 193 (1997)
[arXiv:hep-th/9703217].

\bibitem{0111092}
A.~Hashimoto and N.~Itzhaki,
JHEP {\bf 0201}, 028 (2002)
[arXiv:hep-th/0111092].

\bibitem{0101145}
K.~Hashimoto and K.~Krasnov,
Phys.\ Rev.\ D {\bf 64}, 046007 (2001)
[arXiv:hep-th/0101145].

\bibitem{0102174}
K.~Hashimoto and S.~Hirano,
Phys.\ Rev.\ D {\bf 65}, 026006 (2002) [arXiv:hep-th/0102174].

\bibitem{0202079}
K.~Hashimoto and S.~Nagaoka,
Phys.\ Rev.\ D {\bf 66}, 026001 (2002) [arXiv:hep-th/0202079].

\bibitem{0204203}
K.~Hashimoto,
JHEP {\bf 0207}, 035 (2002)
[arXiv:hep-th/0204203].

\bibitem{0209232}
K.~Hashimoto and N.~Sakai,
JHEP {\bf 0212}, 064 (2002)
[arXiv:hep-th/0209232].

\bibitem{0211090}
K.~Hashimoto, P.~M.~Ho and J.~E.~Wang,
Phys.\ Rev.\ Lett.\  {\bf 90}, 141601 (2003)
[arXiv:hep-th/0211090].

\bibitem{0303172}
K.~Hashimoto, P.~M.~Ho, S.~Nagaoka and J.~E.~Wang,
Phys.\ Rev.\ D {\bf 68}, 026007 (2003)
[arXiv:hep-th/0303172].

\bibitem{0303204}
K.~Hashimoto and S.~Nagaoka,
JHEP {\bf 0306}, 034 (2003)
[arXiv:hep-th/0303204].

\bibitem{0404237}
K.~Hashimoto and S.~Terashima,
JHEP {\bf 0406}, 048 (2004)
[arXiv:hep-th/0404237].


\bibitem{0009105}
H.~Hata and S.~Shinohara,
JHEP {\bf 0009}, 035 (2000) [arXiv:hep-th/0009105].

\bibitem{0101162}
H.~Hata and S.~Teraguchi,
JHEP {\bf 0105}, 045 (2001)
[arXiv:hep-th/0101162].

\bibitem{0108150}
H.~Hata and T.~Kawano,
JHEP {\bf 0111}, 038 (2001) [arXiv:hep-th/0108150].

\bibitem{0111034}
H.~Hata and S.~Moriyama,
JHEP {\bf 0201}, 042 (2002)
[arXiv:hep-th/0111034].

\bibitem{0201177}
H.~Hata, S.~Moriyama and S.~Teraguchi,
JHEP {\bf 0202}, 036 (2002)
[arXiv:hep-th/0201177].

\bibitem{0206208}
H.~Hata and S.~Moriyama,
Nucl.\ Phys.\ B {\bf 651}, 3 (2003)
[arXiv:hep-th/0206208].

\bibitem{0208067}
H.~Hata and H.~Kogetsu,
JHEP {\bf 0209}, 027 (2002)
[arXiv:hep-th/0208067].

\bibitem{0305010}
H.~Hata, H.~Kogetsu and S.~Teraguchi,
JHEP {\bf 0402}, 045 (2004)
[arXiv:hep-th/0305010].

\bibitem{0405064}
M.~Headrick, S.~Minwalla and T.~Takayanagi,
Class.\ Quant.\ Grav.\  {\bf 21}, S1539 (2004)
[arXiv:hep-th/0405064].

\bibitem{0008023}
Y.~Hikida, M.~Nozaki and T.~Takayanagi,
Nucl.\ Phys.\ B {\bf 595}, 319 (2001)
[arXiv:hep-th/0008023].

\bibitem{0101211}
Y.~Hikida, M.~Nozaki and Y.~Sugawara,
Nucl.\ Phys.\ B {\bf 617}, 117 (2001)
[arXiv:hep-th/0101211].

\bibitem{thooft}
G.~'t Hooft,
Nucl.\ Phys.\ B {\bf 79}, 276 (1974).

\bibitem{9812135}
P.~Horava,
Adv.\ Theor.\ Math.\ Phys.\  {\bf 2}, 1373 (1999)
[arXiv:hep-th/9812135].

\bibitem{0211127}
W.~H.~Huang,
Phys.\ Lett.\ B {\bf 561}, 153 (2003)
[arXiv:hep-th/0211127].

\bibitem{0407081}
W.~H.~Huang,
JHEP {\bf 0408}, 060 (2004)
[arXiv:hep-th/0407081].

\bibitem{0204031}
Y.~Imamura,
JHEP {\bf 0207}, 042 (2002)
[arXiv:hep-th/0204031].

\bibitem{0004015}
A.~Iqbal and A.~Naqvi,
arXiv:hep-th/0004015.

\bibitem{0008127}
A.~Iqbal and A.~Naqvi,
JHEP {\bf 0101}, 040 (2001)
[arXiv:hep-th/0008127].

\bibitem{ishibashi}
N.~Ishibashi,
Mod.\ Phys.\ Lett.\ A {\bf 4}, 251 (1989).

\bibitem{0206102}
A.~Ishida and S.~Uehara,
Phys.\ Lett.\ B {\bf 544}, 353 (2002)
[arXiv:hep-th/0206102].

\bibitem{0301179}
A.~Ishida and S.~Uehara,
JHEP {\bf 0302}, 050 (2003)
[arXiv:hep-th/0301179].

\bibitem{0111151}
J.~M.~Isidro,
J.\ Geom.\ Phys.\  {\bf 42}, 325 (2002)
[arXiv:hep-th/0111151].

\bibitem{0007078}
D.~P.~Jatkar, G.~Mandal and S.~R.~Wadia,
JHEP {\bf 0009}, 018 (2000)
[arXiv:hep-th/0007078].

\bibitem{0104229}
D.~P.~Jatkar and R.~Vathsan,
JHEP {\bf 0106}, 039 (2001)
[arXiv:hep-th/0104229].

\bibitem{0107075}
D.~P.~Jatkar, S.~Sur and R.~Vathsan,
Phys.\ Lett.\ B {\bf 520}, 391 (2001)
[arXiv:hep-th/0107075].

\bibitem{9302106}
A.~Jevicki,
arXiv:hep-th/9302106.

\bibitem{0007170}
C.~V.~Johnson,
arXiv:hep-th/0007170.

\bibitem{0408049}
C.~V.~Johnson,
arXiv:hep-th/0408049.

\bibitem{0403050}
G.~Jones, A.~Maloney and A.~Strominger,
Phys.\ Rev.\ D {\bf 69}, 126008 (2004)
[arXiv:hep-th/0403050].

\bibitem{0211180}
N.~T.~Jones and S.~H.~H.~Tye,
JHEP {\bf 0301}, 012 (2003)
[arXiv:hep-th/0211180].

\bibitem{0207041}
G.~Jorjadze and G.~Weigt,
arXiv:hep-th/0207041.

\bibitem{0311202}
G.~Jorjadze and G.~Weigt,
Phys.\ Lett.\ B {\bf 581}, 133 (2004)
[arXiv:hep-th/0311202].

\bibitem{math-ph/0308034}
L.~Joukovskaya and Y.~Volovich,
arXiv:math-ph/0308034.

\bibitem{0104143}
H.~Kajiura, Y.~Matsuo and T.~Takayanagi,
JHEP {\bf 0106}, 041 (2001)
[arXiv:hep-th/0104143].

\bibitem{0003211}
K.~Kamimura and J.~Simon,
Nucl.\ Phys.\ B {\bf 585}, 219 (2000)
[arXiv:hep-th/0003211].

\bibitem{0210108}
S.~Kar,
arXiv:hep-th/0210108.

\bibitem{0306132}
J.~L.~Karczmarek, H.~Liu, J.~Maldacena and A.~Strominger,
arXiv:hep-th/0306132.

\bibitem{0409249}
F.~Katsumata, T.~Takahashi and S.~Zeze,
arXiv:hep-th/0409249.

\bibitem{0106103}
H.~Kawai and T.~Kuroki,
Phys.\ Lett.\ B {\bf 518}, 294 (2001)
[arXiv:hep-th/0106103].

\bibitem{9912274}
T.~Kawano and T.~Takahashi,
hep-th/9912274.

\bibitem{9905195}
C.~Kennedy and A.~Wilkins,
Phys.\ Lett.\ B {\bf 464}, 206 (1999)
[arXiv:hep-th/9905195].

\bibitem{0301076}
C.~j.~Kim, H.~B.~Kim, Y.~b.~Kim and O.~K.~Kwon,
JHEP {\bf 0303}, 008 (2003) [arXiv:hep-th/0301076].

\bibitem{0304180}
C.~j.~Kim, Y.~b.~Kim and C.~O.~Lee,
JHEP {\bf 0305}, 020 (2003)
[arXiv:hep-th/0304180].

\bibitem{0305092}
C.~Kim, Y.~Kim, O.~K.~Kwon and C.~O.~Lee,
JHEP {\bf 0311}, 034 (2003)
[arXiv:hep-th/0305092].

\bibitem{0307184}
C.~j.~Kim, Y.~b.~Kim, O.~K.~Kwon and P.~Yi,
JHEP {\bf 0309}, 042 (2003)
[arXiv:hep-th/0307184].

\bibitem{0404163}
C.~Kim, Y.~Kim and O.~K.~Kwon,
JHEP {\bf 0405}, 020 (2004)
[arXiv:hep-th/0404163].

\bibitem{0204191}
H.~Kim,
JHEP {\bf 0301}, 080 (2003)
[arXiv:hep-th/0204191].

\bibitem{0110124}
I.~Kishimoto,
JHEP {\bf 0112}, 007 (2001)
[arXiv:hep-th/0110124].

\bibitem{0112169}
I.~Kishimoto and K.~Ohmori,
JHEP {\bf 0205}, 036 (2002)
[arXiv:hep-th/0112169].

\bibitem{0205275}
I.~Kishimoto and T.~Takahashi,
Prog.\ Theor.\ Phys.\  {\bf 108}, 591 (2002)
[arXiv:hep-th/0205275].

\bibitem{0012081}
M.~Kleban, A.~Lawrence and S.~Shenker,
hep-th/0012081.

\bibitem{9108019}
I.~R.~Klebanov,
arXiv:hep-th/9108019.

\bibitem{9210105}
I.~R.~Klebanov and A.~Pasquinucci,
arXiv:hep-th/9210105.

\bibitem{0305159}
I.~R.~Klebanov, J.~Maldacena and N.~Seiberg,
JHEP {\bf 0307}, 045 (2003)
[arXiv:hep-th/0305159].

\bibitem{0209186}
A.~Kling, O.~Lechtenfeld, A.~D.~Popov and S.~Uhlmann,
Phys.\ Lett.\ B {\bf 551}, 193 (2003)
[arXiv:hep-th/0209186].

\bibitem{0212335}
A.~Kling, O.~Lechtenfeld, A.~D.~Popov and S.~Uhlmann,
Fortsch.\ Phys.\  {\bf 51}, 775 (2003)
[arXiv:hep-th/0212335].

\bibitem{0004106}
J.~Kluson,
Phys.\ Rev.\ D {\bf 62}, 126003 (2000) [arXiv:hep-th/0004106].

\bibitem{0009189}
J.~Kluson,
JHEP {\bf 0011}, 016 (2000)
[arXiv:hep-th/0009189].

\bibitem{0102063}
J.~Kluson,
JHEP {\bf 0103}, 018 (2001)
[arXiv:hep-th/0102063].

\bibitem{0103079}
J.~Kluson,
Phys.\ Rev.\ D {\bf 64}, 126006 (2001)
[arXiv:hep-th/0103079].

\bibitem{0208028}
J.~Kluson,
arXiv:hep-th/0208028.

\bibitem{0310066}
J.~Kluson,
JHEP {\bf 0311}, 068 (2003)
[arXiv:hep-th/0310066].

\bibitem{0312086}
J.~Kluson,
JHEP {\bf 0401}, 019 (2004)
[arXiv:hep-th/0312086].

\bibitem{0401236}
J.~Kluson,
JHEP {\bf 0402}, 024 (2004)
[arXiv:hep-th/0401236].

\bibitem{0403124}
J.~Kluson,
JHEP {\bf 0406}, 021 (2004)
[arXiv:hep-th/0403124].

\bibitem{KPZ}
V.~G.~Knizhnik, A.~M.~Polyakov and A.~B.~Zamolodchikov,
Mod.\ Phys.\ Lett.\ A {\bf 3}, 819 (1988).

\bibitem{0409044}
S.~Kobayashi, T.~Asakawa and S.~Matsuura,
arXiv:hep-th/0409044.

\bibitem{0212055}
A.~Koshelev,
arXiv:hep-th/0212055.

\bibitem{kost-sam1}
V.~A.~Kostelecky and S.~Samuel,
Phys.\ Lett.\ B {\bf 207}, 169 (1988).

\bibitem{kost-sam2}
V.~A.~Kostelecky and S.~Samuel,
Nucl.\ Phys.\ B {\bf 336}, 263 (1990).

\bibitem{kost-pot1}
V.~A.~Kostelecky and R.~Potting,
Phys.\ Lett.\ B {\bf 381}, 89 (1996) [arXiv:hep-th/9605088].

\bibitem{0008252}
V.~A.~Kostelecky and R.~Potting,
Phys.\ Rev.\ D {\bf 63}, 046007 (2001) [arXiv:hep-th/0008252].

\bibitem{0104090}
T.~Krajewski and M.~Schnabl,
JHEP {\bf 0108}, 002 (2001)
[arXiv:hep-th/0104090].

\bibitem{0010016}
P.~Kraus, A.~Rajaraman and S.~H.~Shenker,
Nucl.\ Phys.\ B {\bf 598}, 169 (2001)
[arXiv:hep-th/0010016].

\bibitem{0012198}
P.~Kraus and F.~Larsen,
Phys.\ Rev.\ D {\bf 63}, 106004 (2001)
[arXiv:hep-th/0012198].

\bibitem{0401003}
K.~R.~Kristjansson and L.~Thorlacius,
Class.\ Quant.\ Grav.\  {\bf 21}, S1359 (2004)
[arXiv:hep-th/0401003].

\bibitem{0204144}
M.~Kruczenski, R.~C.~Myers and A.~W.~Peet,
JHEP {\bf 0205}, 039 (2002)
[arXiv:hep-th/0204144].

\bibitem{0305229}
O.~K.~Kwon and P.~Yi,
arXiv:hep-th/0305229.

\bibitem{0009148}
D.~Kutasov, M.~Marino and G.~W.~Moore,
JHEP {\bf 0010}, 045 (2000) [arXiv:hep-th/0009148].

\bibitem{0010108}
D.~Kutasov, M.~Marino and G.~W.~Moore,
arXiv:hep-th/0010108.

\bibitem{0304045}
D.~Kutasov and V.~Niarchos,
Nucl.\ Phys.\ B {\bf 666}, 56 (2003)
[arXiv:hep-th/0304045].

\bibitem{0405058}
D.~Kutasov,
arXiv:hep-th/0405058.

\bibitem{0408073}
D.~Kutasov,
arXiv:hep-th/0408073.

\bibitem{0002061}
N.~D.~Lambert and I.~Sachs,
JHEP {\bf 0003}, 028 (2000)
[arXiv:hep-th/0002061].

\bibitem{0006122}
N.~D.~Lambert and I.~Sachs,
JHEP {\bf 0008}, 024 (2000)
[arXiv:hep-th/0006122].

\bibitem{0104218}
N.~D.~Lambert and I.~Sachs,
JHEP {\bf 0106}, 060 (2001) [arXiv:hep-th/0104218].

\bibitem{0208217}
N.~D.~Lambert and I.~Sachs,
Phys.\ Rev.\ D {\bf 67}, 026005 (2003) [arXiv:hep-th/0208217].

\bibitem{0303139}
N.~Lambert, H.~Liu and J.~Maldacena,
arXiv:hep-th/0303139.

\bibitem{0010181}
F.~Larsen,
Int.\ J.\ Mod.\ Phys.\ A {\bf 16}, 650 (2001)
[arXiv:hep-th/0010181].

\bibitem{0212248}
F.~Larsen, A.~Naqvi and S.~Terashima,
JHEP {\bf 0302}, 039 (2003)
[arXiv:hep-th/0212248].

\bibitem{0303035}
F.~Leblond and A.~W.~Peet,
JHEP {\bf 0304}, 048 (2003) [arXiv:hep-th/0303035].

\bibitem{0204155}
O.~Lechtenfeld, A.~D.~Popov and S.~Uhlmann,
Nucl.\ Phys.\ B {\bf 637}, 119 (2002)
[arXiv:hep-th/0204155].

\bibitem{0210221}
H.~w.~Lee and W.~S.~l'Yi,
J.\ Korean Phys.\ Soc.\  {\bf 43}, 676 (2003)
[arXiv:hep-th/0210221].

\bibitem{0105115}
T.~Lee,
Phys.\ Rev.\ D {\bf 64}, 106004 (2001)
[arXiv:hep-th/0105115].

\bibitem{0105264}
T.~Lee,
Phys.\ Lett.\ B {\bf 520}, 385 (2001)
[arXiv:hep-th/0105264].

\bibitem{0109032}
T.~Lee, K.~S.~Viswanathan and Y.~Yang,
J.\ Korean Phys.\ Soc.\  {\bf 42}, 34 (2003)
[arXiv:hep-th/0109032].

\bibitem{leigh}
R. Leigh, Mod. Phys. Lett. {\bf A4} (1989) 2767.

\bibitem{9905006}
A.~Lerda and R.~Russo,
Int.\ J.\ Mod.\ Phys.\ A {\bf 15}, 771 (2000)
[arXiv:hep-th/9905006].

\bibitem{9303067}
K.\ Li and E.\ Witten,
Phys.\ Rev.\  {\bf D48}, 853 (1993)
[hep-th/9303067].

\bibitem{0010058}
M.~Li,
Nucl.\ Phys.\ B {\bf 602}, 201 (2001)
[arXiv:hep-th/0010058].

\bibitem{LIAN}
B.~H.~Lian and G.~J.~Zuckerman,
Phys.\ Lett.\ B {\bf 254}, 417 (1991);
Phys.\ Lett.\ B {\bf 266}, 21 (1991).

\bibitem{9604156}
G.~Lifschytz,
Phys.\ Lett.\ B {\bf 388}, 720 (1996)
[arXiv:hep-th/9604156].

\bibitem{9704051}
U.~Lindstrom and R.~von Unge,
Phys.\ Lett.\ B {\bf 403}, 233 (1997) [arXiv:hep-th/9704051].

\bibitem{9910159}
U.~Lindstrom, M.~Zabzine and A.~Zheltukhin,
JHEP {\bf 9912}, 016 (1999)
[hep-th/9910159].

\bibitem{0101213}
U.~Lindstrom and M.~Zabzine,
JHEP {\bf 0103}, 014 (2001)
[arXiv:hep-th/0101213].

\bibitem{0403147}
J.~X.~Lu and S.~Roy,
Phys.\ Lett.\ B {\bf 599}, 313 (2004)
[arXiv:hep-th/0403147].

\bibitem{0409019}
J.~X.~Lu and S.~Roy,
arXiv:hep-th/0409019.

\bibitem{0003124}
J.~Majumder and A.~Sen,
JHEP {\bf 0006}, 010 (2000) [arXiv:hep-th/0003124].

\bibitem{9711200}
J.~M.~Maldacena,
Adv.\ Theor.\ Math.\ Phys.\  {\bf 2}, 231 (1998)
[Int.\ J.\ Theor.\ Phys.\  {\bf 38}, 1113 (1999)]
[arXiv:hep-th/9711200].

\bibitem{0108100}
J.~M.~Maldacena, G.~W.~Moore and N.~Seiberg,
JHEP {\bf 0111}, 062 (2001)
[arXiv:hep-th/0108100].

\bibitem{0302146}
A.~Maloney, A.~Strominger and X.~Yin,
arXiv:hep-th/0302146.

\bibitem{0008214}
G.~Mandal and S.~Rey,
Phys.\ Lett.\ {\bf B495}, 193 (2000)
[hep-th/0008214].

\bibitem{0011094}
G.~Mandal and S.~R.~Wadia,
Nucl.\ Phys.\ B {\bf 599}, 137 (2001)
[arXiv:hep-th/0011094].

\bibitem{0312192}
G.~Mandal and S.~R.~Wadia,
arXiv:hep-th/0312192.

\bibitem{MANDELSTAM}
S.~Mandelstam,
Phys.\ Rev.\ D {\bf 11}, 3026 (1975).

\bibitem{0103089}
M.~Marino,
JHEP {\bf 0106}, 059 (2001)
[arXiv:hep-th/0103089].

\bibitem{0112231}
M.~Marino and R.~Schiappa,
J.\ Math.\ Phys.\  {\bf 44}, 156 (2003)
[arXiv:hep-th/0112231].

\bibitem{0101199}
E.~J.~Martinec and G.~W.~Moore,
arXiv:hep-th/0101199.

\bibitem{0212059}
E.~J.~Martinec and G.~W.~Moore,
arXiv:hep-th/0212059.

\bibitem{0306295}
L.~Martucci and P.~J.~Silva,
JHEP {\bf 0308}, 026 (2003)
[arXiv:hep-th/0306295].

\bibitem{0203071}
P.~Matlock, R.~C.~Rashkov, K.~S.~Viswanathan and Y.~Yang,
Phys.\ Rev.\ D {\bf 66}, 026004 (2002)
[arXiv:hep-th/0203071].

\bibitem{0211286}
P.~Matlock,
Phys.\ Rev.\ D {\bf 67}, 086002 (2003)
[arXiv:hep-th/0211286].

\bibitem{0009002}
Y.~Matsuo,
Phys.\ Lett.\ B {\bf 499}, 223 (2001)
[arXiv:hep-th/0009002].

\bibitem{0107058}
A.~Mazumdar, S.~Panda and A.~Perez-Lorenzana,
Nucl.\ Phys.\ B {\bf 614}, 101 (2001)
[arXiv:hep-ph/0107058].

\bibitem{0304224}
J.~McGreevy and H.~Verlinde,
arXiv:hep-th/0304224.

\bibitem{0305194}
J.~McGreevy, J.~Teschner and H.~Verlinde,
arXiv:hep-th/0305194.

\bibitem{0206212}
T.~Mehen and B.~Wecht,
JHEP {\bf 0302}, 058 (2003)
[arXiv:hep-th/0206212].

\bibitem{0105246}
Y.~Michishita,
Nucl.\ Phys.\ B {\bf 614}, 26 (2001)
[arXiv:hep-th/0105246].

\bibitem{0011079}
M.~Mihailescu, I.~Y.~Park and T.~A.~Tran,
Phys.\ Rev.\ D {\bf 64}, 046006 (2001)
[arXiv:hep-th/0011079].

\bibitem{0008231}
J.~A.~Minahan and B.~Zwiebach,
JHEP {\bf 0009}, 029 (2000) [arXiv:hep-th/0008231].

\bibitem{0009246}
J.~A.~Minahan and B.~Zwiebach,
hep-th/0009246.

\bibitem{0011226}
J.~A.~Minahan and B.~Zwiebach,
JHEP {\bf 0102}, 034 (2001) [arXiv:hep-th/0011226].

\bibitem{0102071}
J.~A.~Minahan,
JHEP {\bf 0103}, 028 (2001)
[arXiv:hep-th/0102071].

\bibitem{0105312}
J.~A.~Minahan,
arXiv:hep-th/0105312.

\bibitem{0203108}
J.~A.~Minahan,
JHEP {\bf 0205}, 024 (2002)
[arXiv:hep-th/0203108].

\bibitem{0205098}
J.~A.~Minahan,
JHEP {\bf 0207}, 030 (2002) [arXiv:hep-th/0205098].

\bibitem{UTTG-16-91}
D.~Minic, J.~Polchinski and Z.~Yang,
Nucl.\ Phys.\ B {\bf 369}, 324 (1992).

\bibitem{0002237}
N.~Moeller and W.~Taylor,
Nucl.\ Phys.\ B {\bf 583}, 105 (2000) [arXiv:hep-th/0002237].

\bibitem{0005036}
N.~Moeller, A.~Sen and B.~Zwiebach,
JHEP {\bf 0008}, 039 (2000) [arXiv:hep-th/0005036].

\bibitem{0008101}
N.~Moeller,
arXiv:hep-th/0008101.

\bibitem{0207107}
N.~Moeller and B.~Zwiebach,
arXiv:hep-th/0207107.

\bibitem{0304213}
N.~Moeller and M.~Schnabl,
JHEP {\bf 0401}, 011 (2004)
[arXiv:hep-th/0304213].

\bibitem{MOORESEI}
G.~W.~Moore and N.~Seiberg,
Int.\ J.\ Mod.\ Phys.\ A {\bf 7}, 2601 (1992).

\bibitem{9912279}
G.~W.~Moore and E.~Witten,
JHEP {\bf 0005}, 032 (2000)
[arXiv:hep-th/9912279].

\bibitem{0111069}
G.~Moore and W.~Taylor,
JHEP {\bf 0201}, 004 (2002)
[arXiv:hep-th/0111069].

\bibitem{0304018}
G.~W.~Moore,
arXiv:hep-th/0304018.

\bibitem{0011002}
S.~Moriyama and S.~Nakamura,
Phys.\ Lett.\ B {\bf 506}, 161 (2001)
[arXiv:hep-th/0011002].

\bibitem{0001066}
S.~Mukhi, N.~V.~Suryanarayana and D.~Tong,
JHEP {\bf 0003}, 015 (2000)
[arXiv:hep-th/0001066].

\bibitem{0009101}
S.~Mukhi and N.~V.~Suryanarayana,
JHEP {\bf 0011}, 006 (2000)
[arXiv:hep-th/0009101].

\bibitem{0107087}
S.~Mukhi and N.~V.~Suryanarayana,
arXiv:hep-th/0107087.

\bibitem{0101014}
P.~Mukhopadhyay and A.~Sen,
JHEP {\bf 0102}, 017 (2001)
[arXiv:hep-th/0101014].

\bibitem{0110136}
P.~Mukhopadhyay,
JHEP {\bf 0112}, 025 (2001)
[arXiv:hep-th/0110136].

\bibitem{0208142}
P.~Mukhopadhyay and A.~Sen,
JHEP {\bf 0211}, 047 (2002)
[arXiv:hep-th/0208142].

\bibitem{0208094}
S.~Mukohyama,
Phys.\ Rev.\ D {\bf 66}, 123512 (2002)
[arXiv:hep-th/0208094].

\bibitem{0309017}
K.~Nagami,
JHEP {\bf 0401}, 005 (2004)
[arXiv:hep-th/0309017].

\bibitem{0312149}
K.~Nagami,
Phys.\ Lett.\ B {\bf 591}, 187 (2004)
[arXiv:hep-th/0312149].

\bibitem{0005114}
M.~Naka, T.~Takayanagi and T.~Uesugi,
JHEP {\bf 0006}, 007 (2000)
[arXiv:hep-th/0005114].

\bibitem{9402156}
M.~Natsuume and J.~Polchinski,
Nucl.\ Phys.\ B {\bf 424}, 137 (1994)
[arXiv:hep-th/9402156].

\bibitem{0408157}
D.~Nemeschansky and V.~Yasnov,
arXiv:hep-th/0408157.

\bibitem{0401066}
V.~Niarchos,
Phys.\ Rev.\ D {\bf 69}, 106009 (2004)
[arXiv:hep-th/0401066].

\bibitem{0102085}
K.~Ohmori,
arXiv:hep-th/0102085.

\bibitem{0104230}
K.~Ohmori,
JHEP {\bf 0105}, 035 (2001)
[arXiv:hep-th/0104230].

\bibitem{0106068}
K.~Ohmori,
JHEP {\bf 0108}, 011 (2001)
[arXiv:hep-th/0106068].

\bibitem{0204138}
K.~Ohmori,
JHEP {\bf 0204}, 059 (2002)
[arXiv:hep-th/0204138].

\bibitem{0208009}
K.~Ohmori,
Nucl.\ Phys.\ B {\bf 648}, 94 (2003)
[arXiv:hep-th/0208009].

\bibitem{0305103}
K.~Ohmori,
arXiv:hep-th/0305103.

\bibitem{0306096}
K.~Ohmori,
Phys.\ Rev.\ D {\bf 69}, 026008 (2004)
[arXiv:hep-th/0306096].

\bibitem{0207004}
K.~Ohta and T.~Yokono,
Phys.\ Rev.\ D {\bf 66}, 125009 (2002)
[arXiv:hep-th/0207004].

\bibitem{0301095}
N.~Ohta,
Phys.\ Lett.\ B {\bf 558}, 213 (2003)
[arXiv:hep-th/0301095].

\bibitem{0204012}
Y.~Okawa,
JHEP {\bf 0207}, 003 (2002)
[arXiv:hep-th/0204012].

\bibitem{0310264}
Y.~Okawa,
arXiv:hep-th/0310264.

\bibitem{0311115}
Y.~Okawa,
arXiv:hep-th/0311115.

\bibitem{0201149}
T.~Okuda,
Nucl.\ Phys.\ B {\bf 641}, 393 (2002)
[arXiv:hep-th/0201149].

\bibitem{0208196}
T.~Okuda and S.~Sugimoto,
Nucl.\ Phys.\ B {\bf 647}, 101 (2002)
[arXiv:hep-th/0208196].

\bibitem{0010028}
K.~Okuyama,
Phys.\ Lett.\ B {\bf 499}, 167 (2001)
[arXiv:hep-th/0010028].

\bibitem{0111087}
K.~Okuyama,
JHEP {\bf 0201}, 043 (2002)
[arXiv:hep-th/0111087].

\bibitem{0201015}
K.~Okuyama,
JHEP {\bf 0201}, 027 (2002)
[arXiv:hep-th/0201015].

\bibitem{0201136}
K.~Okuyama,
JHEP {\bf 0203}, 050 (2002)
[arXiv:hep-th/0201136].

\bibitem{0304108}
K.~Okuyama,
JHEP {\bf 0305}, 005 (2003)
[arXiv:hep-th/0304108].

\bibitem{0308172}
K.~Okuyama,
JHEP {\bf 0309}, 053 (2003) [arXiv:hep-th/0308172].

\bibitem{9904153}
K.~Olsen and R.~J.~Szabo,
Nucl.\ Phys.\ B {\bf 566}, 562 (2000)
[arXiv:hep-th/9904153].

\bibitem{9907140}
K.~Olsen and R.~J.~Szabo,
Adv.\ Theor.\ Math.\ Phys.\  {\bf 3}, 889 (1999)
[arXiv:hep-th/9907140].

\bibitem{0403283}
A.~Parodi,
arXiv:hep-th/0403283.

\bibitem{0405125}
K.~Peeters and M.~Zamaklar,
arXiv:hep-th/0405125.

\bibitem{9612215}
V.~Periwal,
arXiv:hep-th/9612215.

\bibitem{0006223}
V.~Periwal,
JHEP {\bf 0007}, 041 (2000)
[arXiv:hep-th/0006223].

\bibitem{POLCH}
J.~Polchinski,
Nucl.\ Phys.\ B {\bf 362}, 125 (1991).

\bibitem{9404008}
J.~Polchinski and L.~Thorlacius,
Phys.\ Rev.\ D {\bf 50}, 622 (1994) [arXiv:hep-th/9404008].

\bibitem{9407031}
J. Polchinski, Phys. Rev. {\bf D50} (1994) 6041 [hep-th/9407031].

\bibitem{9510017}
J.~Polchinski,
Phys.\ Rev.\ Lett.\  {\bf 75}, 4724 (1995)
[arXiv:hep-th/9510017].

\bibitem{9611050}
J.~Polchinski,
arXiv:hep-th/9611050.

\bibitem{polbook}
J.~Polchinski, String theory. Vol. 1 and 2, Cambridge Univ. Press,
1998.

\bibitem{polyakov}
A.~M.~Polyakov,
JETP Lett.\  {\bf 20}, 194 (1974)
[Pisma Zh.\ Eksp.\ Teor.\ Fiz.\  {\bf 20}, 430 (1974)].

\bibitem{poly1}
A.~M.~Polyakov,
Mod.\ Phys.\ Lett.\ A {\bf 6}, 635 (1991).

\bibitem{0204172}
R.~Potting and J.~Raeymaekers,
JHEP {\bf 0206}, 002 (2002)
[arXiv:hep-th/0204172].

\bibitem{pty}
C.~R.~Preitschopf, C.~B.~Thorn and S.~A.~Yost,
Nucl.\ Phys.\  {\bf B337}, 363 (1990).

\bibitem{0112202}
R.~Rashkov and K.~S.~Viswanathan,
arXiv:hep-th/0112202.

\bibitem{0201229}
R.~Rashkov and K.~S.~Viswanathan,
arXiv:hep-th/0201229.

\bibitem{0006240}
L.~Rastelli and B.~Zwiebach,
JHEP {\bf 0109}, 038 (2001)
[arXiv:hep-th/0006240].

\bibitem{0012251}
L.~Rastelli, A.~Sen and B.~Zwiebach,
Adv.\ Theor.\ Math.\ Phys.\  {\bf 5}, 353 (2002)
[arXiv:hep-th/0012251].

\bibitem{0102112}
L.~Rastelli, A.~Sen and B.~Zwiebach,
Adv.\ Theor.\ Math.\ Phys.\  {\bf 5}, 393 (2002)
[arXiv:hep-th/0102112].

\bibitem{0105058}
L.~Rastelli, A.~Sen and B.~Zwiebach,
JHEP {\bf 0111}, 035 (2001)
[arXiv:hep-th/0105058].

\bibitem{0105168}
L.~Rastelli, A.~Sen and B.~Zwiebach,
JHEP {\bf 0111}, 045 (2001)
[arXiv:hep-th/0105168].

\bibitem{0106010}
L.~Rastelli, A.~Sen and B.~Zwiebach,
arXiv:hep-th/0106010.

\bibitem{0111153}
L.~Rastelli, A.~Sen and B.~Zwiebach,
JHEP {\bf 0202}, 034 (2002)
[arXiv:hep-th/0111153].

\bibitem{0111281}
L.~Rastelli, A.~Sen and B.~Zwiebach,
JHEP {\bf 0203}, 029 (2002)
[arXiv:hep-th/0111281].

\bibitem{9811237}
A.~Recknagel and V.~Schomerus,
Nucl.\ Phys.\ B {\bf 545}, 233 (1999) [arXiv:hep-th/9811237].

\bibitem{9903139}
A.~Recknagel and V.~Schomerus,
Fortsch.\ Phys.\  {\bf 48}, 195 (2000)
[arXiv:hep-th/9903139].

\bibitem{0003110}
A.~Recknagel, D.~Roggenkamp and V.~Schomerus,
Nucl.\ Phys.\ B {\bf 588}, 552 (2000)
[arXiv:hep-th/0003110].

\bibitem{0301049}
S.~J.~Rey and S.~Sugimoto,
Phys.\ Rev.\ D {\bf 67}, 086008 (2003)
[arXiv:hep-th/0301049].

\bibitem{0303133}
S.~J.~Rey and S.~Sugimoto,
arXiv:hep-th/0303133.

\bibitem{0205198}
S.~Roy,
JHEP {\bf 0208}, 025 (2002)
[arXiv:hep-th/0205198].

\bibitem{0010007}
S.~Ryang,
arXiv:hep-th/0010007.

\bibitem{0307078}
Phys.\ Lett.\ B {\bf 573}, 181 (2003)
[arXiv:hep-th/0307078].

\bibitem{0010034}
M.~Schnabl,
JHEP {\bf 0011}, 031 (2000)
[arXiv:hep-th/0010034].

\bibitem{0011238}
M.~Schnabl,
hep-th/0011238.

\bibitem{0201095}
M.~Schnabl,
JHEP {\bf 0301}, 004 (2003)
[arXiv:hep-th/0201095].

\bibitem{0202139}
M.~Schnabl,
Nucl.\ Phys.\ B {\bf 649}, 101 (2003)
[arXiv:hep-th/0202139].

\bibitem{9903205}
V.~Schomerus,
JHEP {\bf 9906}, 030 (1999)
[arXiv:hep-th/9903205].

\bibitem{0306026}
V.~Schomerus,
arXiv:hep-th/0306026.

\bibitem{9908144}
J.~H.~Schwarz,
arXiv:hep-th/9908144.

\bibitem{9908091}
J.~H.~Schwarz,
Class.\ Quant.\ Grav.\  {\bf 17}, 1245 (2000)
[arXiv:hep-th/9908091].

\bibitem{9908142}
N.~Seiberg and E.~Witten,
JHEP {\bf 9909}, 032 (1999)
[arXiv:hep-th/9908142].

\bibitem{0008013}
N.~Seiberg,
hep-th/0008013.

\bibitem{9805019}
A.~Sen,
JHEP {\bf 9808}, 010 (1998)
[arXiv:hep-th/9805019].

\bibitem{9805170}
A.~Sen,
JHEP {\bf 9808}, 012 (1998) [arXiv:hep-th/9805170].

\bibitem{9808141}
A.~Sen,
JHEP {\bf 9809}, 023 (1998)
[arXiv:hep-th/9808141].

\bibitem{9809111}
A.~Sen,
JHEP {\bf 9810}, 021 (1998)
[arXiv:hep-th/9809111].

\bibitem{9812031}
A.~Sen,
JHEP {\bf 9812}, 021 (1998) [arXiv:hep-th/9812031].

\bibitem{9902105}
A.~Sen,
Int.\ J.\ Mod.\ Phys.\ A {\bf 14}, 4061 (1999)
[arXiv:hep-th/9902105].

\bibitem{9904207}
A.~Sen,
arXiv:hep-th/9904207.

\bibitem{9909062}
A.~Sen,
JHEP {\bf 9910}, 008 (1999) [arXiv:hep-th/9909062].

\bibitem{9911116}
A.~Sen,
JHEP {\bf 9912}, 027 (1999) [arXiv:hep-th/9911116].

\bibitem{9912249}
A.~Sen and B.~Zwiebach,
JHEP {\bf 0003}, 002 (2000) [arXiv:hep-th/9912249].

\bibitem{0007153}
A.~Sen and B.~Zwiebach,
JHEP {\bf 0010}, 009 (2000)
[hep-th/0007153].

\bibitem{0009038}
A.~Sen,
JHEP {\bf 0011}, 035 (2000)
[arXiv:hep-th/0009038].

\bibitem{0009090}
A.~Sen,
JHEP {\bf 0012}, 001 (2000)
[arXiv:hep-th/0009090].

\bibitem{0010240}
A.~Sen,
J.\ Math.\ Phys.\  {\bf 42}, 2844 (2001) [arXiv:hep-th/0010240].

\bibitem{0203211}
A.~Sen,
JHEP {\bf 0204}, 048 (2002) [arXiv:hep-th/0203211].

\bibitem{0203265}
A.~Sen,
JHEP {\bf 0207}, 065 (2002) [arXiv:hep-th/0203265].

\bibitem{0204143}
A.~Sen,
Mod.\ Phys.\ Lett.\ A {\bf 17}, 1797 (2002)
[arXiv:hep-th/0204143].

\bibitem{0207105}
A.~Sen,
JHEP {\bf 0210}, 003 (2002) [arXiv:hep-th/0207105].

\bibitem{0209122}
A.~Sen,
Int.\ J.\ Mod.\ Phys.\ A {\bf 18}, 4869 (2003)
[arXiv:hep-th/0209122].

\bibitem{0303057}
A.~Sen,
Phys.\ Rev.\ D {\bf 68}, 066008 (2003) [arXiv:hep-th/0303057].

\bibitem{0305011}
A.~Sen,
Phys.\ Rev.\ D {\bf 68}, 106003 (2003)
[arXiv:hep-th/0305011].

\bibitem{0306137}
A.~Sen,
Phys.\ Rev.\ Lett.\  {\bf 91}, 181601 (2003)
[arXiv:hep-th/0306137].

\bibitem{0308068}
A.~Sen,
Mod.\ Phys.\ Lett.\ A {\bf 19}, 841 (2004)
[arXiv:hep-th/0308068].

\bibitem{0312003}
A.~Sen,
JHEP {\bf 0403}, 070 (2004)
[arXiv:hep-th/0312003].

\bibitem{0312153}
A.~Sen,
arXiv:hep-th/0312153.

\bibitem{0402157}
A.~Sen,
JHEP {\bf 0405}, 076 (2004)
[arXiv:hep-th/0402157].

\bibitem{0403200}
A.~Sen,
JHEP {\bf 0408}, 034 (2004)
[arXiv:hep-th/0403200].

\bibitem{0408064}
A.~Sen,
arXiv:hep-th/0408064.

\bibitem{SENWAD}
A.~M.~Sengupta and S.~R.~Wadia,
Int.\ J.\ Mod.\ Phys.\ A {\bf 6}, 1961 (1991).

\bibitem{9303143}
S.L.\ Shatashvili,
Phys.\ Lett.\  {\bf B311}, 83 (1993)
[hep-th/9303143].

\bibitem{9311177}
S.L.\ Shatashvili,
hep-th/9311177.

\bibitem{0105076}
S.~L.~Shatashvili,
arXiv:hep-th/0105076.

\bibitem{siegel-zwie}
W.~Siegel and B.~Zwiebach,
Nucl.\ Phys.\ B {\bf 288}, 332 (1987).

\bibitem{0310138}
M.~Smedback,
arXiv:hep-th/0310138.

\bibitem{0007217}
C.~Sochichiu,
JHEP {\bf 0008}, 026 (2000)
[arXiv:hep-th/0007217].

\bibitem{9807138}
M.~Srednicki,
JHEP {\bf 9808}, 005 (1998)
[arXiv:hep-th/9807138].

\bibitem{9310006}
C.~R.~Stephens, G.~'t Hooft and B.~F.~Whiting,
Class.\ Quant.\ Grav.\  {\bf 11}, 621 (1994)
[arXiv:gr-qc/9310006].

\bibitem{0209090}
A.~Strominger,
arXiv:hep-th/0209090.

\bibitem{0307034}
Y.~Sugawara,
JHEP {\bf 0308}, 008 (2003)
[arXiv:hep-th/0307034].

\bibitem{0205085}
S.~Sugimoto and S.~Terashima,
JHEP {\bf 0207}, 025 (2002) [arXiv:hep-th/0205085].

\bibitem{9409089}
L.~Susskind,
J.\ Math.\ Phys.\  {\bf 36}, 6377 (1995)
[arXiv:hep-th/9409089].

\bibitem{9805114}
L.~Susskind and E.~Witten,
arXiv:hep-th/9805114.

\bibitem{0101002}
T.~Suyama,
arXiv:hep-th/0101002.

\bibitem{0102192}
T.~Suyama,
Prog.\ Theor.\ Phys.\  {\bf 106}, 1017 (2001)
[arXiv:hep-th/0102192].

\bibitem{0012210}
T.~Takayanagi, S.~Terashima and T.~Uesugi,
JHEP {\bf 0103}, 019 (2001)
[arXiv:hep-th/0012210].

\bibitem{0103021}
T.~Takayanagi,
Nucl.\ Phys.\ B {\bf 603}, 259 (2001)
[arXiv:hep-th/0103021].

\bibitem{0106142}
T.~Takayanagi,
Phys.\ Lett.\ B {\bf 519}, 137 (2001)
[arXiv:hep-th/0106142].

\bibitem{0307083}
T.~Takayanagi and N.~Toumbas,
JHEP {\bf 0307}, 064 (2003)
[arXiv:hep-th/0307083].

\bibitem{0107046}
T.~Takahashi and S.~Tanimoto,
Prog.\ Theor.\ Phys.\  {\bf 106}, 863 (2001)
[arXiv:hep-th/0107046].

\bibitem{0202133}
T.~Takahashi and S.~Tanimoto,
JHEP {\bf 0203}, 033 (2002)
[arXiv:hep-th/0202133].

\bibitem{0302182}
T.~Takahashi,
Nucl.\ Phys.\ B {\bf 670}, 161 (2003)
[arXiv:hep-th/0302182].

\bibitem{0304261}
T.~Takahashi and S.~Zeze,
Prog.\ Theor.\ Phys.\  {\bf 110}, 159 (2003)
[arXiv:hep-th/0304261].

\bibitem{0402196}
T.~Takayanagi,
JHEP {\bf 0405}, 063 (2004)
[arXiv:hep-th/0402196].

\bibitem{0403156}
G.~Tasinato, I.~Zavala, C.~P.~Burgess and F.~Quevedo,
JHEP {\bf 0404}, 038 (2004)
[arXiv:hep-th/0403156].

\bibitem{0001201}
W.~Taylor,
Nucl.\ Phys.\ B {\bf 585}, 171 (2000)
[arXiv:hep-th/0001201].

\bibitem{0008033}
W.~Taylor,
JHEP {\bf 0008}, 038 (2000)
[arXiv:hep-th/0008033].

\bibitem{0208149}
W.~Taylor,
JHEP {\bf 0303}, 029 (2003) [arXiv:hep-th/0208149].

\bibitem{0301094}
W.~Taylor,
arXiv:hep-th/0301094.

\bibitem{0311017}
W.~Taylor and B.~Zwiebach,
arXiv:hep-th/0311017.

\bibitem{0101087}
S.~Terashima,
JHEP {\bf 0105}, 059 (2001)
[arXiv:hep-th/0101087].

\bibitem{0104176}
S.~Terashima and T.~Uesugi,
JHEP {\bf 0105}, 054 (2001)
[arXiv:hep-th/0104176].

\bibitem{0104158}
J.~Teschner,
Class.\ Quant.\ Grav.\  {\bf 18}, R153 (2001)
[arXiv:hep-th/0104158].

\bibitem{0303150}
J.~Teschner,
arXiv:hep-th/0303150.

\bibitem{9512062}
P.~K.~Townsend,
Phys.\ Lett.\ B {\bf 373}, 68 (1996)
[arXiv:hep-th/9512062].

\bibitem{0101125}
L.~S.~Tseng,
Phys.\ Rev.\ D {\bf 64}, 126004 (2001)
[arXiv:hep-th/0101125].

\bibitem{0011033}
A.~A.~Tseytlin,
hep-th/0011033.

\bibitem{0302125}
T.~Uesugi,
arXiv:hep-th/0302125.

\bibitem{9905034}
I.~V.~Vancea,
Int.\ J.\ Mod.\ Phys.\ A {\bf 16}, 4429 (2001)
[arXiv:hep-th/9905034].

\bibitem{0104099}
K.~S.~Viswanathan and Y.~Yang,
Phys.\ Rev.\ D {\bf 64}, 106007 (2001)
[arXiv:hep-th/0104099].

\bibitem{0107098}
K.~S.~Viswanathan and Y.~Yang,
Phys.\ Rev.\ D {\bf 65}, 066001 (2002)
[arXiv:hep-th/0107098].

\bibitem{math-ph/0306018}
V.~S.~Vladimirov and Y.~I.~Volovich,
Theor.\ Math.\ Phys.\  {\bf 138}, 297 (2004)
[Teor.\ Mat.\ Fiz.\  {\bf 138}, 355 (2004)]
[arXiv:math-ph/0306018].

\bibitem{0207089}
J.~E.~Wang,
JHEP {\bf 0210}, 037 (2002)
[arXiv:hep-th/0207089].

\bibitem{osft}
E.~Witten,
Nucl.\ Phys.\ B {\bf 268}, 253 (1986).

\bibitem{inverse}
E.~Witten,
Nucl.\ Phys.\ B {\bf 276} (1986) 291.

\bibitem{9108004}
E.~Witten,
Nucl.\ Phys.\ B {\bf 373}, 187 (1992)
[arXiv:hep-th/9108004].

\bibitem{9201056}
E.~Witten and B.~Zwiebach,
Nucl.\ Phys.\ B {\bf 377}, 55 (1992) [arXiv:hep-th/9201056].

\bibitem{9208027}
E.\ Witten,
Phys.\ Rev.\  {\bf D46}, 5467 (1992)
[hep-th/9208027].

\bibitem{9210065}
E.\ Witten,
Phys.\ Rev.\  {\bf D47}, 3405 (1993)
[hep-th/9210065].

\bibitem{9810188}
E.~Witten,
JHEP {\bf 9812}, 019 (1998) [arXiv:hep-th/9810188].

\bibitem{9802150}
E.~Witten,
Adv.\ Theor.\ Math.\ Phys.\  {\bf 2}, 253 (1998)
[arXiv:hep-th/9802150].

\bibitem{0006071}
E.~Witten,
arXiv:hep-th/0006071.

\bibitem{0007175}
E.~Witten,
Int.\ J.\ Mod.\ Phys.\ A {\bf 16}, 693 (2001)
[arXiv:hep-th/0007175].

\bibitem{0209197}
H.~t.~Yang,
JHEP {\bf 0211}, 007 (2002)
[arXiv:hep-th/0209197].

\bibitem{0406023}
H.~t.~Yang,
JHEP {\bf 0409}, 002 (2004)
[arXiv:hep-th/0406023].

\bibitem{0402027}
H.~U.~Yee and P.~Yi,
arXiv:hep-th/0402027.

\bibitem{9901159}
P.~Yi,
Nucl.\ Phys.\ B {\bf 550}, 214 (1999) [arXiv:hep-th/9901159].

\bibitem{0305249}
W.~S.~L'Yi,
arXiv:hep-th/0305249.

\bibitem{9912255}
T.~Yoneya,
Nucl.\ Phys.\ B {\bf 576}, 219 (2000)
[arXiv:hep-th/9912255].

\bibitem{9506136}
A.~B.~Zamolodchikov and A.~B.~Zamolodchikov,
Nucl.\ Phys.\ B {\bf 477}, 577 (1996)
[arXiv:hep-th/9506136].

\bibitem{0101152}
A.~B.~Zamolodchikov and A.~B.~Zamolodchikov,
arXiv:hep-th/0101152.

\bibitem{9705241}
B.~Zwiebach,
Annals Phys.\  {\bf 267}, 193 (1998) [arXiv:hep-th/9705241], and
references therein.

\bibitem{0008227}
B.\ Zwiebach,
JHEP {\bf 0009}, 028 (2000)
[hep-th/0008227].

\bibitem{0010190}
B.~Zwiebach,
arXiv:hep-th/0010190.

\end{thebibliography}
\end{document}